\documentclass[12pt,a4paper,fleqn]{article}
\usepackage[utf8]{inputenc}
\usepackage{authblk}
\usepackage{rotating}
\usepackage{geometry}
\usepackage{xcolor}
\usepackage{natbib}
\usepackage{graphicx}
\usepackage{pdflscape}
\usepackage{appendix}
\usepackage{setspace}
\usepackage{makecell}
\usepackage{float}

\usepackage{colortbl}

\providecommand{\shadecell}{\cellcolor[rgb]{0.9, 0.9, 0.9}}

\usepackage{threeparttable}
\usepackage{booktabs}

\usepackage{amssymb}
\usepackage{amsbsy}
\usepackage{bm}
\usepackage{amsmath}
\usepackage{amsthm}
\usepackage{graphicx}
\usepackage{mathtools}

\usepackage{enumitem}
\newlist{steps}{enumerate}{1}
\setlist[steps, 1]{label = Step \arabic*:}

\usepackage{dcolumn}
\newcolumntype{d}[1]{D{.}{.}{#1}}



\usepackage{caption}
\usepackage{subcaption}

\usepackage{fancyvrb}
\definecolor{nblue}{HTML}{000660}
\usepackage[colorlinks=true,urlcolor=nblue,linkcolor=nblue,citecolor=nblue]{hyperref}
\VerbatimFootnotes

\title{\LARGE \textbf{Integration or fragmentation? A closer look at euro area financial markets}}
\author[1]{\MakeUppercase{Martin Feldkircher}}
\author[2]{\MakeUppercase{Karin Klieber}\thanks{Corresponding author: Martin Feldkircher. \textit{Email}: \href{mailto:martin.feldkircher@da-vienna.ac.at}{martin.feldkircher@da-vienna.ac.at}. We would like to thank Julia H\"{o}ftberger for research assistance and Niko Hauzenberger and Michael Pfarrhofer for helpful comments and suggestions. The views expressed in this work do not necessarily reflect those of the Oesterreichische Nationalbank or the Eurosystem.}}
\affil[1]{\textit{Vienna School of International Studies}}
\affil[2]{\textit{Oesterreichische Nationalbank}}

\begin{document}

\maketitle\thispagestyle{empty}\normalsize\vspace*{-2em}\small

\begin{center}
\begin{minipage}{0.8\textwidth}
\noindent\small This paper examines the degree of integration at euro area financial markets. To that end, we  estimate overall and country-specific integration indices based on a panel vector-autoregression with factor stochastic volatility. Our results indicate a  more heterogeneous bond market compared to the market for lending rates. At both markets, the global financial crisis and the sovereign debt crisis led to a severe decline in financial integration, which fully recovered since then. We furthermore identify countries that deviate from their peers either by responding differently to crisis events or by taking on different roles in the spillover network. The latter analysis reveals two set of countries, namely a main body of countries that receives and transmits spillovers and a second, smaller group of spillover absorbing economies. Finally, we demonstrate by estimating an augmented Taylor rule that euro area short-term interest rates are positively linked to the level of integration on the bond market.
\\[1em]
\end{minipage}
\end{center}

\onehalfspacing\normalsize\renewcommand{\thepage}{\arabic{page}}

\newpage

\section{Introduction}

The creation of the European monetary union sparked interest in the effects of the common currency on euro area financial markets. Without exchange rate risks and based on the law of one price theory, economists predicted a convergence of interest rates, credit flows as well as corporate and sovereign bonds. The early literature on financial market connectedness hence focused on the degree of cross-country variation of interest rates and government bond yields and indeed found evidence for a surge in integration \citep{Baele2004, Gill2014}. With the onset of the global financial crisis (GFC) and especially during the subsequent European sovereign debt crisis (GFC), investors started to price debt from euro area periphery countries significantly higher than that of their peers from core countries. As a consequence, integration plummeted and a new series of studies started to examine the degree of fragmentation rather than integration and potential consequences for monetary policy \citep{Horvath2018, Candelon2022, Costola2022}.

Fragmentation endangers monetary policy transmission either directly through differences in bank lending rates or indirectly through distinct effects of monetary policy on euro area assets such as bonds and stocks and through the home-bias of banks in holding these assets. In the current high inflation environment, after two years of pandemic and amidst massive geopolitical tensions, these problems are aggravated by a significant trade-off the European Central Bank (ECB) faces, namely between ensuring price stability and guarding the common monetary policy area and its currency, the euro. Against this backdrop, this chapter analyses re-assesses empirically the degree of connectedness in euro area financial markets.

In what follows, we focus on two segments of the financial market, government bond yields and lending rates. Our analysis is complemented by investigating intra-euro area money flows. Each market is analyzed separately by employing state-of-the-art time series models that are geared towards using large-scale cross-sectional data and can account for potential drifts in coefficients and / or residual volatilities. To quantify the degree of connectedness in each market, we rely on out-of-sample forecast error variance decompositions and the Diebold-Yilmaz (DY) spillover index \citep{diebold2009measuring,Diebold2014}. In this setting, spillovers can be thought of as either arising directly through cross-border asset holdings of financial institutions or indirectly through the exposure to common, external shocks. It is hence a more encompassing measure of integration compared to correlation analysis \citep{Fernandez2016} and the out-of-sample approach renders our measure potentially faster to respond to crisis periods \citep{Buse2019}. 

Our main results show a sharp decrease in integration for government bond yields and long-term lending rates during the GFC, reaching a trough at the end of the ESDC. The decoupling was particularly pronounced for German and Dutch bonds. Using network analysis, we reveal a particular role for Austria, Belgium and Germany; countries that receive significantly more spillovers than they transmit. Finally and by estimating Taylor rules, we show that our integration measures are significantly related to euro area short-term interest rates. This holds especially true for integration at the sovereign bond market, indicating that the ECB is more reluctant to raise interest rates in case integration is low.

The chapter is structured as follows: In the next section, we provide a literature review of studies dealing with financial connectedness. Section \ref{sec:data} describe the data and stylized facts on the financial markets we consider. Section \ref{sec:eco} outlines the empirical framework we use to estimate the DY spillover index. Section \ref{sec:results} contains our main results, while section \ref{sec:concl} concludes.

\section{Literature review}\label{sec:lit}

The literature investigating \textit{fragmentation in euro area financial markets} is large. One strand of the literature focuses on a particular segment of financial markets, namely whether government bond yields move in tandem or not. In a perfectly integrated market, without exchange rate risks, and with the Maastricht criteria ensuring macroeconomic convergence, bond prices should tend to equalize. 
That said, there is ample evidence that in periods of global financial stress, a flight to safety motive of international investors reveal a different pricing of euro area debt. Viewed from an international perspective, this flight to quality can also be explained by the international risk-taking channel of \citet{Bruno2015} which predicts a general retrenchment of risky assets in case funding costs increase. There are a few recent studies looking at fragmentation of euro area bond yields. \citet{Candelon2022} demonstrate that fragmentation risks preceded the outbreak of the euro area sovereign debt crisis, especially so between core and periphery countries, and re-emerged during the recent COVID crisis. In a similar vein \citet{Antonakakis2013} find evidence for large spillovers (integration) within euro area periphery countries on the one hand and euro area core countries on the other hand, whereas spillovers between these two groups are negligible. Using a similar econometric framework, \citet{Fernandez2016} assess spillovers in the European sovereign bond market and relate them to macrofundamental variables.

Another part of the literature looks at financial integration in a broader sense. A lack of connectivity or fragmentation can arise due to distinct investment opportunities, institutional differences between market segments as well as information barriers that increase cross-border transaction costs. In a monetary union, these factors should not be pronounced. Still there is ample evidence of fragmentation \citep{Battistini2014, Gagnon2020, Mayordomo2015}. To highlight a few relevant and recent contributions, \citet{ArceAlfaro2022} show that  bank lending rates and credit volumes in the euro area are driven by a common factor. The importance of the common factor, however,  started to decline significantly with the onset of the GFC. In a similar vein, \citet{Gagnon2020} find evidence for fragmentation which can be partially offset by euro area (unconventional) monetary policy. \citet{Pungulescu2013} investigate convergence for various financial market segments including the bond, stock and interest rate market in the European Union. Their analysis reveals a partial convergence of these markets within the group of old EU member states and separately for those that joined after 2000. However, and in line with \citet{ArceAlfaro2022}, convergence and hence financial integration has started to reverse since the years after the GFC. \citet{Hoffmann2020} develop a composite indicator of financial integration drawing on price-based and quantity-based data for the money, bond, equity and credit market and find a significant, positive relationship between financial integration and economic growth. Focusing on equity prices,  \citet{Vides2018} find evidence that stock markets of the four major European countries are cointegrated, but this relationship is not robust to crisis episodes such as the GFC or the ESDC. \citet{Horny2018} study a large sample of corporate bonds from selected, major euro area countries and demonstrate that the ESDC has been a turning point as regards financial integration, especially so between German, Italian and Spanish corporate bonds. \citet{Horvath2018} and more recently \citet{Hoffmann2020} reach a similar conclusion, namely  that fragmentation has peaked around the ESDC.

\section{Data \& descriptive statistics}\label{sec:data}

We collect monthly data for different segments of the financial market as well as within-euro area money flows. In accordance with the existing literature on integration, our main focus is on 10-year government bond yields. These time series are available for the longest time period and start in 1994m01. We further include data on credit markets as advocated in \citet{Baele2004}. More specifically, we include data on short- and long-term lending rates to households and non-financial corporations. Both lending rates are constructed by aggregating different maturity structures using corresponding new loan volumes as weights. 
These datasets are available from 2003m01 onward. Last, we include data on TARGET2 balances, i.e., the net flow of money between euro area countries, which relates to trade and financial integration. Here, the data starts in 2007m01. All data are available until 2022m05 and taken from the ECB's statistical data warehouse. An exception to this are government bond yields data that stem from the OECD's main economic indicator data base. 

\begin{figure}[htbp!]
\caption{Financial market segments in the euro area.}\label{fig:int}
        \captionsetup[subfigure]{aboveskip=0pt,belowskip=-5pt}
	\begin{subfigure}{0.45\textwidth}\caption{Government bond yields (core)}
		\centering\includegraphics[scale=0.3]{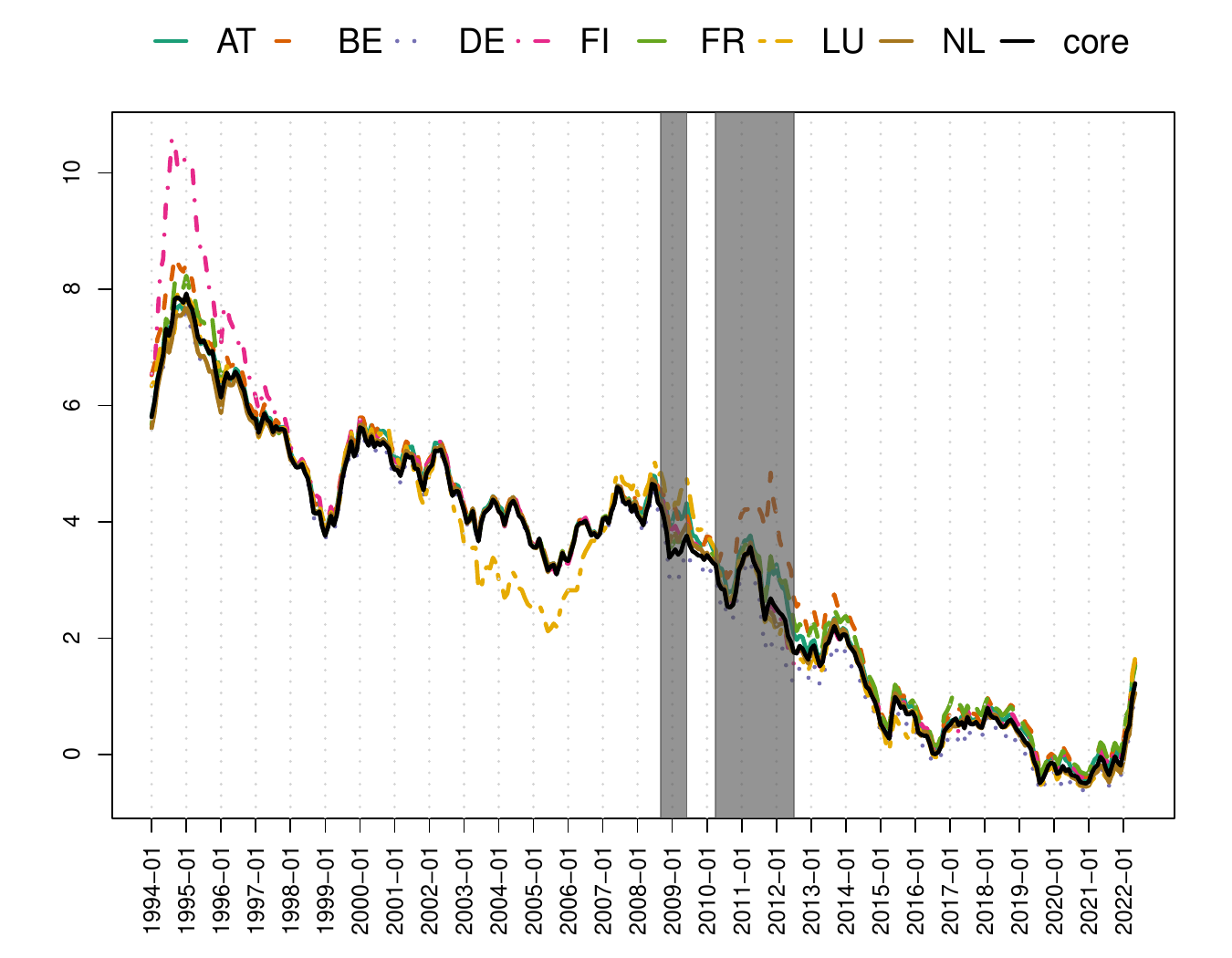}
	\end{subfigure}
	\begin{subfigure}{0.45\textwidth}\caption{Government bond yields (periphery)}
		\centering\includegraphics[scale=0.3]{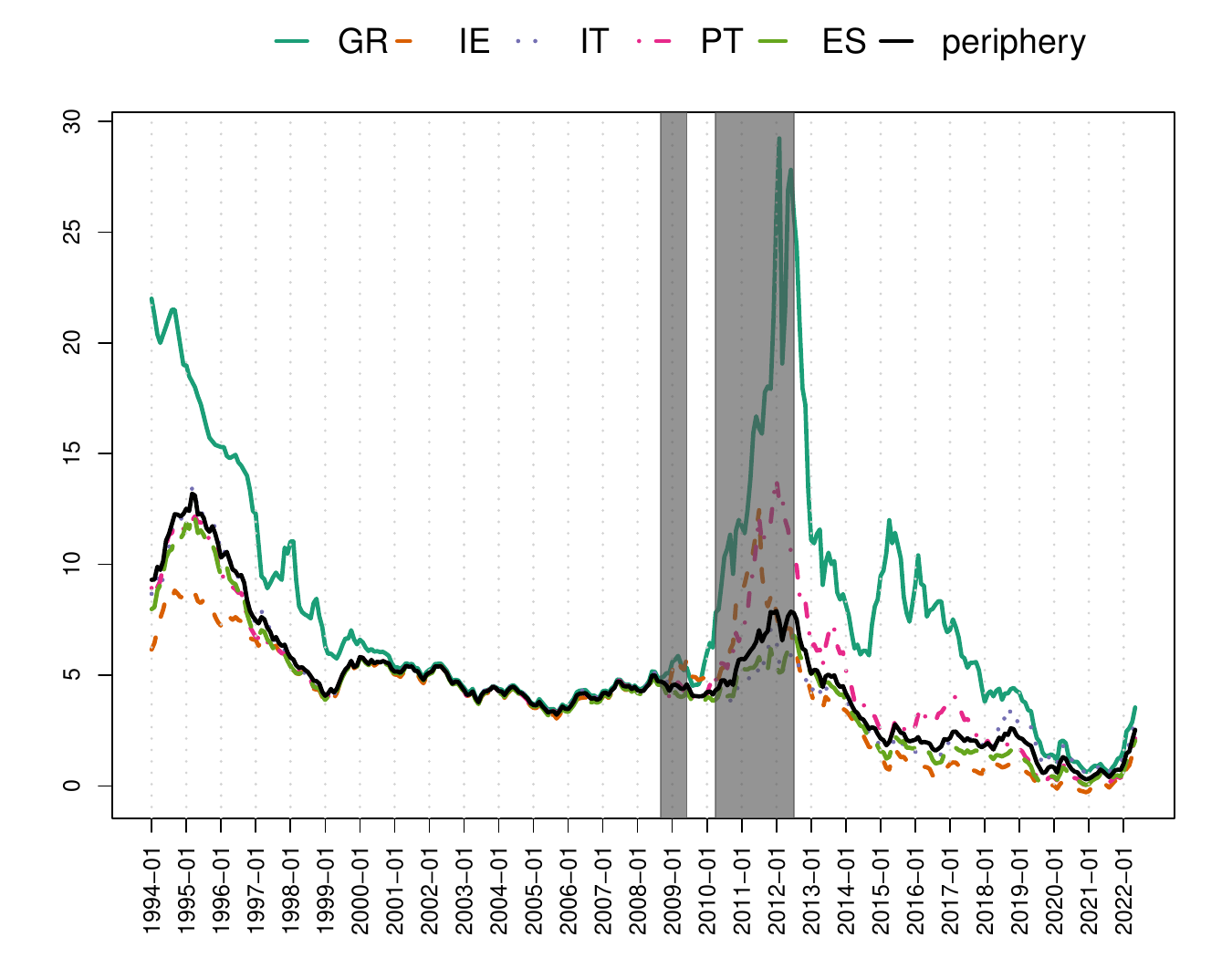}
	\end{subfigure}\\
	\vspace{0.5cm}\\
	\begin{subfigure}{0.45\textwidth}\caption{Borrow long (core)}
		\centering\includegraphics[scale=0.3]{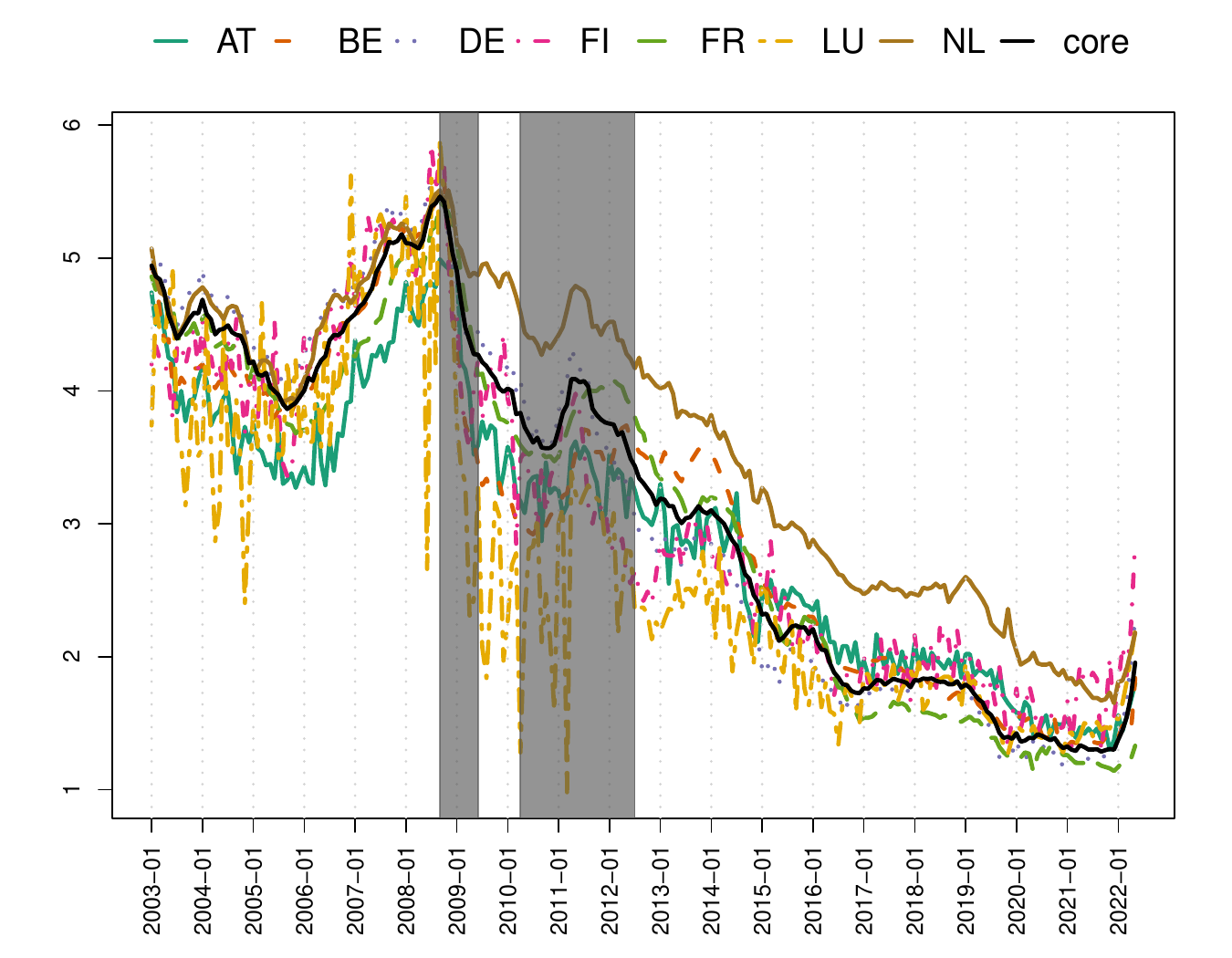}
	\end{subfigure}
		\begin{subfigure}{0.45\textwidth}\caption{Borrow long (periphery)}
		\centering\includegraphics[scale=0.3]{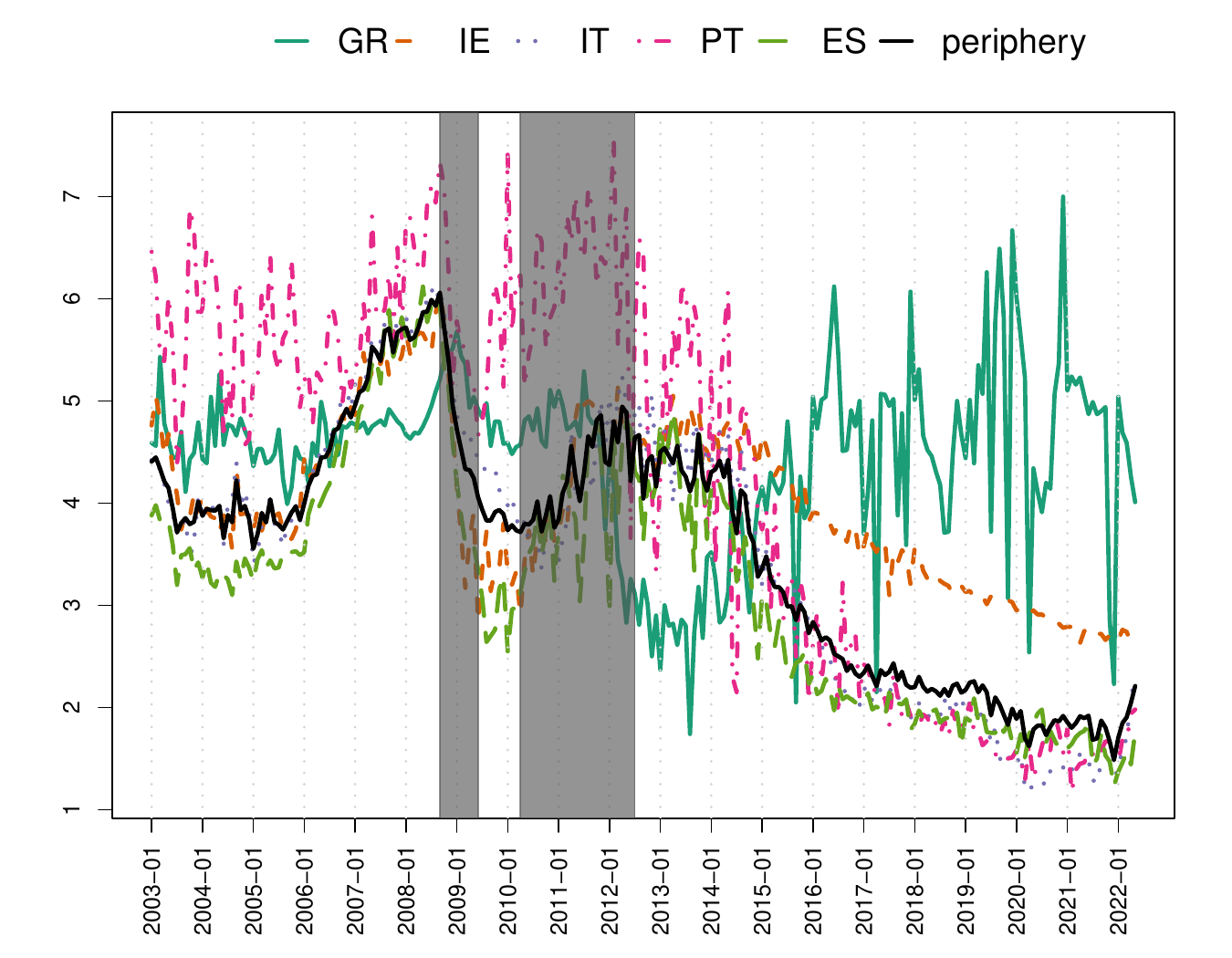}
	\end{subfigure}\\
		\vspace{0.5cm}\\
	\begin{subfigure}{0.45\textwidth}\caption{Borrow short (core)}
		\centering\includegraphics[scale=0.3]{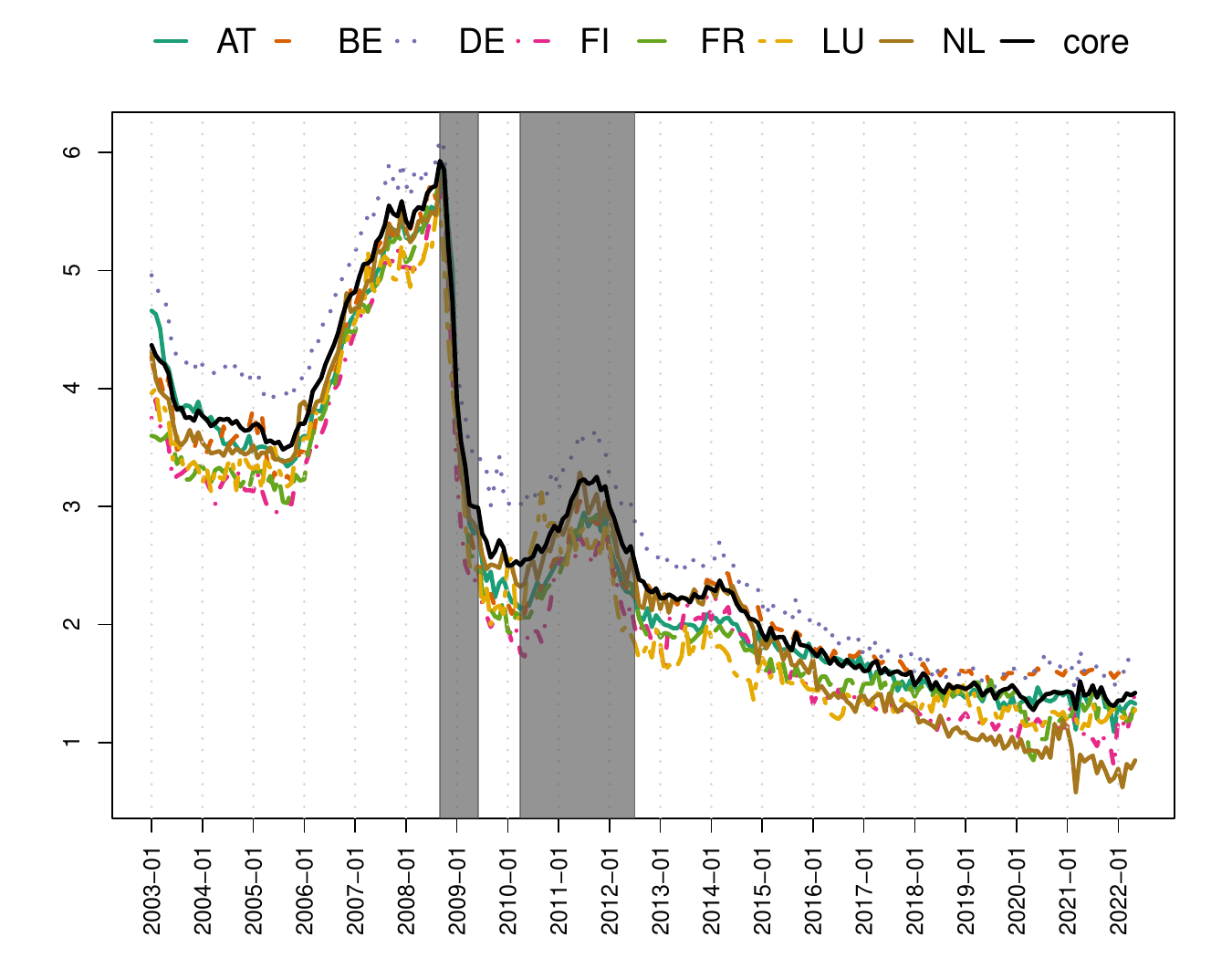}
	\end{subfigure}
	\begin{subfigure}{0.45\textwidth}\caption{Borrow short (periphery)}
		\centering\includegraphics[scale=0.3]{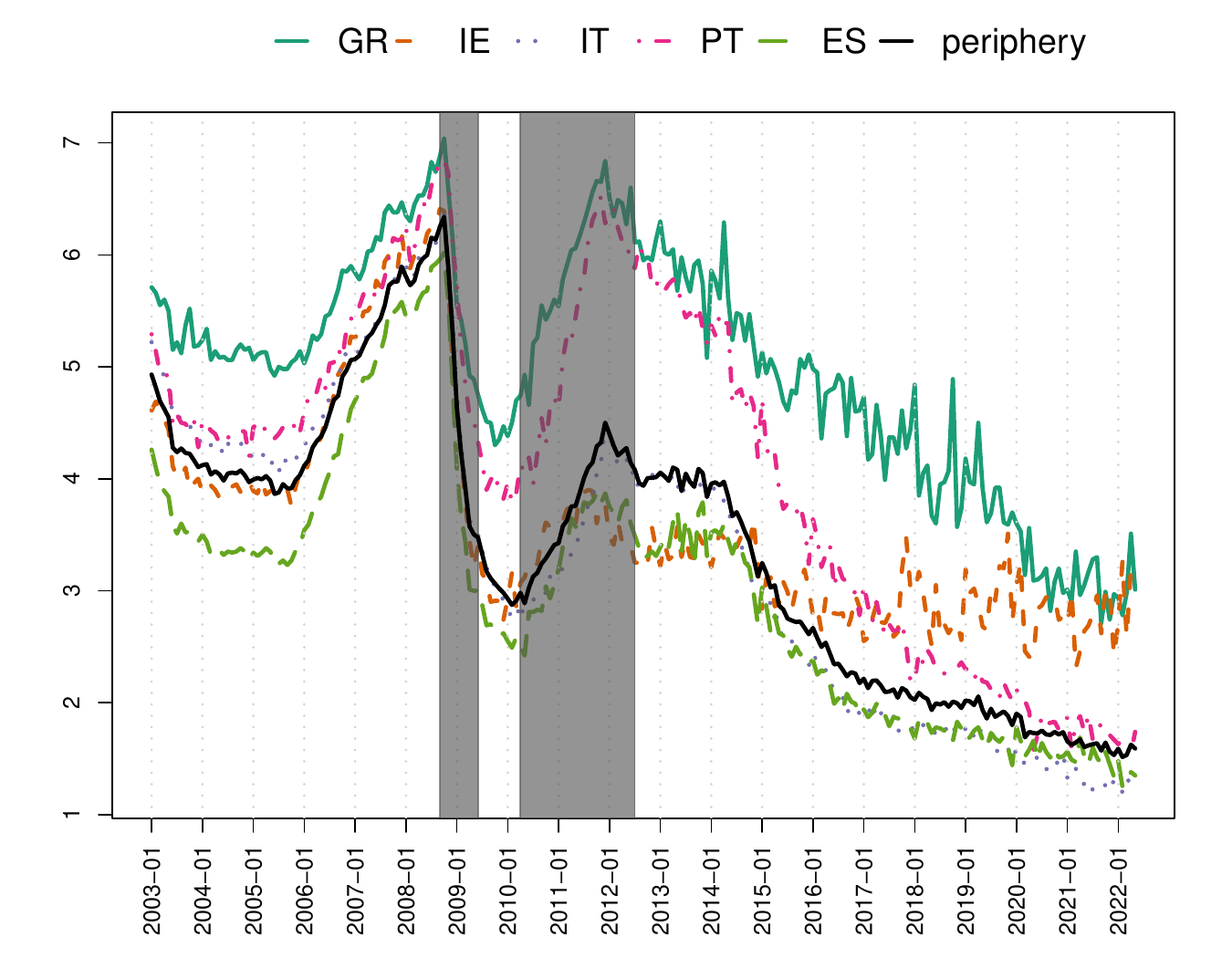}
	\end{subfigure}\\
	\vspace{0.5cm}\\
	\begin{subfigure}{0.45\textwidth}\caption{TARGET2 (core)}
		\centering\includegraphics[scale=0.3]{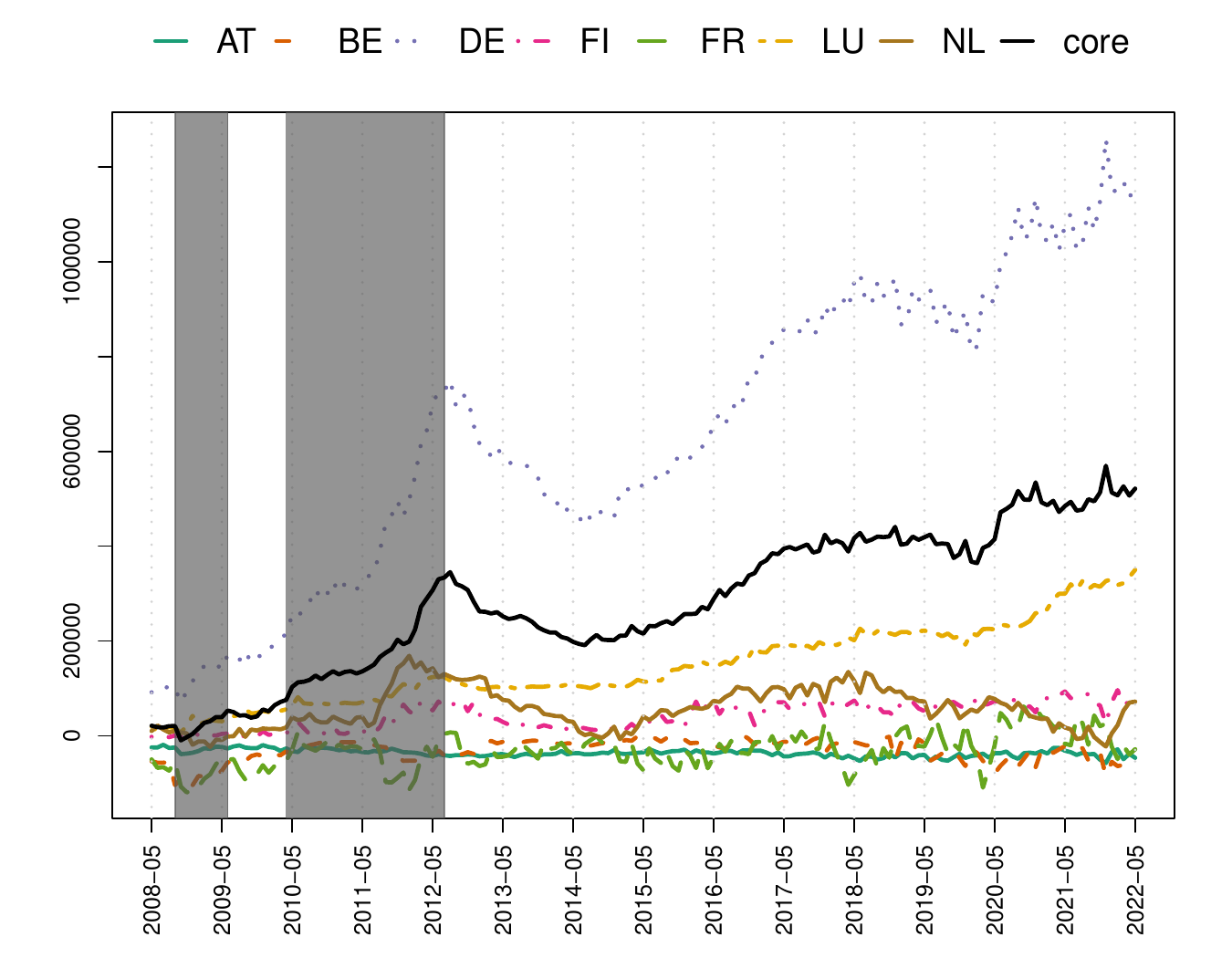}
	\end{subfigure}
		\begin{subfigure}{0.45\textwidth}\caption{TARGET2 (periphery)}
		\centering\includegraphics[scale=0.3]{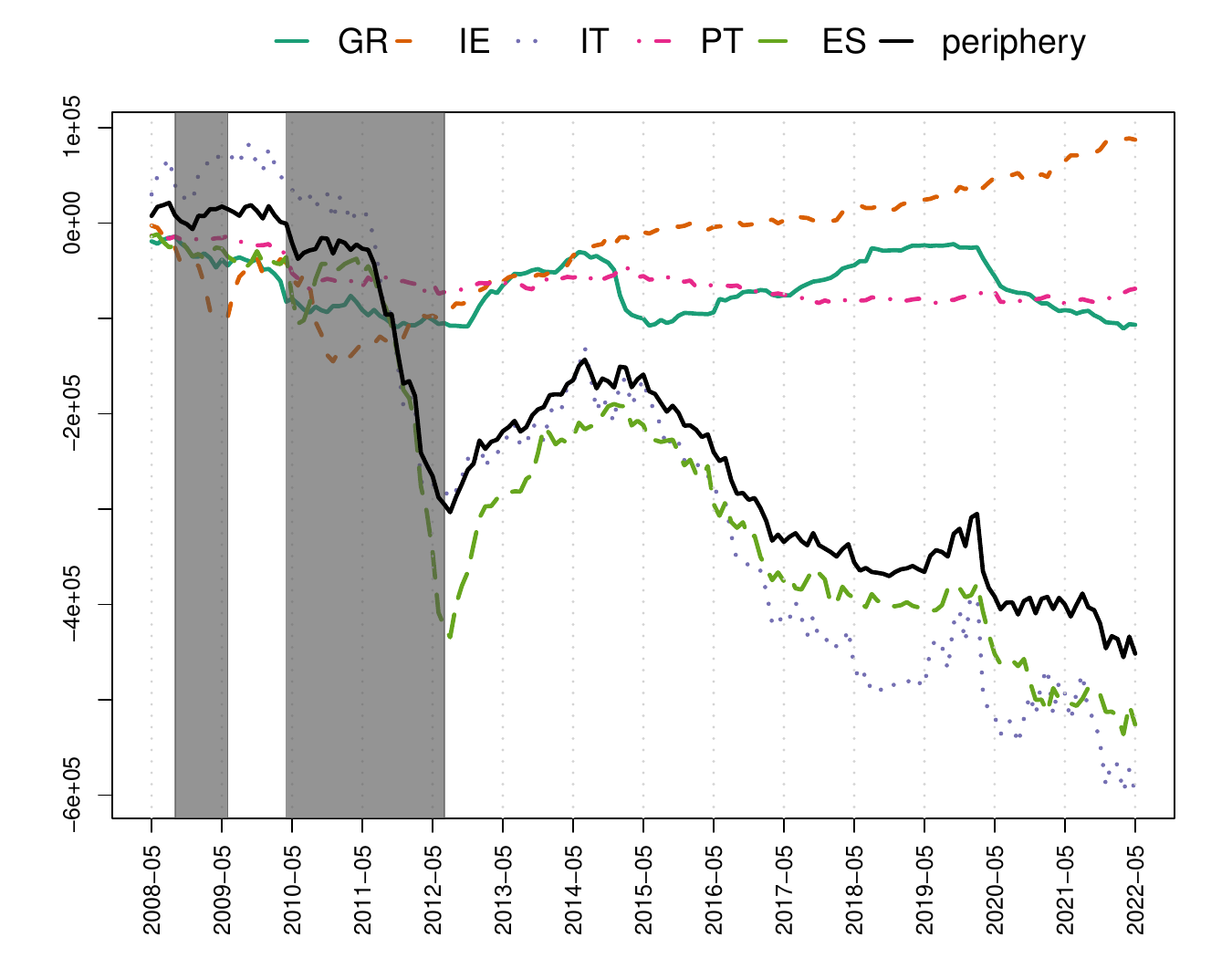}
	\end{subfigure}\\
	\begin{minipage}{14cm}~\\
\scriptsize \emph{Note:} The graph shows government bond yields, long- and short-term borrowing costs as well as TARGET2 balances for different country groups. Regional aggregates calculated with PPP weights. In light grey, we depict the periods of the GFC and the ESDC. \end{minipage}%
\end{figure}

The country coverage spans euro area \textbf{core} countries (Austria, Belgium, Germany, Finland, France, Luxembourg and the Netherlands) and  \textbf{periphery} countries (Greece, Ireland, Italy, Portugal, Spain). In \autoref{fig:int} we display the data for core and periphery countries separately and include regional means based on a set of rolling purchasing power parity (PPP) weights. In the graphs, we highlight the period of the GFC (2008m09 to 2009m06) and the ESDC \citep[2010m04 to 2012m07, see][]{Candelon2022}.

We see that government bond yields have been on a steady downward trend for both core and periphery euro area countries. A striking difference emerges during the ESDC. Here, spreads shot up significantly in the group of periphery countries, while they remained comparably stable in the core group. This implies a pronounced widening of government bond yields between the two groups. In the latest part of the sample we see a reversal of the downward trend and an increase in yields. Long-term borrowing costs are tied to government bond yields. They can differ across countries due to different market structures of the respective banking system. For the core countries, long-term lending rate tend to show parallel movements in government bond yields: they are on a general downward trend throughout the sample period which is reversed in the most recent part of our sample. The variation tends to be large within the group of periphery countries. While the regional mean for periphery countries follows the trend of their peers from the core region, long-term borrowing costs significantly deviate in Portugal and Greece. The latter countries show a steady upward movement in lending rates, which contrasts both the development of lending rates of other euro area countries as well as the trend seen in Greek government bond yields. Short-term borrowing costs show a less diverse picture. First, for euro area core countries, the variation around the regional mean is not very pronounced indicating a similar cross-sectional development of short-term lending rates. Second, the overall downward trend is also evident in lending rates. This trend flattens with the beginning of the quantitative easing (QE) program and the binding of the zero lower bound. Short-term lending rates are again more volatile for periphery countries, with the Greek and Portuguese credit markets showing the largest deviations from their peers. In general, also here, a steady decline in rates is evident. Last, we investigate the development of TARGET2 balances. These are expressed in absolute terms, such that their movement is not obscured by developments of a normalizing metric (such as GDP). Here we see a steady increase of positive German net balances, which only declines in the aftermath of the ESDC as well as around 2018. The remaining core countries show a comparably more gradual development of TARGET2 balances. Most periphery countries tend to be importers of goods and services from other euro area countries, which implies a negative TARGET2 balance. This applies especially so to Spain and Italy and to a lesser degree to Portugal. By contrast, Irish balances are positive and on a continuous upward trend. For more details, we provide summary statistics in \autoref{tbl:desc} in the appendix.

\section{Econometric framework}\label{sec:eco}
Our main analysis rests on the DY spillover index \citep{diebold2009measuring, Diebold2014}, which in turn is based on decompositions of forecast error variances. For that purpose, we estimate large systems of equations that can account for relations in the variables or through the error structure. 

The set of models we consider comprise a Bayesian vector autoregression (BVAR) model with and without time-varying parameters, a global VAR model \citep[GVAR, ][]{pesaran2004gvar,feldkircher2016gvar} with and without time-varying parameters as well as a panel VAR (PVAR) with a factor stochastic volatility specification. These models are evaluated using a forecast exercise as well as examining the DY spillover index, both of which are provided in Appendix \ref{sec:robust}. Based on these considerations we opt for the PVAR model with stochastic volatility since it provides competitive forecasts and is computationally the least costly of all provided alternatives \citep{huber2018bayesian,koop2019forecasting,feldkircher2022approximate}.

\subsection{The panel vector autoregressive (PVAR) model}\label{sec:pvar}
In this section, we describe the PVAR model with factor stochastic volatility. 
Let $\bm y_{it}$ denote a $M$-dimensional vector of endogenous variables for country $i = 1,\dots,N$ at time $t = 1,\dots,T$. Moreover, we introduce a $M$-dimensional vector $\bm y^{\ast}_{it}$ capturing the other countries' variables for which we exclude the $i^{th}$ subvector from $\bm y_t$ and define $\bm y^{\ast}_{it} = (\bm y'_{1t}, \dots, \bm y'_{i-1t},\bm y'_{i+1t},\dots,\bm y'_{Nt})'$. The PVAR model is then given by
\begin{equation}
    \bm y_{it} = \bm \alpha_{it} + \sum^P_{p=1} \bm A_{ip,t} \bm y_{it-p} + \sum^P_{p=1} \bm B_{ip,t} \bm y^{\ast}_{it-p} + \bm \epsilon_{it}
\end{equation}
where $\bm \alpha_t$ is a $M$-dimensional intercept vector and $\bm A_{ip,t}$ and  $\bm B_{ip,t}$ denote the $M \times 1$ vectors of coefficients. $P$ denotes the number of lags, which is set to $2$ in our applications. The error term $\bm \epsilon_{it}$ is assumed to be normally distributed with zero mean and a time-varying variance-covariance matrix $\bm \Sigma_{it}$.

The PVAR approach leaves the weights on the linkages between countries unrestricted. That is, we do not a priori impose restrictions on the relationship between foreign variables, e.g., via trade weights. Rather we estimate the nature and extent of linkages between countries and divide our set of countries into different clusters. In particular, we assume that the domestic coefficients follow a mixture of Gaussian distributions to deal with parameter homogeneity across countries and avoid overfitting \citep[see ][]{huber2018bayesian}.

We let $\bm C_i = (\bm \alpha_i, \bm A_{i1}, \dots, \bm A_{iP})$ and define the mixture distribution for $\bm c_i = \text{vec}(\bm C_i)$ as
\begin{equation}\label{eq:mixture1}
    p(\bm c_i | \bm \nu, \bm \mu_1, \dots, \bm \mu_G, \bm V) = \sum_{g=1}^G \nu_g \mathnormal{f}_\mathcal{N} (\bm c_i|\bm \mu_g, \bm V).
\end{equation}

The vector of component weights $\bm \nu = (\nu_1, \dots, \nu_G)'$ satisfies two conditions, $\sum_{g=1}^G \nu_g = 1$ and $\nu_g \geq 0$. We denote $\mathnormal{f}_\mathcal{N}$ as the density of the multivariate Gaussian distribution with the component-specific mean vector $\bm \mu_g$ of dimension $m = M(Mp+1) \times 1$ and the common variance-covariance matrix $\bm V$ of dimension $m \times m$. Introducing the binary indicator $\delta_i$ (for $i = 1, \dots, N$) allows us to rewrite \autoref{eq:mixture1} as
\begin{equation*}
    p(\bm c_i | \delta_i=g, \bm \mu_g, \bm V) = \mathnormal{f}_\mathcal{N} (\bm c_i|\bm \mu_g, \bm V)
\end{equation*}
where $\text{Pr}(\delta_i=g) = \nu_g$. This auxiliary representation allows to interpret the averages of the posterior draws of $\delta_i$ as the probability of country $i$ belonging to a specific cluster of countries. Furthermore, it facilitates the estimation of a mixture model \citep[see, ][]{malsiner2016model}. 

We use a weakly informative prior on the common variances and apply a set of independent inverted Gamma priors on the main diagonal elements of $\bm V$ in form of:
\begin{equation*}
    v_j \sim \mathcal{G}(a_0,a_1), \quad \text{for } j = 1, \dots, m
\end{equation*}
where $a_0$ and $a_1$ are hyperparameters with small values ($a_0 = a_1 = 0.01$).

For each $\bm \mu_g$ we assume that it follows a common distribution with mean $\bm \mu_0$ and a diagonal variance-covariance matrix $\bm Q_0$ and specify a hierarchical prior of the following form
\begin{equation*}
    \bm \mu_g | \bm \mu_0, \bm Q_0 \sim \mathcal{N}(\bm \mu_0, \bm Q_0), \quad \text{for } g = 1, \dots, G
\end{equation*}
where 
\begin{align*}
    \bm Q_0 &= \bm \Lambda \bm R_0 \bm \Lambda \\
    \bm \Lambda &= \text{diag}(\sqrt{\lambda_1}, \dots \sqrt{\lambda_m}) \\
    \lambda_j &\sim \mathcal{G}(b_0,b_1) \\
    \bm \mu_0 &\sim \mathcal{N}(\bm m_0, \bm M_0).
\end{align*}
We define the additional scaling matrix $\bm R_0 = \text{diag}(R^2_1, \dots, R^2_m)$ with the main diagonal elements $R^2_j$ denoting the range of $c = (c_1, \dots, c_N)$ along the $j^{th}$ dimension. The hyperparameters $b_0$ and $b_1$ of the Gamma prior on $\lambda_j$ are set to $b_0 = b_1 = 1/2$. $\bm m_0$ is defined as the median over the columns of $\bm c$ and $\bm M_0^{-1}=\bm 0$

Moreover and following \cite{malsiner2016model}, we endogenously select the number of mixture components via a symmetric Dirichlet prior on the mixture component weights $\bm \nu$ as well as a Gamma prior on the intensity parameter of the Dirichlet distribution $p_0$. Formally, this boils down to
\begin{equation*}
    \bm \nu \sim \mathcal{D}ir(p_0, \dots, p_0)
\end{equation*}
and 
\begin{equation*}
    p_0 \sim \mathcal{G} (c_0, c_0G).
\end{equation*}
where $c_0$ is a hyperparameter controlling the variance of the prior and set to $c_0 = 10$ \citep{malsiner2016model, huber2018bayesian}.

For achieving parsimony with respect to the coefficients in $\bm B_{ip,t}$, capturing the dynamic interdependencies between countries, we implement a Normal-Gamma shrinkage prior with a local and a global shrinkage component. We stack the matrices $\bm B_{ip}$ for $i = 1, \dots, N$ to $\bm B_i = (\bm B_{i1}, \dots, \bm B_{iP})'$ and put a normally distributed prior on each element $b_{ij}$ in vec($\bm B_i$):
\begin{equation}
    b_{ij} | \gamma_{ij}, \phi_{i} \sim \mathcal{N} \left(0, \frac{2\gamma^2_{ij}}{\phi_{i}}\right), \quad \gamma_{ij} \sim \mathcal{G}(\vartheta_i, \vartheta_i), \quad \phi_{i} \sim \mathcal{G}(d_0,d_1)
\end{equation}
for $j = 1, \dots, k = PM^2(N-1)$ and $i = 1, \dots, N$. The degree of shrinkage is driven by the global scaling parameters $d_0$ and $d_1$, which we set to $d_0 = d_1 = 0.01$. The country-specific shrinkage parameters is given by $\gamma_i$ and controls the degree of dynamic interdependence between a given country and all other countries. We follow \cite{Huber2019} and choose $\vartheta_i = 0.1$.

\subsection{The treatment of the error variance}
We assume that the error term is a Gaussian vector white noise process featuring a time-varying variance-covariance matrix; the same assumption applies also to the models used in the robustness check. To achieve parsimony for our high-dimensional time-varying variance-covariance matrix, we assume a factor stochastic volatility structure on $\bm \Sigma_{it}$ \citep{kastner2019sparse,kastner2020sparse}. Following \cite{aguilar2000bayesian} and stacking the vectors of error terms of the different countries in a $K=MN$ vector $\bm \epsilon_t = (\bm \epsilon_{1t}, \dots, \bm \epsilon_{Nt})'$ we define
\begin{align}
    \bm \Sigma_t &= \bm L \bm H_t \bm L' + \bm \Omega_t\\
    \bm \epsilon_t &= \bm L \bm f_t + \bm \eta_t, \quad \bm f_t \sim \mathcal{N}(\bm 0, \bm H_t), \quad \bm \eta_t \sim \mathcal{N}(\bm 0, \bm \Omega_t)
\end{align}
where $\bm f_t$ is a $d$-dimensional vector of common factors ($d \ll K$) with matrix $\bm H_t = \text{diag}(\text{exp}(h_{1t}, \dots, h_{qt}))$ capturing the time-varying variances. $\bm L$ is a $K \times d$ matrix of factor loadings and $\bm \eta_t$ is a vector of idiosyncratic shocks with time-varying variance-covariance matrix $\bm \Omega_t = \text{diag}(\text{exp}(\omega_{1t}), \dots, \text{exp}(\omega_{Kt}))$.
The diagonal elements of $\bm H_t$ and $\bm \Omega_t$ are assumed to follow independent AR(1) processes given by
\begin{align*}
    h_{jt} &= \rho_{hj} h_{jt-1} + \zeta_{jt}, \quad \zeta_{jt} \sim \mathcal{N}(0, \sigma_{hj}), \quad \text{for } j = 1, \dots, d\\
    \omega_{jt} &= \varpi_{\omega j} + \rho_{\omega j} (\omega_{jt-1} - \varpi_{\omega j}) + \xi_{jt}, \quad \xi_{jt} \sim \mathcal{N}(0, \sigma_{\omega j}), \quad \text{for } j = 1, \dots K
\end{align*}
with the state-equation innovation variances denoted by $\sigma^2_{hj}$ and $\sigma^2_{\omega j}$, the autoregressive parameters $\rho_{hj}$ and $\rho_{\omega j}$ and the unconditional mean of the log-volatility $\varpi_{\omega j}$.

The priors concerning the factor stochastic volatility specification are chosen according to \cite{kastner2019sparse}. In particular, we impose a row-wise Normal-Gamma shrinkage prior on the factor loadings in $\bm L$. We let $l_{ij}$ denote the elements in $\bm L$ and specify the prior as
\begin{equation}
    l_{ij}|\varphi_{ij}, \varrho_{i} \sim \mathcal{N}(0, 2 \varphi^2_{ij} / \varrho_{i}), \quad \varphi^2_{ij} \sim \mathcal{G}(\psi_l, \psi_l), \quad \varrho_i \sim \mathcal{G}(s_0, s_1)
\end{equation}
with the hyperparameters $\psi_l = 0.1$, $s_0 = s_1 = 1$. Furthermore, we choose $\varpi_{\omega j} \sim \mathcal{N}(0,10)$ on the unconditional mean, $\sigma^2_{\omega j} \sim \mathcal{G}(1/2,1/2)$ and $\sigma^2_{h j} \sim \mathcal{G}(1/2,1/2)$ on the process innovation variances and $(\rho_{\omega j}+1)/2 \sim \mathcal{B}(10,3)$ and $(\rho_{h j}+1)/2 \sim \mathcal{B}(10,3)$ on the autoregressive parameters of the state equation of the factor stochastic volatility specification.

\subsection{Posterior simulation}
The posterior simulation consists of four main steps which we repeat $10,000$ times, discarding the first $4,000$ draws as burn-ins. We briefly sketch the algorithm below \citep[for more details we refer to][]{huber2018bayesian}.

\begin{enumerate}
    \item In the first step, we simulate the VAR coefficients as well as the stochastic volatility component and the factor loadings for the stochastic volatility process. The VAR coefficients are drawn from a multivariate Gaussian distribution with posterior moments taking well-known forms. To simulate the parameters concerning the factor stochastic volatility process we use the \texttt{factorstochvol} R-package of \citet{kastner2019sparse}. 
    
    \item Next, we focus on obtaining the quantities for the mixture model specification. This involves drawing the mixture probabilities $\bm \nu$ from a Dirichlet distribution, the regime indicators $\delta_i$ from a multinomial distribution and the group-specific means $\bm \mu_g$ from a multivariate Gaussian distribution. The corresponding mean and variance-covariance matrix of $\bm \mu_g$ are obtained through simulating $\lambda_j$ from a GIG distribution and $\bm \mu_0$ from a Gaussian distribution. The common variance-covariance matrix $\bm V$ is constructed by drawing its elements from an inverse Gamma distribution. Finally, $p_0$, which refers to the intensity parameter of the Dirichlet prior, is obtained using a random walk Metropolis Hastings step.

    \item In the next step we draw the shrinkage parameters for the coefficients associated with the dynamic interdependencies between countries $\bm B_{ip,t}$. This is achieved by sampling the global and local shrinkage parameters from a Gamma distributed posterior and a generalized inverse Gaussian (GIG) distribution, respectively.

    \item The last step deals with the fact that the proposed PVAR approach is statistically not identified. To avoid the so-called label switching problem we apply the random permutation sampler of \citet{fruhwirth2001markov}.
\end{enumerate}

\section{Financial connectedness}\label{sec:results}

In this section, we present our empirical results. The basis of our analysis is the estimation of a generalized forecast error variance decomposition (GFEVD). The GFEVD is independent of the  ordering of the variables in the system \citep[see, ][]{koop1996gfevd,pesaran1998gfevd} and can be used to calculate an overall spillover index as well as country-specific measures of integration. More specifically, the $H$-step ahead GFEVD is given by 

\begin{equation}\label{equ:fevd}
    \theta_{ij}(H)=\frac{\sigma_{jj}^{-1}\sum_{h=0}^{H-1}\left(\bm e_i' \bm A_h\bm \Sigma_t \bm e_j\right)^2}{\sum_{h=0}^{H-1}\left(e_i' \bm A_h \bm \Sigma_t \bm e_j\right)}
\end{equation}

with $\bm \Sigma_t$ denoting the variance-covariance matrix of the error vector $\bm \epsilon$, $\sigma_{jj}$ the standard deviation of the error term for the $j^{th}$ equation, $\bm A_h$ the matrix corresponding to the moving average representation of the VAR and $\bm e_i$ a selection vector with unity as the $i^{th}$ element and zero otherwise. The element $\theta_{ij}(H)$ of the $N \times N$ GFEVD matrix $\bm \theta$ indicates the share of shocks to equation $j$ that explain forecast error variance of variable $i$.

In general, this contribution can be large because there is a direct connection between financial assets in country $j$ and $i$ through, e.g., intra-euro area holdings and operations of euro area banks. Alternatively, both $j$ and $i$ can be exposed to the same external shock, such as a reversal in global investors' risk sentiment or an overall economic downturn.\footnote{\citet{Eickmeier2023} examine cross-country differences of a broad set of economies to a common global financial stress shock.} In both cases, our analysis would reveal a high spillover index (i.e., a comparably low role for domestic as opposed to non-domestic shocks) and hence point to a high degree of financial integration.\footnote{\citet{Forbes2002} define contagion as periods of short-lived high correlation, whereas integration is defined by persistent periods of high cross-asset correlation. Our approach sidesteps this discussion by not focusing on direct correlations of the underlying assets but on the GFEVD, which also benefits from taking asymmetries into account.} 

From a modeling perspective, the assumptions about the specification of the error variances, both regarding the cross-country variances and whether variances can drift over time, can have a potentially large influence on the GFEVD and hence, the spillover analysis. In particular, assuming a homoskedastic variance could lead to a situation with changes in variances being falsely attributed to changes in variances of other equations / countries and hence lead to distorted inference of spillovers. We address this potential problem by choosing a very general specification by opting for the PVAR model with factor stochastic volatility. 

Summing up the contributions from all other variables for each country separately leads to the "directional" spillover index

\begin{equation}
    S_{i\leftarrow j}(H)=\frac{\sum_{j=1,j\not i}^{D}\theta_{ij}(H)}{D}\times 100
\end{equation}

with $D=\sum_{i,j=1}\theta_{ij}(H)$. This index indicates the importance of non-domestic shocks for each country. 

The overall spillover index is then finally defined as

\begin{equation}
    S(H)=\frac{\sum_{i,j=1,i\not j}{D}\theta_{ij}(H)}{D}\times 100
\end{equation}

In contrast to \citet{Diebold2014}, we calculate the GFEVDs based on an expanding window, meaning, we estimate the underlying PVAR model using data up to the first period of the hold-out sample and repeat this process until we reach the last period in our hold-out. This implies that our measure is based on a truly out-of-sample exercise and hence is potentially able to respond to crisis periods more quickly than common in-sample techniques \citep{Buse2019}. 

\subsection{Overall spillovers}\label{sec:tot}

We first assess the overall degree of spillovers for the financial variables considered in this study. These are calculated using a series of estimates from the PVAR model based on an expanding window of observations as explained in the previous section. The results are depicted in \autoref{fig:DYindex_data} below. The graph shows the total spillover index for the 1-month horizon with the grey shaded areas denoting the periods of the GFC and the ESDC.

\begin{figure}[!htbp]
\caption{Overall spillover index for the 1-month forecast horizon. \label{fig:DYindex_data}}

\begin{minipage}{0.49\textwidth}
\centering
\small \textit{Government bond yields $(S_{gb}(1))$}
\end{minipage}
\begin{minipage}{0.49\textwidth}
\centering
\small \textit{Borrow long $(S_{bl}(1))$}
\end{minipage}

\begin{minipage}{0.49\textwidth}
\centering
\includegraphics[scale=.3]{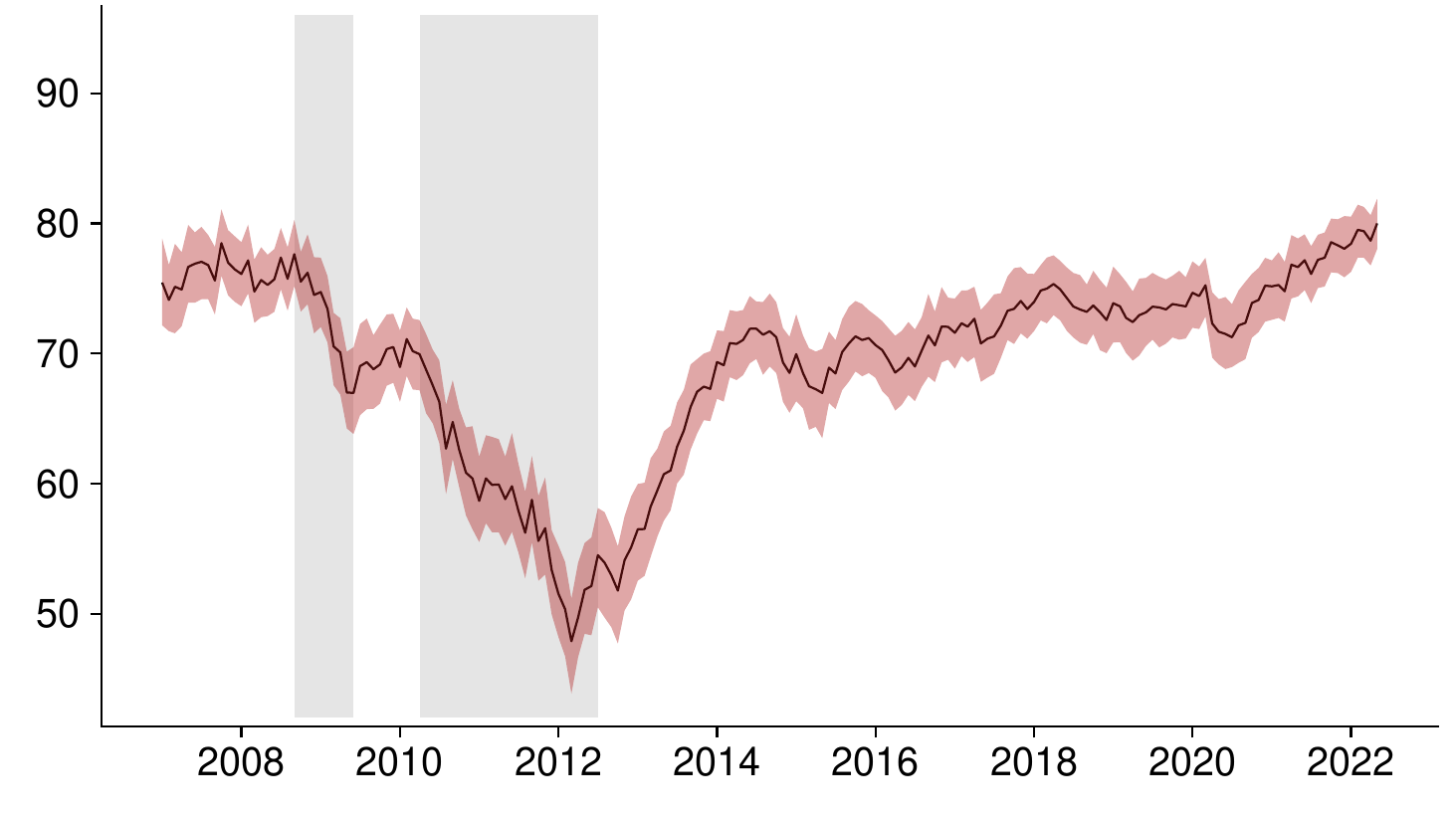}
\end{minipage}
\begin{minipage}{0.49\textwidth}
\centering
\includegraphics[scale=.3]{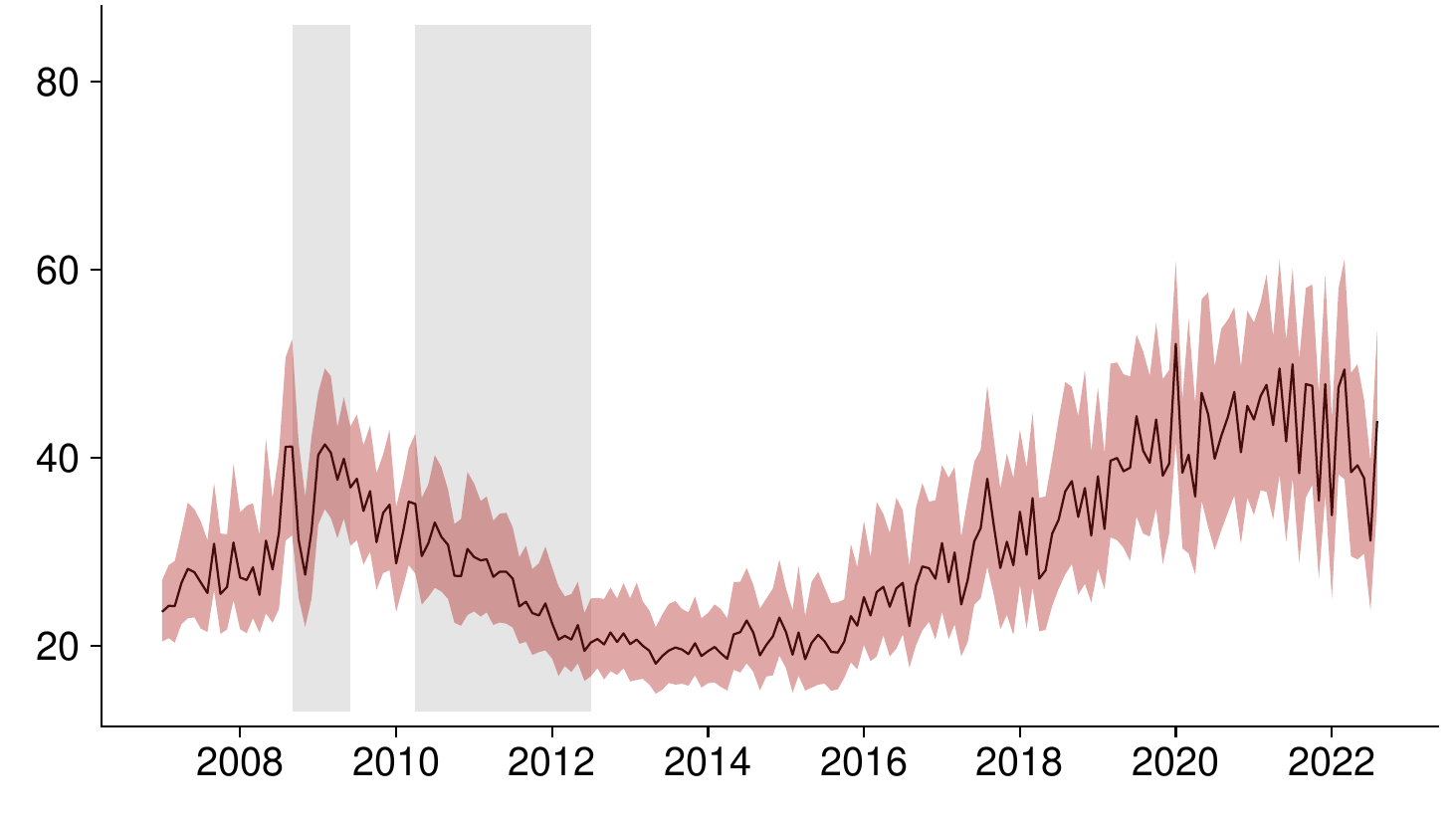}
\end{minipage}

\begin{minipage}{0.49\textwidth}
\centering
\small \textit{Borrow short $(S_{bs}(1))$}
\end{minipage}
\begin{minipage}{0.49\textwidth}
\centering
\small \textit{TARGET2 $(S_{t2}(1))$}
\end{minipage}

\begin{minipage}{0.49\textwidth}
\centering
\includegraphics[scale=.3]{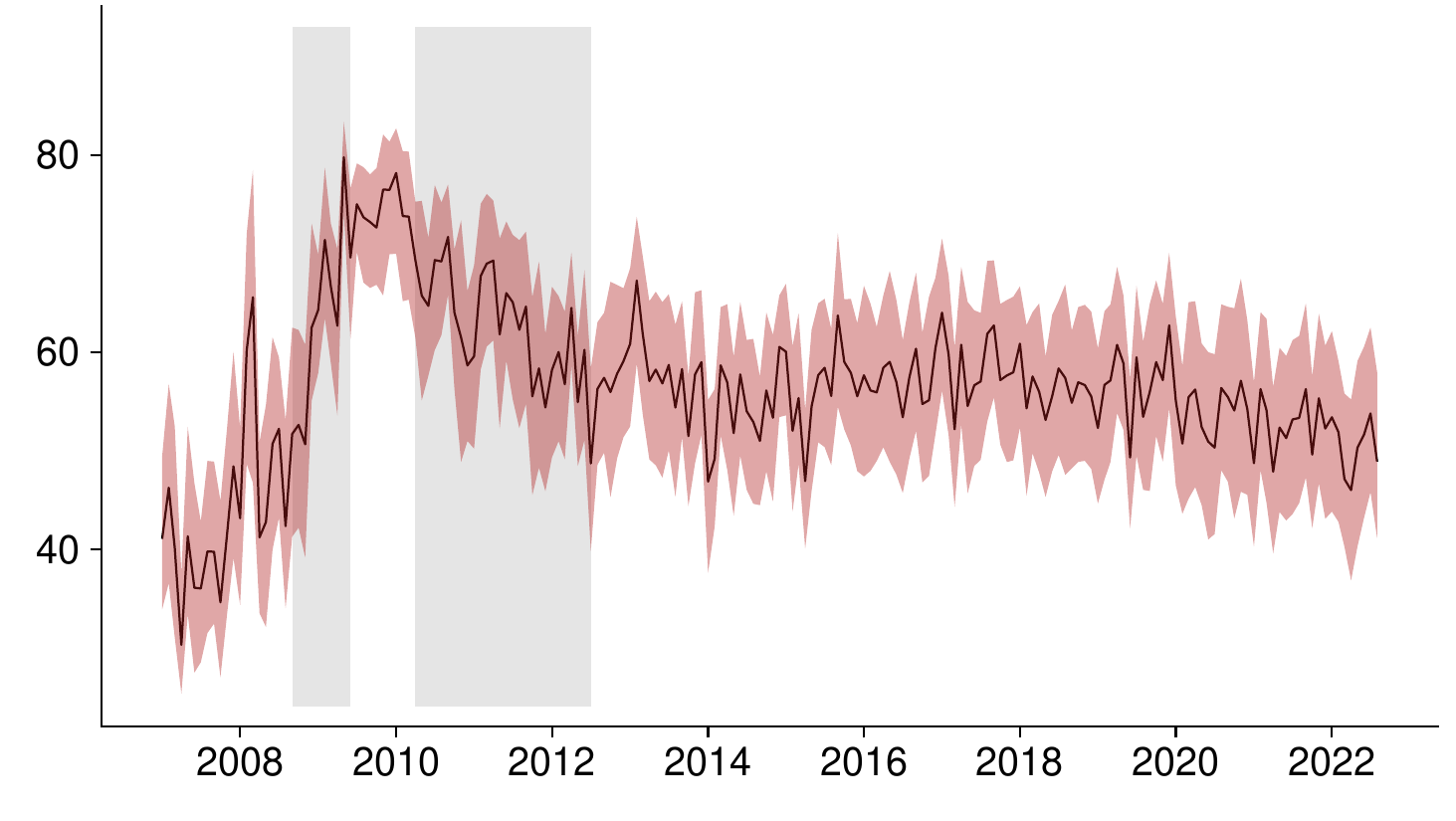}
\end{minipage}
\begin{minipage}{0.49\textwidth}
\centering
\includegraphics[scale=.3]{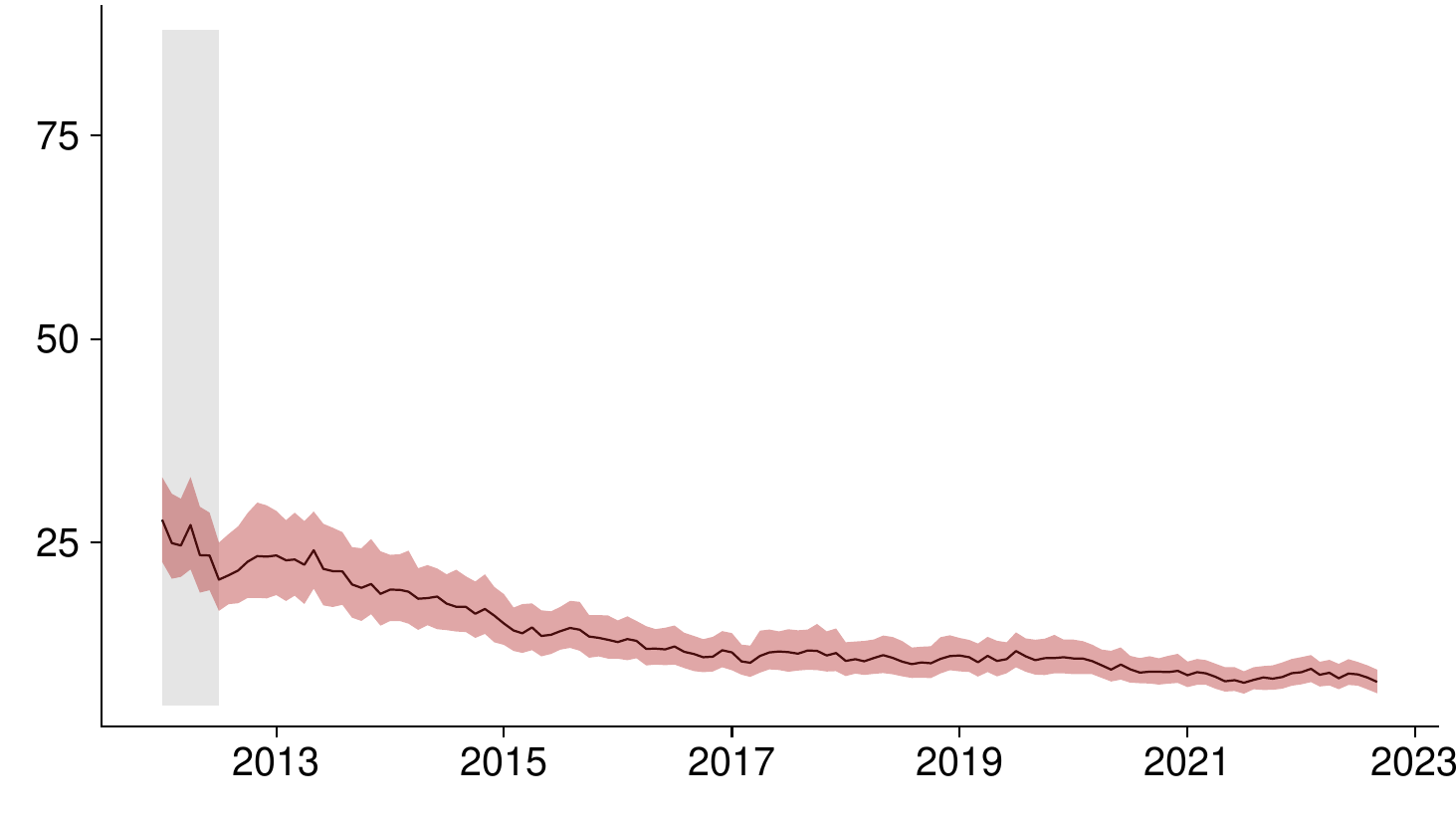}
\end{minipage}

\begin{minipage}{\textwidth}
\vspace{2pt}
\scriptsize \emph{Note:} This index indicates the share of spillovers across countries (averaged over all countries) according to \cite{diebold2009measuring} and is estimated based on an expanding window. The solid line is the posterior median alongside the $68\%$ posterior credible set. The grey shaded area depicts the period of the ESDC.
\end{minipage}
\end{figure}

We start with analyzing the sovereign bond market ($gb$). Sovereign bond yields, corrected for differences in country-specific risk premia, should be closely related  in a credible monetary union due to common interest and inflation expectations. 
Spillovers in the bond market started to decline significantly with the onset of the GFC. Not surprisingly, the trough is reached at the height of the ESDC, during which a sharp break in macrofundamentals and sovereign bond yields has been observed \citep{Horny2018}. From 2012 onward, integration increased markedly leading to a full recovery of financial connectedness in the latest part of our sample compared to the period prior to the ESDC. Our results do not indicate a re-emergence of fragmentation during the COVID crisis.  By contrast, the pandemic stimulus package seems to have created a new upward tick of connectivity, as evidenced by the increase of the total spillover measure after 2020. 

We next analyze integration on the credit market. Euro area lending rates might be related due to common monetary policy or cross-country banking links. \citet{Emter2019} show that a deterioration of asset quality in a source country can lead to a retrenchment of banking flows to other euro area countries. Through the "home bias" of commercial banks, lower asset quality can also be related to developments on the sovereign bond market. The total spillover index for the long-term borrowing costs follow a similar trend as the one for government bond yields. In general, a decrease of connectedness has set in at the height of the GFC and the trough has been reached at the end of the ESDC. Compared to connectedness of government bond yields, the upward trend of integration starts at a later date. More precisely and coinciding with the launch of the asset purchase programs, connectedness starts to increase sharply until the end of our sample reaching similar levels to before the ESDC. Easing financing conditions for households and firms has been one of the major targets of quantitative easing programs. 

The picture changes when looking at short-term borrowing costs. Here, the policy rate cuts by the ECB in response to the GFC have increased connectivity. The degree of connectivity declined, once the policy rates reached the zero lower bound and different tools of monetary policy have been used. Last, we analyze spillovers between TARGET2 balances to investigate a variable that reflects trade imbalances within the euro area. The hold-out starts in 2012 and indicates a gradual decline in connectivity. This implies that we observe a low and persistent degree of connectivity from 2015 onward. 

\begin{figure}[!htbp]
\caption{Overall spillover index. Comparison for the 1-month (red) and 12-months (blue) forecast horizons. \label{fig:comp}}

\begin{minipage}{0.49\textwidth}
\centering
\small \textit{Government bond yields}
\end{minipage}
\begin{minipage}{0.49\textwidth}
\centering
\small \textit{TARGET2}
\end{minipage}

\begin{minipage}{0.49\textwidth}
\centering
\includegraphics[scale=.3]{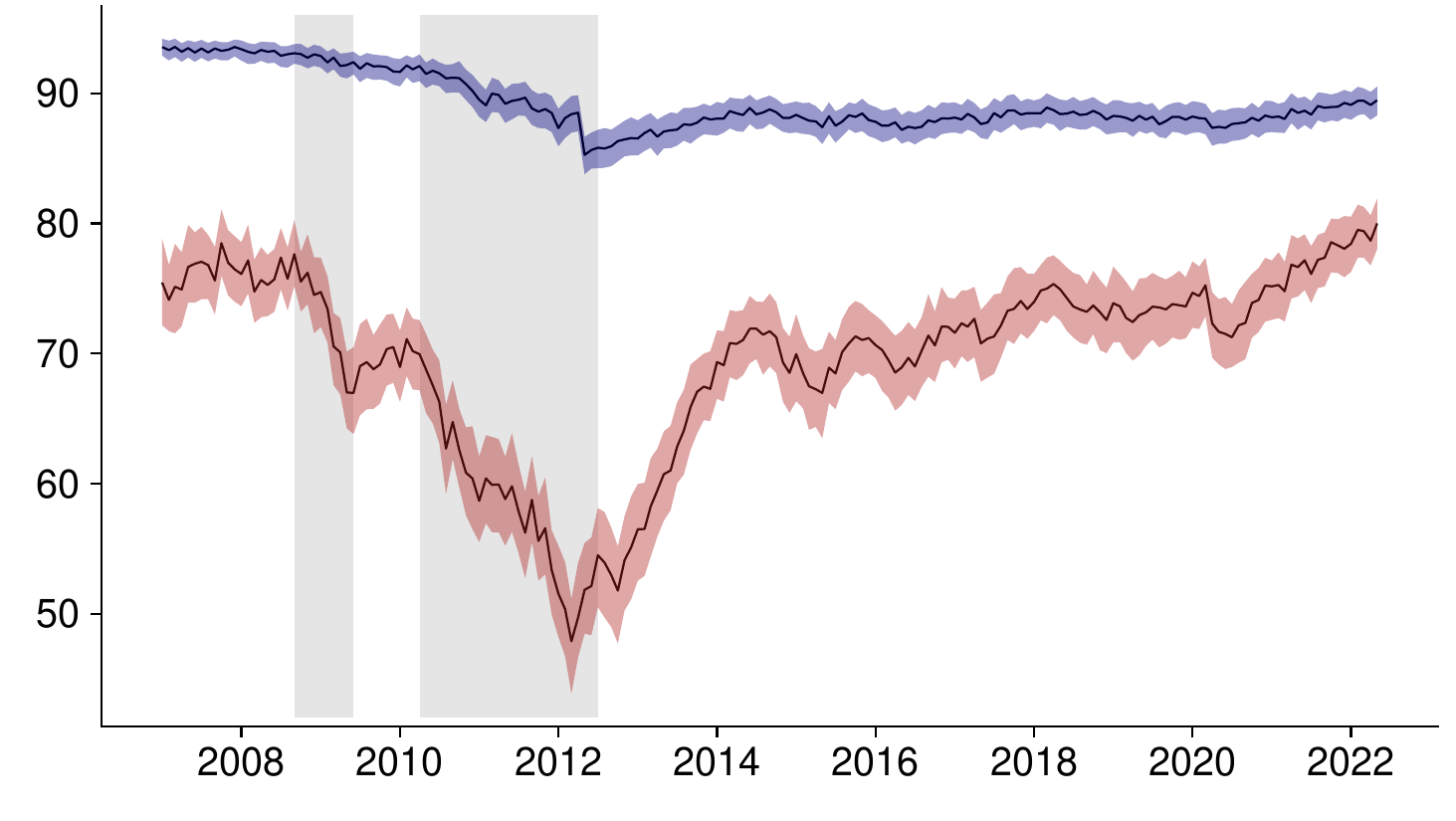}
\end{minipage}
\begin{minipage}{0.49\textwidth}
\centering
\includegraphics[scale=.3]{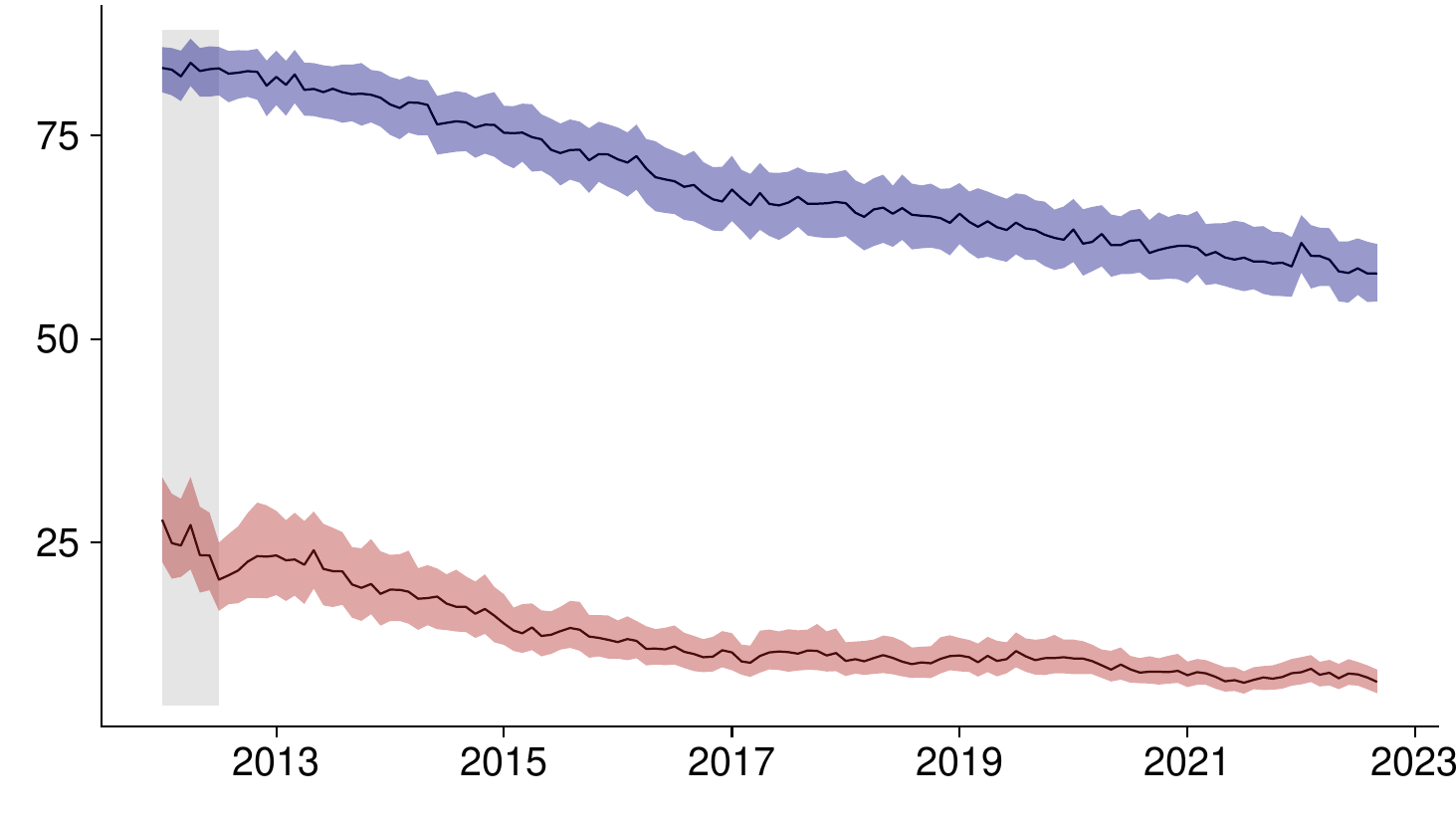}
\end{minipage}

\begin{minipage}{0.49\textwidth}
\centering
\small \textit{Borrow long}
\end{minipage}
\begin{minipage}{0.49\textwidth}
\centering
\small \textit{Borrow short}
\end{minipage}

\begin{minipage}{0.49\textwidth}
\centering
\includegraphics[scale=.3]{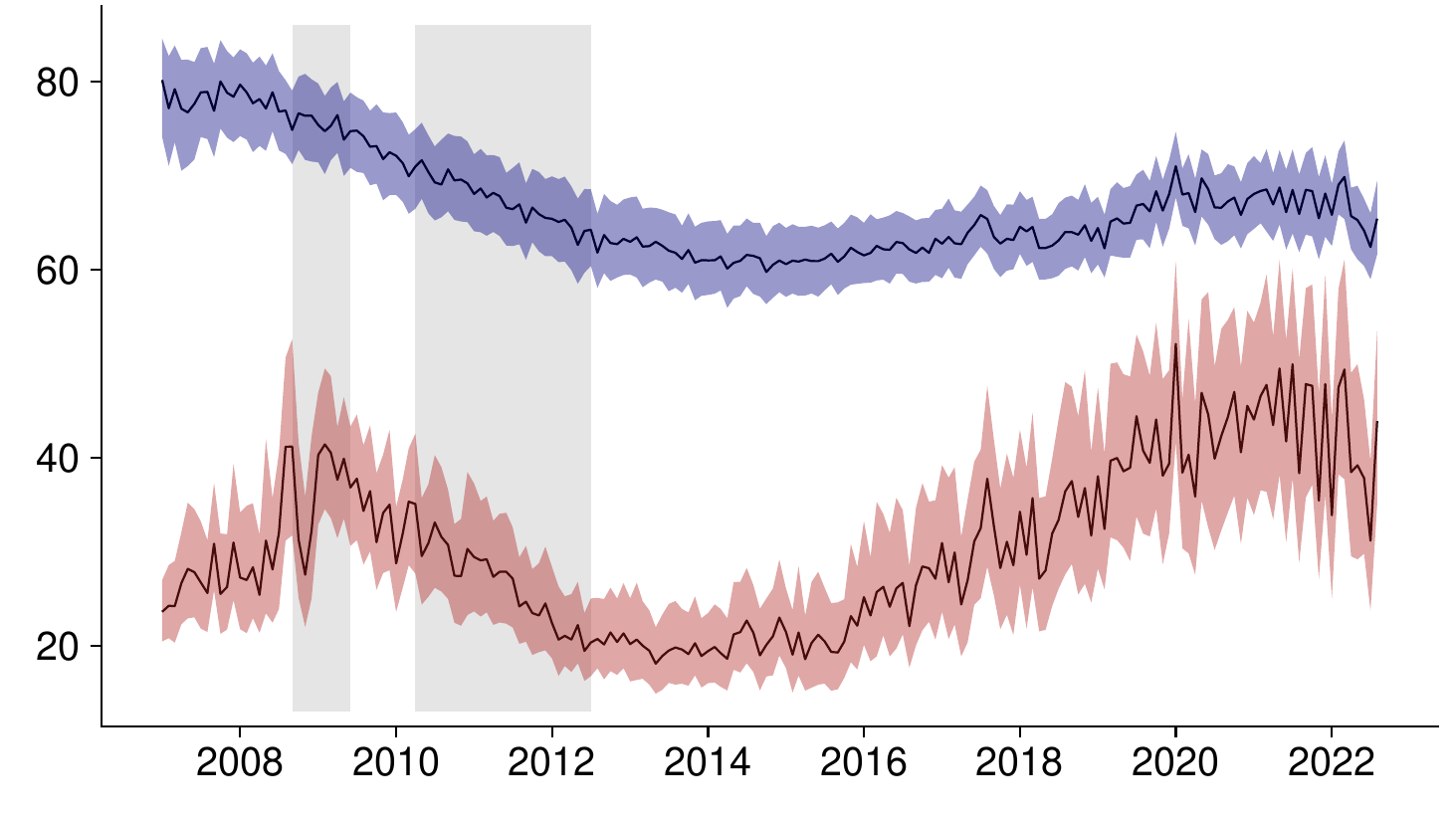}
\end{minipage}
\begin{minipage}{0.49\textwidth}
\centering
\includegraphics[scale=.3]{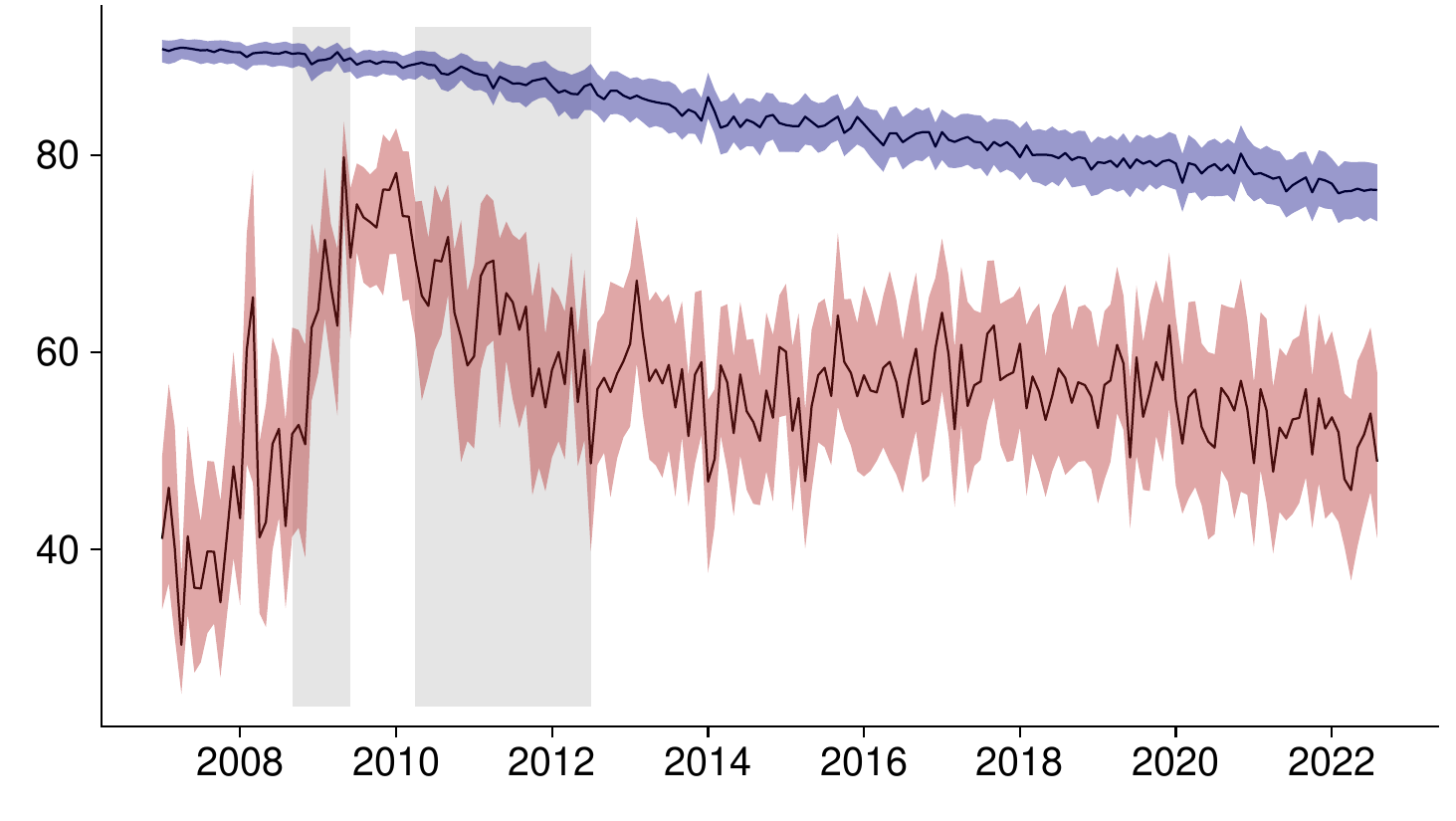}
\end{minipage}

\begin{minipage}{\textwidth}
\vspace{2pt}
\scriptsize \emph{Note:} This index indicates the share of spillovers across countries (averaged over all countries) according to \cite{diebold2009measuring} and is estimated based on an expanding window. The solid line is the posterior median alongside the $68\%$ posterior credible set. The posterior credible set colored in red refers to the 1-month ahead forecast horizon whereas the blue colored bounds indicate the 12-months ahead forecast horizon. The grey shaded area depicts the period of the ESDC.
\end{minipage}
\end{figure}

As a sensitivity check, we include the spillover indices based on a 12-months forecast error variance decomposition in \autoref{fig:comp}. A general observation when using the DY spillover index is that as the forecast horizon increases, and through the lag structure in the underlying VAR model, the share of error variance explained by non-domestic shocks tends to increase.
Looking at \autoref{fig:comp}, we see this level shift of the DY spillover index with higher values of integration being associated with a larger forecast horizon. That said, the measures based on the 12-months forecast horizon tend to show a similar trend as the 1-month based index. An exception to this are long-term borrowing costs. Here, the short-term DY spillover index points to an increase in connectivity, whereas the long-run measure barely changes at the beginning of our sample period. For the interested reader, we provide the full set of results for the 12-months horizon in the appendix.

Summing up, our analysis reveals that financial connectedness in the euro area bond market dropped significantly during the period of the GFC and reached a trough level at the end of the ESDC crisis. In the most recent part of our sample, we see connectedness levels similar to the ones before the ESDC. These results corroborate findings of \citet{Hoffmann2020} who use composite indicators to measure financial integration as well as \citet{Chatziantoniou2021} and \citet{Costola2022} who use daily bond data. Our results also remain qualitatively unchanged if we consider a different forecast horizon as well as different econometric frameworks. Results for the credit market show a similar but less pronounced responsiveness to crisis periods, whereas integration regarding TARGET2 balances is on a steady downward trend.

\subsection{Country-specific spillovers}

In this section, we delve deeper into the cross-country variation of financial connectedness by looking at country-specific spillover indices. This allows us to assess to what degree countries differ from the overall spillover index discussed in the previous section and to see whether spillover indices of some countries behave similarly, especially so during crisis events. A related exercise has been carried out in \citet{Chatziantoniou2021}.

\begin{figure}[htbp!]
\caption{Country-specific spillover index for the 1-month forecast horizon.}\label{fig:DYindex_cc_1}
        \captionsetup[subfigure]{aboveskip=0pt,belowskip=-5pt}
	\begin{subfigure}{0.45\textwidth}\caption{Government bond yields (core)}
		\centering\includegraphics[scale=0.3]{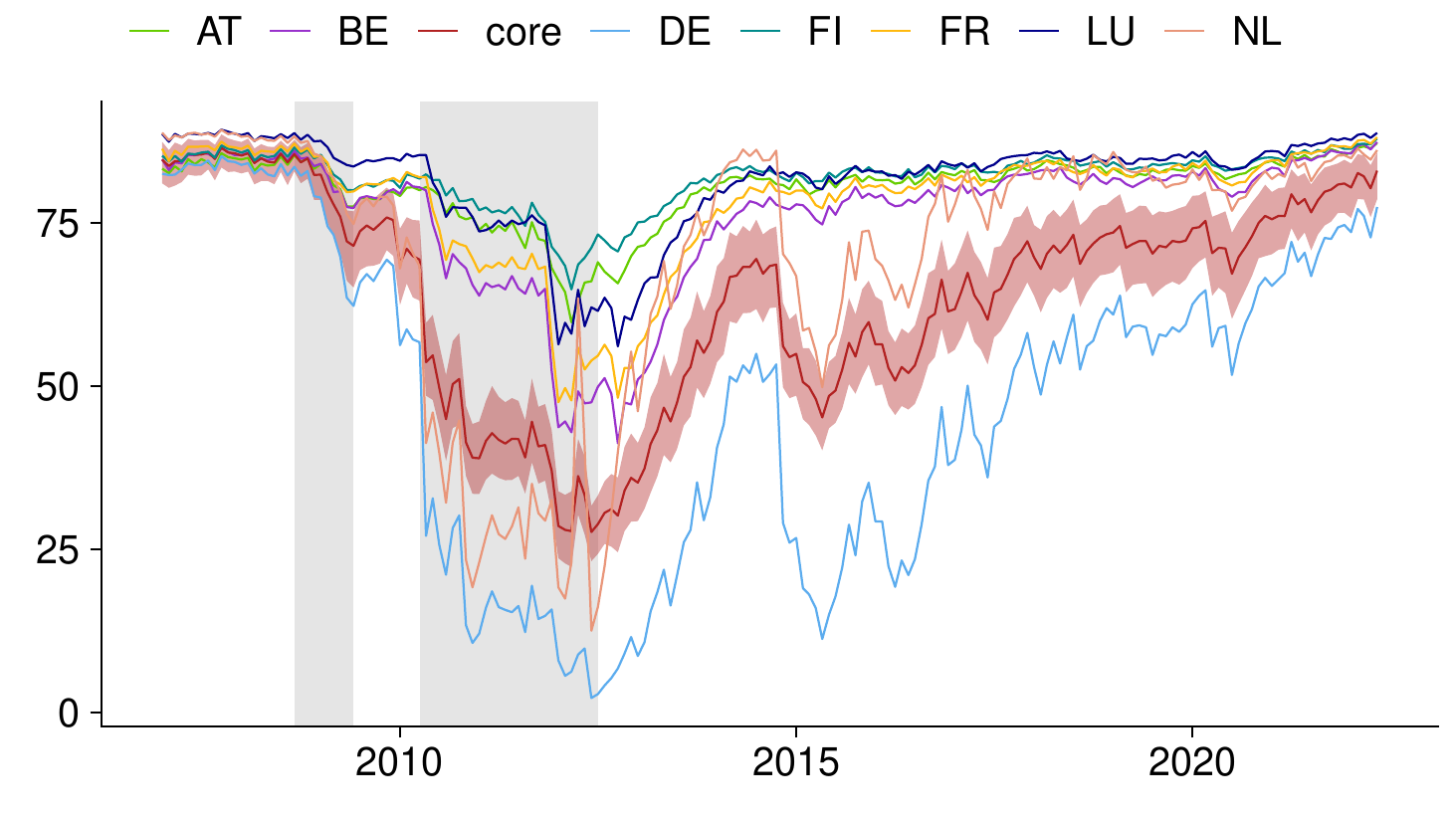}
	\end{subfigure}
	\begin{subfigure}{0.45\textwidth}\caption{Government bond yields (periphery)}
		\centering\includegraphics[scale=0.3]{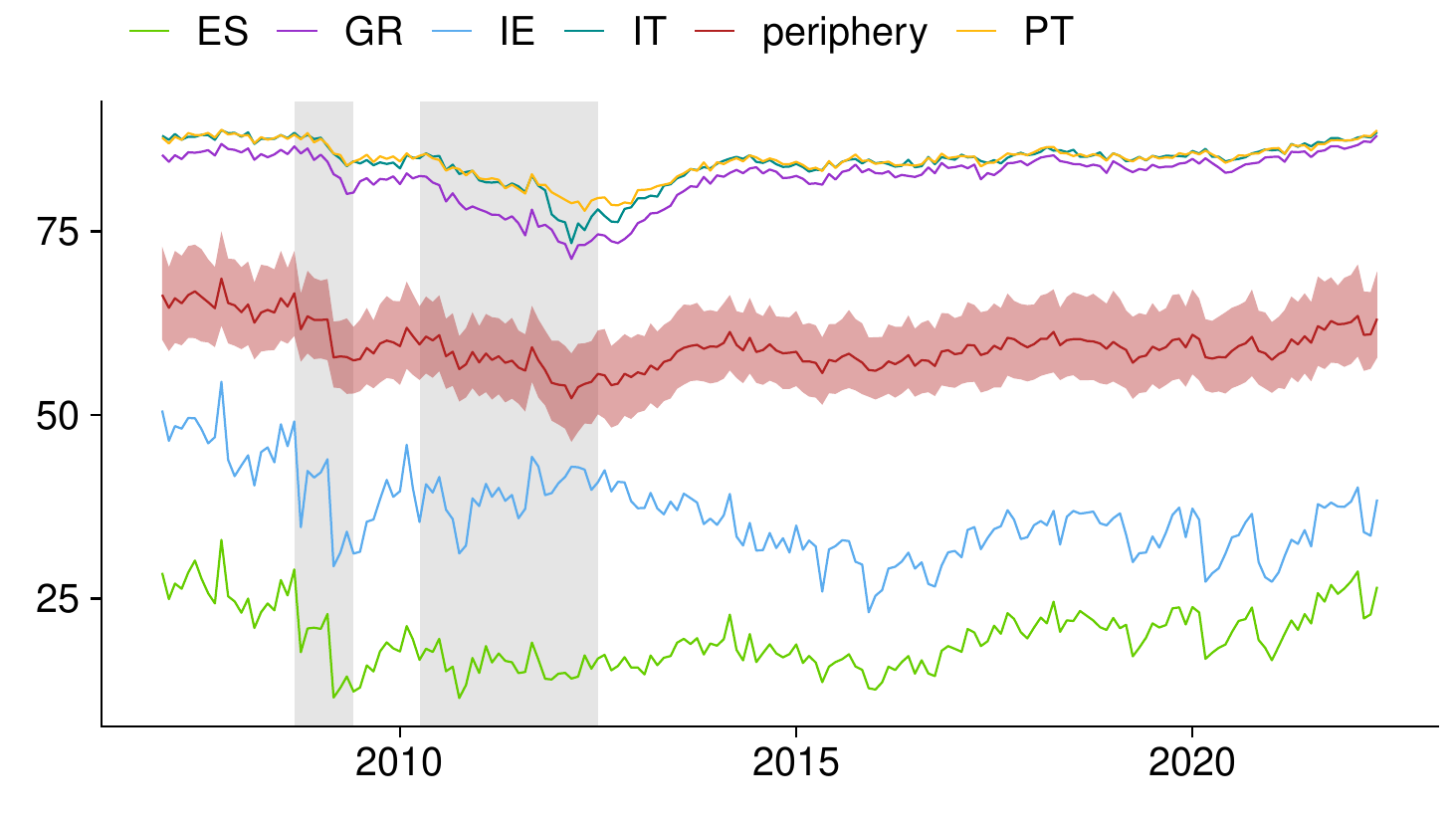}
	\end{subfigure}\\
	\vspace{0.5cm}\\
	\begin{subfigure}{0.45\textwidth}\caption{Borrow long (core)}
		\centering\includegraphics[scale=0.3]{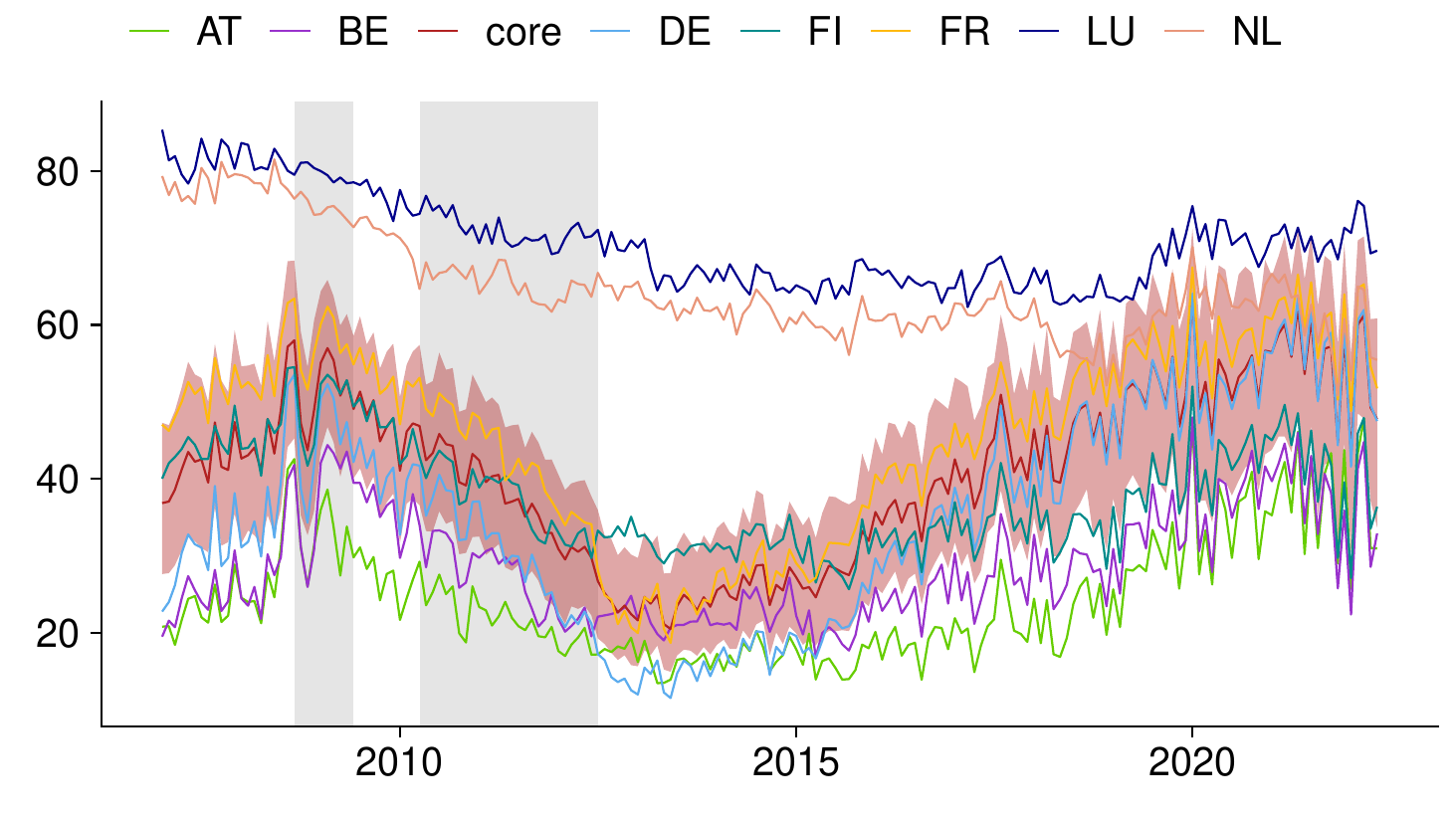}
	\end{subfigure}
		\begin{subfigure}{0.45\textwidth}\caption{Borrow long (periphery)}
		\centering\includegraphics[scale=0.3]{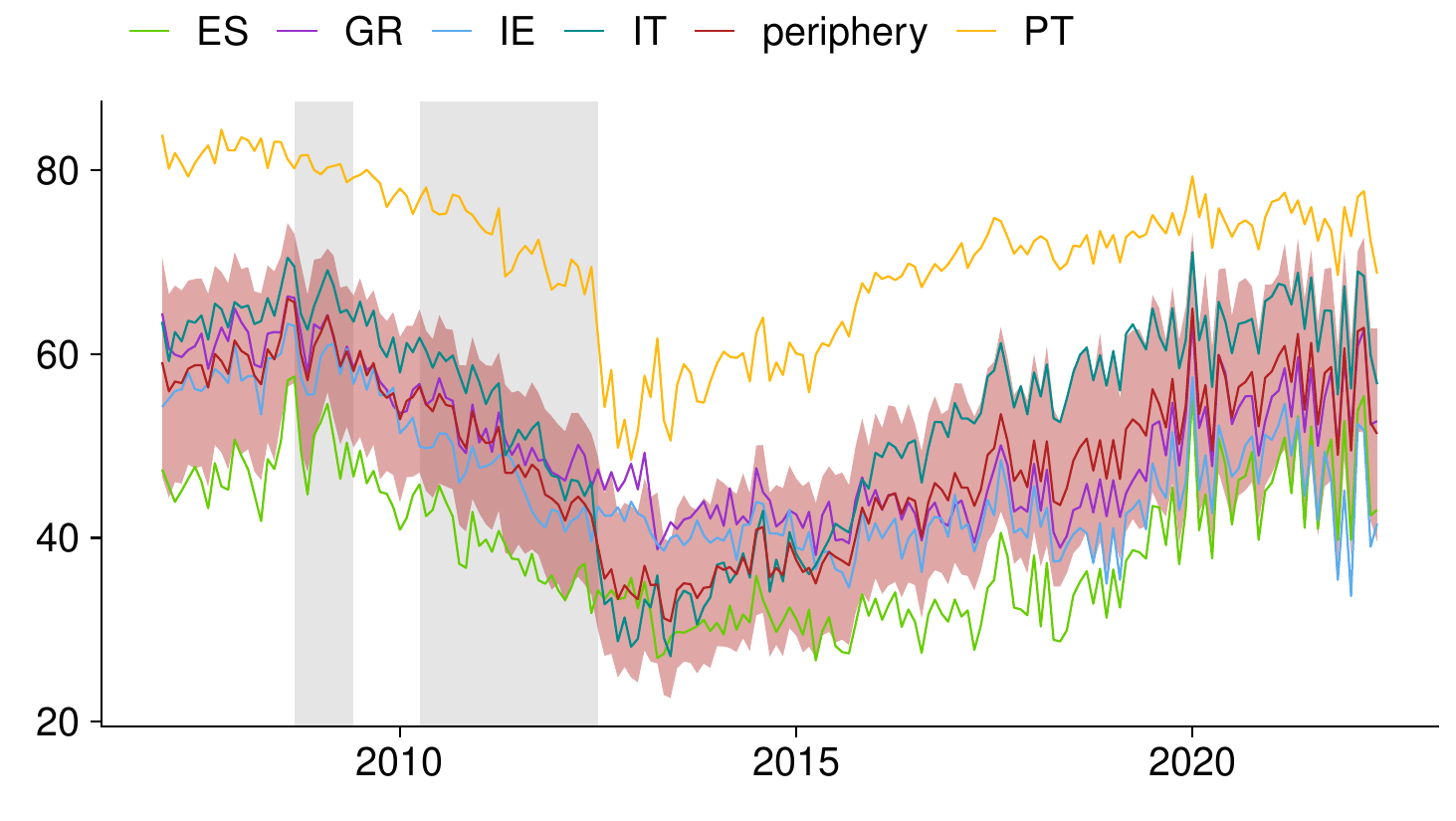}
	\end{subfigure}\\
		\vspace{0.5cm}\\
	\begin{subfigure}{0.45\textwidth}\caption{Borrow short (core)}
		\centering\includegraphics[scale=0.3]{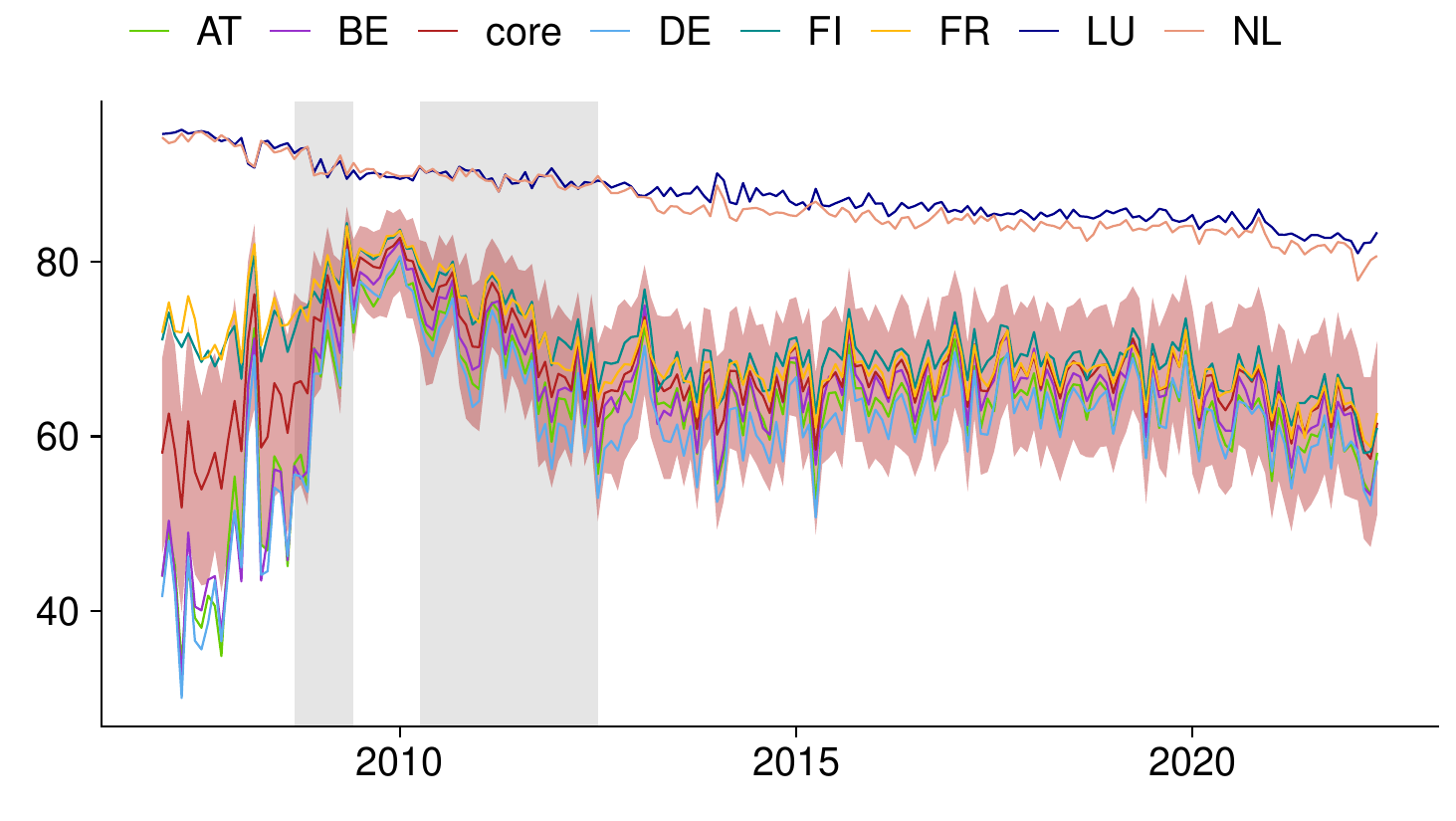}
	\end{subfigure}
	\begin{subfigure}{0.45\textwidth}\caption{Borrow short (periphery)}
		\centering\includegraphics[scale=0.3]{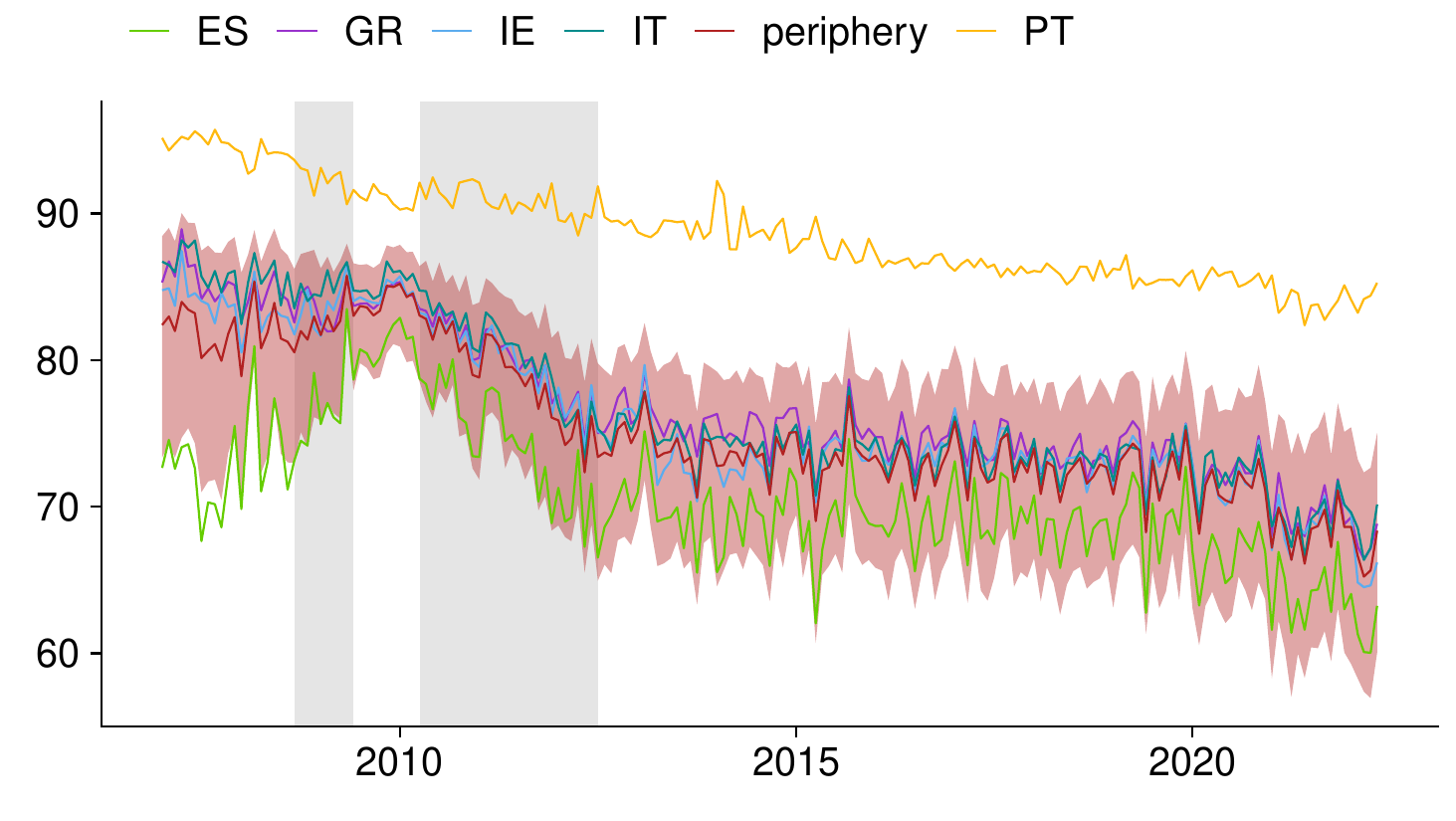}
	\end{subfigure}\\
	\vspace{0.5cm}\\
	\begin{subfigure}{0.45\textwidth}\caption{TARGET2 (core)}
		\centering\includegraphics[scale=0.3]{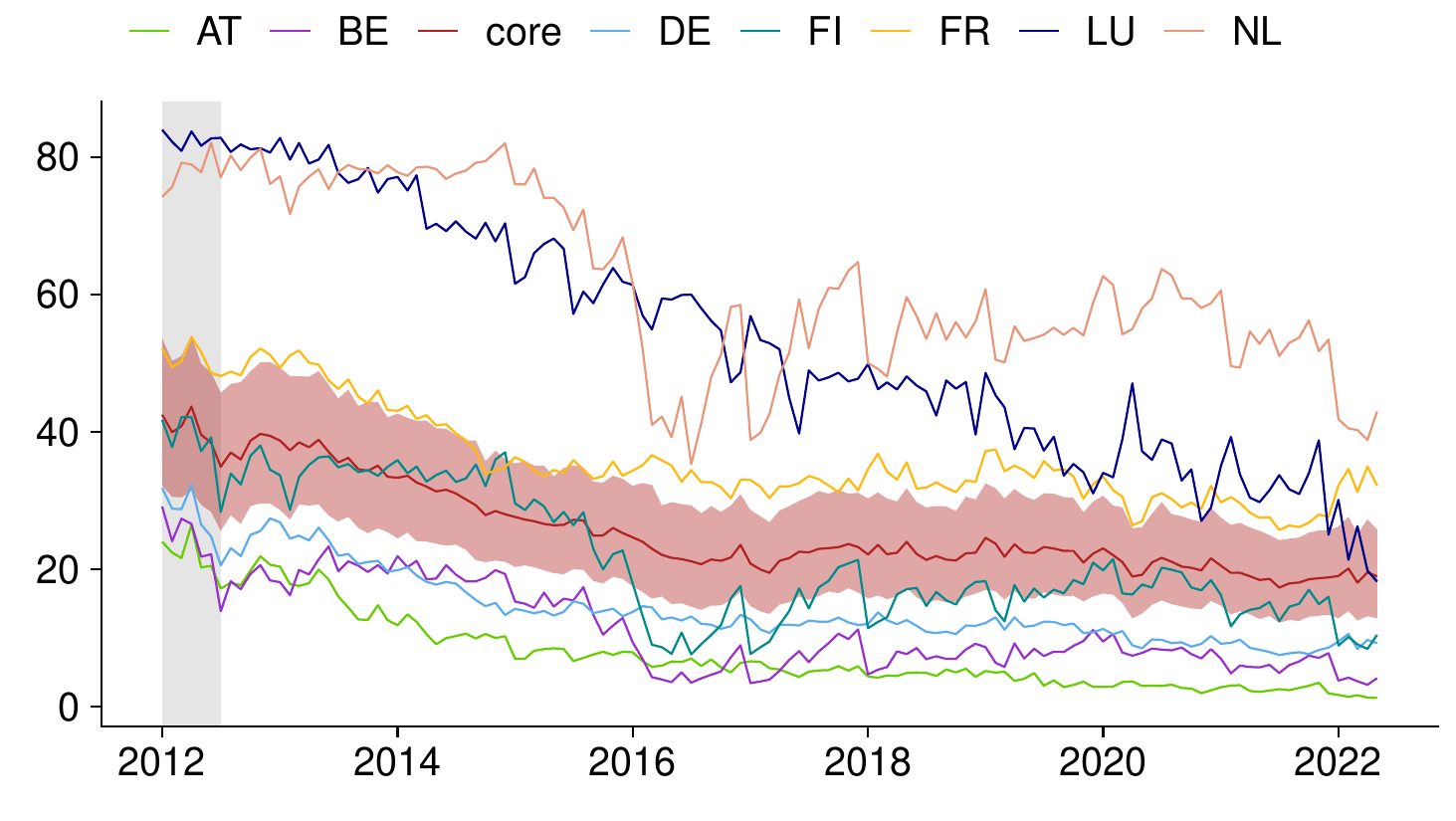}
	\end{subfigure}
		\begin{subfigure}{0.45\textwidth}\caption{TARGET2 (periphery)}
		\centering\includegraphics[scale=0.3]{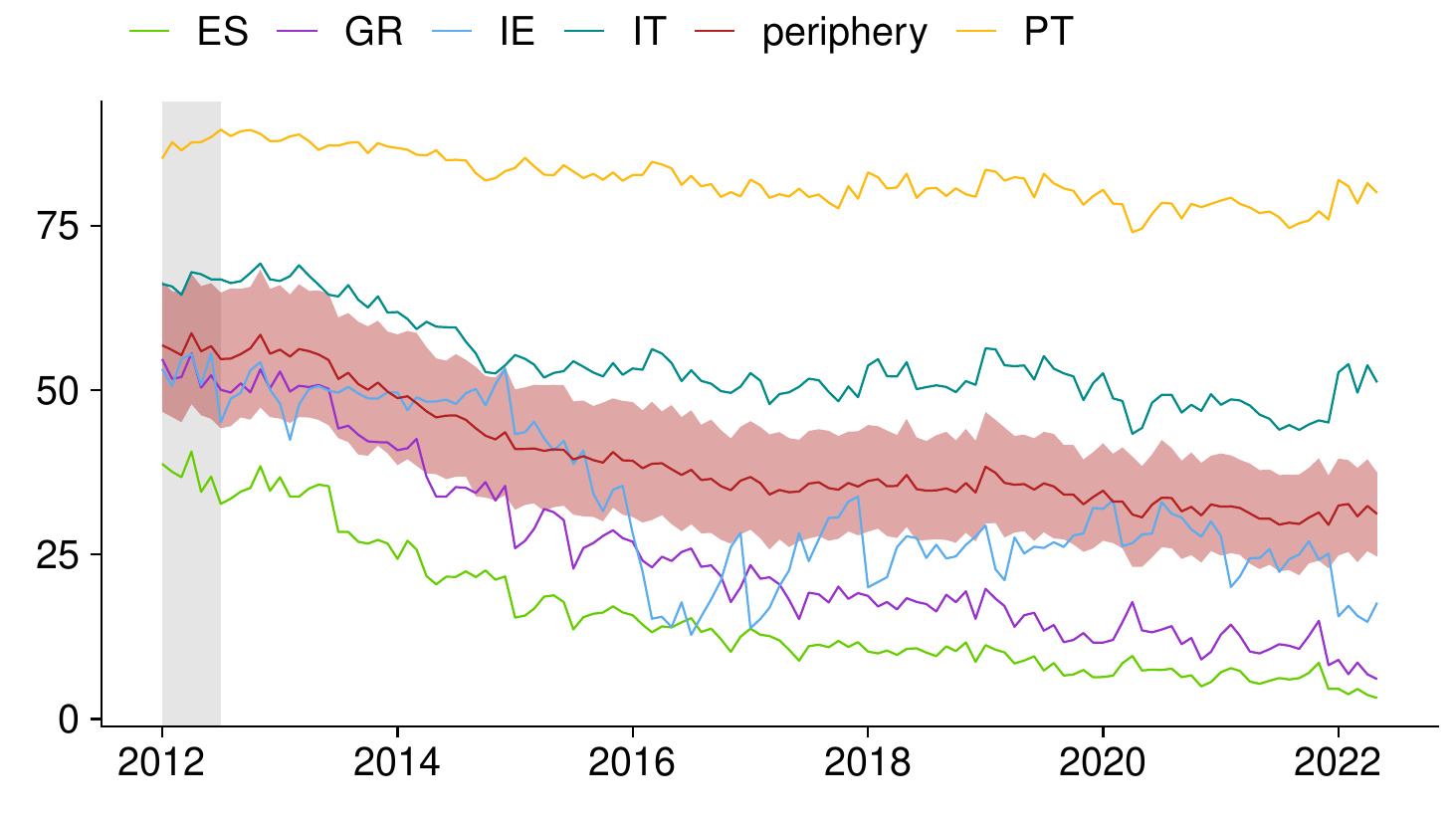}
	\end{subfigure}\\
	\begin{minipage}{14cm}~\\
\scriptsize \emph{Note:} This index indicates the share of spillovers for each country according to \cite{diebold2009measuring} and is estimated based on an expanding window. The solid line is the posterior median alongside the $68\%$ posterior credible set. Regional aggregates are calculated with PPP weights. In light grey, we depict the periods of the GFC and the ESDC. \end{minipage}%
\end{figure}

\autoref{fig:DYindex_cc_1} shows responses for core and periphery countries separately, with credible sets along the respective regional means. The full set of results is available in Figures \autoref{fig:DYindex_detail_1} and \autoref{fig:DYindex_detail_2} in the appendix.

Starting with the bond market, we see some remarkable differences in integration, even for the rather homogeneous group of core countries. More precisely, whereas financial integration plummets for all core countries during the ESDC, the decrease is significantly more pronounced for German bonds. That is, German bonds decoupled from their peers.  Moreover, German (and to a lesser extent Dutch) bonds show a second decrease after the ESDC, which only starts to revert once the ECB started its QE program. This outstanding position of German bonds might be related to their status serving as a "safe haven" during turbulent times. At the end of our sample period, connectedness levels are as high as before the ESDC. A different picture emerges when looking at the periphery countries. Here, one group of countries, i.e., Greece, Portugal and Spain, show a slight decline in financial integration during the ESDC followed by a quick recovery -- similar to some of the core economies. Spain and Ireland, however, deviate markedly from both core and periphery economies by showing a persistently low degree of financial integration with their peers. This low degree of financial connectedness is moreover only modestly affected by crisis episodes.


Looking at the credit market for core economies, we find distinct patterns for both short- and long-term rates for the Netherlands and Luxembourg. In these economies, financial connectedness is firstly, always higher compared to their peers and secondly, persistently so. Hence, in these economies, crisis periods have a comparably small impact on financial integration. 
Looking at the periphery countries, we find that spillovers for Portuguese rates considerably differ from their regional peers. While for long-term rates, the spillover index also declines at the end of the ESDC, it is rather persistent considering short-term rates. For both financial market segments, Portuguese spillover indices lie well above the ones of other periphery countries.


 Last, we turn to the spillover indices for TARGET2 balances.  Looking at the core countries and consistent with our previous results, spillover indices for the Netherlands and Luxembourg consistently lie above indices for the rest of the sample. We also have countries with spillover indices that lie below the regional mean such as Austria and Belgium. Regarding periphery countries, we find an elevated spillover index for Portugal and low indices for Ireland and Greece. The fact that in both groups most countries lie either above or below the credible sets spanned by the regional mean index indicates that we observe a high degree of cross-country variation for TARGET2 balances. This is in contrast to spillover indices for the credit market. 
 
 Summing up, we find that the country-specific spillover indices can deviate significantly from the overall spillover index. This holds especially true for the government bond market and TARGET2 balances, whereas indices for lending rates are more homogeneous. Even here though,  countries like the Netherlands, Luxembourg and Portugal show a significantly different degree of financial integration compared to their peers. This implies a distinct responsiveness of these countries' financial integration to crisis events.

\subsection{A network analysis of financial connectedness using blockmodels}

In this section, we complement our analysis by drawing on network techniques \citep[for recent applications in related contexts see, e.g.,][]{Buse2019, Chen2022}. Whereas in the previous section we examined cross-country differences in how countries' integration changes during crisis events, here we aim to identify countries that play similar roles in the euro area financial network. For that purpose, we  use the generalized blockmodeling approach of \citet{vziberna2014blockmodeling}, implemented in the \verb+R+ package of \citet{Matjasevic2021}. The main idea behind blockmodeling is to analyze relationships or connections in a cross-section of units and to identify units that ``behave" similarly.\footnote{As is common in the literature on blockmodeling, we use as a similarity condition ``regular" as opposed to ``structural" equivalence. Structural equivalence would imply that similar countries not only have to have a similar extent of spillovers received but the countries from which these spillovers are sent would also have to be similar -- a condition that is rarely fulfilled in real world applications.} For example, some countries could act as spillover absorbers receiving more spillovers than transmitting. Other countries could function as gatekeepers, receiving a lot and transmitting a lot 
\citep{IMF2012}. Yet others might be isolated from their peers. In that case there is no ``traffic" from and to other countries implying a truly peripheral position in the network of euro area financial markets.

In what follows, we are going to allow for $i\in 1,\ldots,m_{Cl}=4$ different blocks or clusters of the matrix representation of the financial network. The \textit{within strength} of interactions is reflected in the block density of a particular cluster $Cl_{ij}$ with $i=j$. In our case the interactions between countries are the extent to which countries receive and transmit spillovers, i.e., are connected to the rest of the countries.\footnote{In standard blockmodeling applications of so-called valued networks, one uses directly bilateral data on interactions. In our case, the units are the $N$ euro area countries and the interactions we focus on are the spillover intensities as measured by the GFEVD for the different variables discussed in the previous sections.}   This analysis also allows to look at the  \textit{cross-cluster strength} of interactions, which is denoted by $Cl_{ij}$ and indicates to which extent $Cl_i$ \textit{receives} spillovers from $Cl_j$. 

\begin{figure}[htbp!]
\caption{Clustering of GFEVD using a generalized block model for 1-month forecast horizon.}\label{fig:blockmodel}
\begin{minipage}{1\linewidth}~\\
\centering \textbf{Government bond yields}
\end{minipage}\\
\begin{minipage}[b]{.24\linewidth}
Pre-GFC
\centering \includegraphics[scale=0.22]{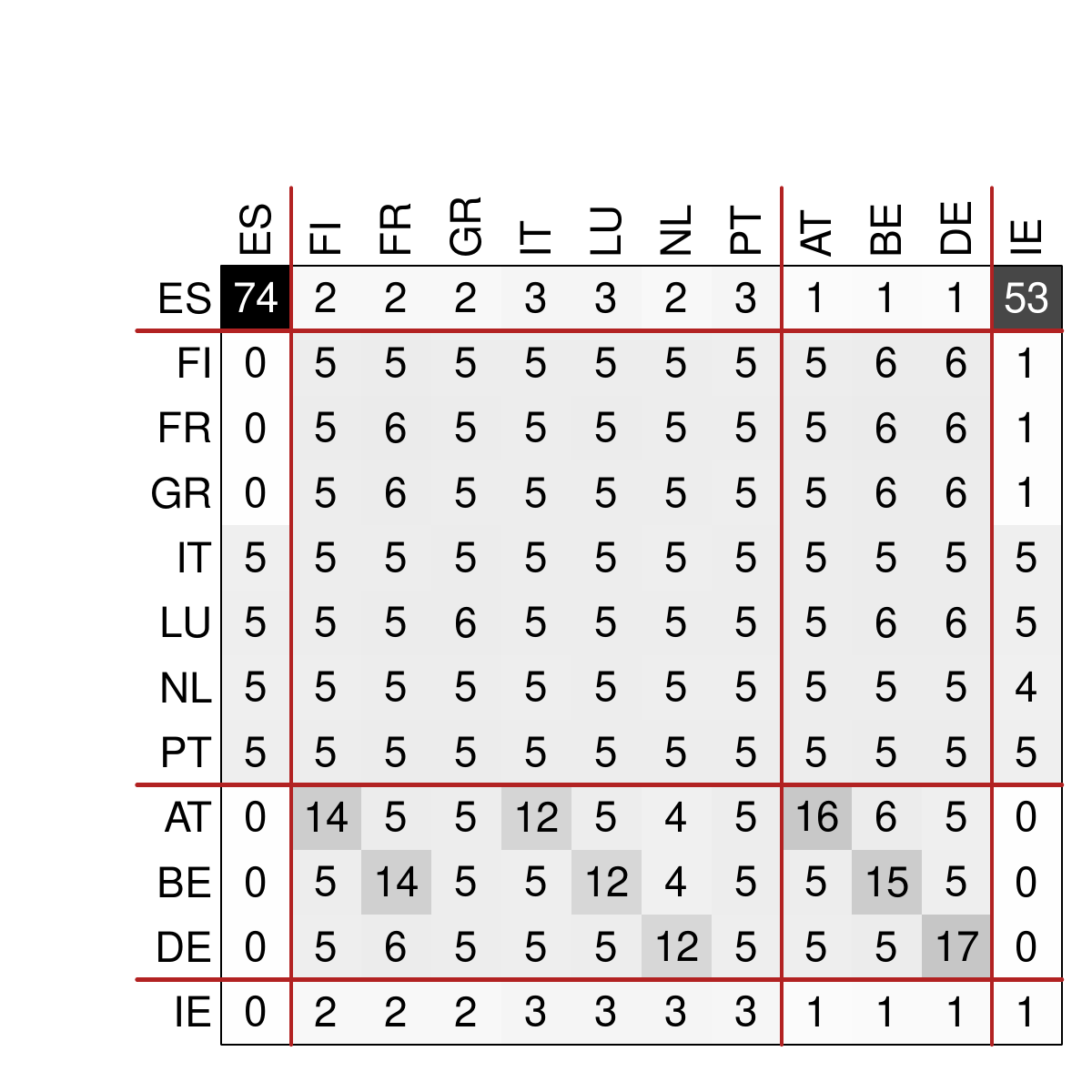}
\end{minipage}%
\begin{minipage}[b]{.24\linewidth}
GFC
\centering \includegraphics[scale=0.22]{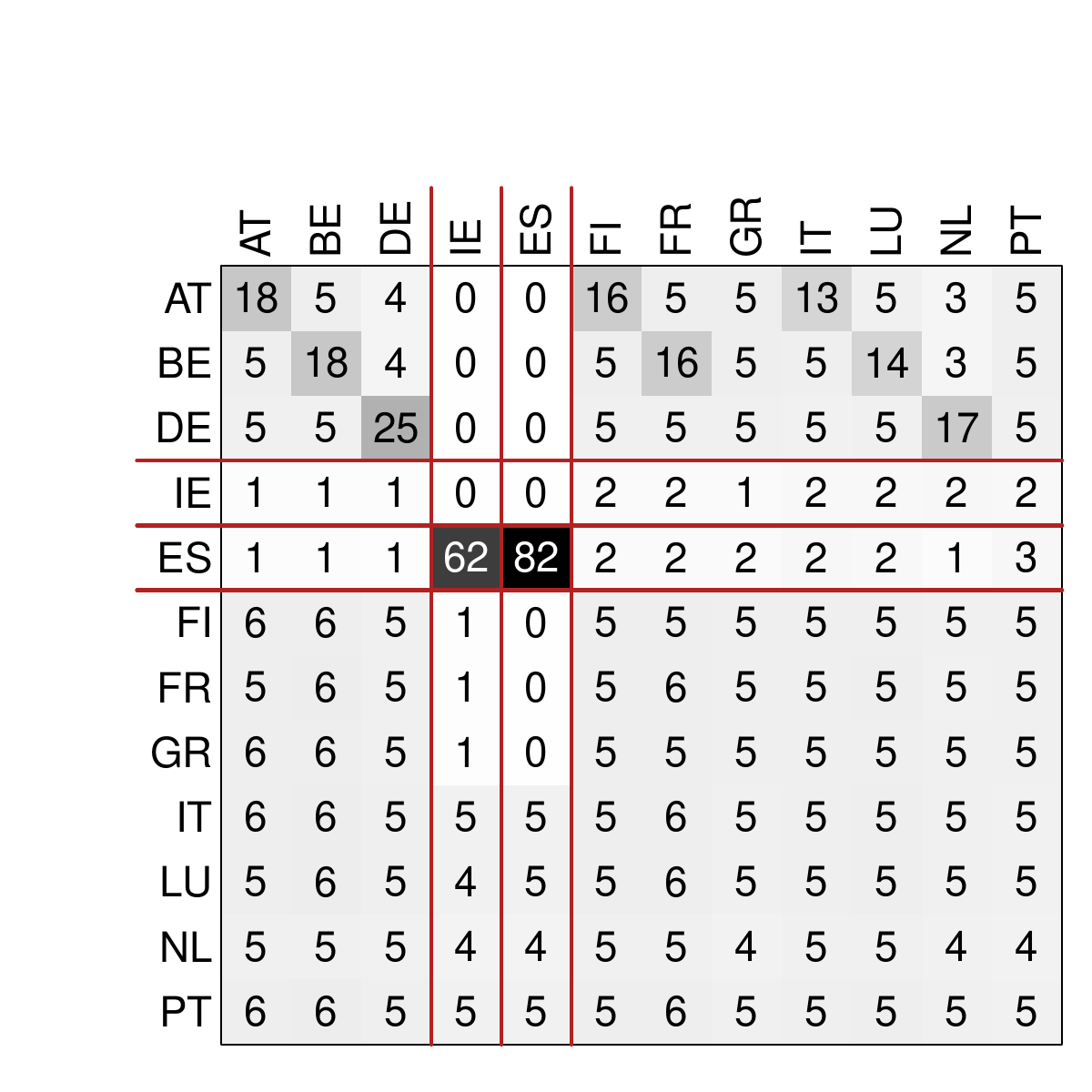}
\end{minipage}
\begin{minipage}[b]{.24\linewidth}
ESDC
\centering \includegraphics[scale=0.22]{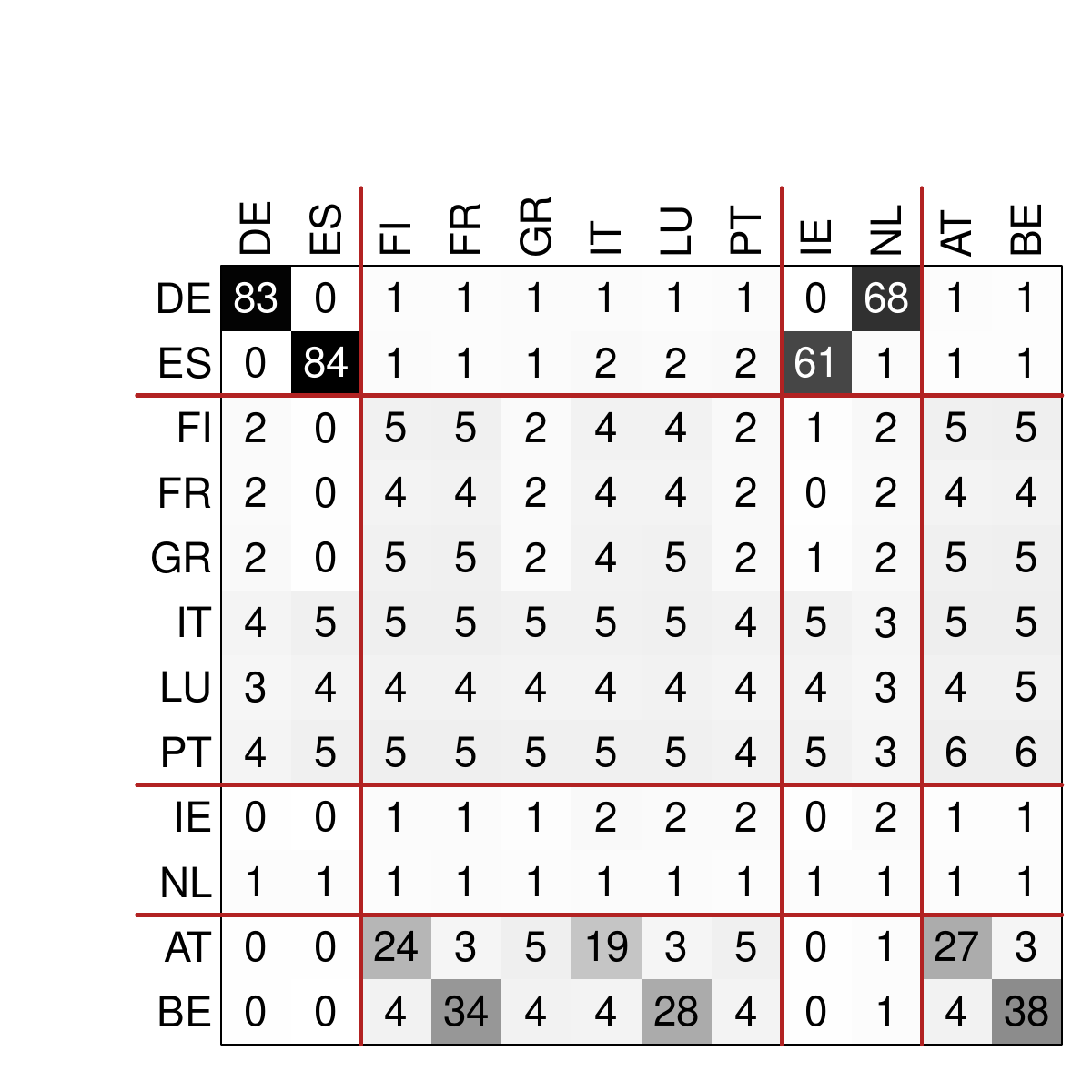}
\end{minipage}
\begin{minipage}[b]{.24\linewidth}
Post-ESDC
\centering \includegraphics[scale=0.22]{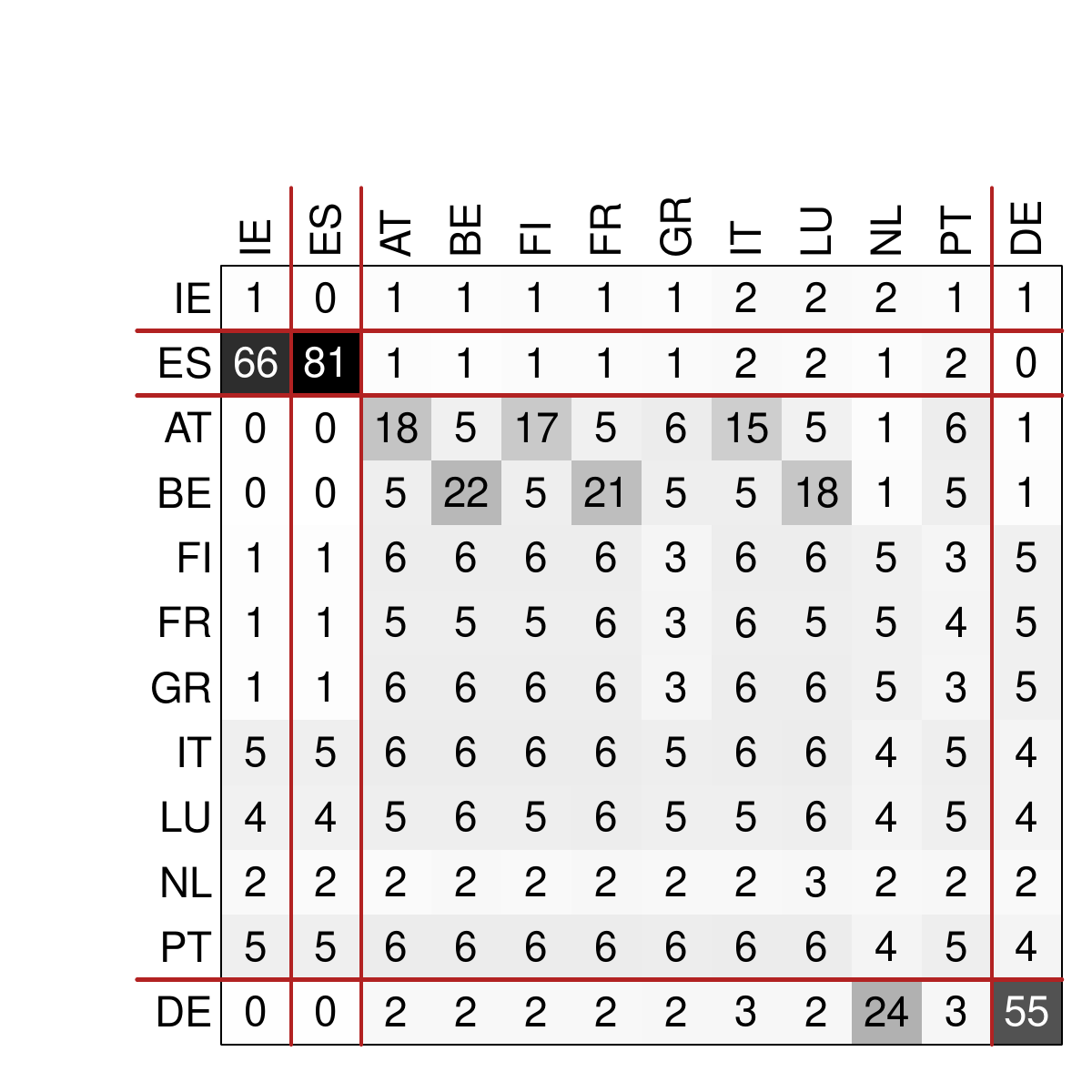}
\end{minipage}\\
\begin{minipage}{1\linewidth}~\\
\centering \textbf{Borrow long}
\end{minipage}\\
\begin{minipage}[b]{.24\linewidth}
Pre-GFC
\centering \includegraphics[scale=0.22]{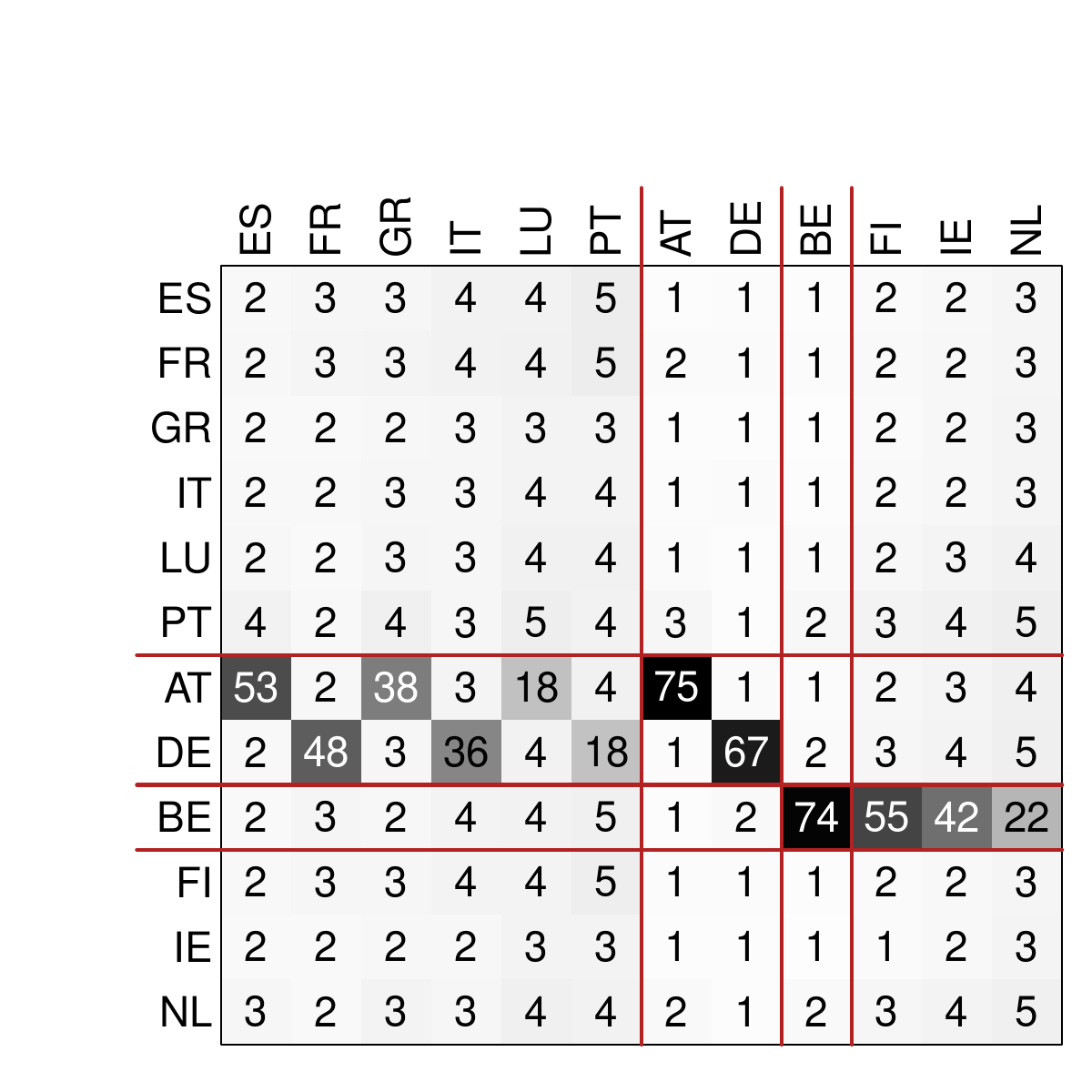}
\end{minipage}%
\begin{minipage}[b]{.24\linewidth}
GFC
\centering \includegraphics[scale=0.22]{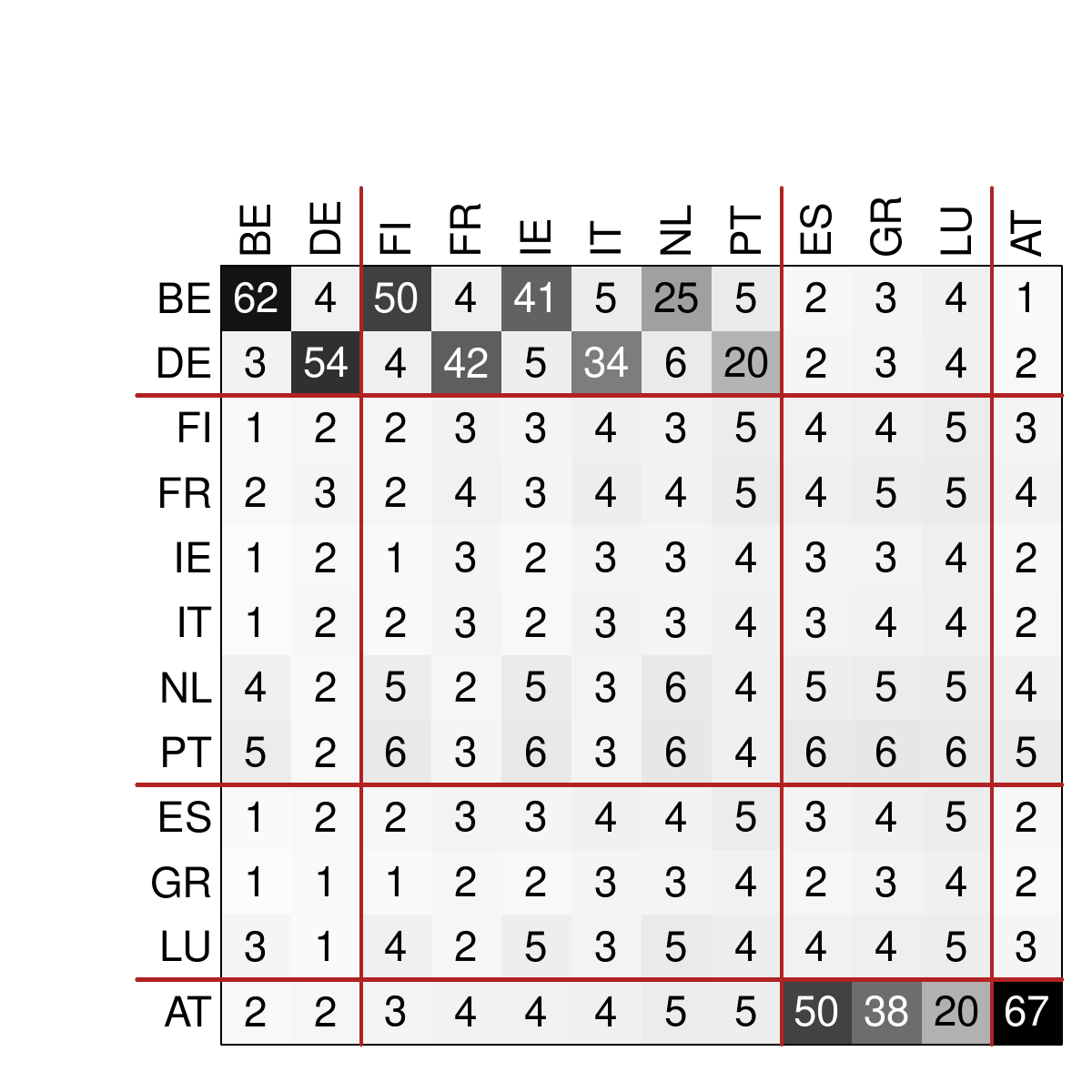}
\end{minipage}
\begin{minipage}[b]{.24\linewidth}
ESDC
\centering \includegraphics[scale=0.22]{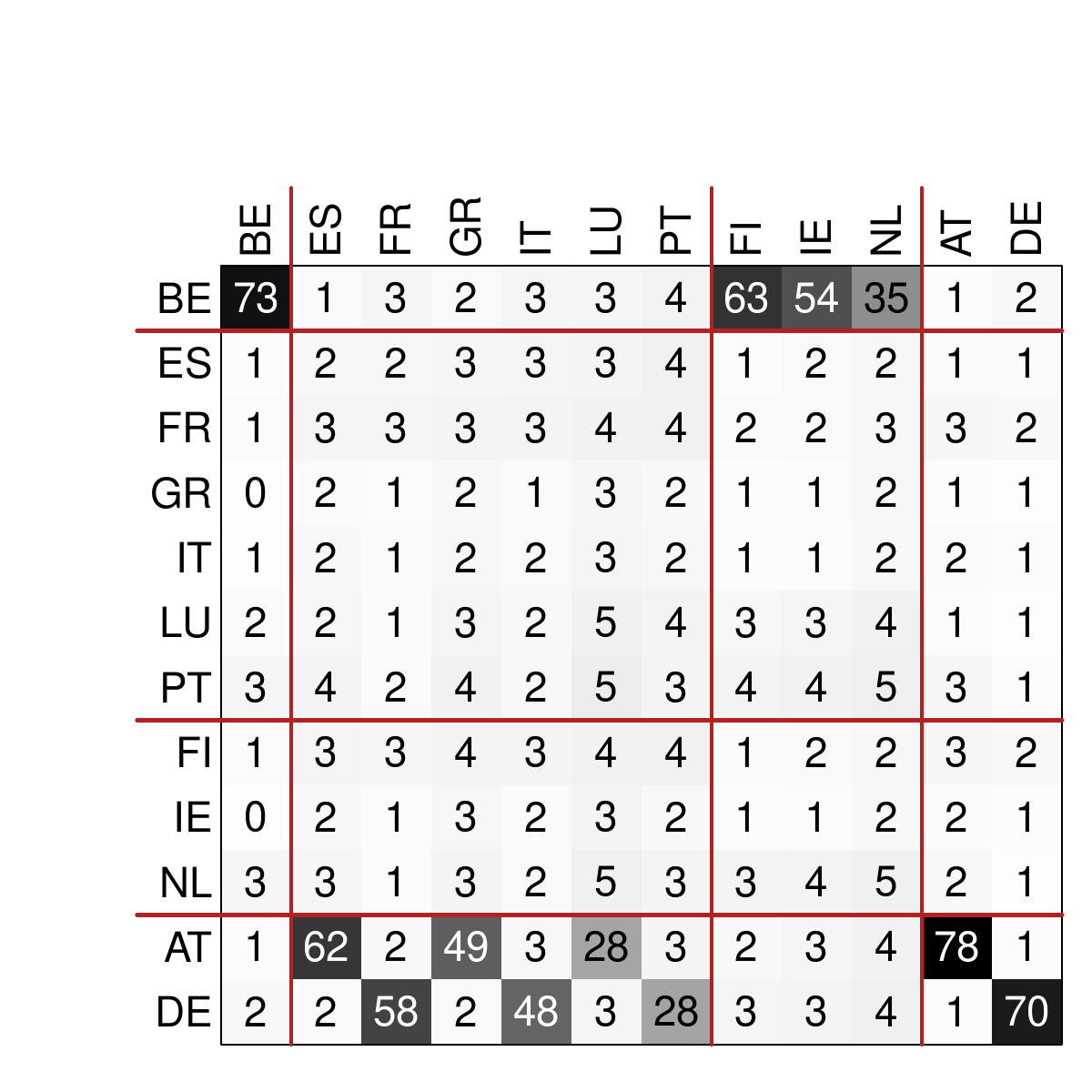}
\end{minipage}
\begin{minipage}[b]{.24\linewidth}
Post-ESDC
\centering \includegraphics[scale=0.22]{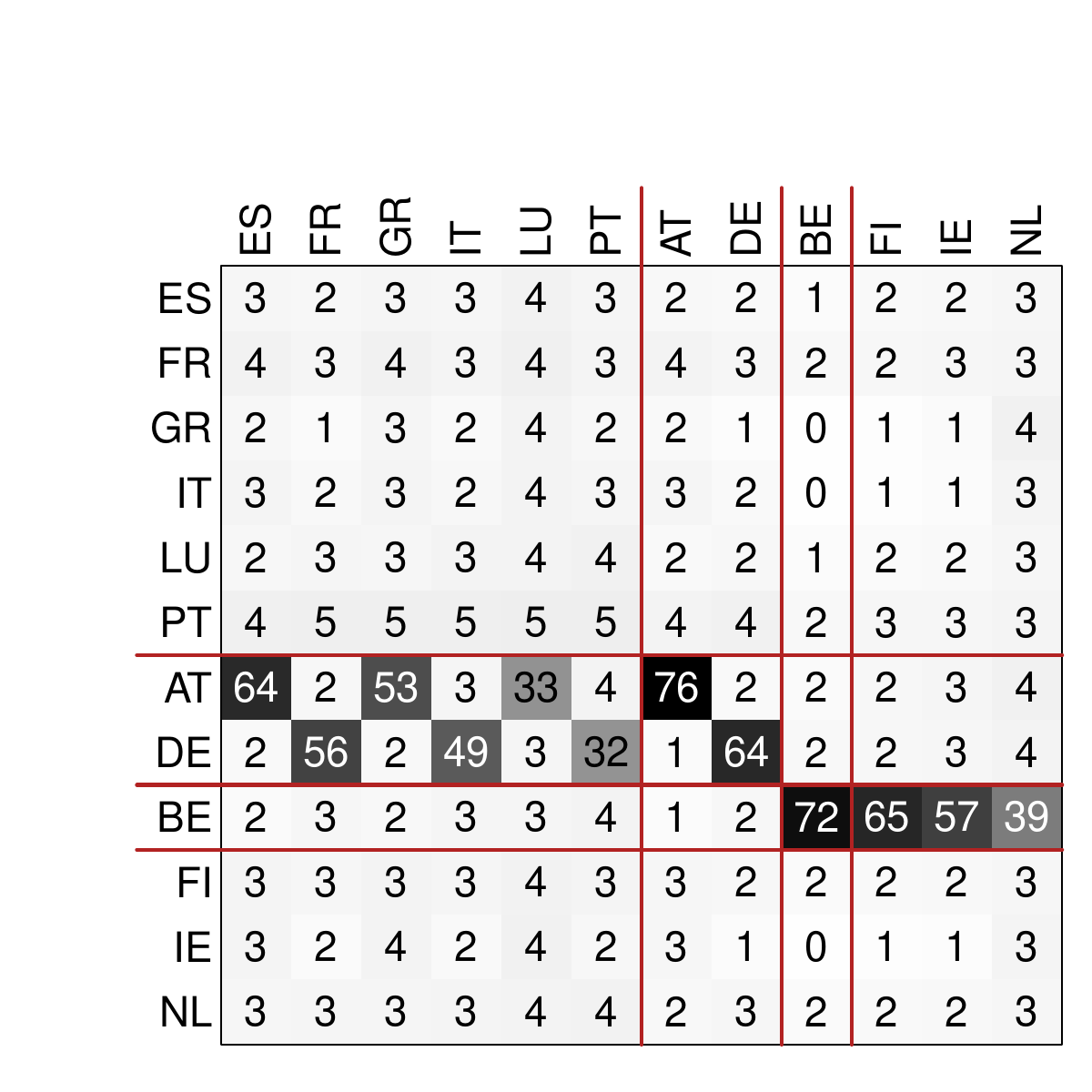}
\end{minipage}\\
\begin{minipage}{1\linewidth}~\\
\centering \textbf{Borrow short}
\end{minipage}\\
\begin{minipage}[b]{.24\linewidth}
Pre-GFC
\centering \includegraphics[scale=0.22]{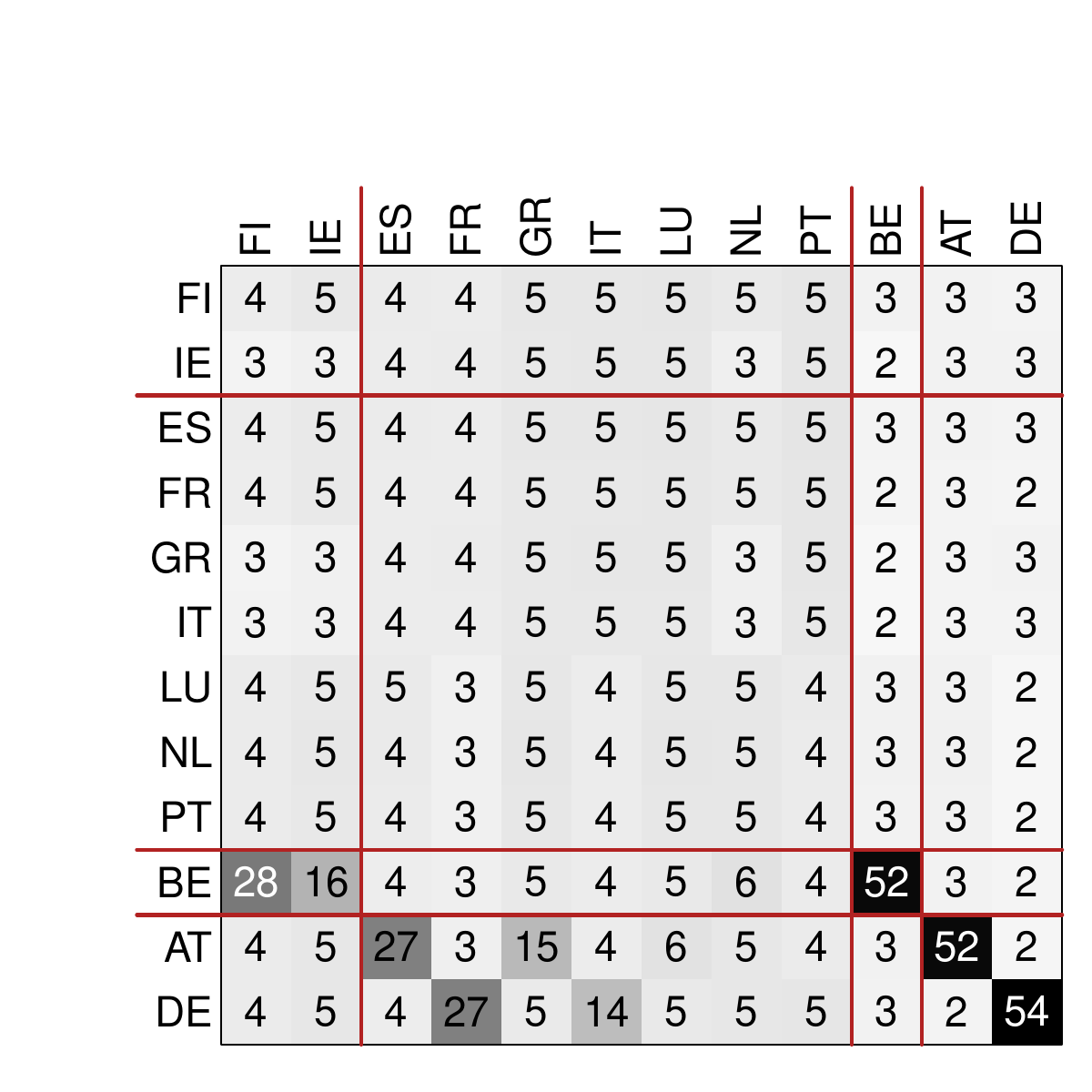}
\end{minipage}%
\begin{minipage}[b]{.24\linewidth}
GFC
\centering \includegraphics[scale=0.22]{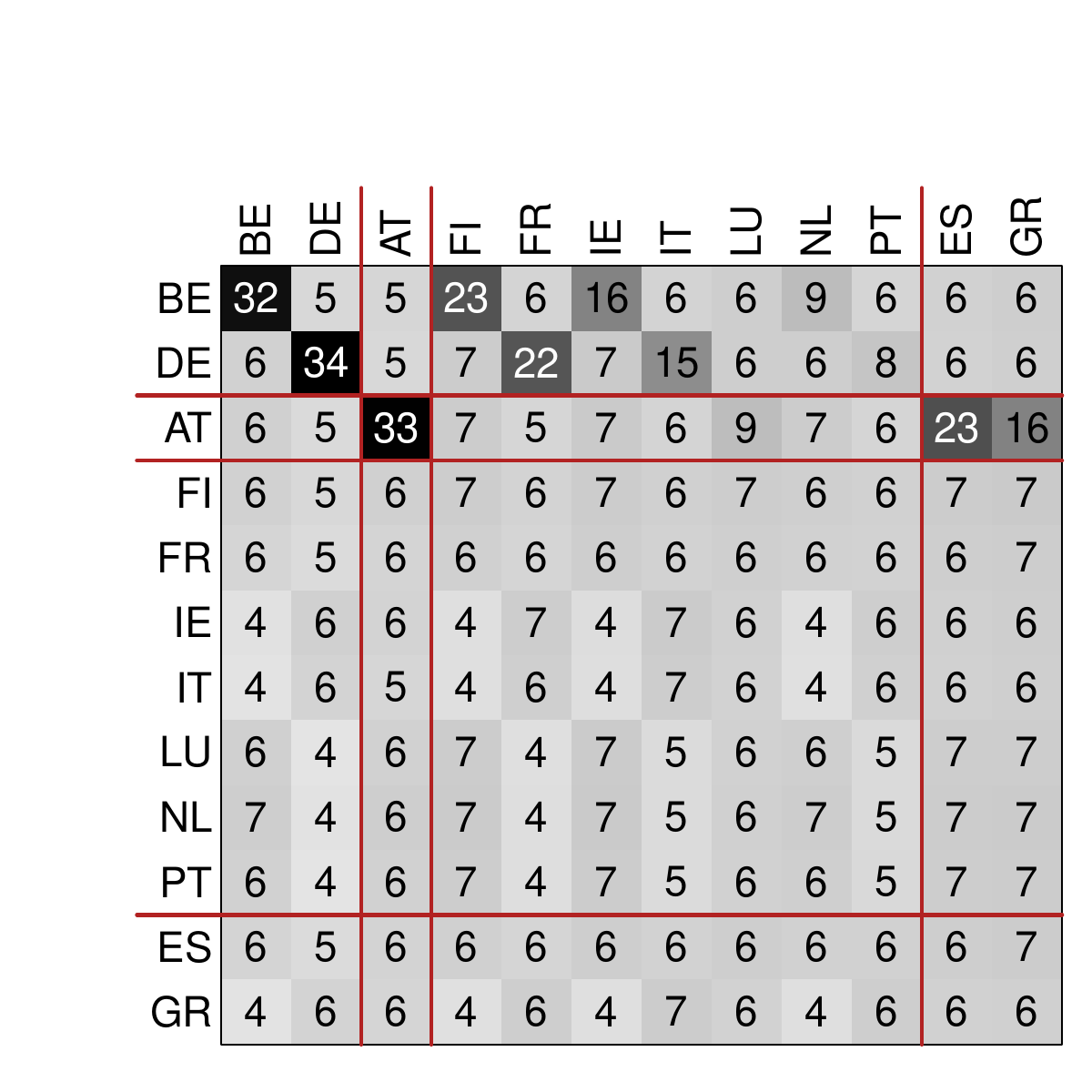}
\end{minipage}
\begin{minipage}[b]{.24\linewidth}
ESDC
\centering \includegraphics[scale=0.22]{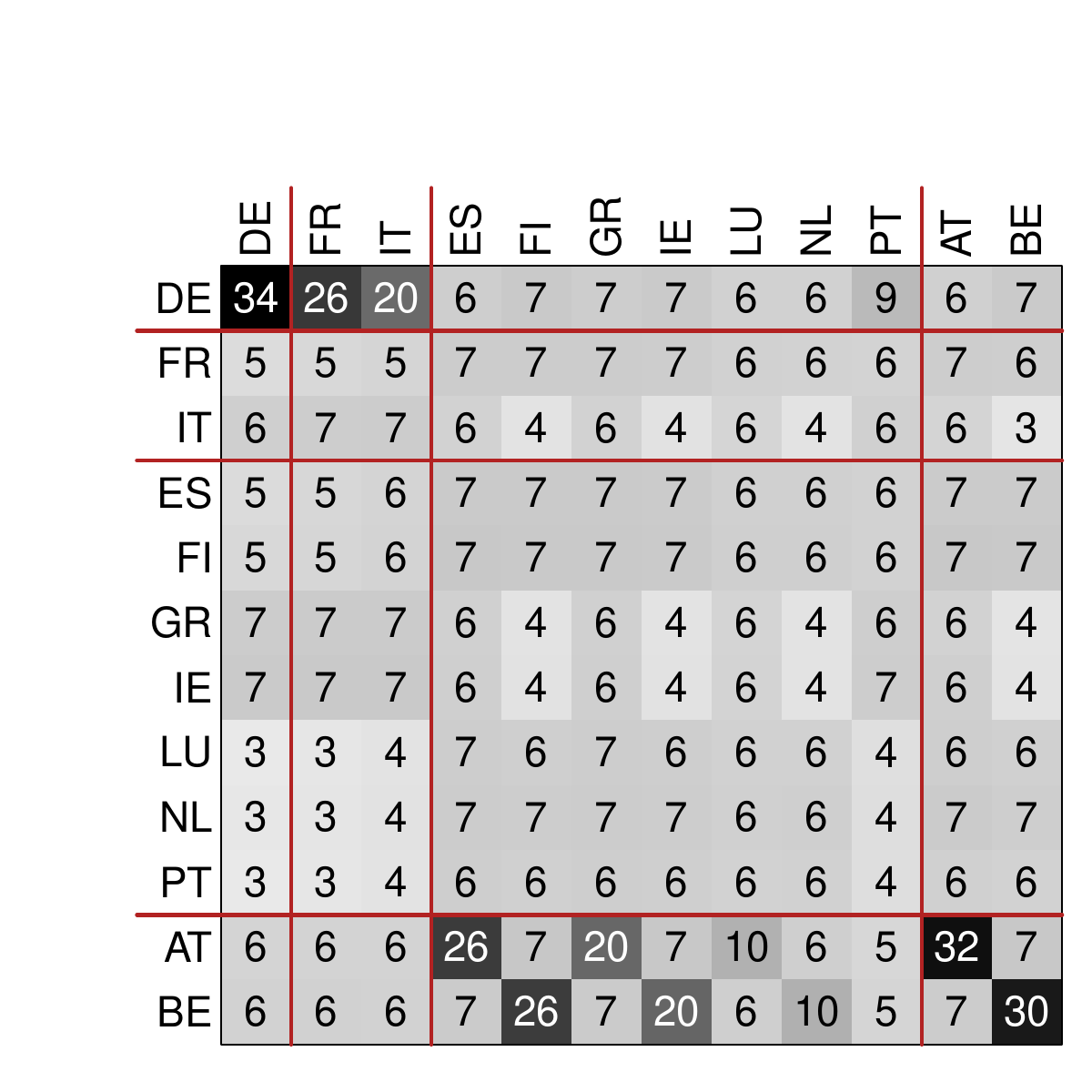}
\end{minipage}
\begin{minipage}[b]{.24\linewidth}
Post-ESDC
\centering \includegraphics[scale=0.22]{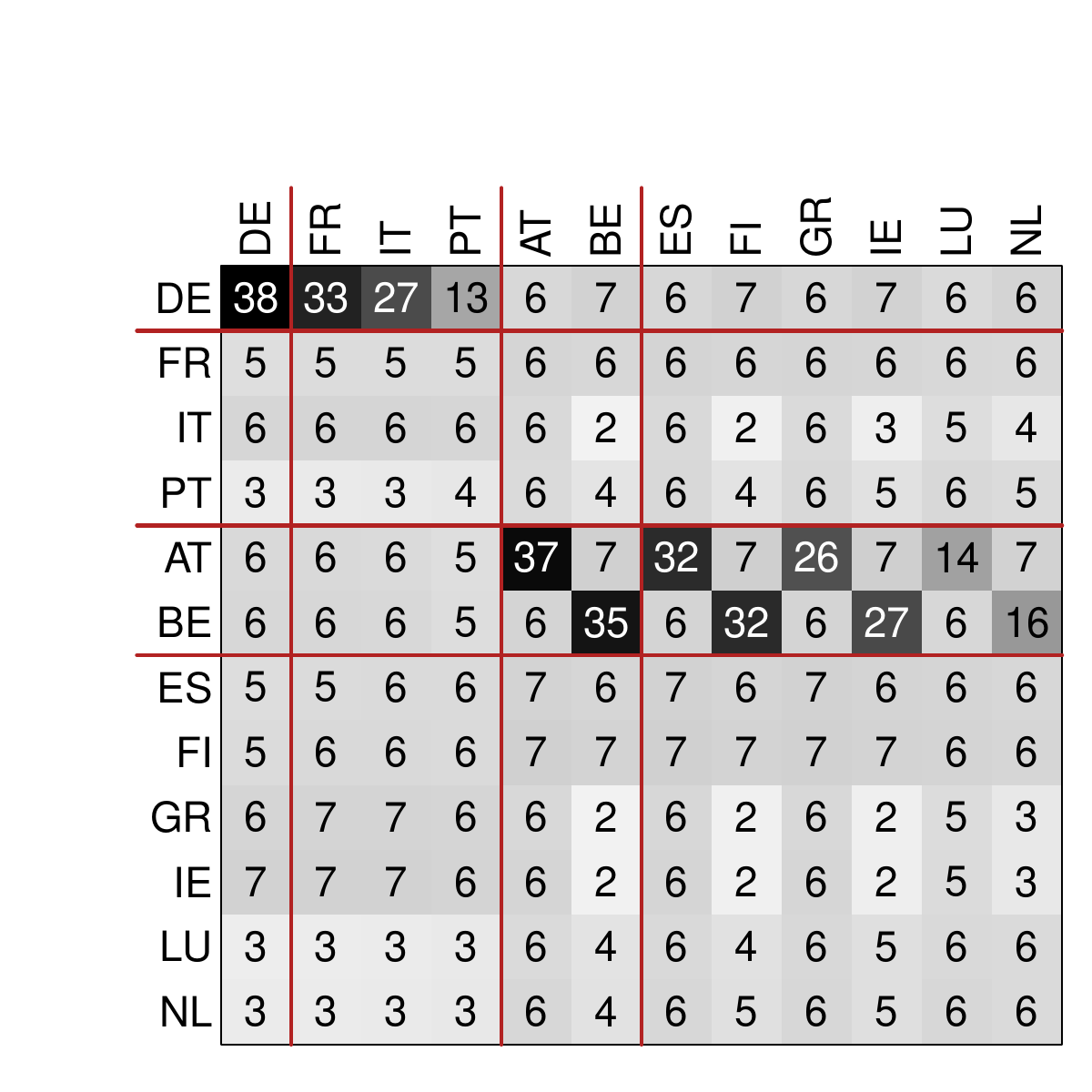}
\end{minipage}\\
\begin{minipage}{1\linewidth}~\\
\centering \textbf{TARGET2}
\end{minipage}\\
\begin{minipage}[b]{.24\linewidth}
\centering 
\end{minipage}%
\begin{minipage}[b]{.24\linewidth}
\centering 
\end{minipage}
\begin{minipage}[b]{.24\linewidth}
ESDC
\centering \includegraphics[scale=0.22]{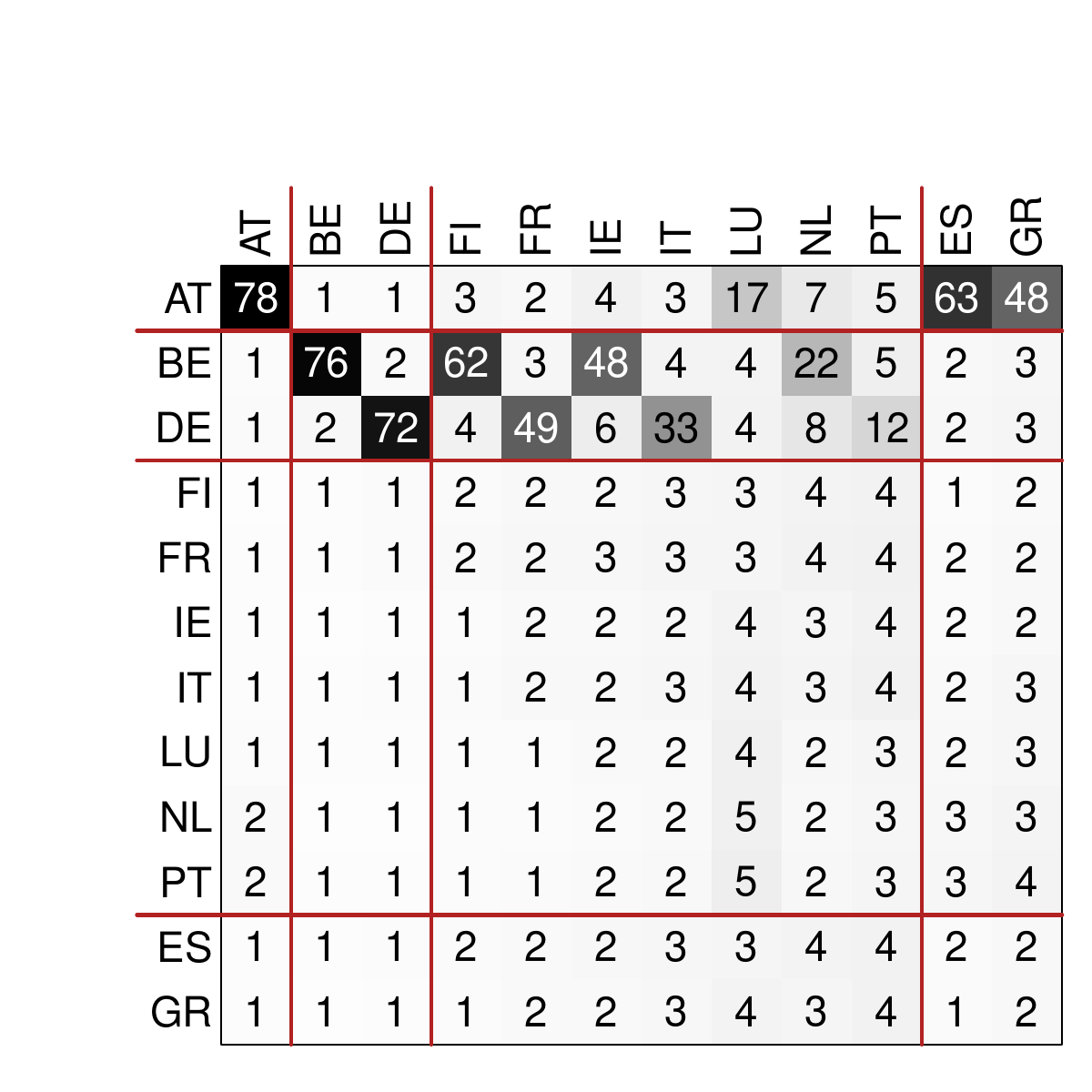}
\end{minipage}
\begin{minipage}[b]{.24\linewidth}
Post-ESDC
\centering \includegraphics[scale=0.22]{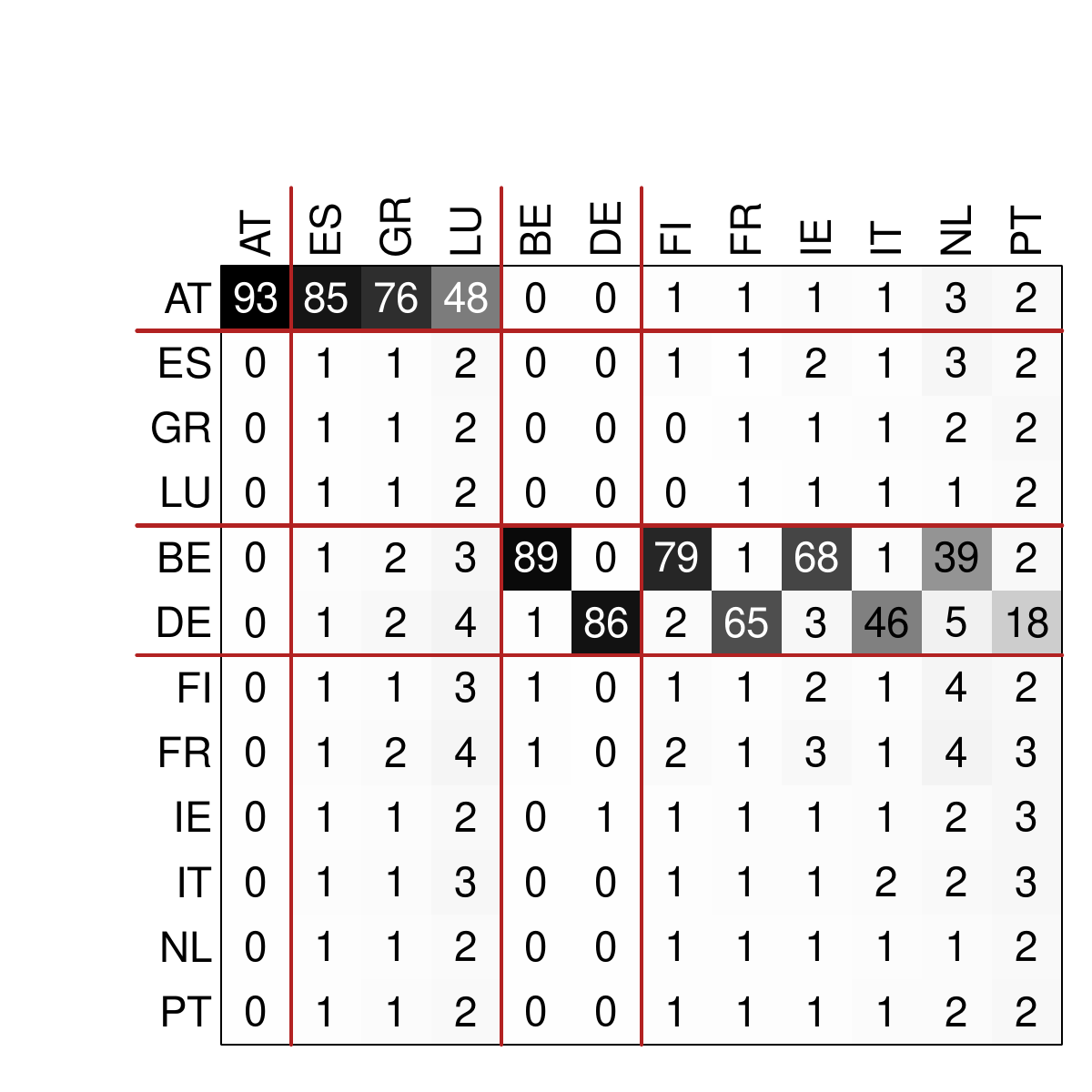}
\end{minipage}\\

\begin{minipage}{\linewidth}
\footnotesize \textit{Note:} The figure shows results of a generalized blockmodel that clusters the GFEVD at the 1-month horizon. The pre-GFC period is defined as running from 2007m01 to 2008-08, the GFC runs from 2008m09 to 2009m06, ESDC from 2010m04 to 2012m07 and the post-ESDC from 2012m08 to 2022m05.
  \end{minipage}%
\end{figure}

In \autoref{fig:blockmodel} we show the best partitions of the GFEVD averaged across different time periods. We start with looking at the government bond market. The patterns for the pre-GFC and the GFC periods are very similar in the sense that we can distinguish three different clusters of countries. First, we have Ireland taking on a very isolated role sending spillovers to Spain, but not receiving spillovers from any of the other countries. Second, we have a group of countries that receive spillovers to a larger extent than they transmit spillovers. We call these countries \textit{absorbing economies}. These constitute Spain on the one hand and Austria, Belgium and Germany on the other hand. The absorbing economies receive spillovers from the third group of countries, i.e., the main group (Finland, France, Greece, Italy, Luxembourg, Netherlands and Portugal). More specifically, Austria receives from Finland and Italy, Belgium from France and Luxembourg and Germany from the Netherlands. On top of that, in the absorbing economies bond yields are to a large extent driven by a strong domestic component (i.e., high values on the diagonal). Absorbing economies also transmit spillovers to the main group, but the extent to which they receive is by far larger. The main group of countries is well integrated with itself and the absorbing economies. This pattern remains remarkably stable over the four sample splits. The size of the main group, however, changes. During the period of the ESDC, the main group shrinks and the network gets more dispersed. This is reminiscent of the de-coupling tendencies already described in the previous section. This tendency reverses in the post-ESDC period. Here, the main group takes on the largest size over the sample splits, and outstanding positions are only taken on by Spain, Ireland and Germany.


Next, we consider the credit market. Compared to the government bond network, the credit market reveals only two type of countries: A main group (Spain, France, Greece, Italy, Luxembourg, Netherlands and Portugal) and a group of absorbing economies (Austria, Belgium and Germany). More precisely, Austria receives spillovers from Spain, Greece and Luxembourg, Germany from France and Italy, and Belgium from Finland and Ireland. On top of that, borrowing rates in the absorbing economies are also strongly determined by a domestic component. Note that our analysis is silent on the specific reason why a country turns out to be an absorber. For Austria and Belgium, it could be related to a retrenchment of cross-border bank claims from these countries by the main group \citep{Emter2019}, whereas reduction in exposure to crisis stricken countries is reflected by the block-densities within the main group. Germany could receive spillovers due to its perception as safe haven. Compared to the bond market, the main group of countries in the long-term borrowing market interacts to a lesser degree with itself and the size of the main group does not change over the periods considered. The pattern for short-term borrowing rates is very similar. The main difference to long-term borrowing rates arises due to higher within spillovers as can be seen by the darker shading of the plots, especially so after the pre-GFC period.


Last, we investigate TARGET2 balances. In line with results for the credit markets, we can identify again two types of countries, i.e., absorbing economies and a main group (Finland, France, Ireland, Italy, Luxembourg, Netherlands and Portugal). The absorbing economies are the same as before, namely Austria receiving from Greece, Spain and Luxembourg, Belgium receiving from Finland and Ireland and the Netherlands as well as Germany receiving from France and Italy. Compared to the credit market, there is much less within interaction in the main group; rather they send spillovers to the absorbing economies. This observation is even more pronounced during the post-ESDC period which reveals a very sparse network showing nearly only interactions from some members of the main group to the absorbing economies.

Summing up, for most markets the cluster analysis reveals two different set of countries, which are spillover absorbing economies who mostly receive but also transmit and a main group of countries. The absorbing economies constitute Austria, Belgium and Germany which, depending on the market under study, receive spillovers from selected members of the main group. On top of that, these countries are characterized by a strong domestic component that can explain movements in the variables considered.  This pattern tends to be stable over different sample splits.

\subsection{An augmented Taylor rule}\label{sec:tr}

In this section, we estimate Taylor rules augmented by our overall spillover indices, estimated in \autoref{sec:tot}. In specifying the Taylor rule, we follow \citet{Carvalho2021} who use a similar specification as in  \citet{Clarida2000}, estimated by simple ordinary least squares (OLS). More specifically, we regress the 3-months euribor ($i_s$) on year-on-year inflation ($Dp_{yoy}$), the IMF's output gap measure ($y_{gap}$), a commodity price index ($pcom$, in logarithms), 2- and 10-year government bond yields  ($gb_{med}$, $gb_{long}$) and months-on-months changes of $M2$ ($m2_{mom}$). The output gap measure stems from the IMF's world economic outlook data base and has been disaggregated from yearly to monthly frequency using the Chow-Lin method \citep{Chow1971} and the unemployment rate as high-frequency indicator. In a robustness exercise, we replace the 3-months euribor with the shadow rate (SSR) of \citet{Krippner2013}, which is a broader measure of the monetary policy stance that can also take on negative values.

We collect all explanatory variables in a data matrix $\mathbf{Z}$ and the spillover indices at the one-month horizon into $\mathbf{S}=\left(S_{gb}(1),S_{bl}(1),S_{bs}(1)\right)$ and estimate:

\begin{equation}
 i_s =\alpha+ \mathbf{S}\gamma+ \mathbf{Z} \beta+\varepsilon
\label{eq:eq1}
\end{equation}

with $\varepsilon$ denoting an i.i.d., white noise process. The results are provided in \autoref{tbl:TR}. Column (1) contains the plain specification put forth in \citet{Clarida2000} excluding our computed spillover indices. This benchmark model shows that euro area interest rates can be reasonably well explained by the suggested variables. More specifically and as expected, interest rates respond positively to the output gap as well as to inflation. Since during the time span considered, the ECB considerably increased its balance sheet through quantitative easing, it is not surprising that the coefficient on money growth turns out to be significantly and positively related to short-term interest rates as well. The coefficient on inflation is not precisely estimated. This finding is in line with evidence from other studies \citep[see e.g.,][]{Gorter2008} and could be related to the weak price growth over the sample period or the cross-country heterogeneity in euro area inflation rates. Different segments of the yield curve as well as the commodity price indicator also significantly impact euro area short-term interest rates.

We next include our total spillover measures $S_{gb}(1)$, $S_{bs}(1)$ and $S_{bl}(1)$ in columns (2) to (4) step by step. The measures for the bond market and  short-term lending rates are statistically significantly related to euro area interest rates. The general fit of the regression based on the adjusted $R^2$ also increases markedly. In column (5), we include all three measures simultaneously. Our results reveal that two out of the three integration measures are statistically significantly related to short-term interest rates, namely $S_{gb}(1)$ and $S_{bs}(1)$. The results imply that interest rates respond positively to high levels of financial integration. The opposite holds true when looking at the integration measure for short-term lending rates. Here, a high degree of integration is associated with lower interest rates. 
In column (6) we include results of a specification with the shadow rate instead of the 3-months euribor as the dependent variable. In line with the main specification, $S_{gb}(1)$ and $S_{bs}(1)$ turn out to be significantly related to movements in the shadow rate. On top of that, also our third measure, $S_{bl}(1)$, shows a significant relationship. 

As a robustness exercise, we estimate the same specifications accounting for interest rate smoothing by including a lag of the dependent variable ($i_{s,t-1}$) on the right hand side of \autoref{eq:eq1}. The estimation is carried again by using OLS, which follows the recommendation put forth in \citet{Carvalho2021}. The results, shown in \autoref{tbl:TRint} of the appendix, reveal the $S_{gb}(1)$ measure of integration as statistically significantly and positively related to $i_s$, while the other measures or not precisely estimated. 

Summing up, we found that measures of integration for the bond market and -- depending on the specification -- for the credit market are significantly related to euro area short-term interests rates. Taken at face value, this implies that the ECB considers integration / fragmentation as a further policy variable when setting interest rates. These results are robust to different measures of short-term interest rates for the euro area as well as Taylor rule specifications that include interest rate smoothing.

\begin{table}[H] \centering 
  \caption{Taylor rule specifications: 2007m1 to 2022m5}\label{tbl:TR}
  \footnotesize
\begin{tabular}{@{\extracolsep{5pt}}lcccccc} 
\\[-1.8ex]\hline 
\hline \\[-1.8ex] 
 & \multicolumn{6}{c}{\textit{Dependent variable:}} \\ 
\cline{2-7} 
\\[-1.8ex] & \multicolumn{5}{c}{$i_s$: euribor3m} & $i_s$: SSR \\ 
\\[-1.8ex] & (1) & (2) & (3) & (4) & (5) & (6)\\ 
\hline \\[-1.8ex] 
  $y_{gap}$ & 0.295$^{***}$ & 0.215$^{***}$ & 0.295$^{***}$ & 0.216$^{***}$ & 0.169$^{***}$ & 0.136$^{***}$ \\ 
  & (0.030) & (0.031) & (0.030) & (0.030) & (0.030) & (0.033) \\ 
  & & & & & & \\ 
 $Dp_{yoy}$ & 0.050 & 0.0004 & 0.023 & 0.086 & 0.026 & 0.132$^{**}$ \\ 
  & (0.061) & (0.057) & (0.064) & (0.055) & (0.054) & (0.059) \\ 
  & & & & & & \\ 
 $m2_{mom}$ & 1.099$^{***}$ & 0.902$^{***}$ & 1.072$^{***}$ & 0.754$^{***}$ & 0.620$^{***}$ & 0.334$^{**}$ \\ 
  & (0.140) & (0.133) & (0.140) & (0.136) & (0.131) & (0.144) \\ 
  & & & & & & \\ 
 $pcom$ & $-$0.019$^{***}$ & $-$0.011$^{**}$ & $-$0.017$^{***}$ & $-$0.024$^{***}$ & $-$0.017$^{***}$ & 0.001 \\ 
  & (0.005) & (0.005) & (0.005) & (0.005) & (0.005) & (0.005) \\ 
  & & & & & & \\ 
 $gb_{med}$ & 0.078$^{**}$ & 0.157$^{***}$ & 0.082$^{**}$ & 0.055$^{*}$ & 0.111$^{***}$ & 0.029 \\ 
  & (0.032) & (0.032) & (0.032) & (0.029) & (0.031) & (0.034) \\ 
  & & & & & & \\ 
 $gb_{long}$ & 0.516$^{***}$ & 0.578$^{***}$ & 0.539$^{***}$ & 0.551$^{***}$ & 0.611$^{***}$ & 1.143$^{***}$ \\ 
  & (0.055) & (0.051) & (0.056) & (0.049) & (0.048) & (0.053) \\ 
  & & & & & & \\ 
 $S_{gb}(1)$ &  & 0.031$^{***}$ &  &  & 0.020$^{***}$ & 0.034$^{***}$ \\ 
  &  & (0.005) &  &  & (0.006) & (0.007) \\ 
  & & & & & & \\ 
$S_{bl}(1)$ &  &  & 0.003 &  & 0.003 & 0.008$^{***}$ \\ 
  &  &  & (0.002) &  & (0.002) & (0.002) \\ 
  & & & & & & \\ 
 $S_{bs}(1)$ &  &  &  & $-$0.023$^{***}$ & $-$0.021$^{***}$ & $-$0.023$^{***}$ \\ 
  &  &  &  & (0.003) & (0.004) & (0.004) \\ 
  & & & & & & \\ 
 Constant & 0.079$^{***}$ & $-$0.089$^{**}$ & 0.063$^{**}$ & 0.193$^{***}$ & 0.063 & $-$0.119$^{**}$ \\ 
  & (0.024) & (0.037) & (0.026) & (0.028) & (0.043) & (0.047) \\ 
  & & & & & & \\ 
\hline \\[-1.8ex] 
Observations & 185 & 185 & 185 & 185 & 185 & 185 \\ 
R$^{2}$ & 0.863 & 0.884 & 0.865 & 0.890 & 0.904 & 0.932 \\ 
Adjusted R$^{2}$ & 0.858 & 0.879 & 0.859 & 0.886 & 0.899 & 0.928 \\ 
Residual Std. Error & 0.006  & 0.006  & 0.006  & 0.005  & 0.005  & 0.006  \\ 
F Statistic & 186.415$^{***}$  & 192.636$^{***}$ & 161.358$^{***}$& 204.963$^{***}$ & 183.321$^{***}$  & 265.438$^{***}$  \\ 
\hline 
\hline \\[-1.8ex] 
\textit{}  & \multicolumn{6}{r}{$^{*}$p$<$0.1; $^{**}$p$<$0.05; $^{***}$p$<$0.01} \\ 
\end{tabular} 
\end{table} 

\section{Conclusions}\label{sec:concl}
In this paper, we investigate the degree of integration at the euro area sovereign bond yield market, the market for lending rates with different maturities as well as within-euro area money flows. To that end we calculate Diebold-Yilmaz (DY) spillover indices based on a panel vector-autoregressive model with factor stochastic volatility. A high degree of spillovers between countries indicates a high degree of integration that can either arise through a similar response to joint external shocks or direct cross-country links.

Our main results are as follows: First, we observe that integration in the markets for government bond yields and long-term lending rates started to decrease with the outbreak of the GFC. A further, sharp drop can be observed during the ESDC, which constitutes the trough in integration for our sample period. This result is in line with the recent literature on government bond fragmentation \citep{Chatziantoniou2021, Hoffmann2020} and is robust to changes in the model specification. In both markets, the level of integration has fully recovered since then, which could be attributed to the accommodative policy stance of the ECB, especially since 2015. Financial integration of short-term borrowing costs and TARGET2 balances is comparably more stable and less prone to crisis events.

Second, we examine cross-country differences of these overall trends by examining similarities of the responsiveness of integration over time. In general, country-specific spillover indices behave very similar in the credit markets. By contrast and for the sovereign bond market, we reveal a severe decoupling of German -- and to a lesser extent Dutch -- government bond yields during the ESDC. German bonds started to re-connect in 2015, the period in which the ECB started its asset purchase programs. A second way of clustering countries is by identifying similar roles they play in the financial market network. This analysis reveals two sets of countries, namely a main body of countries that receives and transmits spillovers and a second, smaller group of spillover absorbing economies comprised of Austria, Belgium and Germany. These countries tend to receive more spillovers than they transmit. This general classification tends to apply to all financial market segments as well as to different time periods considered. 

Last, we model euro area short-term interest rates by a Taylor rule, augmented with the estimated DY integration indices. Our results show that the integration index for the bond market is positively related to euro area short-term interest rates. This implies that the ECB is more likely to increase interest rates, when integration in the bond market is high and less so otherwise. This result is robust to different specification choices of the Taylor rule.

\clearpage
\small{\setstretch{0.85}
\addcontentsline{toc}{section}{References}
\bibliographystyle{cit_econometrica.bst}
\bibliography{lit}}

\newpage

\begin{appendices}
\begin{center}
\LARGE\textbf{Appendices}
\end{center}

\setcounter{equation}{0}
\setcounter{table}{0}
\setcounter{figure}{0}
\renewcommand\theequation{A.\arabic{equation}}
\renewcommand\thetable{A.\arabic{table}}
\renewcommand\thefigure{A.\arabic{figure}}

\section{Additional results}

\begin{table}[htbp!]
\caption{Descriptive statistics}\label{tbl:desc}
\centering
\footnotesize
\begin{tabular*}{\textwidth}{l @{\extracolsep{\fill}} ccccccc}
  \hline
\textbf{Core}  & AT & BE & DE & FI & FR & LU & NL \\ 
  \midrule
  \multicolumn{7}{l}{\textit{Government bond yields:}}\\
Mean & 3.3 & 3.5 & 3 & 3.4 & 3.3 & 3.1 & 3.2 \\ 
  Min & -0.4 & -0.4 & -0.7 & -0.4 & -0.3 & -0.6 & -0.6 \\ 
  Max & 7.7 & 8.6 & 7.6 & 10.6 & 8.2 & 8 & 7.7 \\ 
  SD & 2.2 & 2.2 & 2.3 & 2.5 & 2.2 & 2.3 & 2.2 \\ 
    \multicolumn{7}{l}{\textit{Borrow long:}}\\
  Mean & 3 & 3.1 & 3.1 & 3.1 & 3.1 & 2.7 & 3.7 \\ 
  Min & 1.3 & 1.3 & 1.2 & 1.3 & 1.1 & 1 & 1.6 \\ 
  Max & 5 & 5.6 & 5.6 & 5.9 & 5.4 & 5.9 & 5.5 \\ 
  SD & 1 & 1.2 & 1.4 & 1.2 & 1.3 & 1.2 & 1.1 \\ 
    \multicolumn{7}{l}{\textit{Borrow short:}}\\
  Mean & 2.6 & 2.7 & 3.1 & 2.4 & 2.5 & 2.4 & 2.5 \\ 
  Min & 1.1 & 1.4 & 1.4 & 0.8 & 0.8 & 1.1 & 0.6 \\ 
  Max & 5.9 & 5.8 & 6.1 & 5.7 & 5.8 & 5.5 & 5.8 \\ 
  SD & 1.3 & 1.2 & 1.3 & 1.2 & 1.2 & 1.2 & 1.4 \\ 
    \multicolumn{7}{l}{\textit{TARGET2:}}\\
  Mean & -36150 & -31023 & 640090 & 41069 & -31739 & 147433 & 56286 \\ 
  Min & -57437 & -104233 & 70196 & -3280 & -119079 & 6151 & -23477 \\ 
  Max & -18960 & 1538 & 1260673 & 96524 & 61812 & 350290 & 168883 \\ 
  SD & 7334 & 22743 & 309992 & 28268 & 36696 & 83258 & 44016 \\ 
  \midrule
\textbf{Periphery}   &GR&    IE&    IT    &PT &   ES& &\\
  \midrule
    \multicolumn{7}{l}{\textit{Government bond yields:}}\\
  Mean & 7.8 & 4.1 & 4.5 & 4.9 & 4.2 & - & - \\ 
  Min & 0.6 & -0.3 & 0.6 & 0 & 0 & - & - \\ 
  Max & 29.2 & 12.4 & 13.4 & 13.8 & 12.3 & - & - \\ 
  SD & 5.6 & 2.6 & 2.7 & 3 & 2.6 & - & - \\ 
 \multicolumn{7}{l}{\textit{Borrow long:}}\\
  Mean & 4.4 & 4 & 3.5 & 4.3 & 3.2 & - & - \\ 
  Min & 1.7 & 2.6 & 1.2 & 1.2 & 1.2 & - & - \\ 
  Max & 7 & 6.1 & 6.1 & 7.6 & 6.1 & - & - \\ 
  SD & 0.8 & 0.8 & 1.4 & 1.9 & 1.2 & - & - \\ 
   \multicolumn{7}{l}{\textit{Borrow short:}}\\
  Mean & 5 & 3.6 & 3.3 & 4.1 & 3 & - & - \\ 
  Min & 2.7 & 2.3 & 1.2 & 1.6 & 1.2 & - & - \\ 
  Max & 7 & 6.4 & 6.4 & 6.8 & 6 & - & - \\ 
  SD & 1 & 0.9 & 1.4 & 1.5 & 1.2 & - & - \\ 
   \multicolumn{7}{l}{\textit{TARGET2:}}\\
  Mean & -67173 & -20788 & -260204 & -62627 & -273754 & - & - \\ 
  Min & -110805 & -145185 & -596913 & -84418 & -536135 & - & - \\ 
  Max & -14497 & 88627 & 82169 & -12606 & -11894 & - & - \\ 
  SD & 28947 & 58266 & 205862 & 20009 & 158386 & - & - \\ 
   \bottomrule
\end{tabular*}
\begin{minipage}{16cm}~\\
\footnotesize \emph{Note:} The table shows descriptive statistics for sovereign government bond yields, short- and long-term lending rates and TARGET2 balances. Data for interest rates are in \%, TARGET2 balances are in millions of euro.
\end{minipage}%
\end{table}

\clearpage

\begin{landscape}
\begin{figure}[htbp!]
\caption{Spillover index for each country and the 1-month forecast horizon. \label{fig:DYindex_detail_1}}

\begin{minipage}{\textwidth}
\centering
\small \textit{\textbf{Government bond yields }}
\end{minipage}

\begin{minipage}{0.21\textwidth}
\centering
\scriptsize \textit{AT}
\end{minipage}
\begin{minipage}{0.21\textwidth}
\centering
\scriptsize \textit{BE}
\end{minipage}
\begin{minipage}{0.21\textwidth}
\centering
\scriptsize \textit{DE}
\end{minipage}
\begin{minipage}{0.21\textwidth}
\centering
\scriptsize \textit{FI}
\end{minipage}
\begin{minipage}{0.21\textwidth}
\centering
\scriptsize \textit{FR}
\end{minipage}
\begin{minipage}{0.21\textwidth}
\centering
\scriptsize \textit{LU}
\end{minipage}
\begin{minipage}{0.21\textwidth}
\centering
\scriptsize \textit{NL}
\end{minipage}

\begin{minipage}{0.21\textwidth}
\centering
\includegraphics[scale=.24]{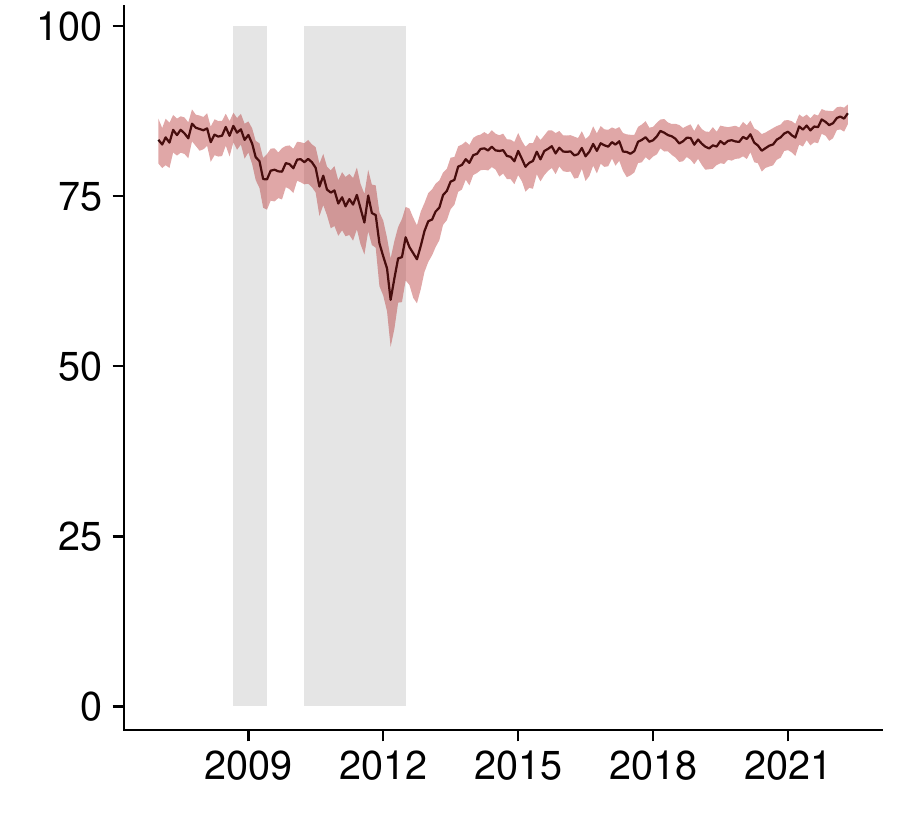}
\end{minipage}
\begin{minipage}{0.21\textwidth}
\centering
\includegraphics[scale=.24]{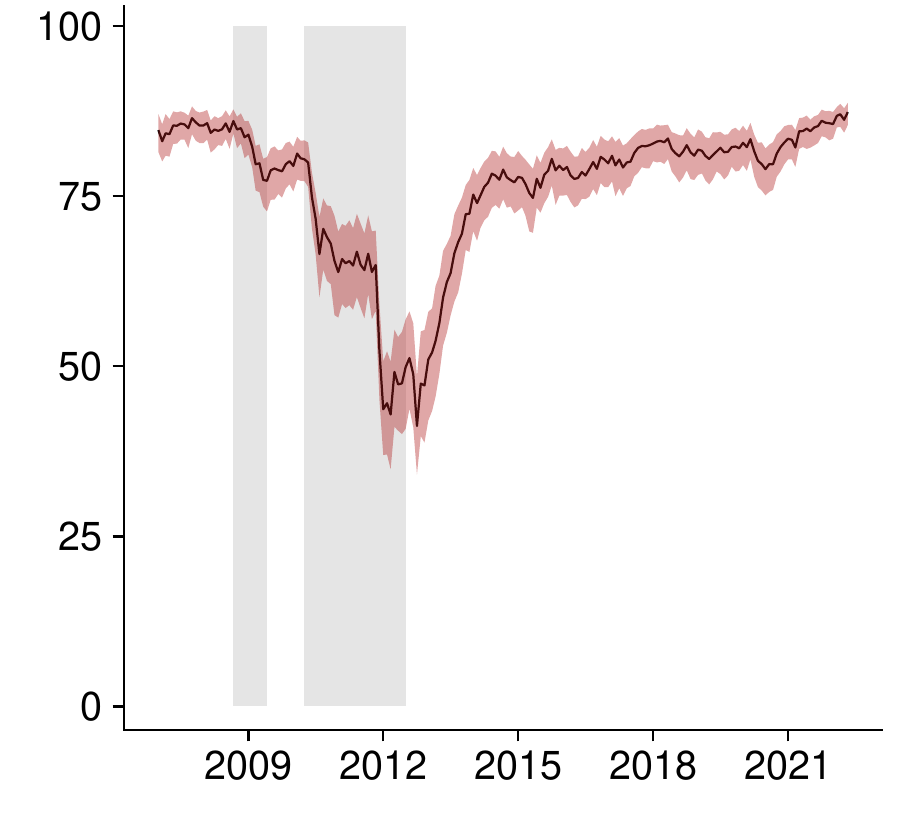}
\end{minipage}
\begin{minipage}{0.21\textwidth}
\centering
\includegraphics[scale=.24]{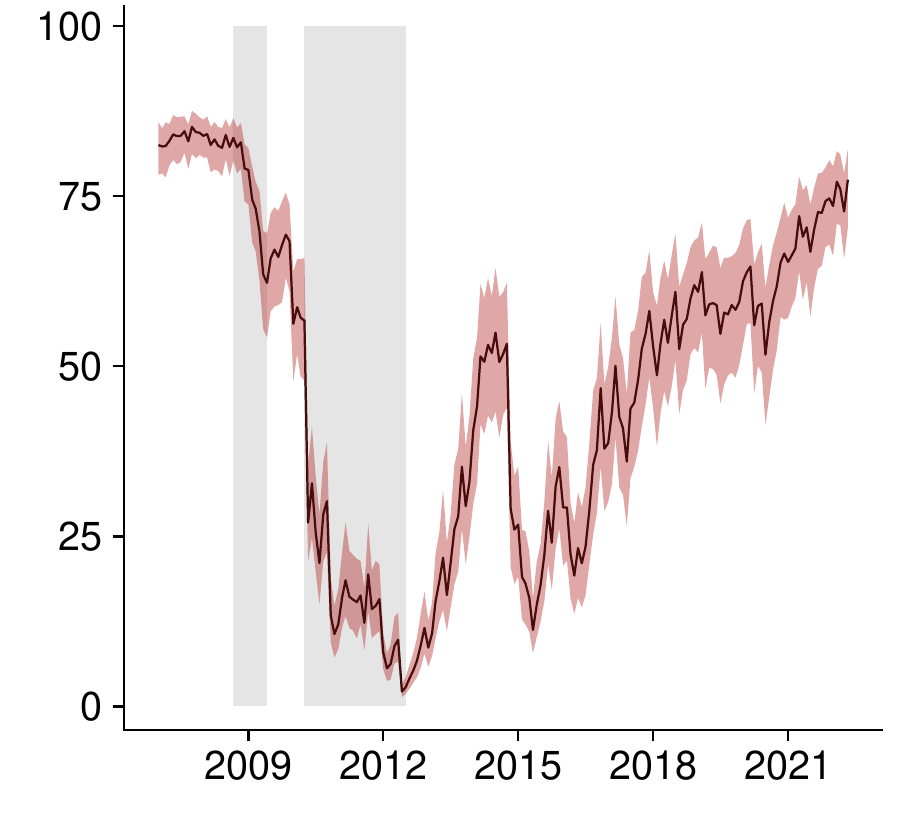}
\end{minipage}
\begin{minipage}{0.21\textwidth}
\centering
\includegraphics[scale=.24]{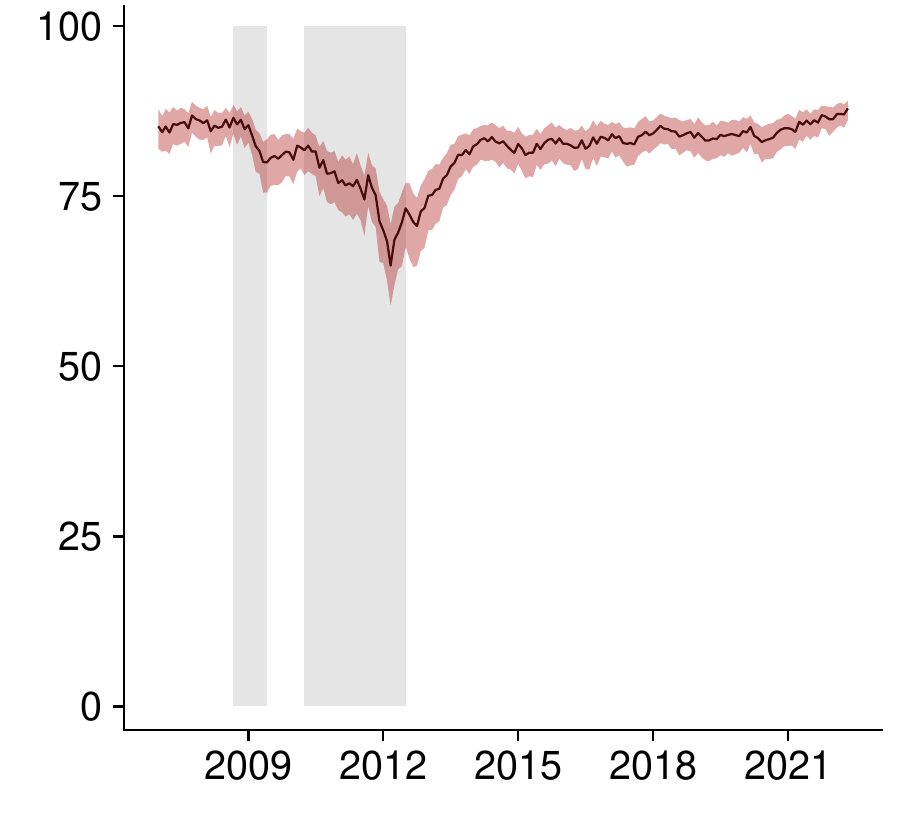}
\end{minipage}
\begin{minipage}{0.21\textwidth}
\centering
\includegraphics[scale=.24]{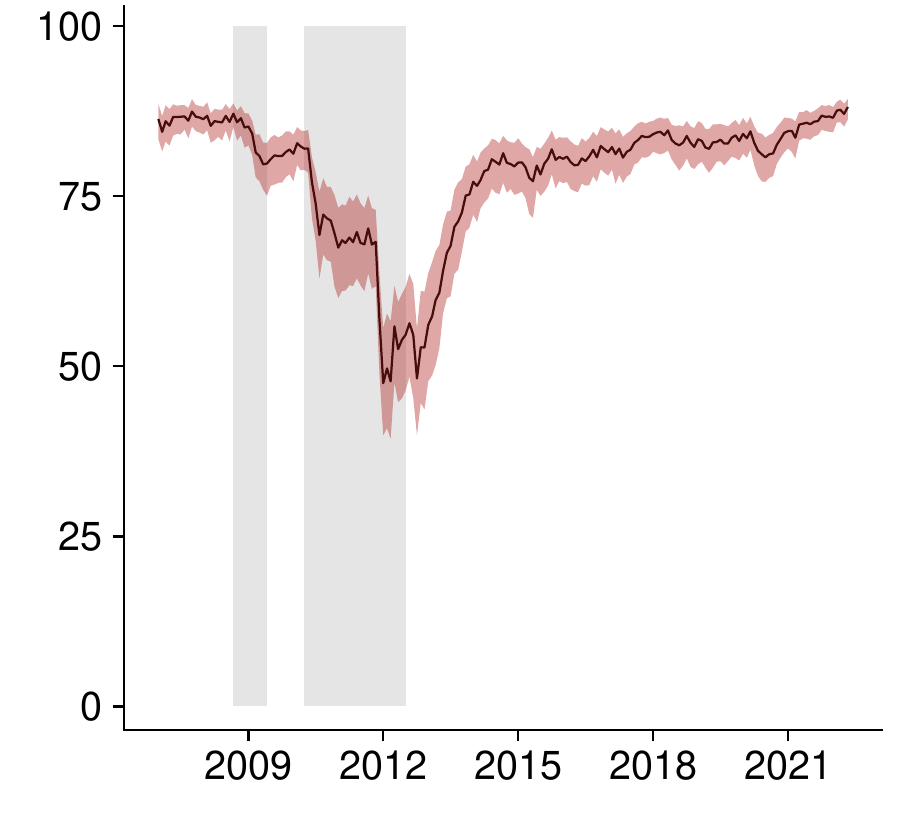}
\end{minipage}
\begin{minipage}{0.21\textwidth}
\centering
\includegraphics[scale=.24]{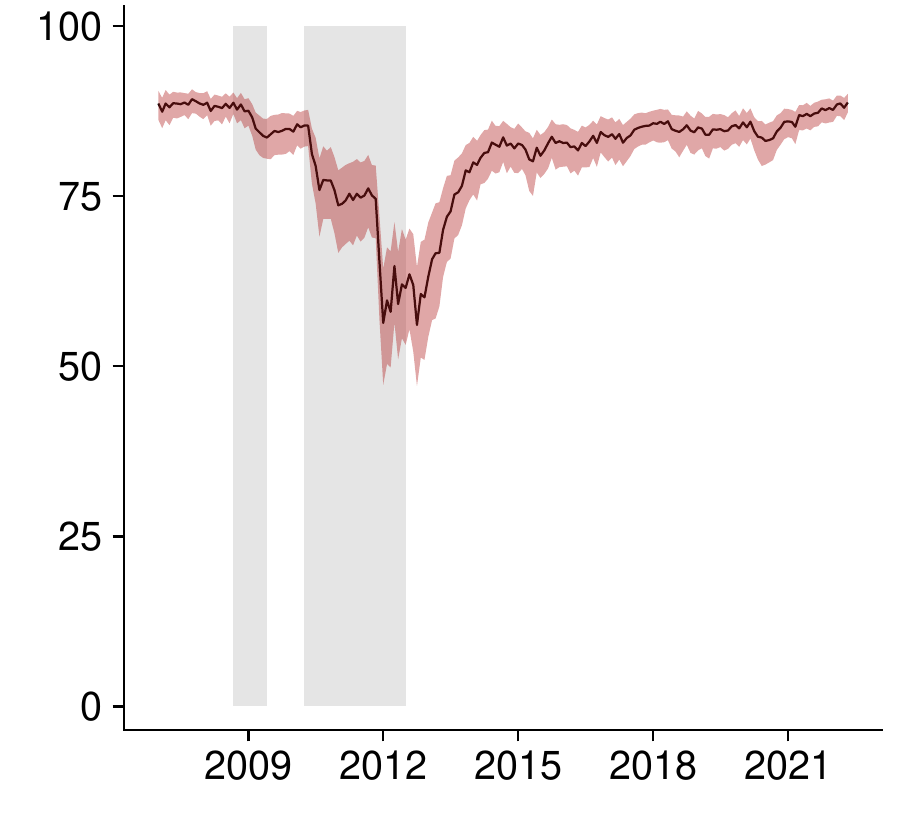}
\end{minipage}
\begin{minipage}{0.21\textwidth}
\centering
\includegraphics[scale=.24]{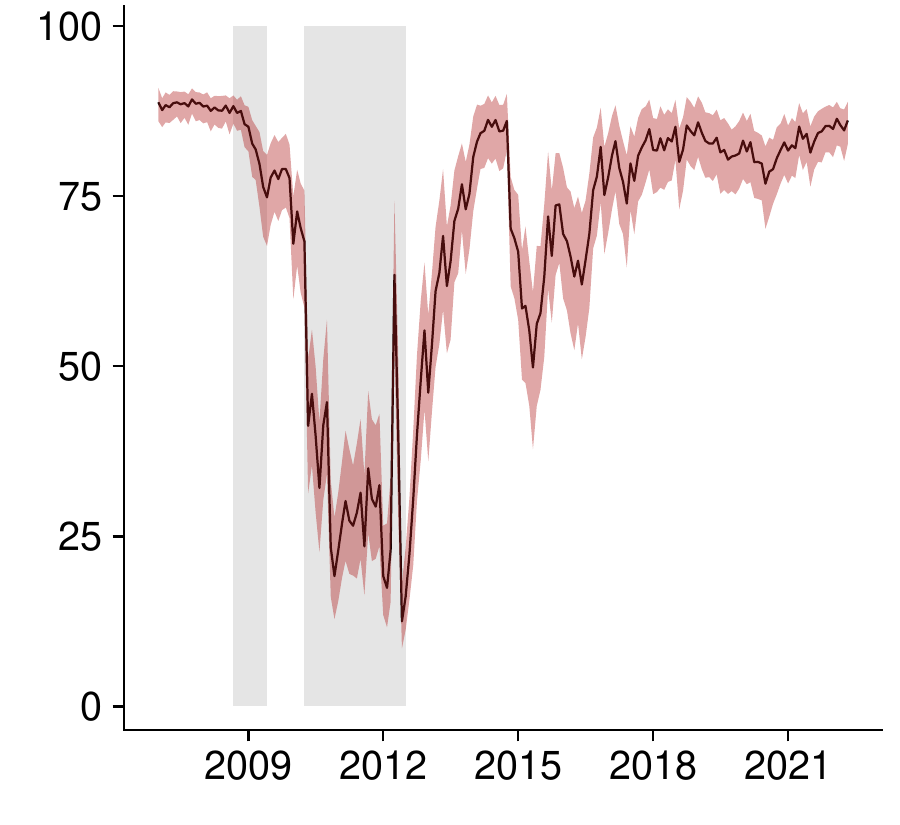}
\end{minipage}

\begin{minipage}{0.24\textwidth}
\centering
\scriptsize \textit{GR}
\end{minipage}
\begin{minipage}{0.24\textwidth}
\centering
\scriptsize \textit{IE}
\end{minipage}
\begin{minipage}{0.24\textwidth}
\centering
\scriptsize \textit{IT}
\end{minipage}
\begin{minipage}{0.24\textwidth}
\centering
\scriptsize \textit{PT}
\end{minipage}
\begin{minipage}{0.24\textwidth}
\centering
\scriptsize \textit{ES}
\end{minipage}

\begin{minipage}{0.24\textwidth}
\centering
\includegraphics[scale=.24]{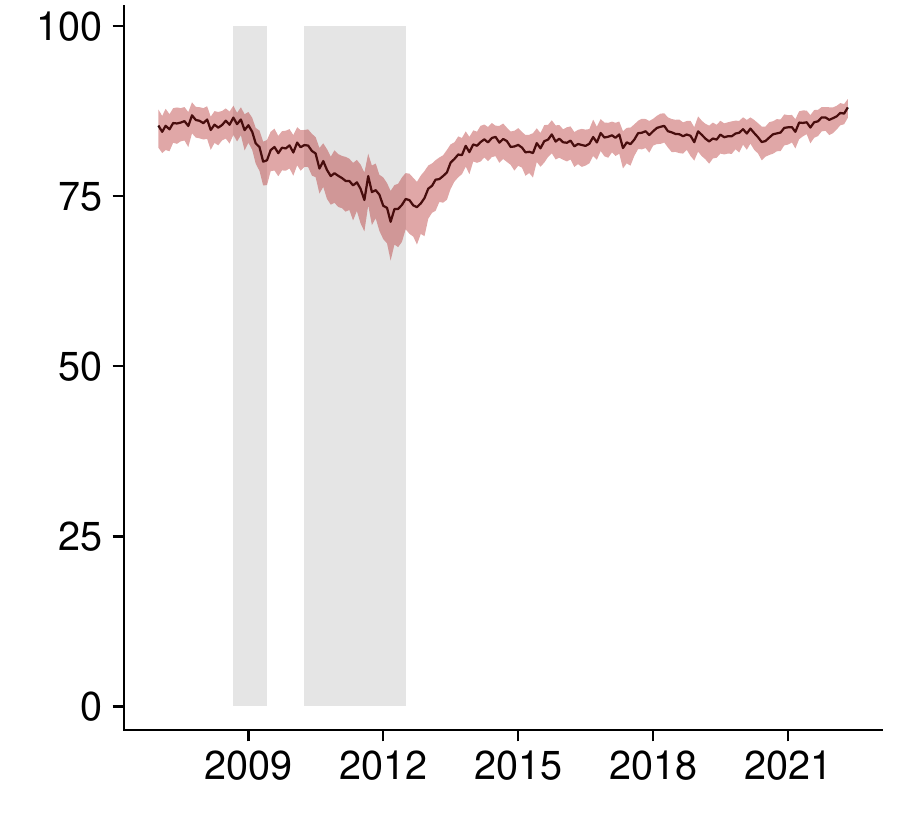}
\end{minipage}
\begin{minipage}{0.24\textwidth}
\centering
\includegraphics[scale=.24]{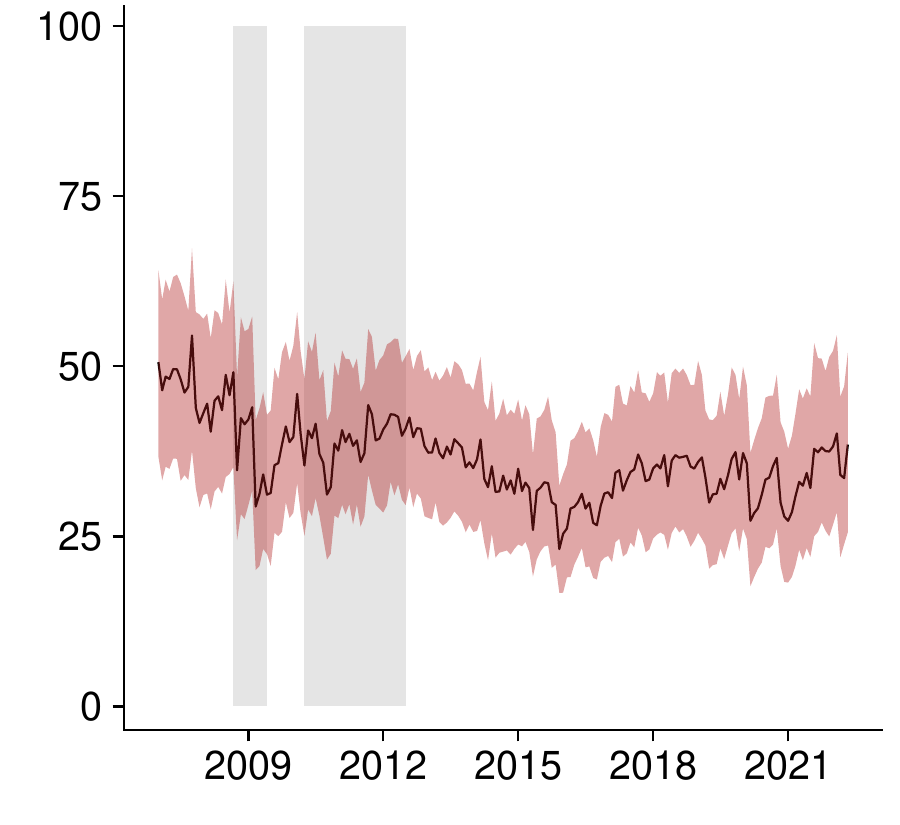}
\end{minipage}
\begin{minipage}{0.24\textwidth}
\centering
\includegraphics[scale=.24]{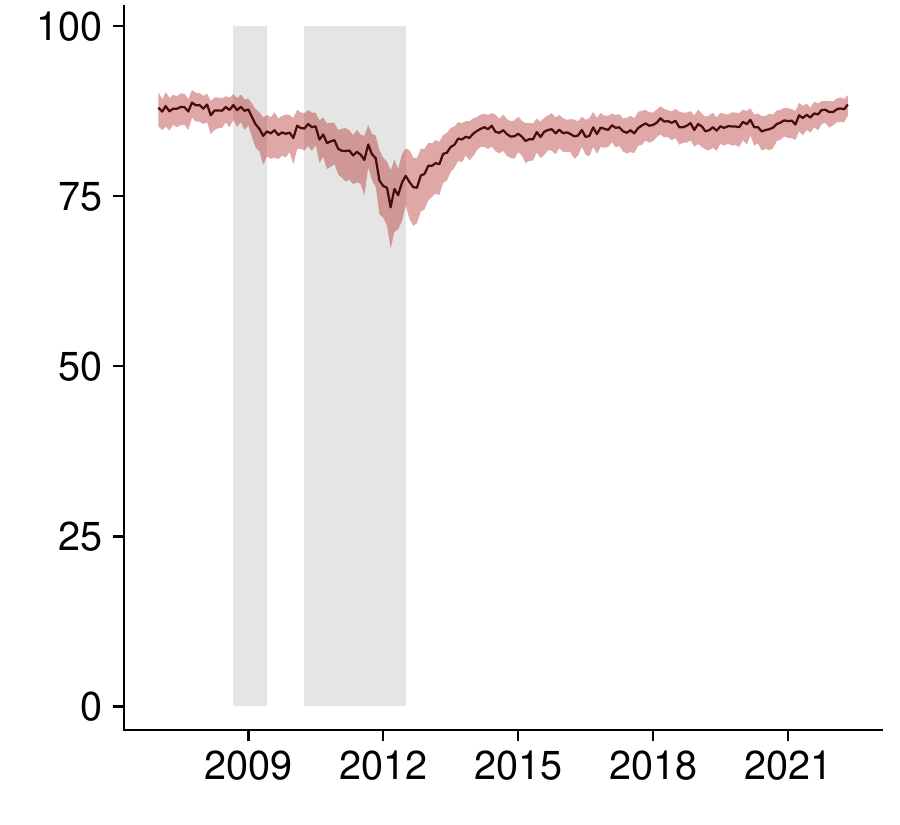}
\end{minipage}
\begin{minipage}{0.24\textwidth}
\centering
\includegraphics[scale=.24]{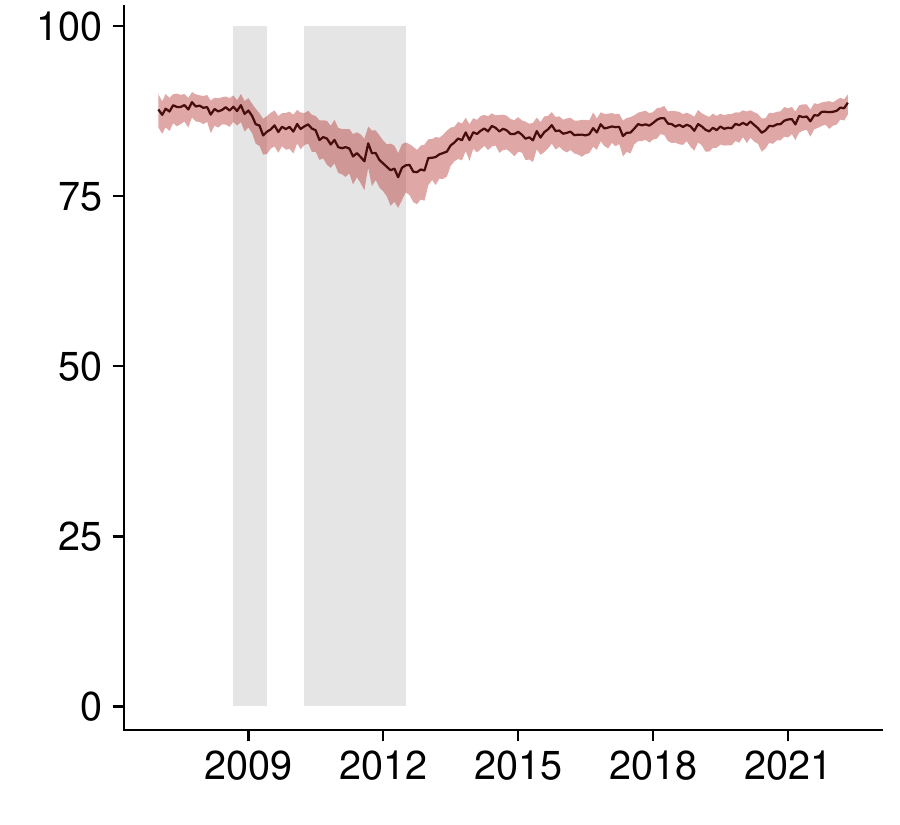}
\end{minipage}
\begin{minipage}{0.24\textwidth}
\centering
\includegraphics[scale=.24]{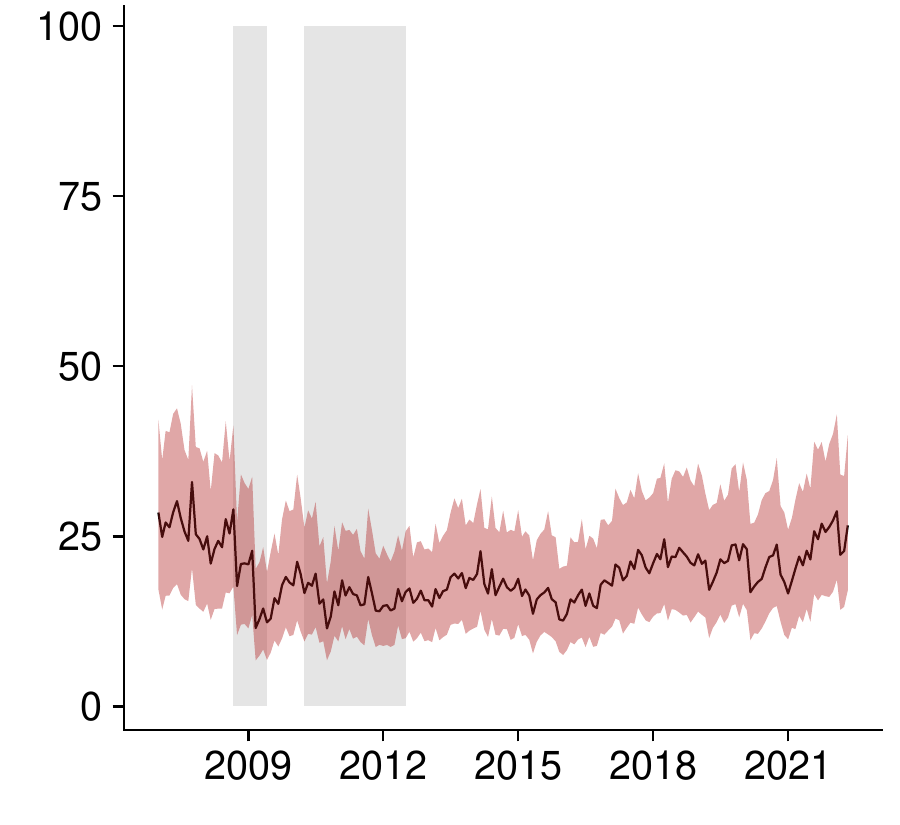}
\end{minipage}

\begin{minipage}{\textwidth}
\centering
\small \textit{\textbf{Borrow long}}
\end{minipage}

\begin{minipage}{0.21\textwidth}
\centering
\scriptsize \textit{AT}
\end{minipage}
\begin{minipage}{0.21\textwidth}
\centering
\scriptsize \textit{BE}
\end{minipage}
\begin{minipage}{0.21\textwidth}
\centering
\scriptsize \textit{DE}
\end{minipage}
\begin{minipage}{0.21\textwidth}
\centering
\scriptsize \textit{FI}
\end{minipage}
\begin{minipage}{0.21\textwidth}
\centering
\scriptsize \textit{FR}
\end{minipage}
\begin{minipage}{0.21\textwidth}
\centering
\scriptsize \textit{LU}
\end{minipage}
\begin{minipage}{0.21\textwidth}
\centering
\scriptsize \textit{NL}
\end{minipage}

\begin{minipage}{0.21\textwidth}
\centering
\includegraphics[scale=.24]{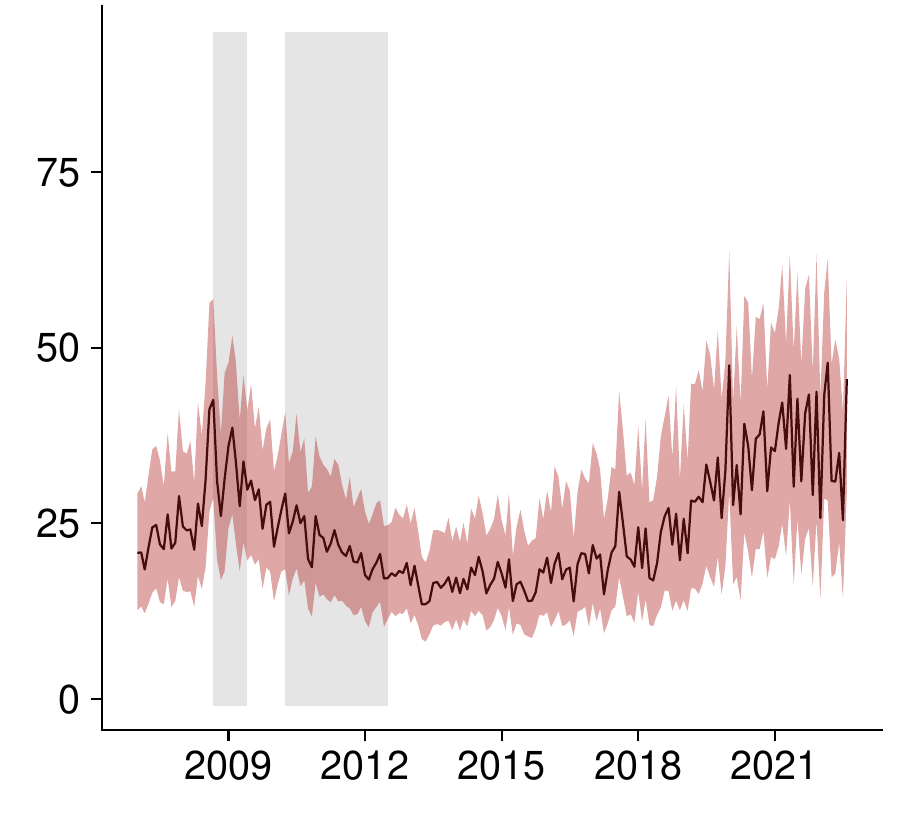}
\end{minipage}
\begin{minipage}{0.21\textwidth}
\centering
\includegraphics[scale=.24]{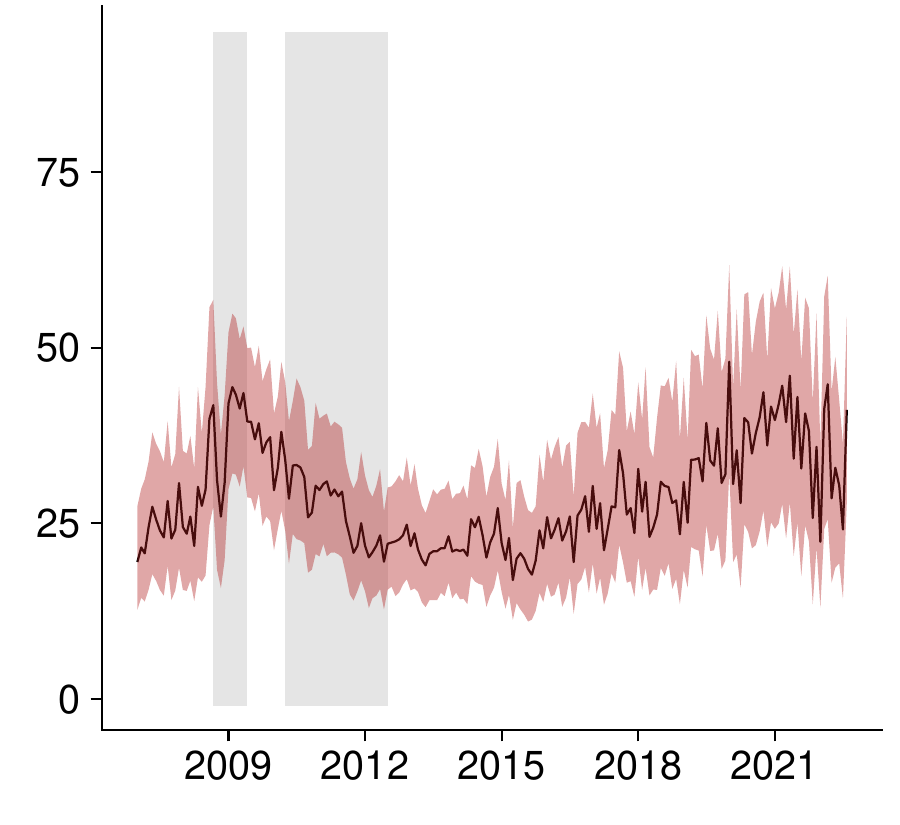}
\end{minipage}
\begin{minipage}{0.21\textwidth}
\centering
\includegraphics[scale=.24]{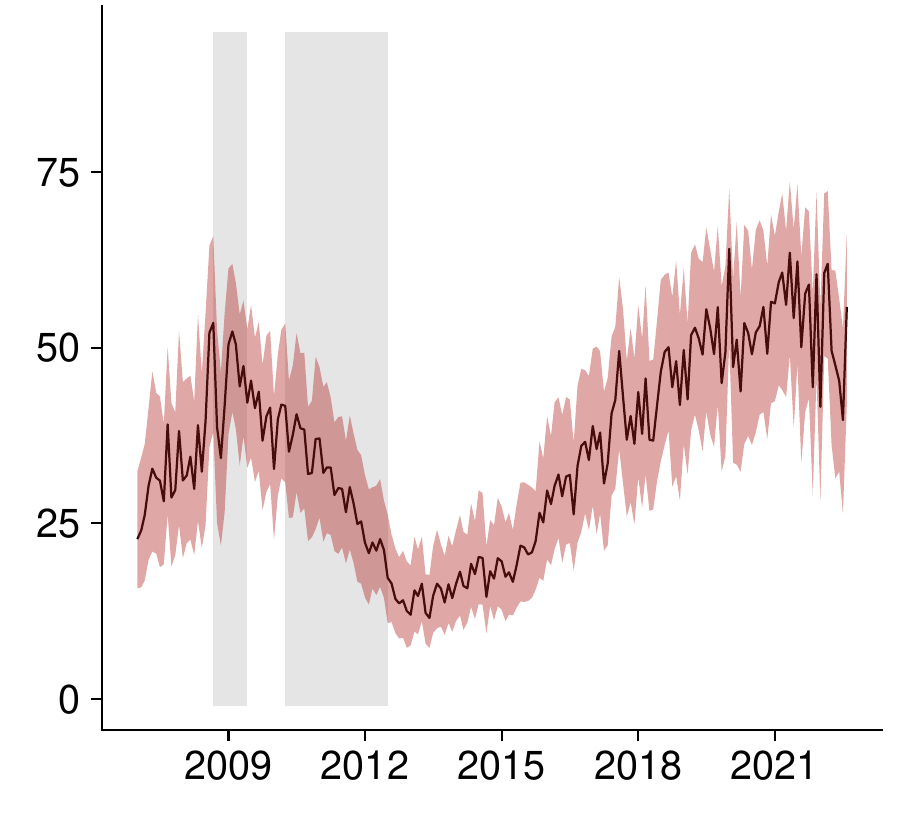}
\end{minipage}
\begin{minipage}{0.21\textwidth}
\centering
\includegraphics[scale=.24]{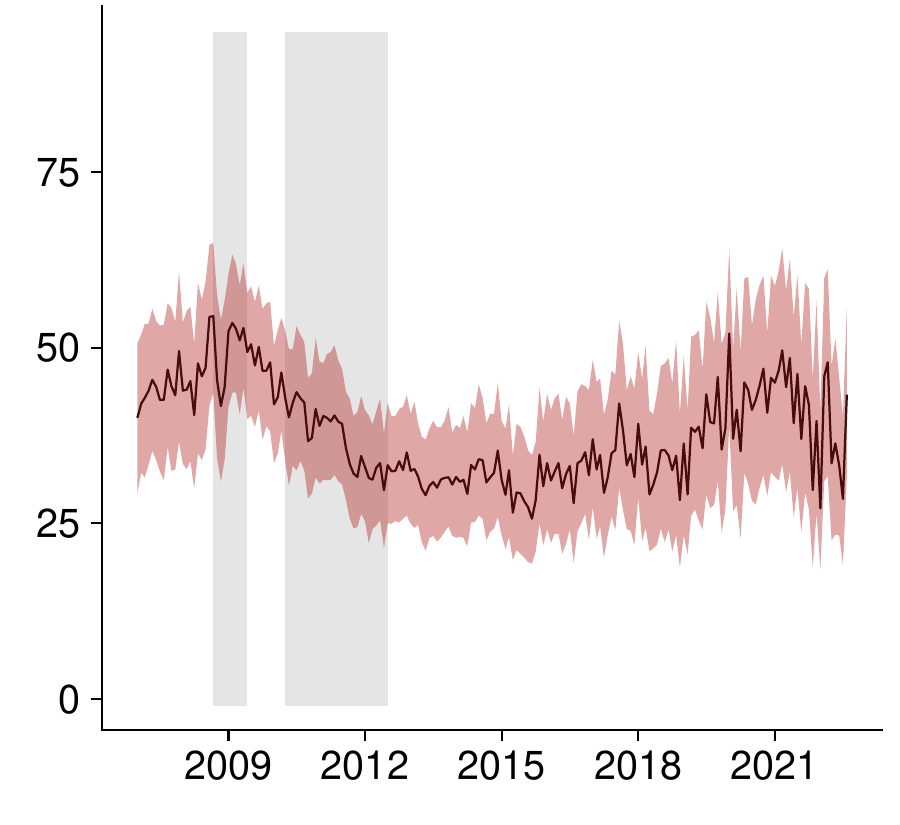}
\end{minipage}
\begin{minipage}{0.21\textwidth}
\centering
\includegraphics[scale=.24]{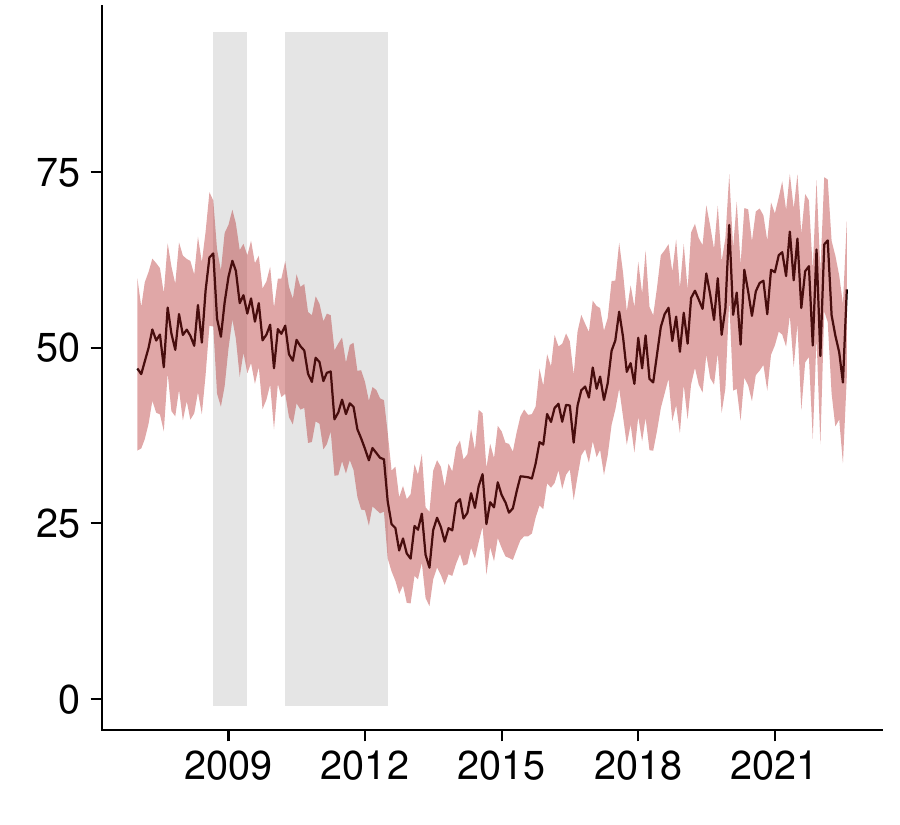}
\end{minipage}
\begin{minipage}{0.21\textwidth}
\centering
\includegraphics[scale=.24]{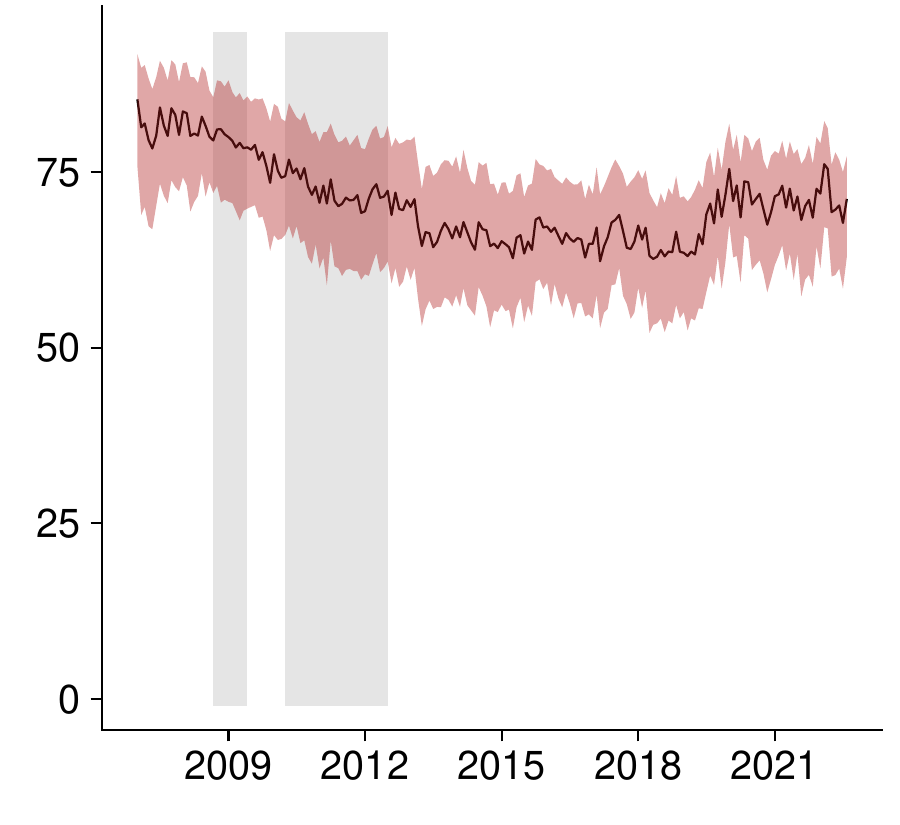}
\end{minipage}
\begin{minipage}{0.21\textwidth}
\centering
\includegraphics[scale=.24]{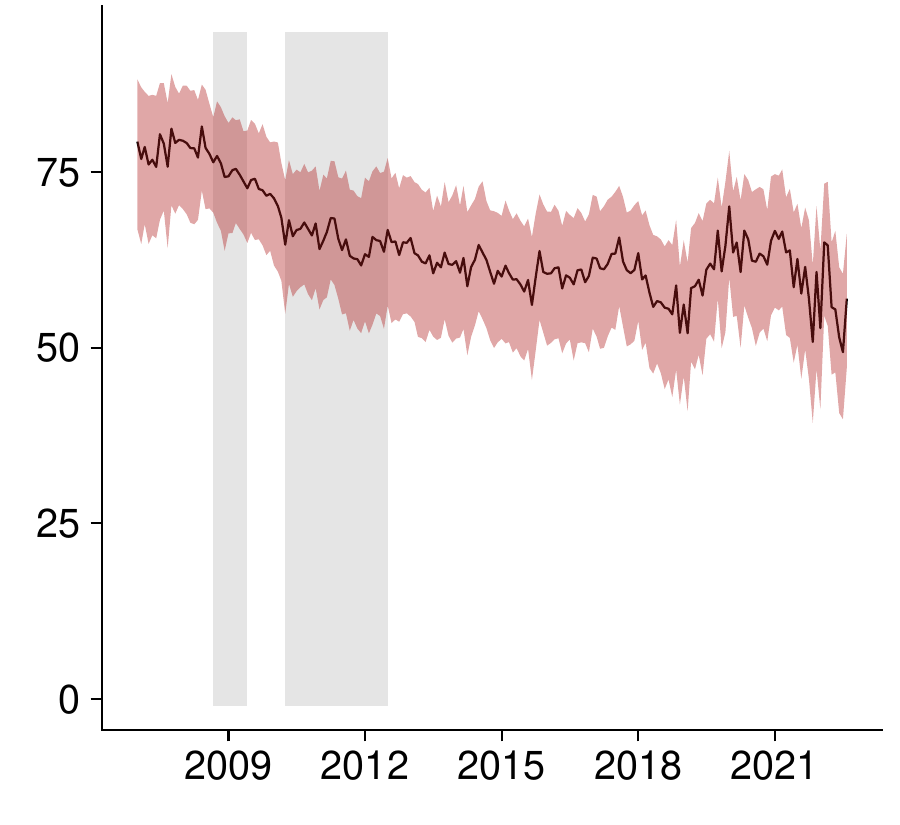}
\end{minipage}

\begin{minipage}{0.24\textwidth}
\centering
\scriptsize \textit{GR}
\end{minipage}
\begin{minipage}{0.24\textwidth}
\centering
\scriptsize \textit{IE}
\end{minipage}
\begin{minipage}{0.24\textwidth}
\centering
\scriptsize \textit{IT}
\end{minipage}
\begin{minipage}{0.24\textwidth}
\centering
\scriptsize \textit{PT}
\end{minipage}
\begin{minipage}{0.24\textwidth}
\centering
\scriptsize \textit{ES}
\end{minipage}

\begin{minipage}{0.24\textwidth}
\centering
\includegraphics[scale=.24]{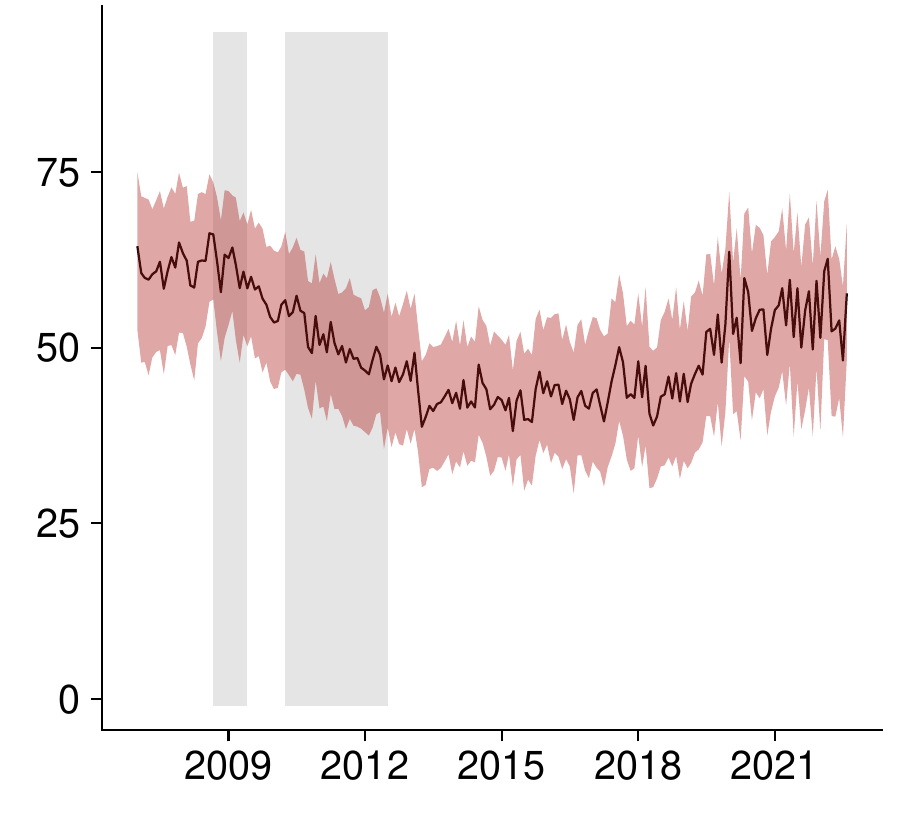}
\end{minipage}
\begin{minipage}{0.24\textwidth}
\centering
\includegraphics[scale=.24]{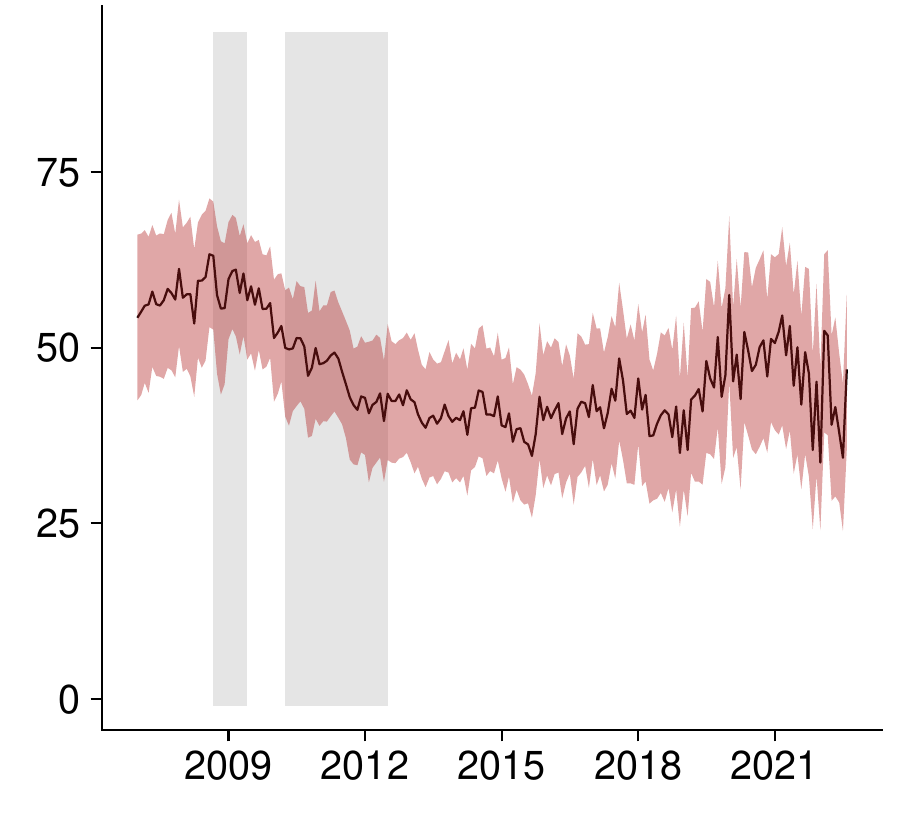}
\end{minipage}
\begin{minipage}{0.24\textwidth}
\centering
\includegraphics[scale=.24]{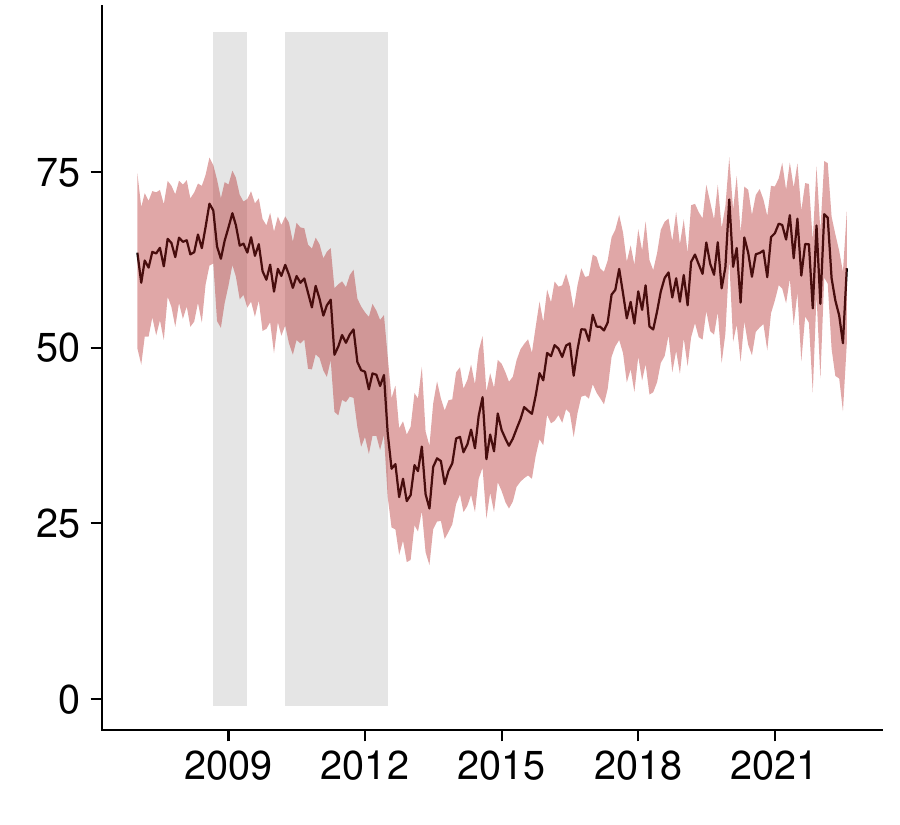}
\end{minipage}
\begin{minipage}{0.24\textwidth}
\centering
\includegraphics[scale=.24]{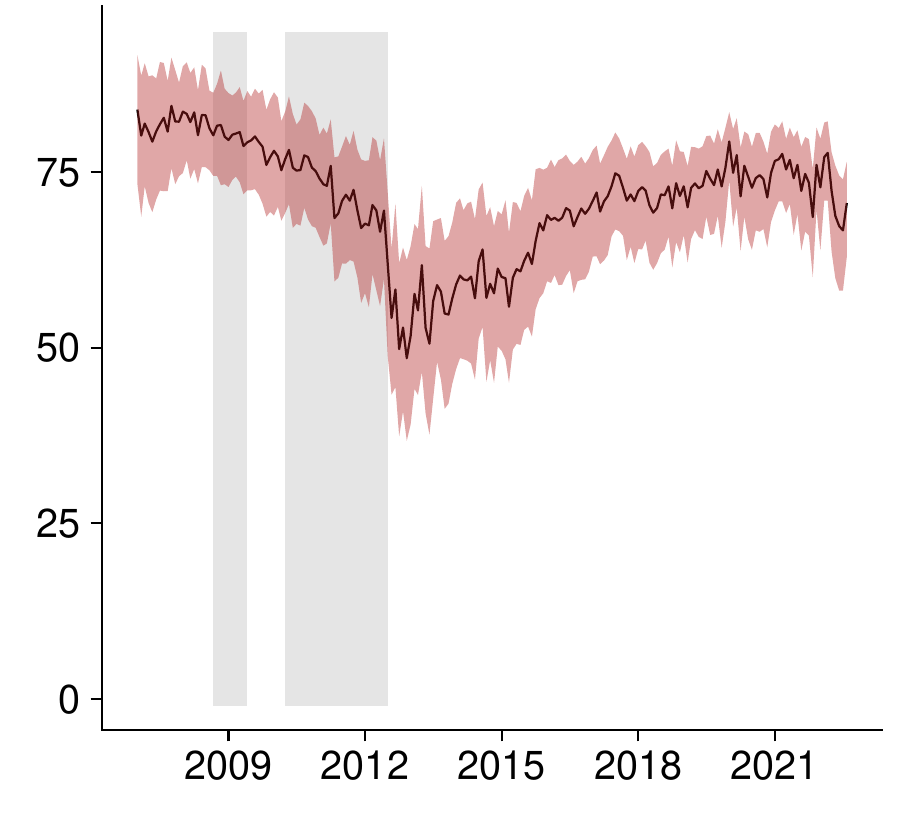}
\end{minipage}
\begin{minipage}{0.24\textwidth}
\centering
\includegraphics[scale=.24]{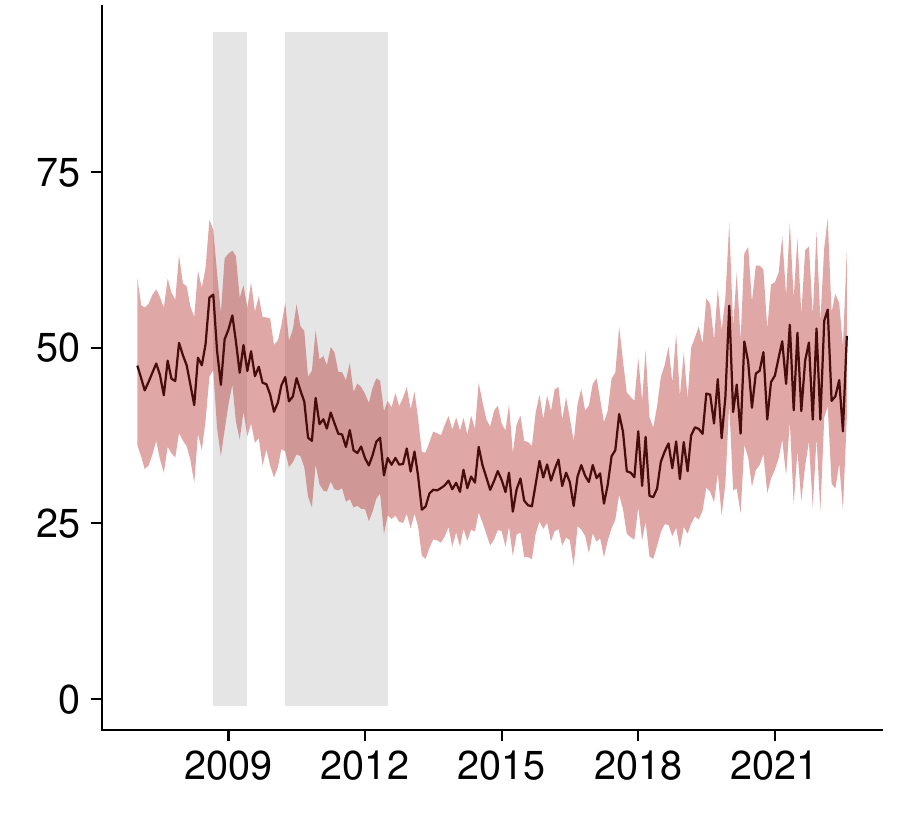}
\end{minipage}

\begin{minipage}{\linewidth}
\vspace{2pt}
\scriptsize \emph{Note:} This index indicates the share of spillovers for each country according to \cite{diebold2009measuring} and is estimated based on an expanding window. The solid line is the posterior median alongside the $68\%$ posterior credible set. The grey shaded area depicts the period of the ESDC.
\end{minipage}
\end{figure}
\end{landscape}

\clearpage

\begin{landscape}
\begin{figure}[htbp!]
\caption{Spillover index for each country and the 1-month forecast horizon. \label{fig:DYindex_detail_2}}

\begin{minipage}{\textwidth}
\centering
\small \textit{\textbf{Borrow short}}
\end{minipage}

\begin{minipage}{0.21\textwidth}
\centering
\scriptsize \textit{AT}
\end{minipage}
\begin{minipage}{0.21\textwidth}
\centering
\scriptsize \textit{BE}
\end{minipage}
\begin{minipage}{0.21\textwidth}
\centering
\scriptsize \textit{DE}
\end{minipage}
\begin{minipage}{0.21\textwidth}
\centering
\scriptsize \textit{FI}
\end{minipage}
\begin{minipage}{0.21\textwidth}
\centering
\scriptsize \textit{FR}
\end{minipage}
\begin{minipage}{0.21\textwidth}
\centering
\scriptsize \textit{LU}
\end{minipage}
\begin{minipage}{0.21\textwidth}
\centering
\scriptsize \textit{NL}
\end{minipage}

\begin{minipage}{0.21\textwidth}
\centering
\includegraphics[scale=.24]{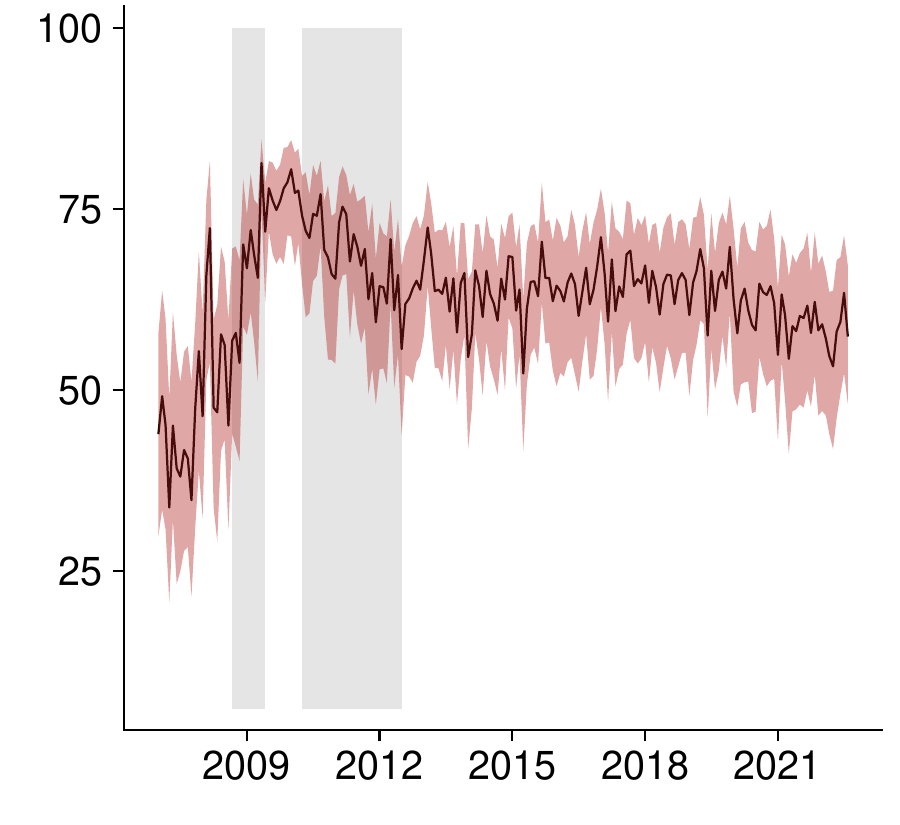}
\end{minipage}
\begin{minipage}{0.21\textwidth}
\centering
\includegraphics[scale=.24]{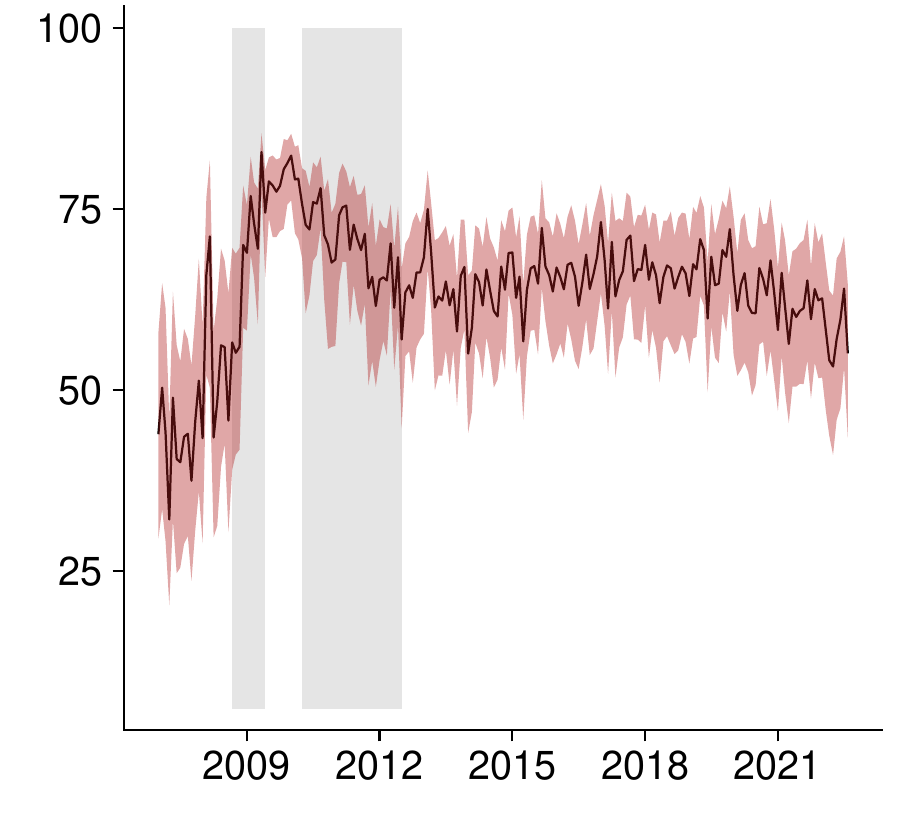}
\end{minipage}
\begin{minipage}{0.21\textwidth}
\centering
\includegraphics[scale=.24]{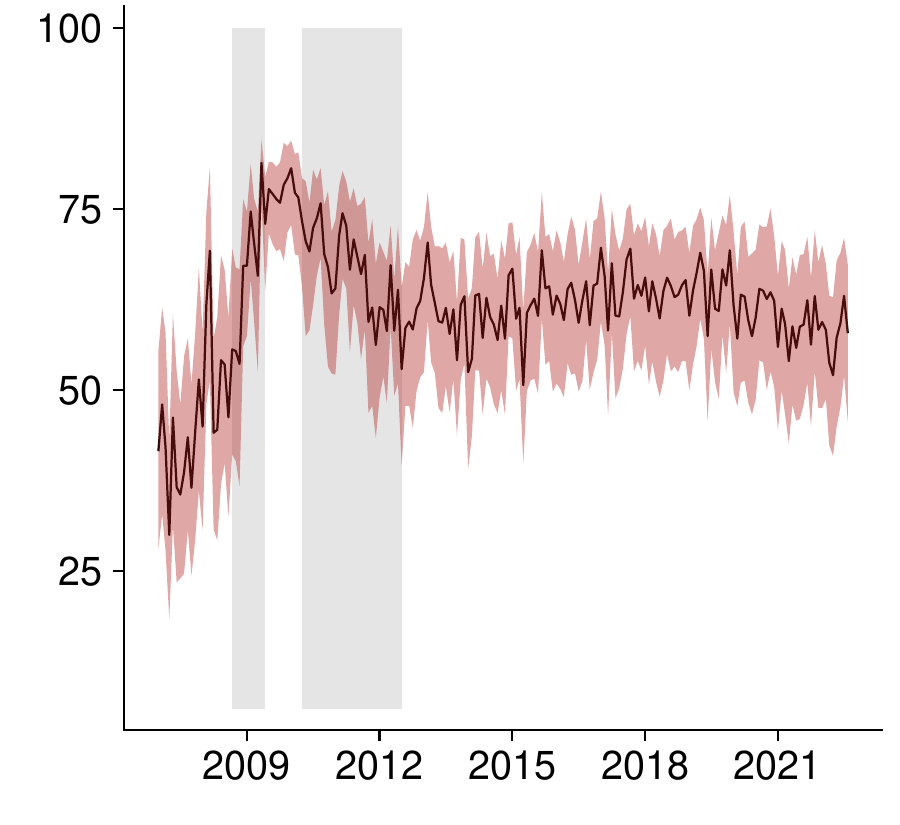}
\end{minipage}
\begin{minipage}{0.21\textwidth}
\centering
\includegraphics[scale=.24]{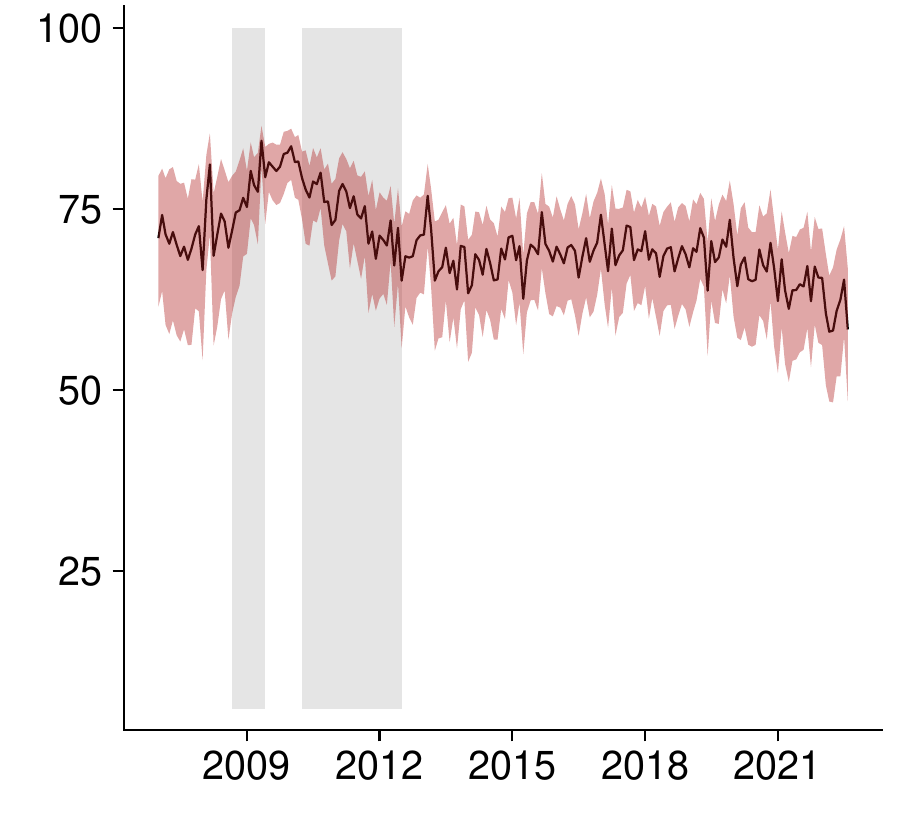}
\end{minipage}
\begin{minipage}{0.21\textwidth}
\centering
\includegraphics[scale=.24]{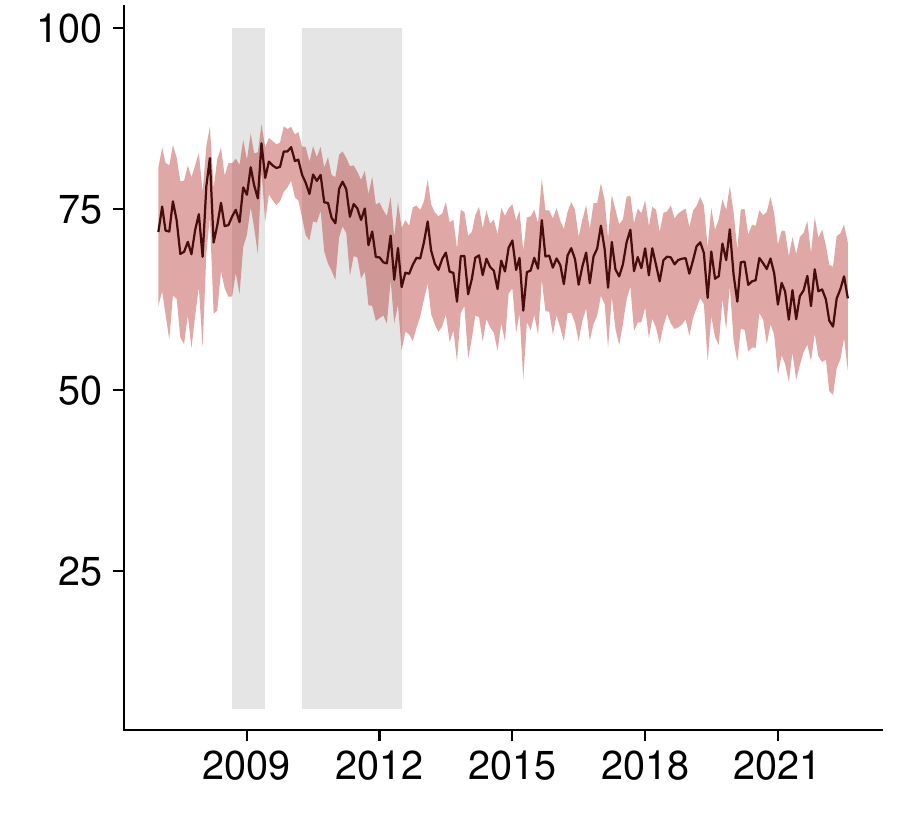}
\end{minipage}
\begin{minipage}{0.21\textwidth}
\centering
\includegraphics[scale=.24]{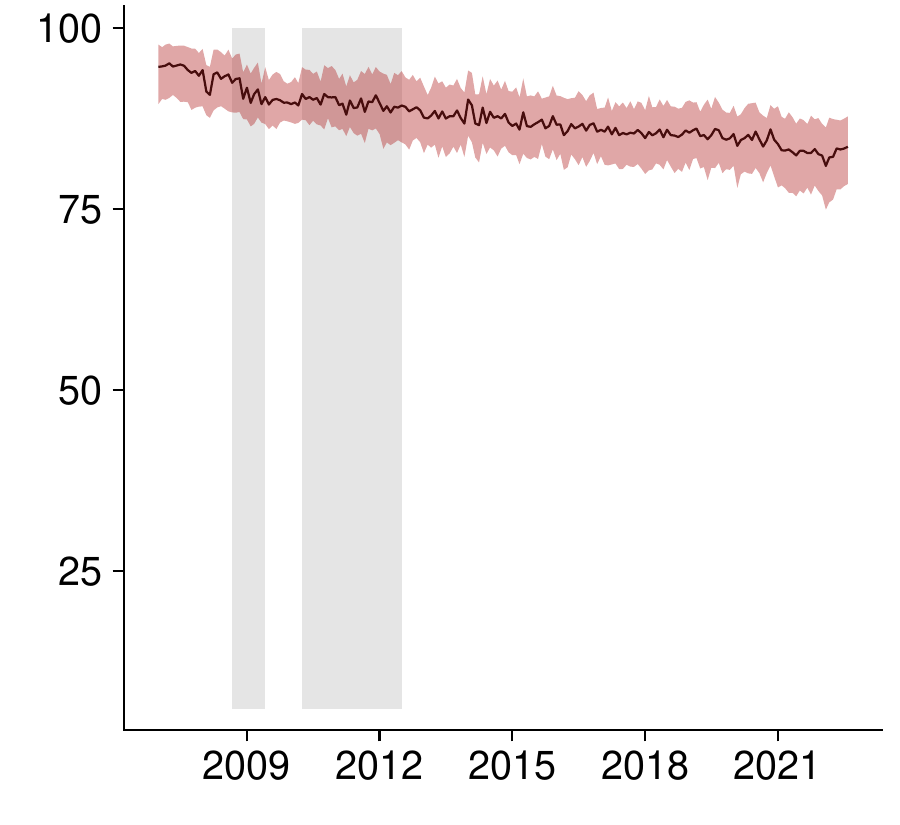}
\end{minipage}
\begin{minipage}{0.21\textwidth}
\centering
\includegraphics[scale=.24]{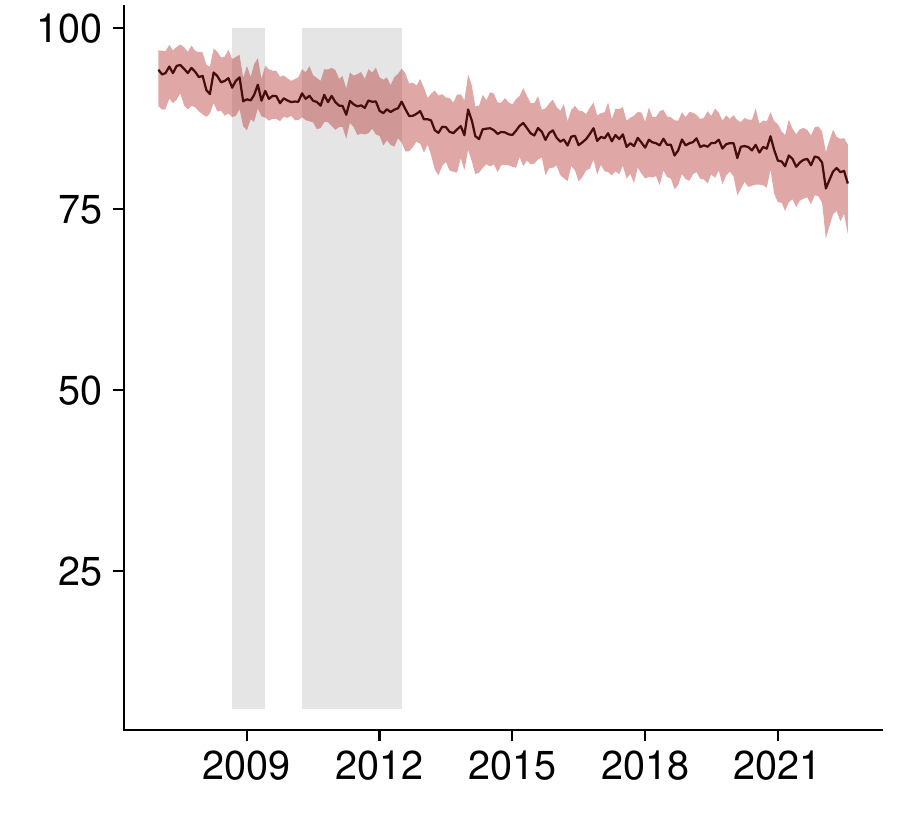}
\end{minipage}

\begin{minipage}{0.24\textwidth}
\centering
\scriptsize \textit{GR}
\end{minipage}
\begin{minipage}{0.24\textwidth}
\centering
\scriptsize \textit{IE}
\end{minipage}
\begin{minipage}{0.24\textwidth}
\centering
\scriptsize \textit{IT}
\end{minipage}
\begin{minipage}{0.24\textwidth}
\centering
\scriptsize \textit{PT}
\end{minipage}
\begin{minipage}{0.24\textwidth}
\centering
\scriptsize \textit{ES}
\end{minipage}

\begin{minipage}{0.24\textwidth}
\centering
\includegraphics[scale=.24]{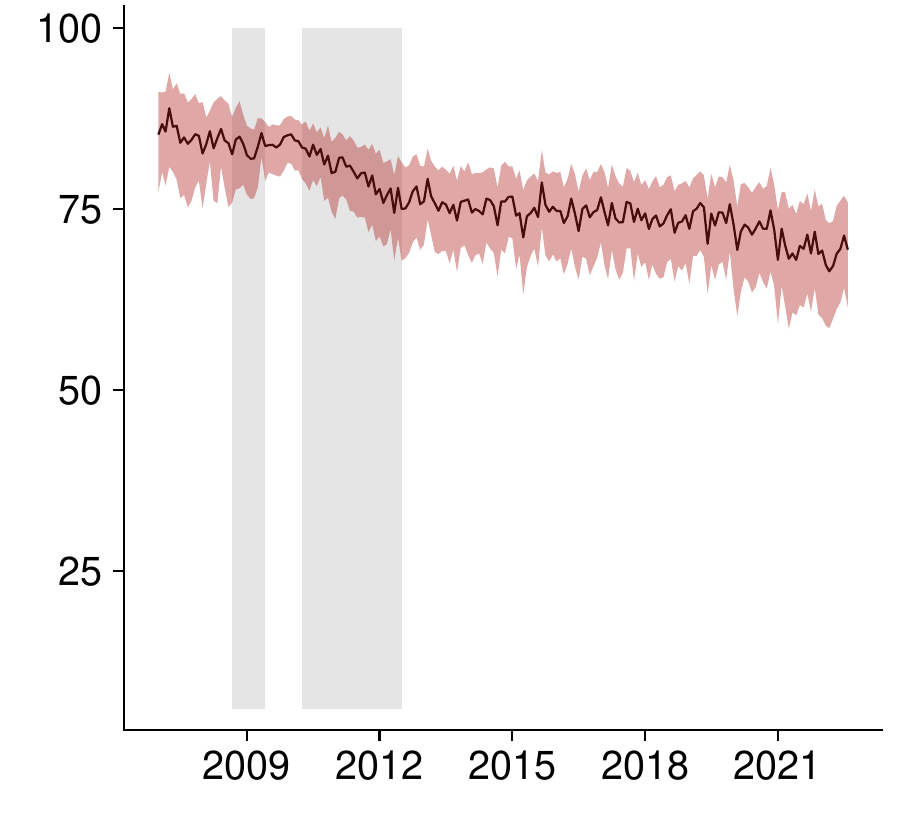}
\end{minipage}
\begin{minipage}{0.24\textwidth}
\centering
\includegraphics[scale=.24]{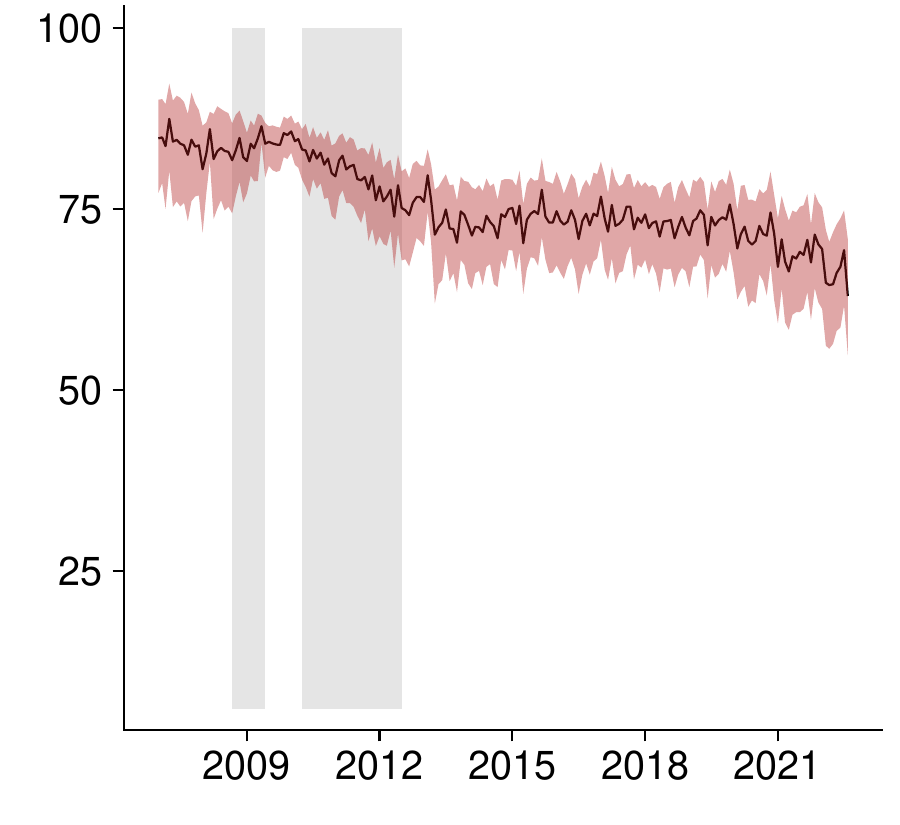}
\end{minipage}
\begin{minipage}{0.24\textwidth}
\centering
\includegraphics[scale=.24]{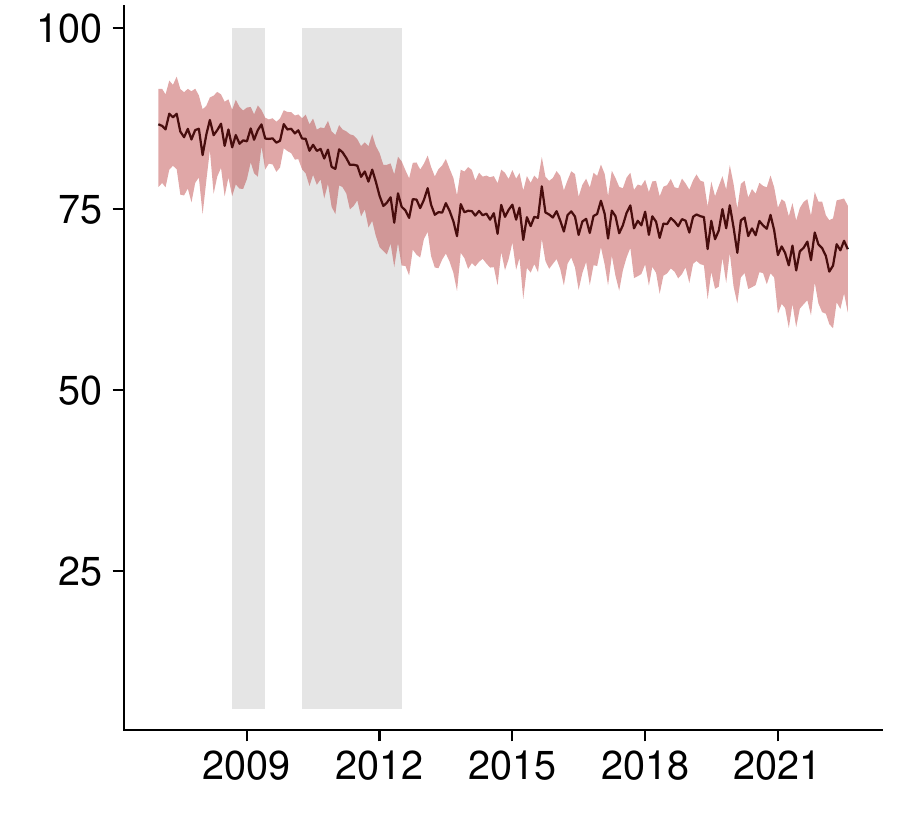}
\end{minipage}
\begin{minipage}{0.24\textwidth}
\centering
\includegraphics[scale=.24]{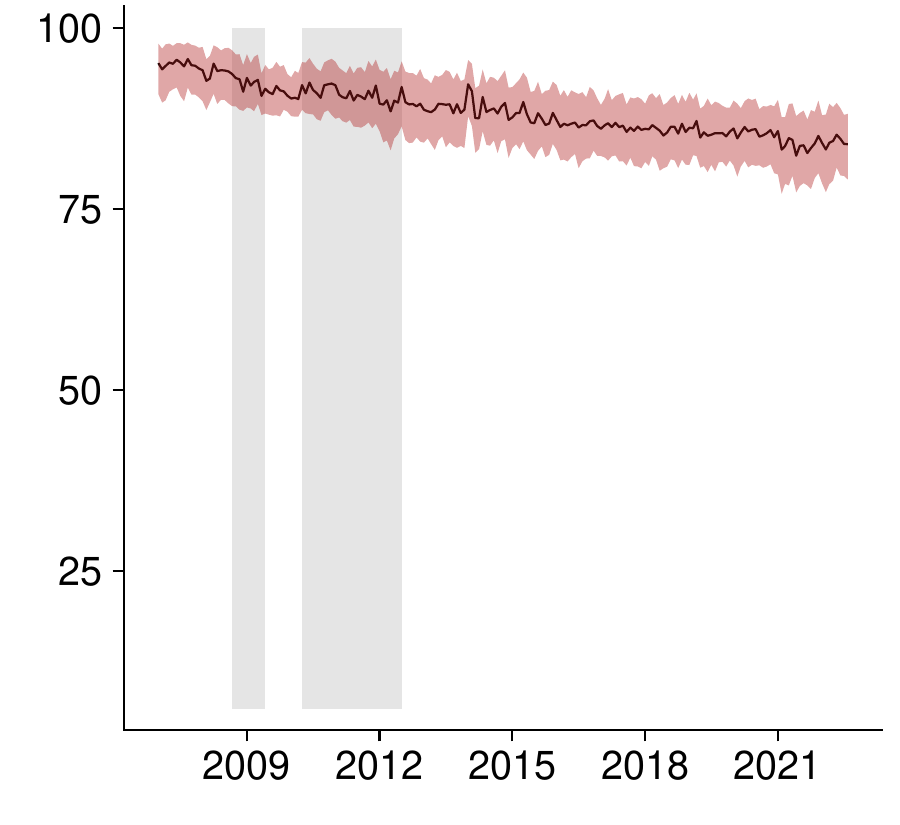}
\end{minipage}
\begin{minipage}{0.24\textwidth}
\centering
\includegraphics[scale=.24]{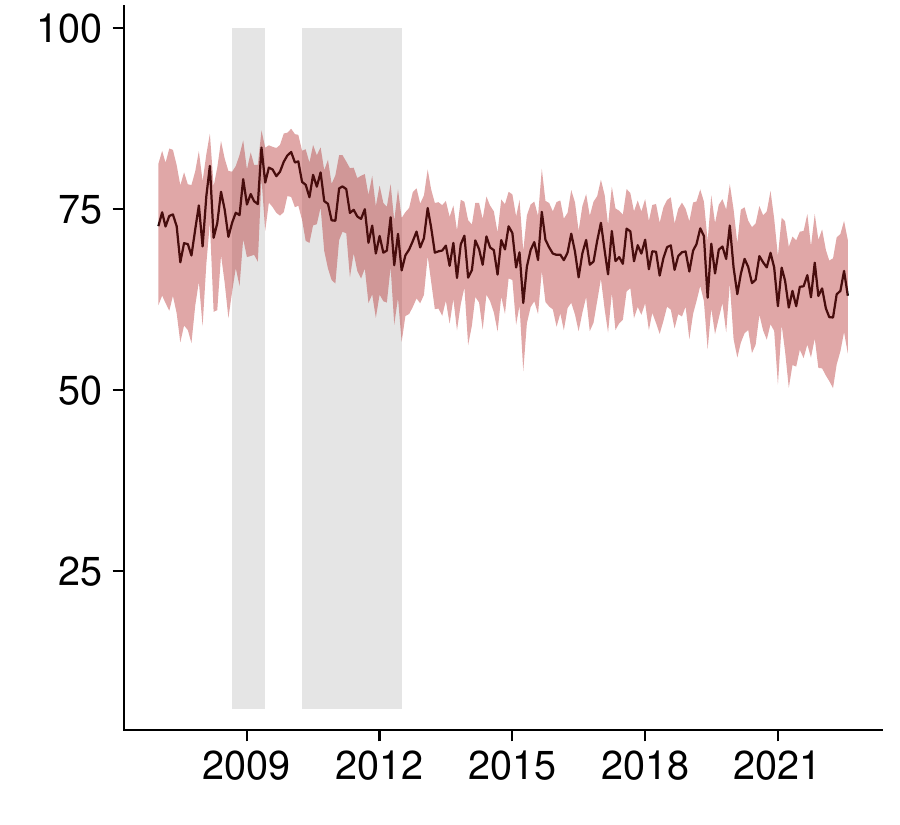}
\end{minipage}

\begin{minipage}{\textwidth}
\centering
\small \textit{\textbf{TARGET2}}
\end{minipage}

\begin{minipage}{0.21\textwidth}
\centering
\scriptsize \textit{AT}
\end{minipage}
\begin{minipage}{0.21\textwidth}
\centering
\scriptsize \textit{BE}
\end{minipage}
\begin{minipage}{0.21\textwidth}
\centering
\scriptsize \textit{DE}
\end{minipage}
\begin{minipage}{0.21\textwidth}
\centering
\scriptsize \textit{FI}
\end{minipage}
\begin{minipage}{0.21\textwidth}
\centering
\scriptsize \textit{FR}
\end{minipage}
\begin{minipage}{0.21\textwidth}
\centering
\scriptsize \textit{LU}
\end{minipage}
\begin{minipage}{0.21\textwidth}
\centering
\scriptsize \textit{NL}
\end{minipage}

\begin{minipage}{0.21\textwidth}
\centering
\includegraphics[scale=.24]{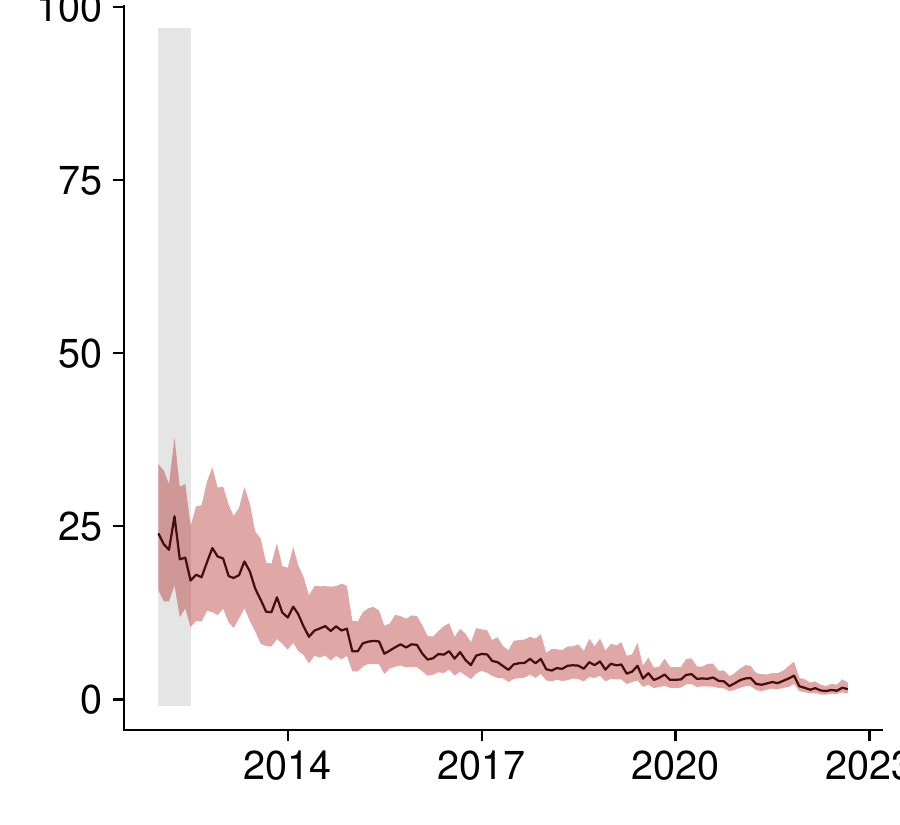}
\end{minipage}
\begin{minipage}{0.21\textwidth}
\centering
\includegraphics[scale=.24]{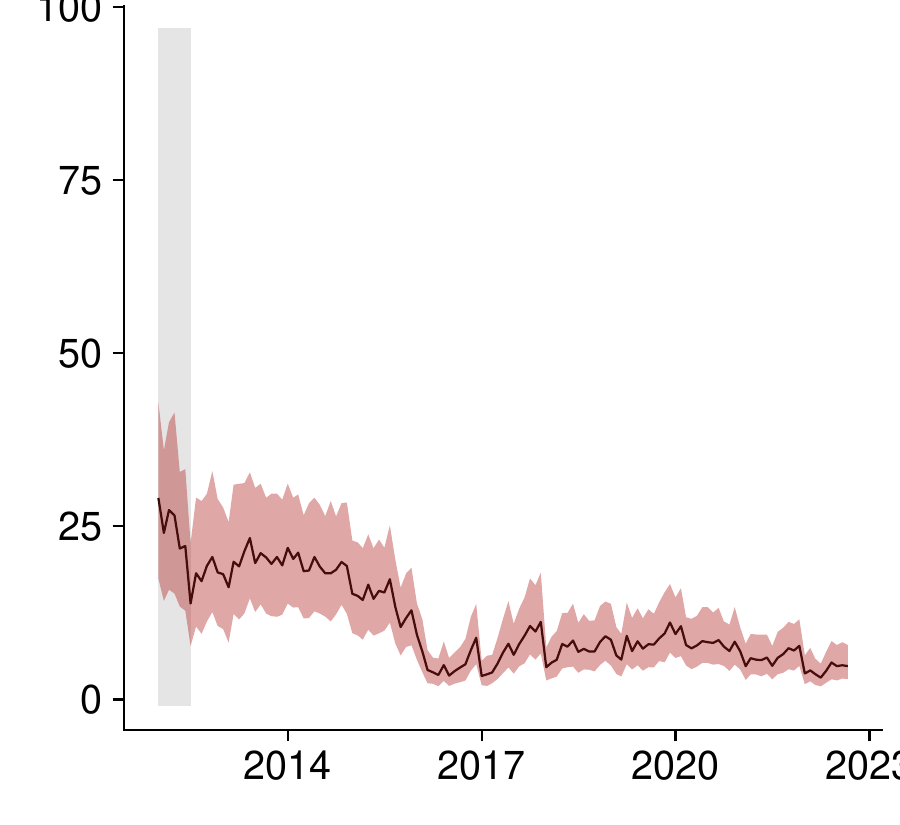}
\end{minipage}
\begin{minipage}{0.21\textwidth}
\centering
\includegraphics[scale=.24]{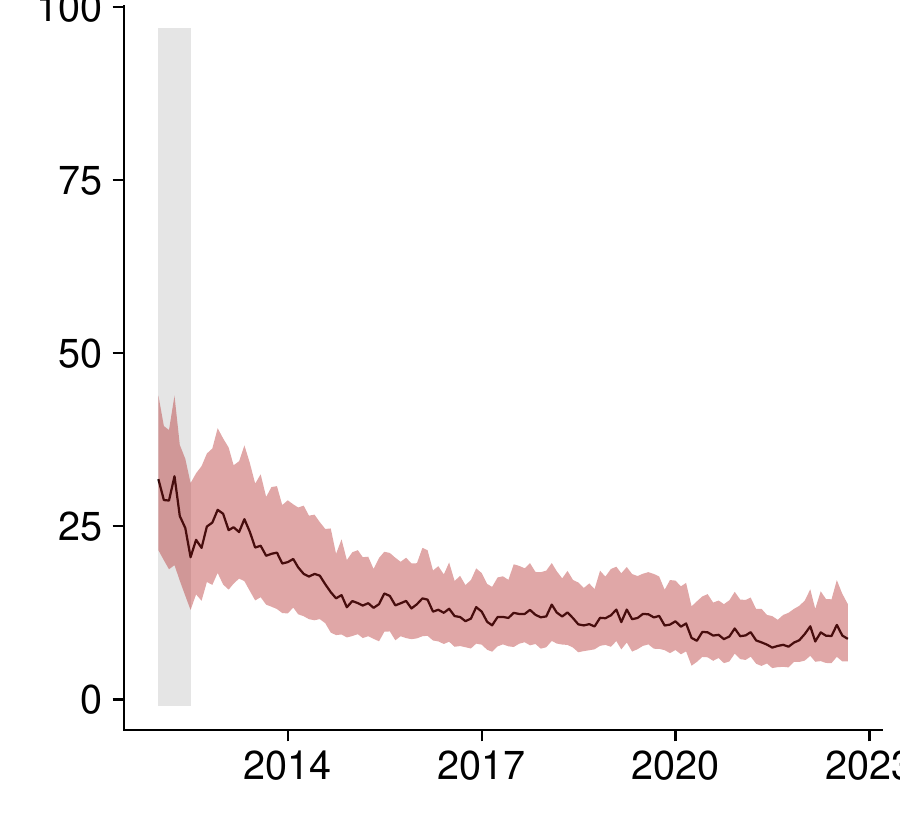}
\end{minipage}
\begin{minipage}{0.21\textwidth}
\centering
\includegraphics[scale=.24]{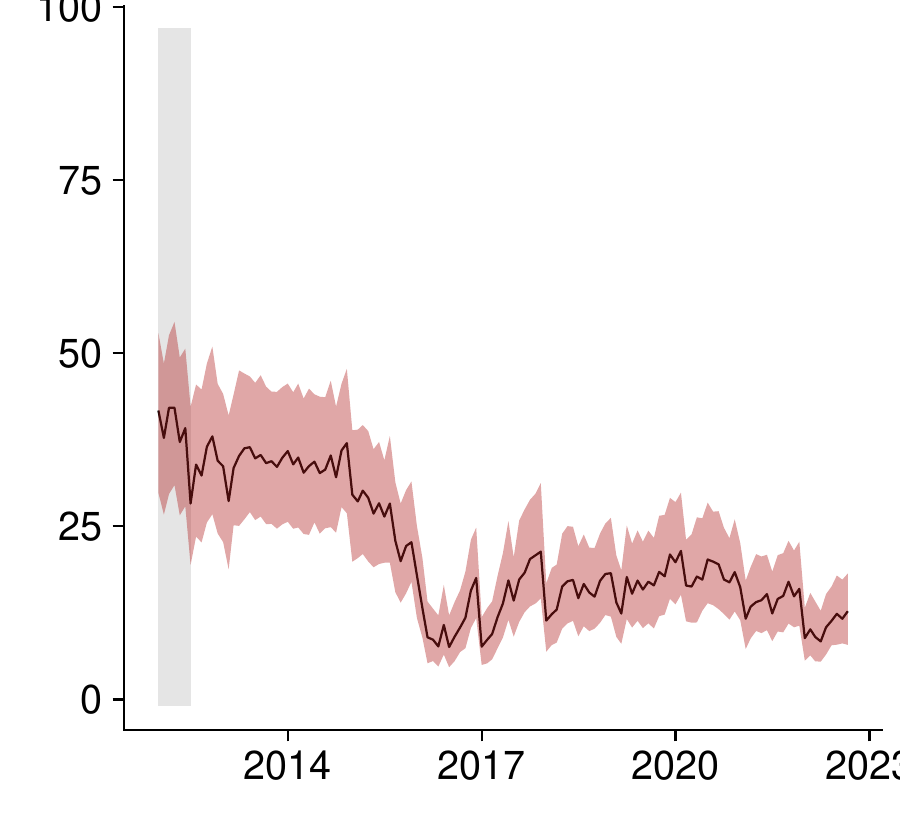}
\end{minipage}
\begin{minipage}{0.21\textwidth}
\centering
\includegraphics[scale=.24]{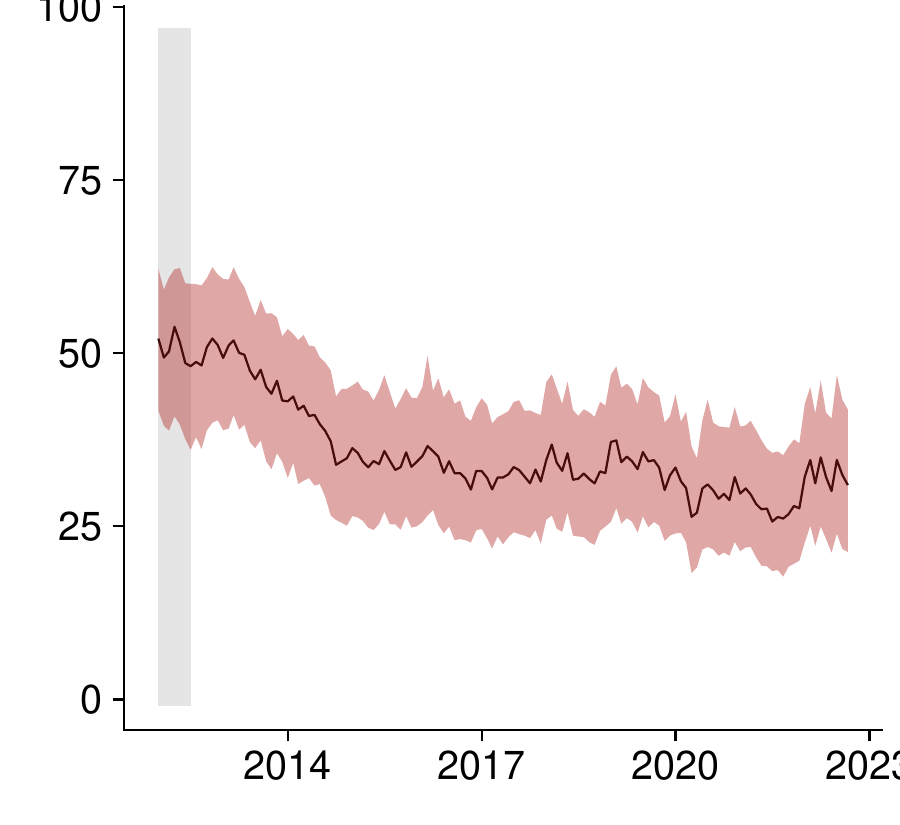}
\end{minipage}
\begin{minipage}{0.21\textwidth}
\centering
\includegraphics[scale=.24]{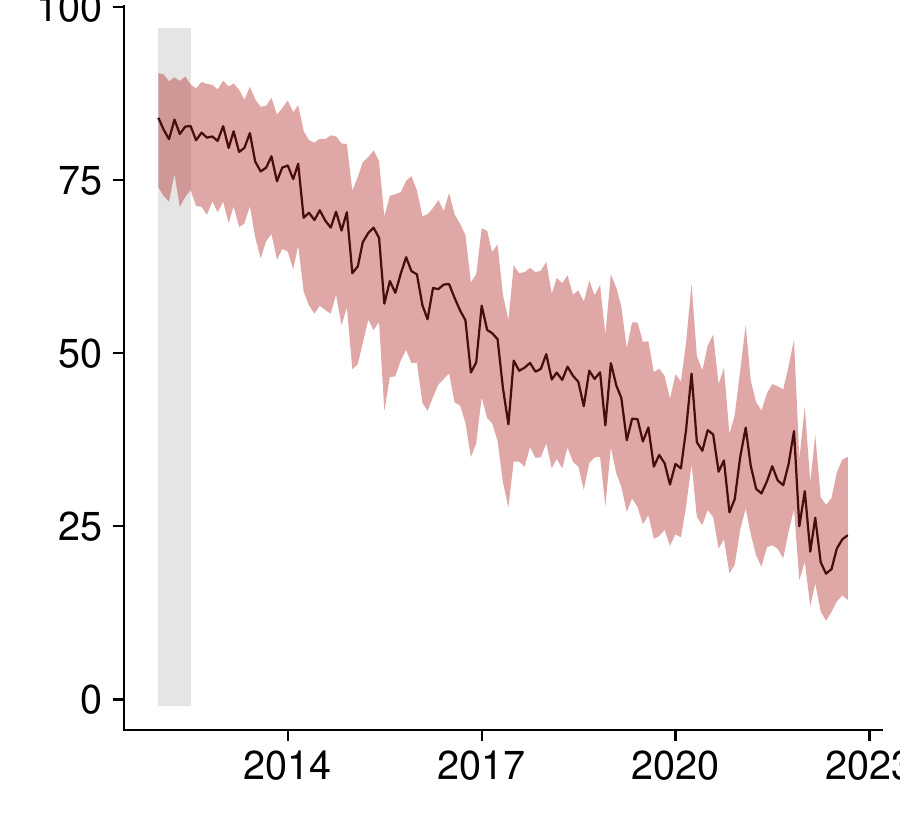}
\end{minipage}
\begin{minipage}{0.21\textwidth}
\centering
\includegraphics[scale=.24]{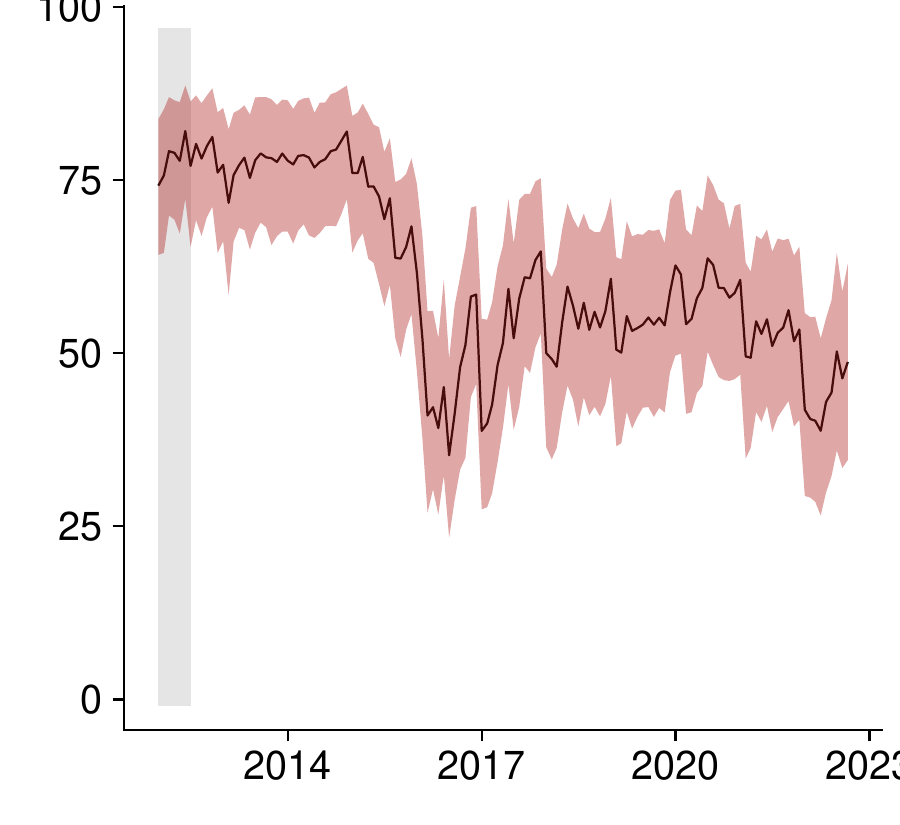}
\end{minipage}

\begin{minipage}{0.24\textwidth}
\centering
\scriptsize \textit{GR}
\end{minipage}
\begin{minipage}{0.24\textwidth}
\centering
\scriptsize \textit{IE}
\end{minipage}
\begin{minipage}{0.24\textwidth}
\centering
\scriptsize \textit{IT}
\end{minipage}
\begin{minipage}{0.24\textwidth}
\centering
\scriptsize \textit{PT}
\end{minipage}
\begin{minipage}{0.24\textwidth}
\centering
\scriptsize \textit{ES}
\end{minipage}

\begin{minipage}{0.24\textwidth}
\centering
\includegraphics[scale=.24]{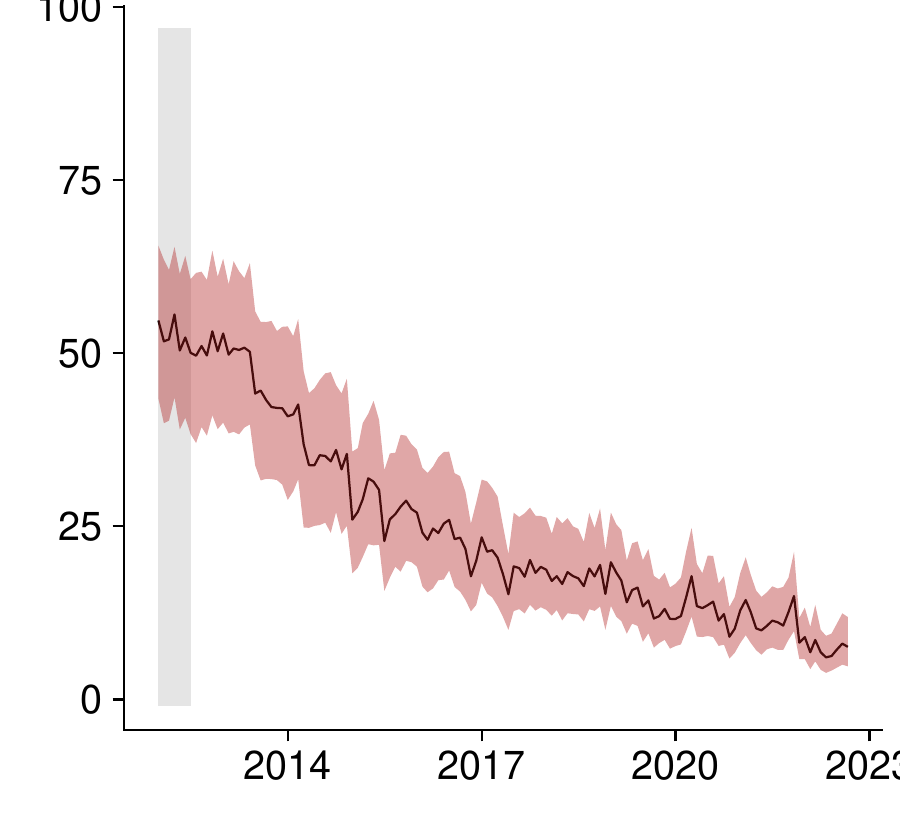}
\end{minipage}
\begin{minipage}{0.24\textwidth}
\centering
\includegraphics[scale=.24]{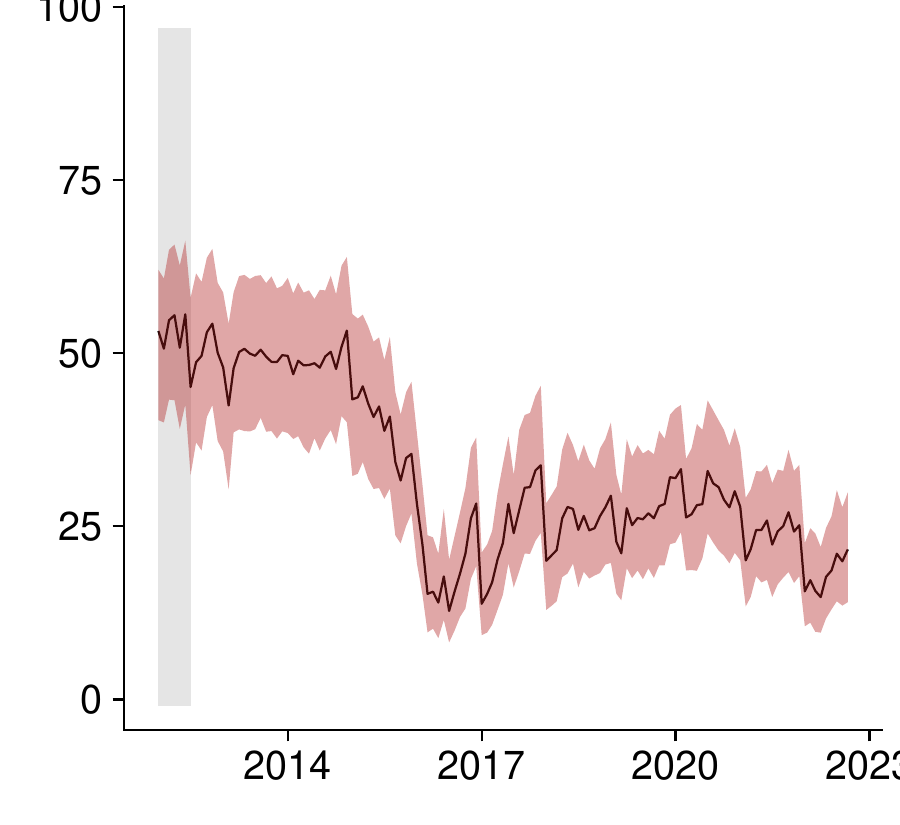}
\end{minipage}
\begin{minipage}{0.24\textwidth}
\centering
\includegraphics[scale=.24]{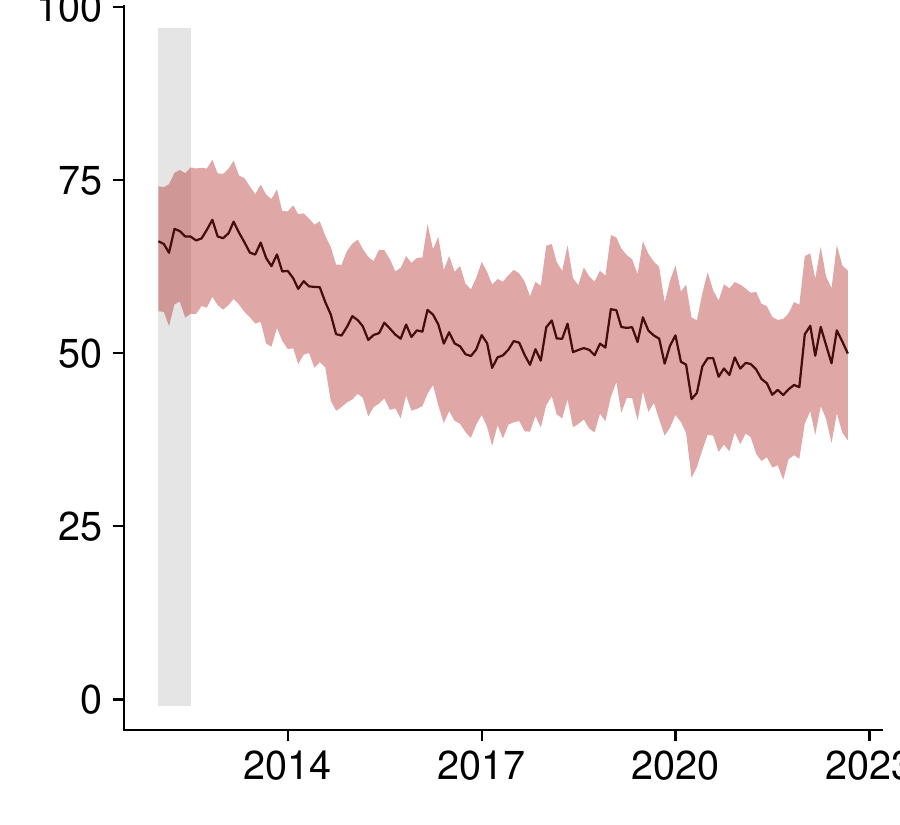}
\end{minipage}
\begin{minipage}{0.24\textwidth}
\centering
\includegraphics[scale=.24]{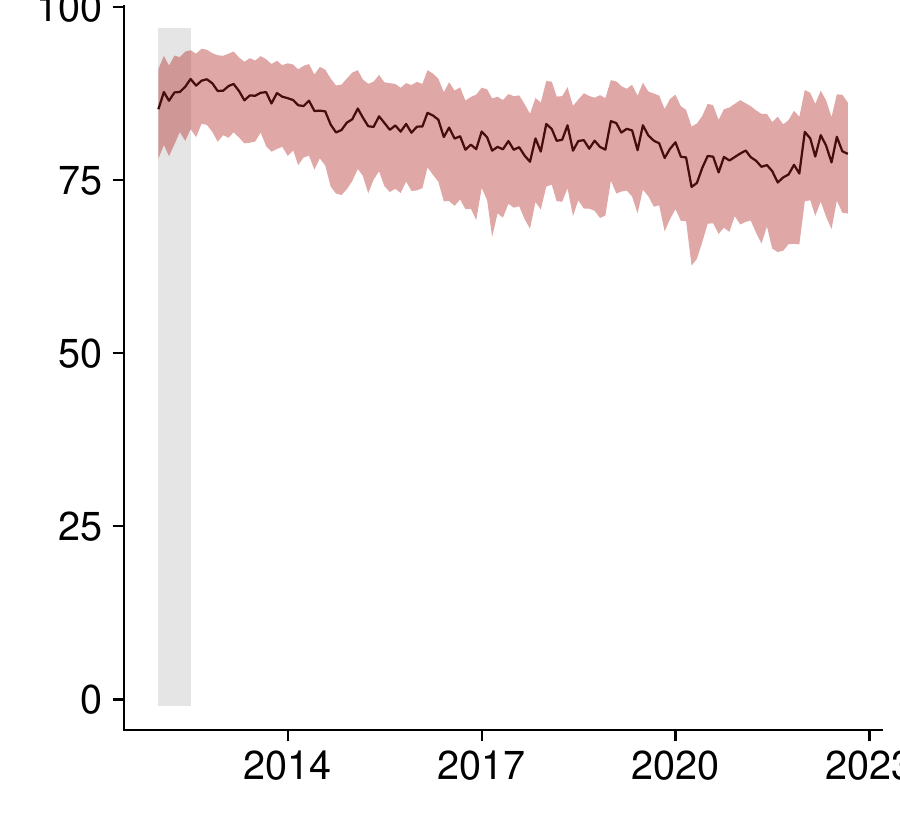}
\end{minipage}
\begin{minipage}{0.24\textwidth}
\centering
\includegraphics[scale=.24]{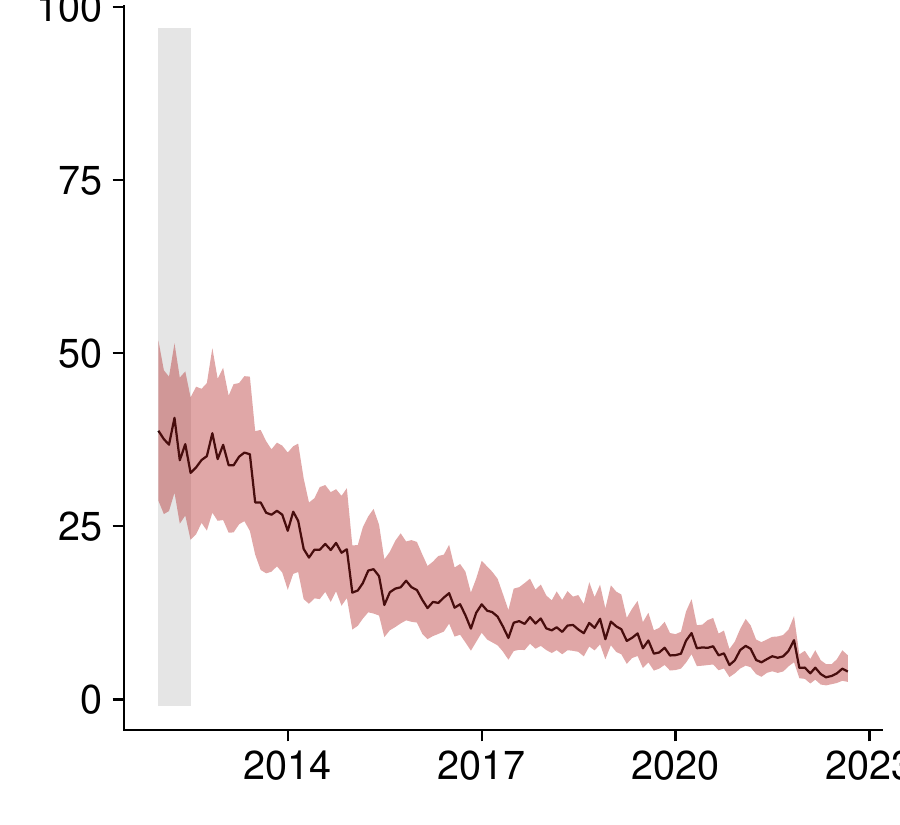}
\end{minipage}

\begin{minipage}{\linewidth}
\vspace{2pt}
\scriptsize \emph{Note:} This index indicates the share of spillovers for each country according to \cite{diebold2009measuring} and is estimated based on an expanding window. The solid line is the posterior median alongside the $68\%$ posterior credible set. The grey shaded area depicts the period of the ESDC.
\end{minipage}
\end{figure}
\end{landscape}

\clearpage

\begin{landscape}
\begin{figure}[htbp!]
\caption{Spillover index for each country and the 12-months forecast horizon. \label{fig:DYindex_detail_12}}

\begin{minipage}{\textwidth}
\centering
\small \textit{\textbf{Government bond yields}}
\end{minipage}

\begin{minipage}{0.21\textwidth}
\centering
\scriptsize \textit{AT}
\end{minipage}
\begin{minipage}{0.21\textwidth}
\centering
\scriptsize \textit{BE}
\end{minipage}
\begin{minipage}{0.21\textwidth}
\centering
\scriptsize \textit{DE}
\end{minipage}
\begin{minipage}{0.21\textwidth}
\centering
\scriptsize \textit{FI}
\end{minipage}
\begin{minipage}{0.21\textwidth}
\centering
\scriptsize \textit{FR}
\end{minipage}
\begin{minipage}{0.21\textwidth}
\centering
\scriptsize \textit{LU}
\end{minipage}
\begin{minipage}{0.21\textwidth}
\centering
\scriptsize \textit{NL}
\end{minipage}

\begin{minipage}{0.21\textwidth}
\centering
\includegraphics[scale=.24]{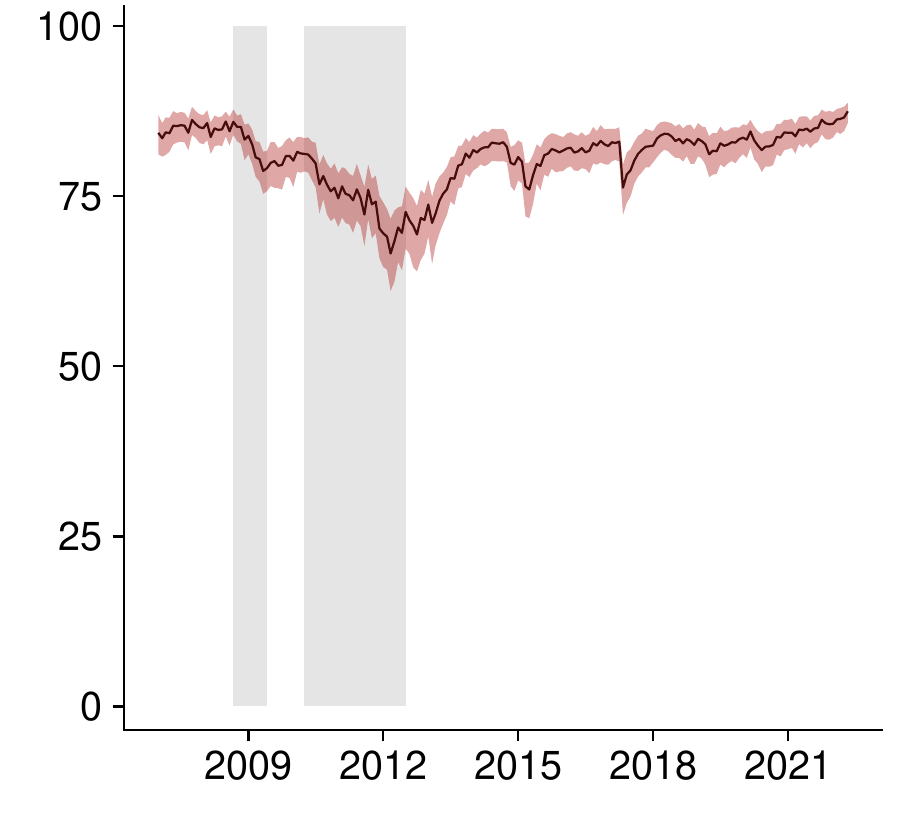}
\end{minipage}
\begin{minipage}{0.21\textwidth}
\centering
\includegraphics[scale=.24]{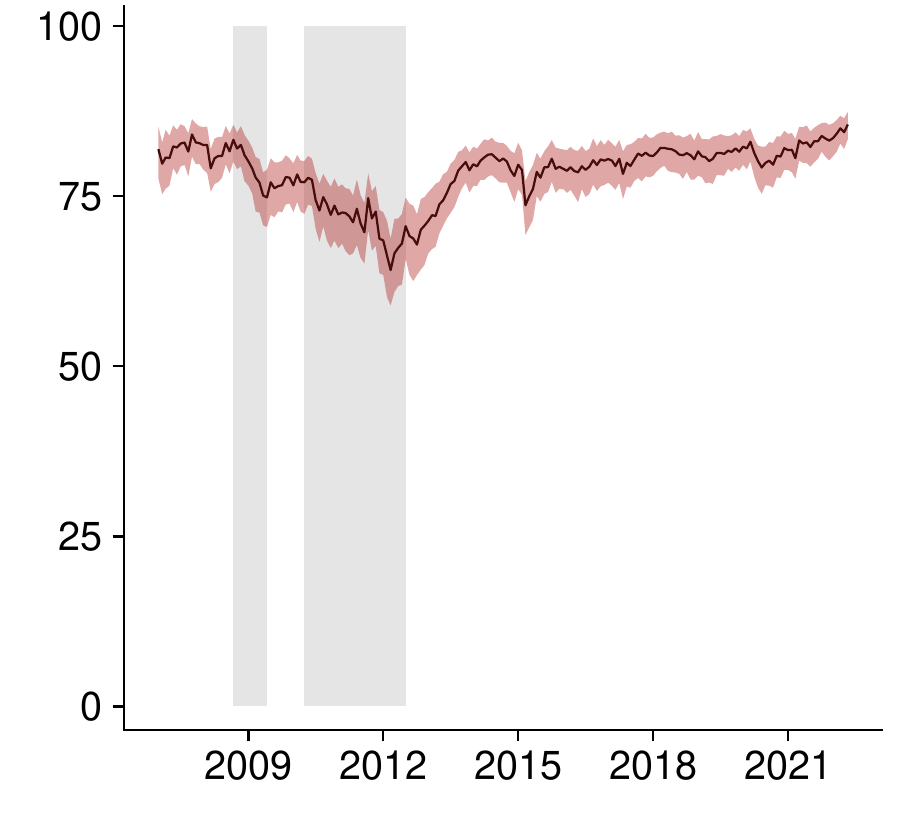}
\end{minipage}
\begin{minipage}{0.21\textwidth}
\centering
\includegraphics[scale=.24]{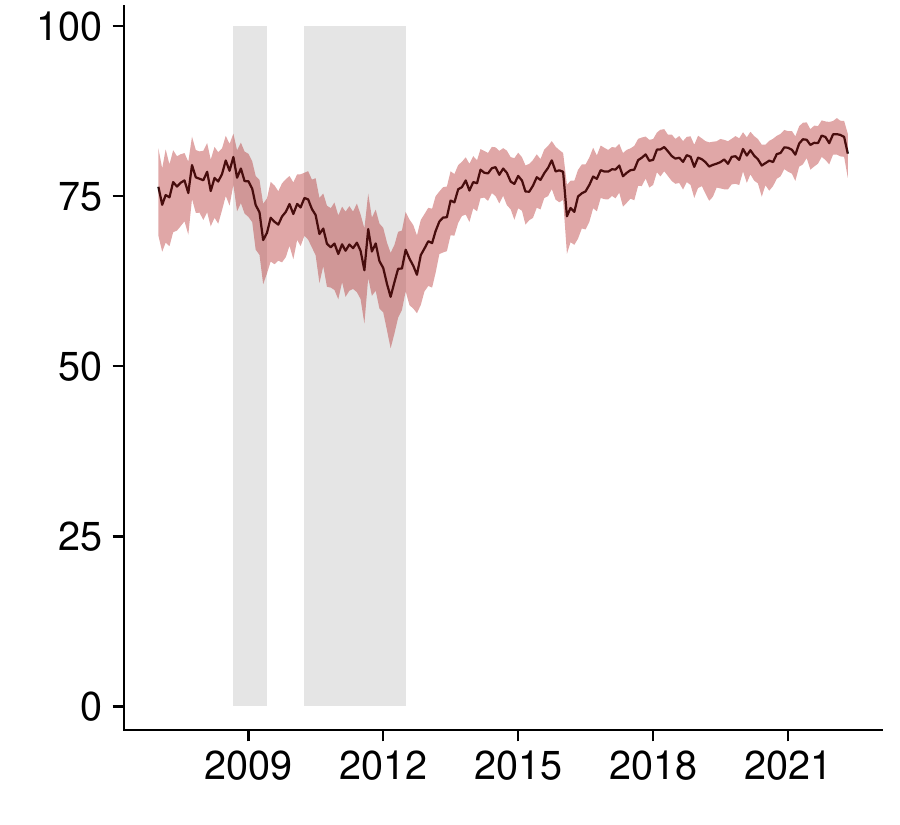}
\end{minipage}
\begin{minipage}{0.21\textwidth}
\centering
\includegraphics[scale=.24]{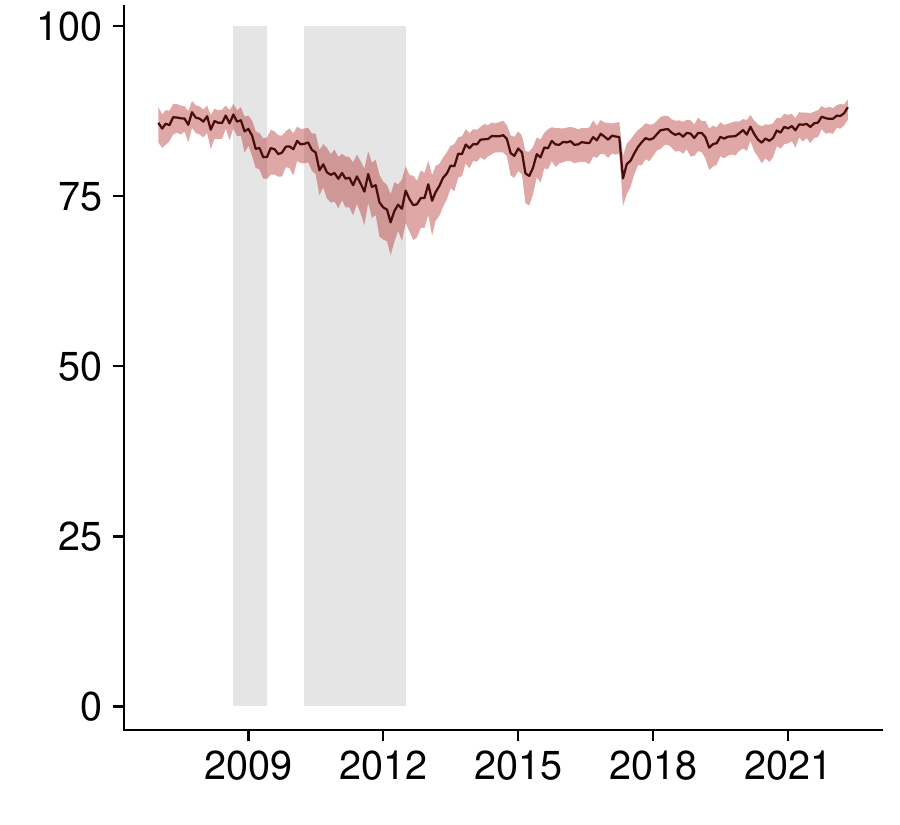}
\end{minipage}
\begin{minipage}{0.21\textwidth}
\centering
\includegraphics[scale=.24]{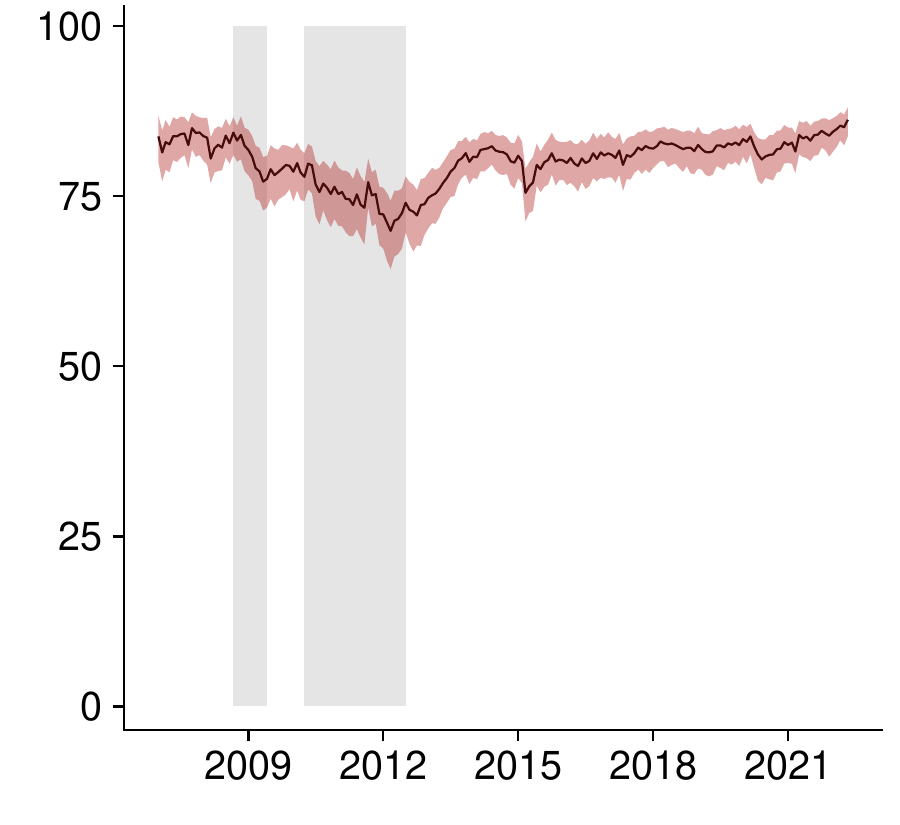}
\end{minipage}
\begin{minipage}{0.21\textwidth}
\centering
\includegraphics[scale=.24]{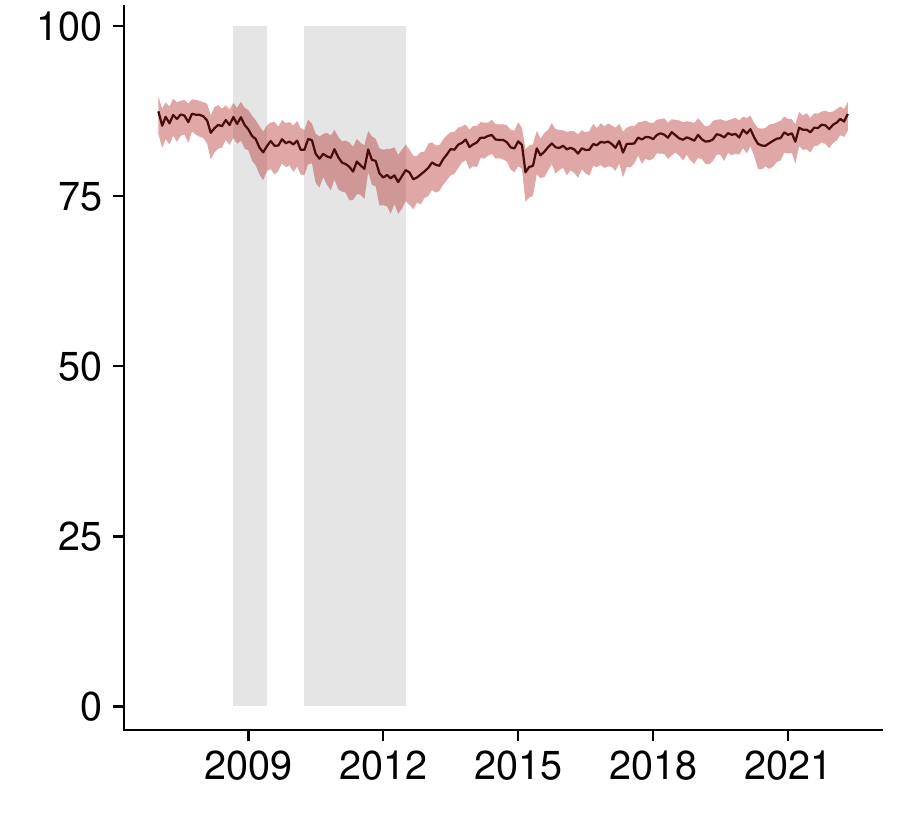}
\end{minipage}
\begin{minipage}{0.21\textwidth}
\centering
\includegraphics[scale=.24]{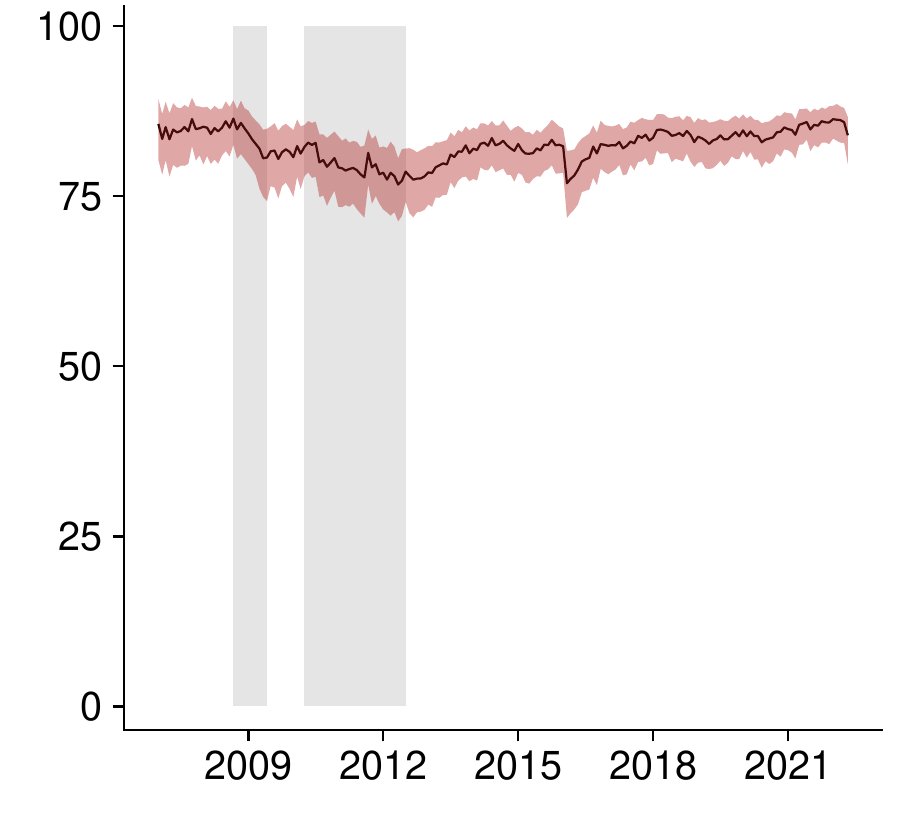}
\end{minipage}

\begin{minipage}{0.24\textwidth}
\centering
\scriptsize \textit{GR}
\end{minipage}
\begin{minipage}{0.24\textwidth}
\centering
\scriptsize \textit{IE}
\end{minipage}
\begin{minipage}{0.24\textwidth}
\centering
\scriptsize \textit{IT}
\end{minipage}
\begin{minipage}{0.24\textwidth}
\centering
\scriptsize \textit{PT}
\end{minipage}
\begin{minipage}{0.24\textwidth}
\centering
\scriptsize \textit{ES}
\end{minipage}

\begin{minipage}{0.24\textwidth}
\centering
\includegraphics[scale=.24]{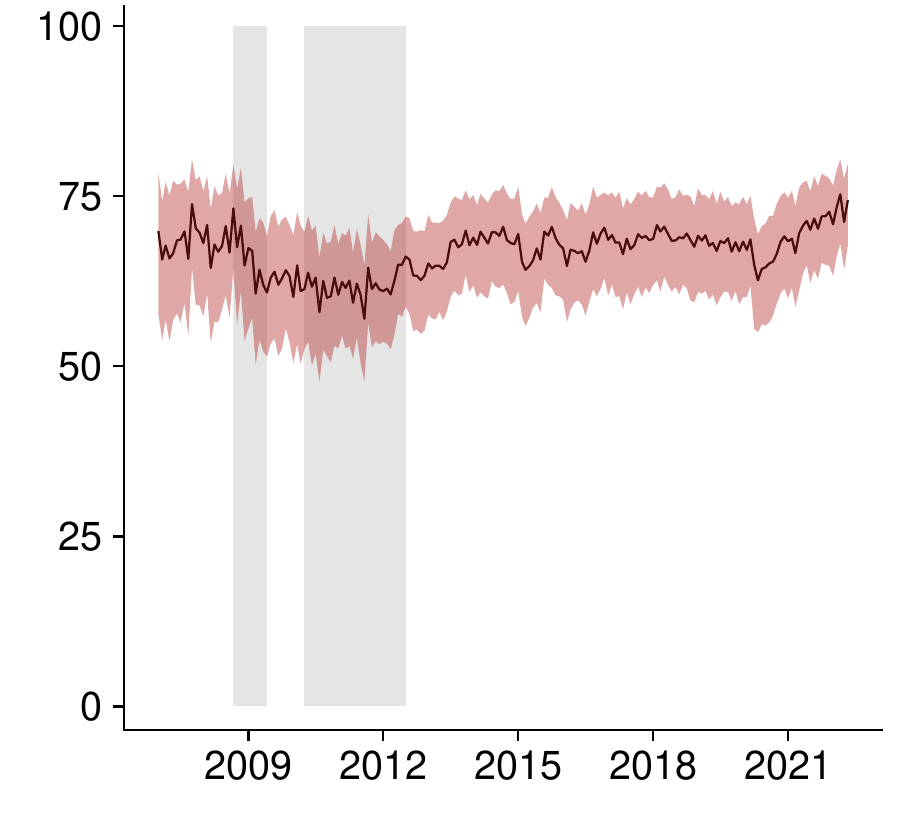}
\end{minipage}
\begin{minipage}{0.24\textwidth}
\centering
\includegraphics[scale=.24]{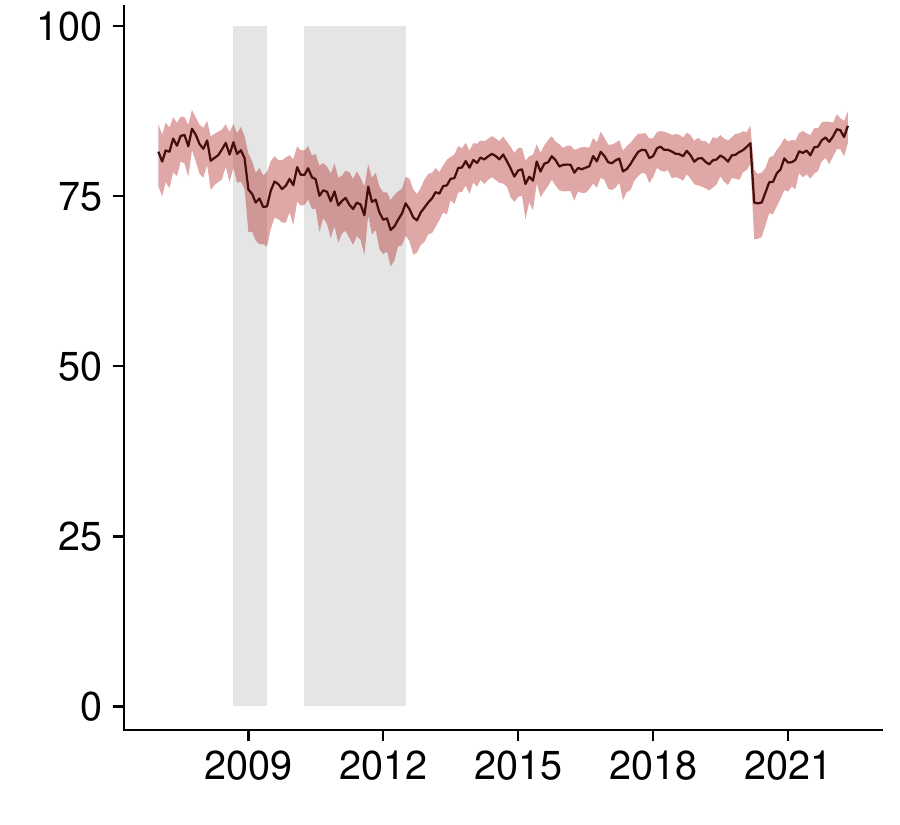}
\end{minipage}
\begin{minipage}{0.24\textwidth}
\centering
\includegraphics[scale=.24]{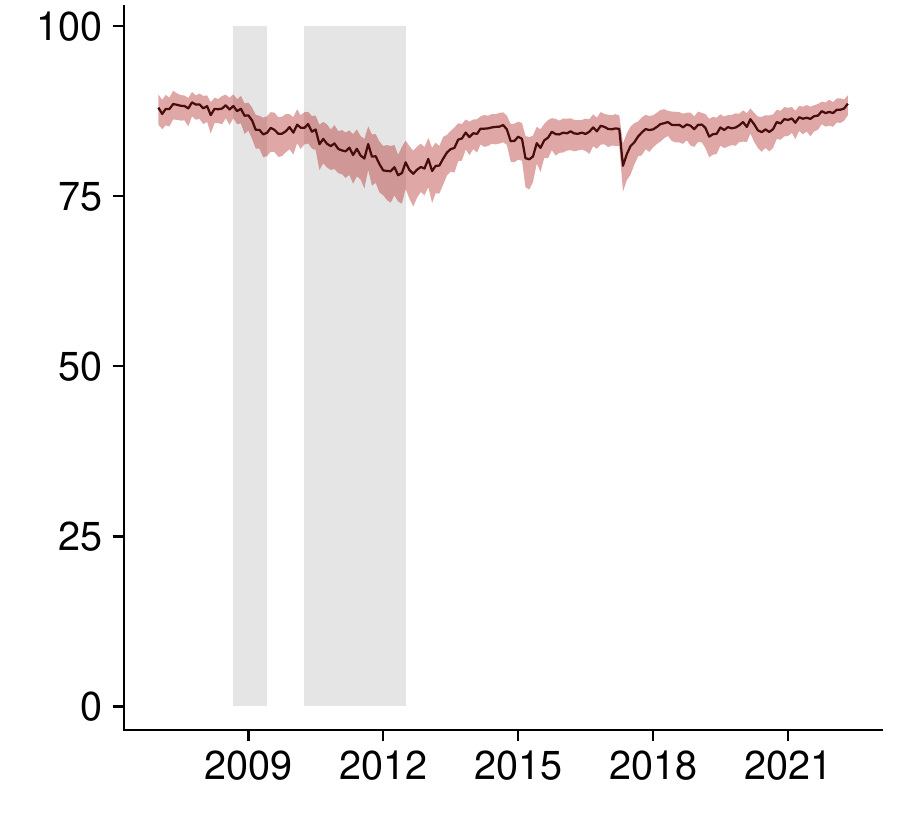}
\end{minipage}
\begin{minipage}{0.24\textwidth}
\centering
\includegraphics[scale=.24]{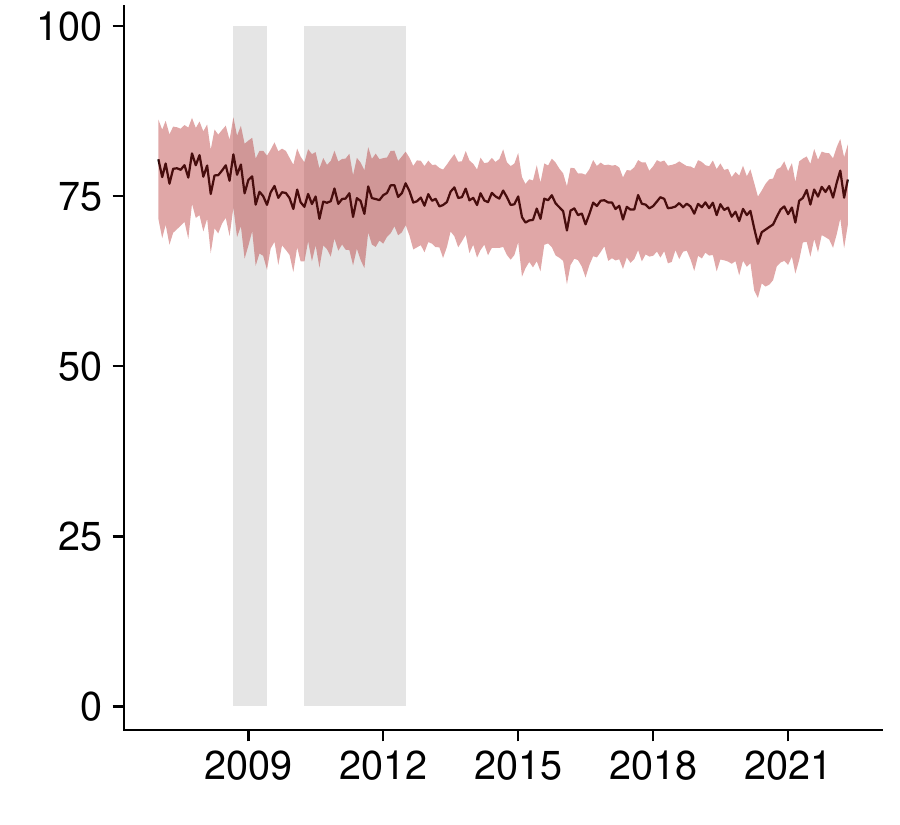}
\end{minipage}
\begin{minipage}{0.24\textwidth}
\centering
\includegraphics[scale=.24]{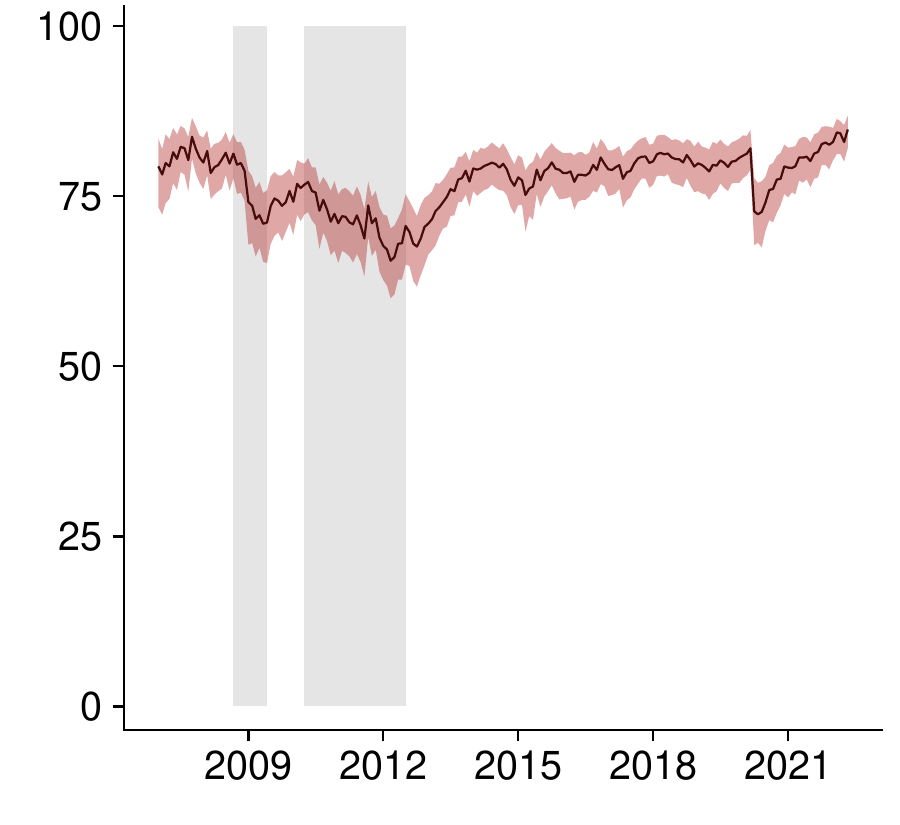}
\end{minipage}

\begin{minipage}{\textwidth}
\centering
\small \textit{\textbf{Borrow long}}
\end{minipage}

\begin{minipage}{0.21\textwidth}
\centering
\scriptsize \textit{AT}
\end{minipage}
\begin{minipage}{0.21\textwidth}
\centering
\scriptsize \textit{BE}
\end{minipage}
\begin{minipage}{0.21\textwidth}
\centering
\scriptsize \textit{DE}
\end{minipage}
\begin{minipage}{0.21\textwidth}
\centering
\scriptsize \textit{FI}
\end{minipage}
\begin{minipage}{0.21\textwidth}
\centering
\scriptsize \textit{FR}
\end{minipage}
\begin{minipage}{0.21\textwidth}
\centering
\scriptsize \textit{LU}
\end{minipage}
\begin{minipage}{0.21\textwidth}
\centering
\scriptsize \textit{NL}
\end{minipage}

\begin{minipage}{0.21\textwidth}
\centering
\includegraphics[scale=.24]{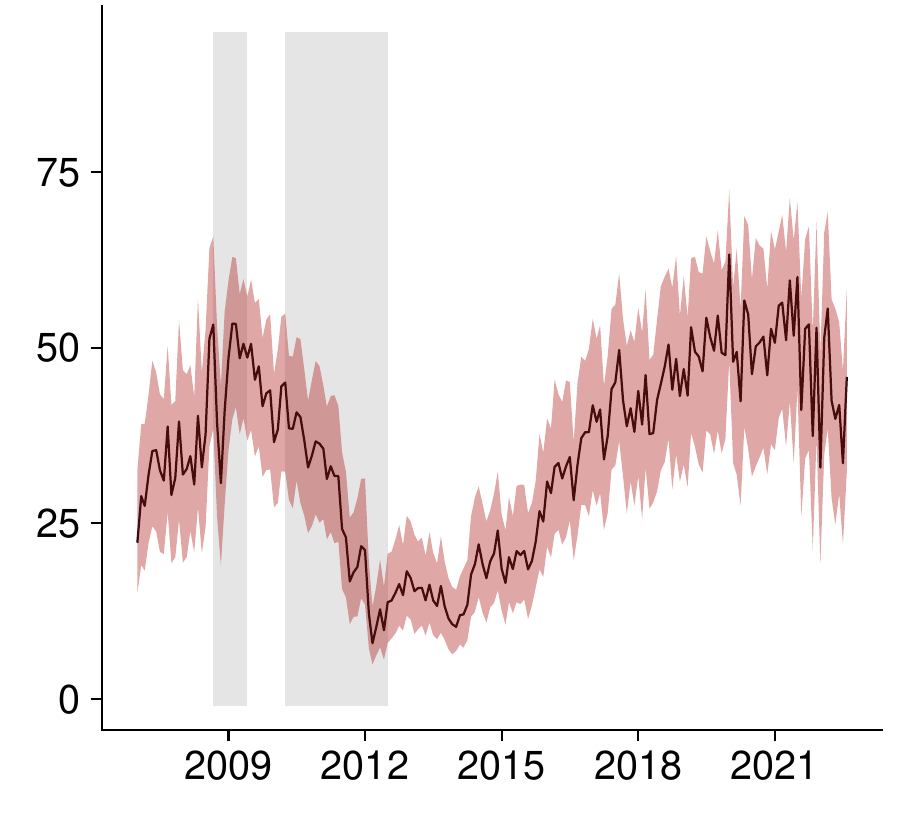}
\end{minipage}
\begin{minipage}{0.21\textwidth}
\centering
\includegraphics[scale=.24]{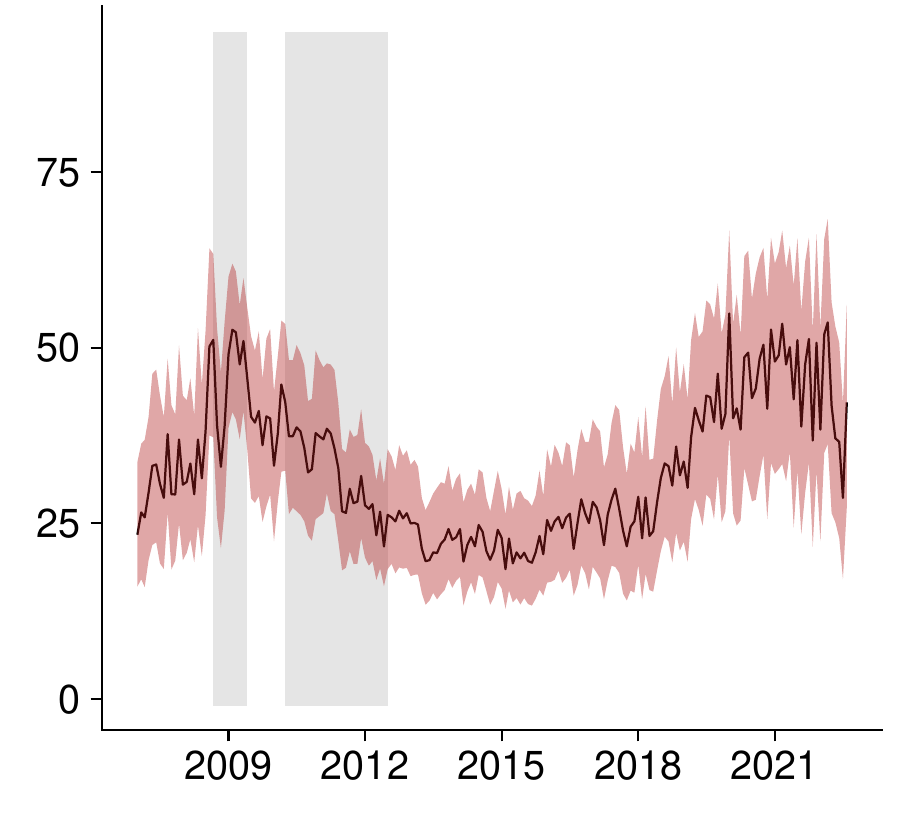}
\end{minipage}
\begin{minipage}{0.21\textwidth}
\centering
\includegraphics[scale=.24]{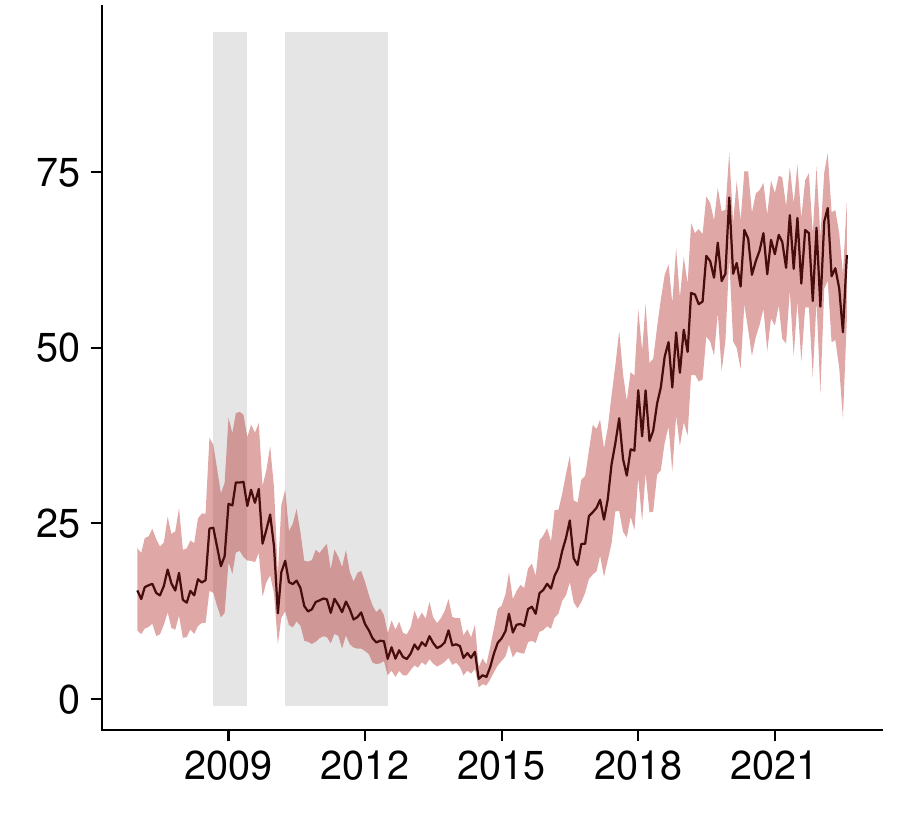}
\end{minipage}
\begin{minipage}{0.21\textwidth}
\centering
\includegraphics[scale=.24]{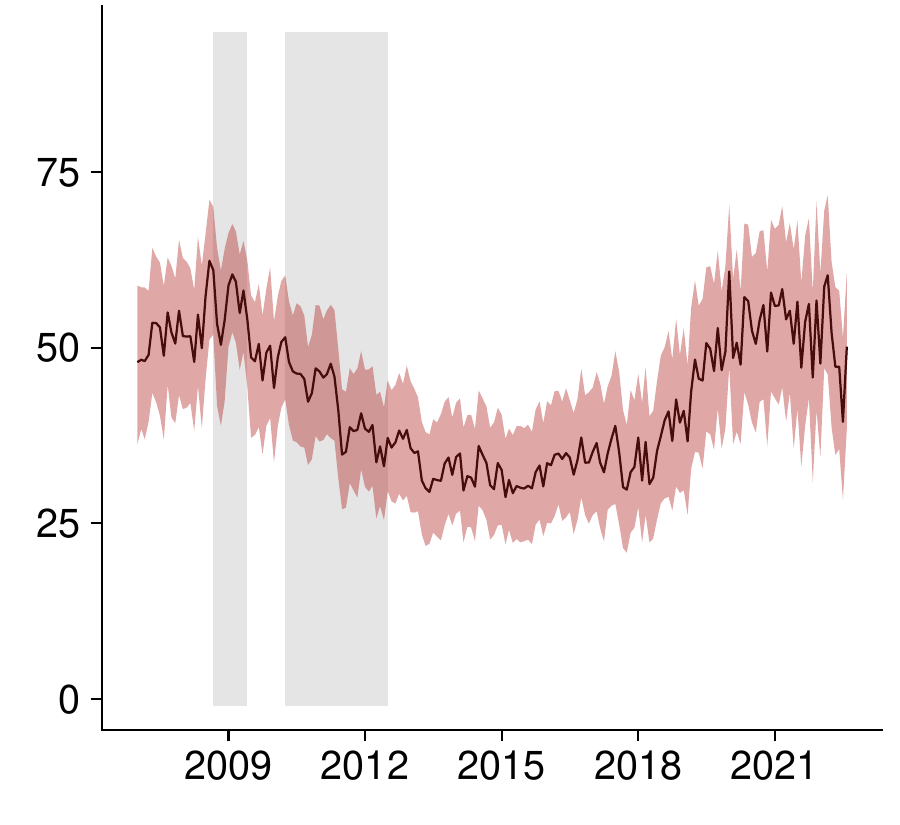}
\end{minipage}
\begin{minipage}{0.21\textwidth}
\centering
\includegraphics[scale=.24]{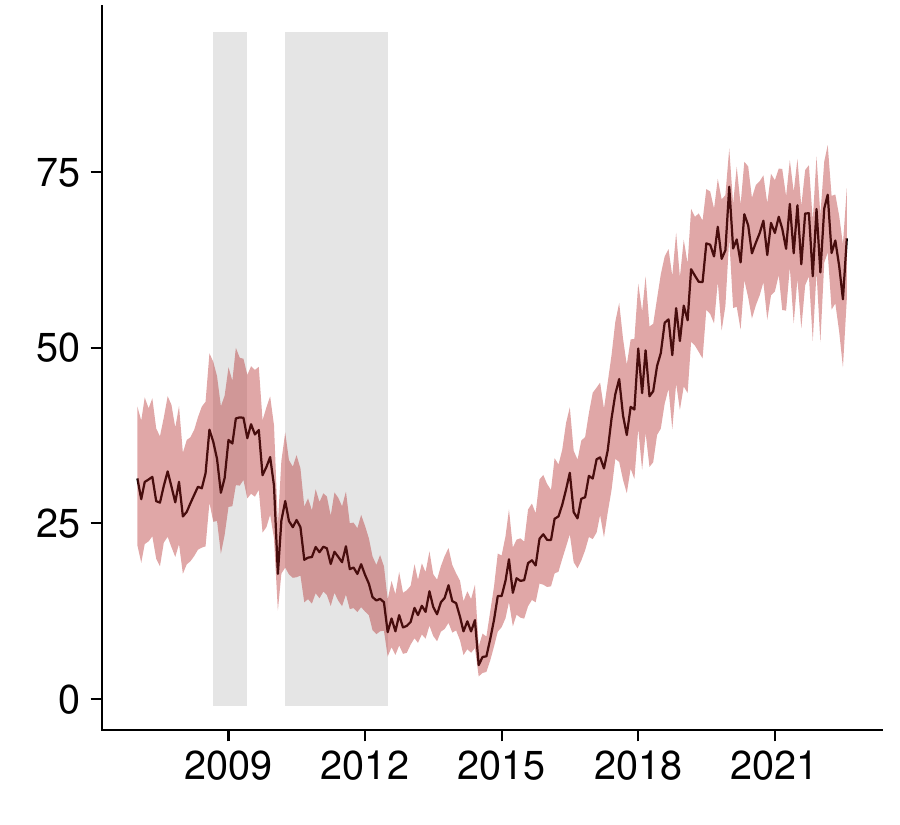}
\end{minipage}
\begin{minipage}{0.21\textwidth}
\centering
\includegraphics[scale=.24]{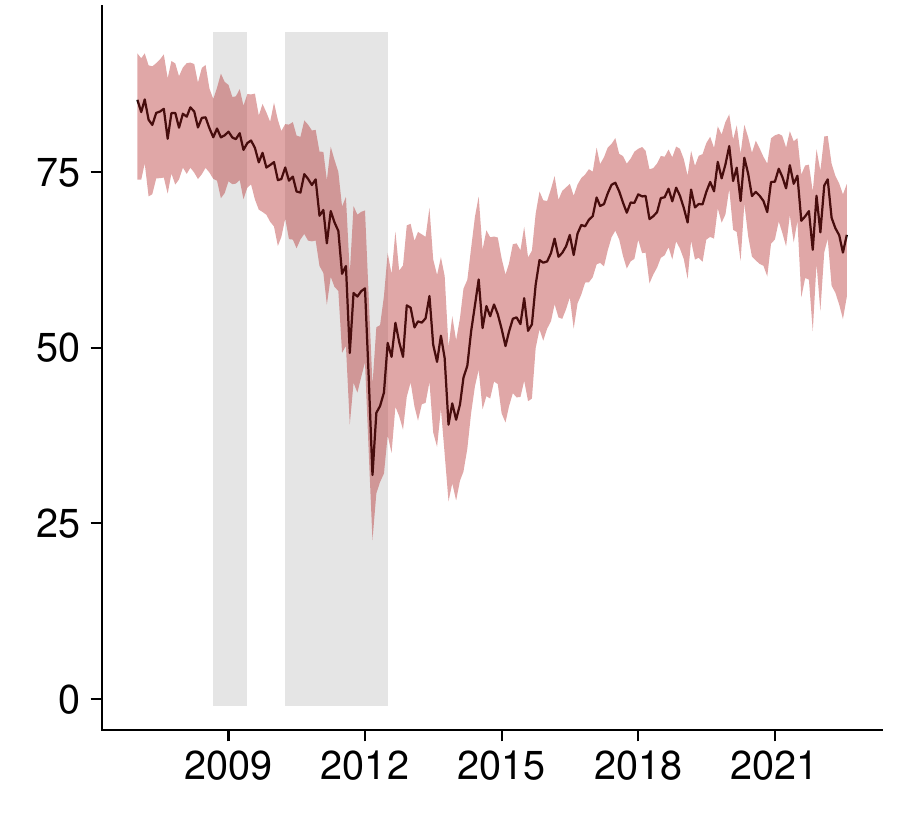}
\end{minipage}
\begin{minipage}{0.21\textwidth}
\centering
\includegraphics[scale=.24]{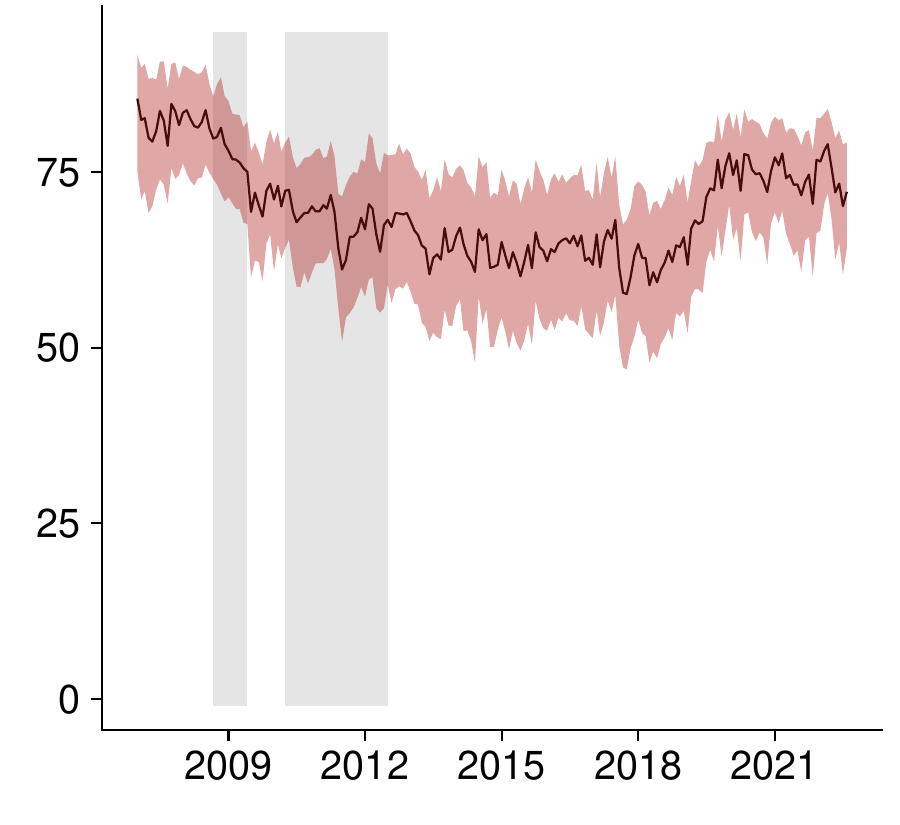}
\end{minipage}

\begin{minipage}{0.24\textwidth}
\centering
\scriptsize \textit{GR}
\end{minipage}
\begin{minipage}{0.24\textwidth}
\centering
\scriptsize \textit{IE}
\end{minipage}
\begin{minipage}{0.24\textwidth}
\centering
\scriptsize \textit{IT}
\end{minipage}
\begin{minipage}{0.24\textwidth}
\centering
\scriptsize \textit{PT}
\end{minipage}
\begin{minipage}{0.24\textwidth}
\centering
\scriptsize \textit{ES}
\end{minipage}

\begin{minipage}{0.24\textwidth}
\centering
\includegraphics[scale=.24]{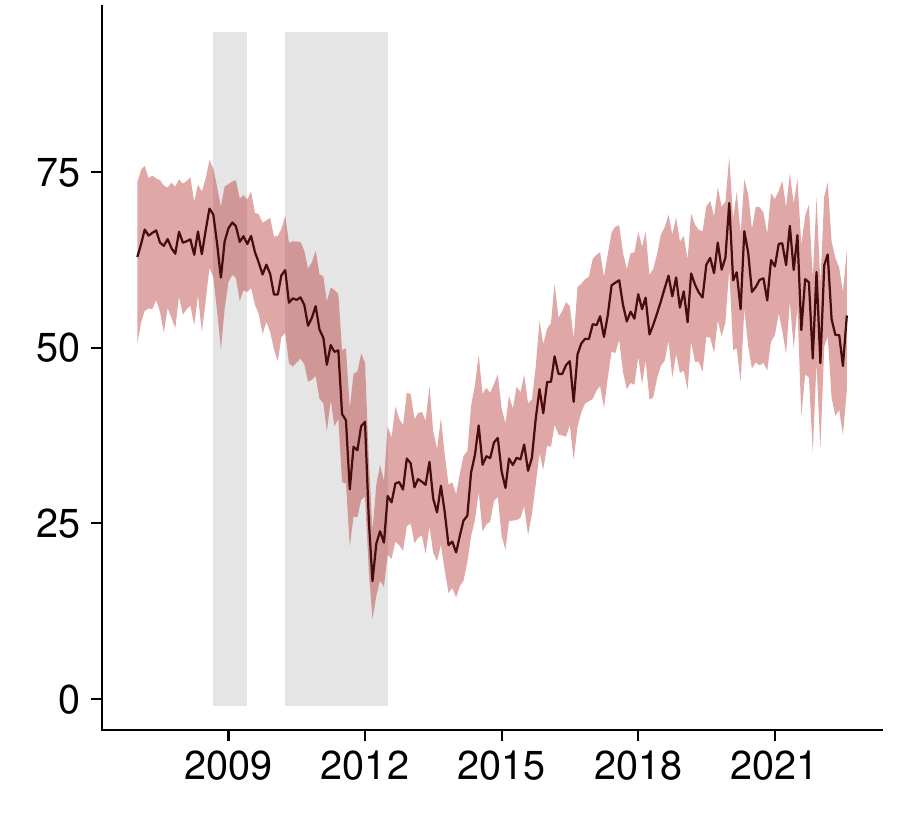}
\end{minipage}
\begin{minipage}{0.24\textwidth}
\centering
\includegraphics[scale=.24]{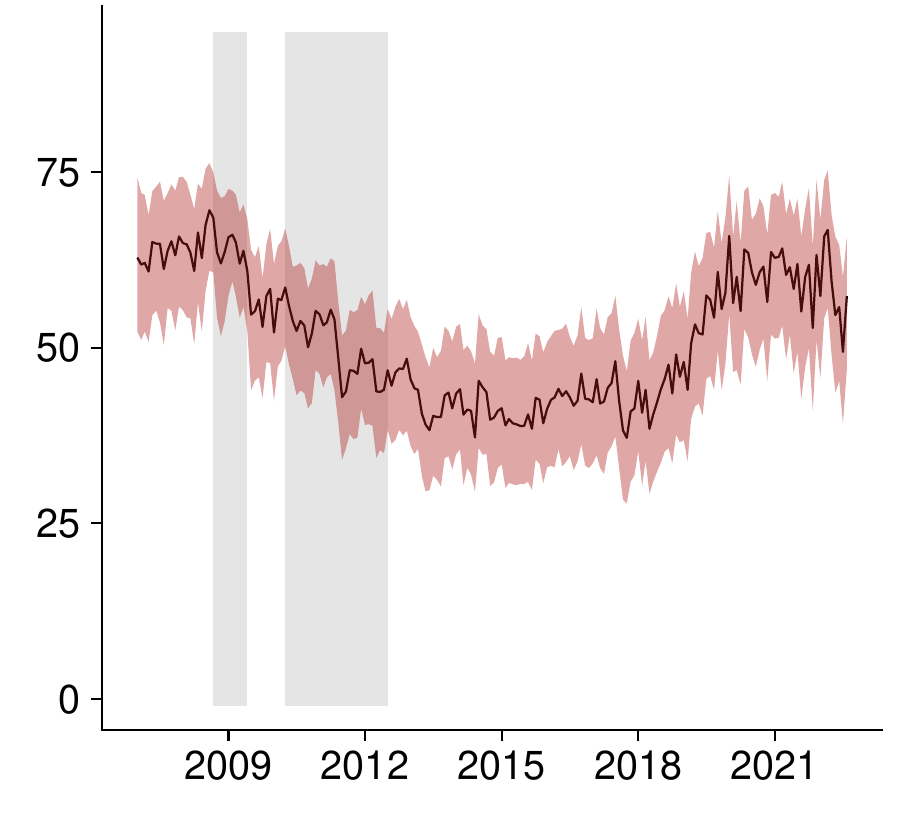}
\end{minipage}
\begin{minipage}{0.24\textwidth}
\centering
\includegraphics[scale=.24]{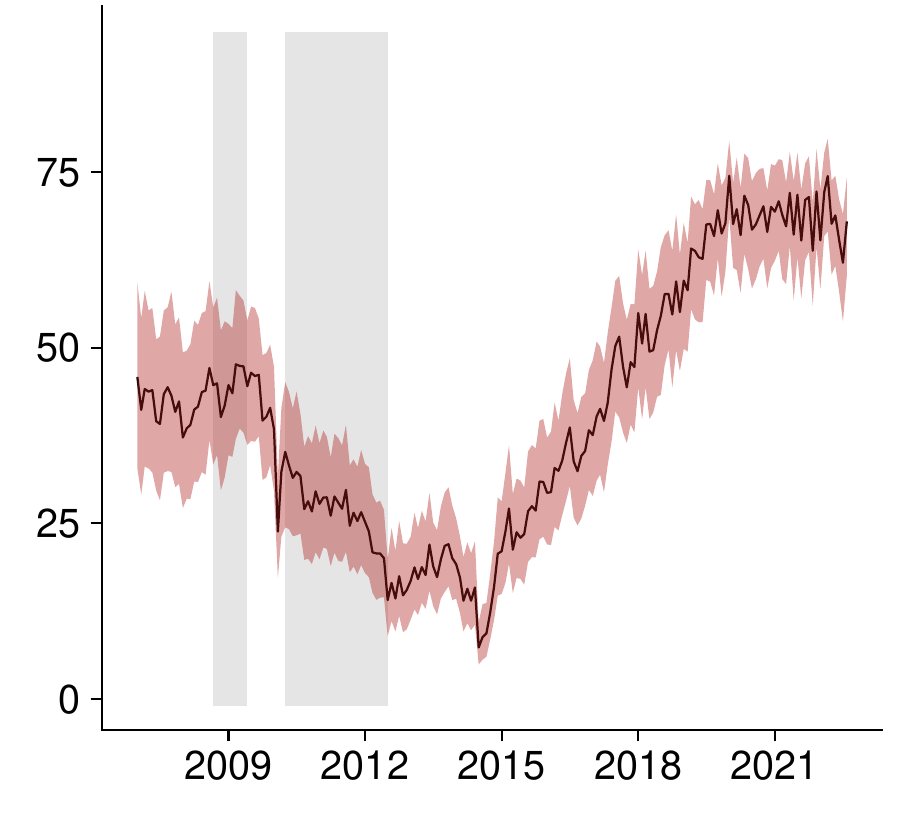}
\end{minipage}
\begin{minipage}{0.24\textwidth}
\centering
\includegraphics[scale=.24]{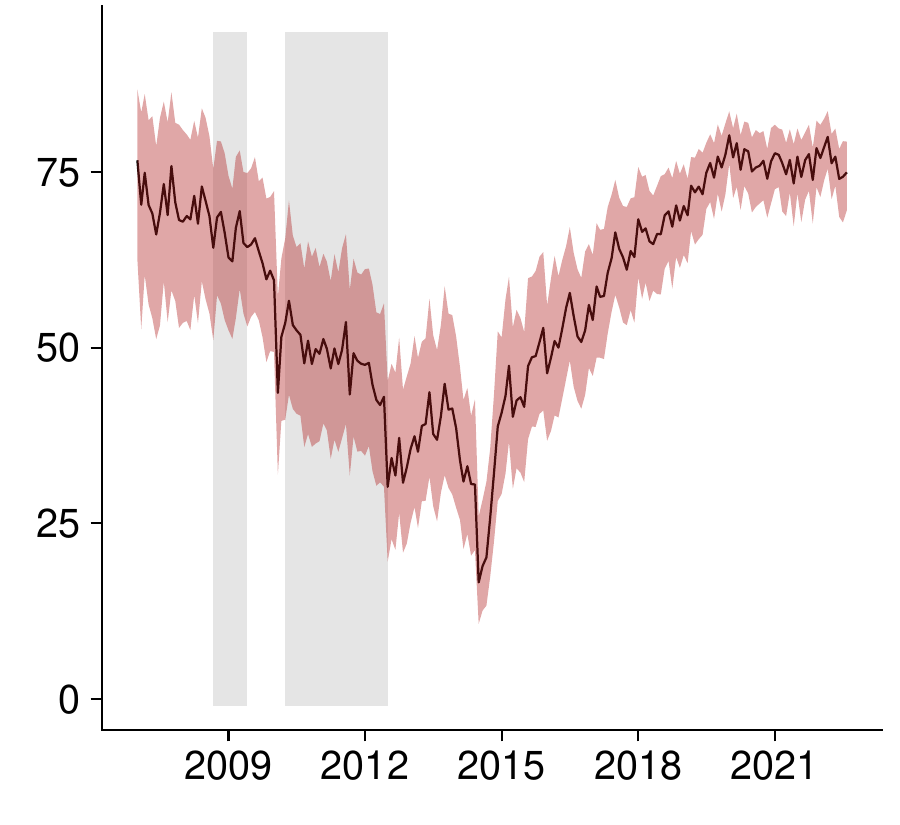}
\end{minipage}
\begin{minipage}{0.24\textwidth}
\centering
\includegraphics[scale=.24]{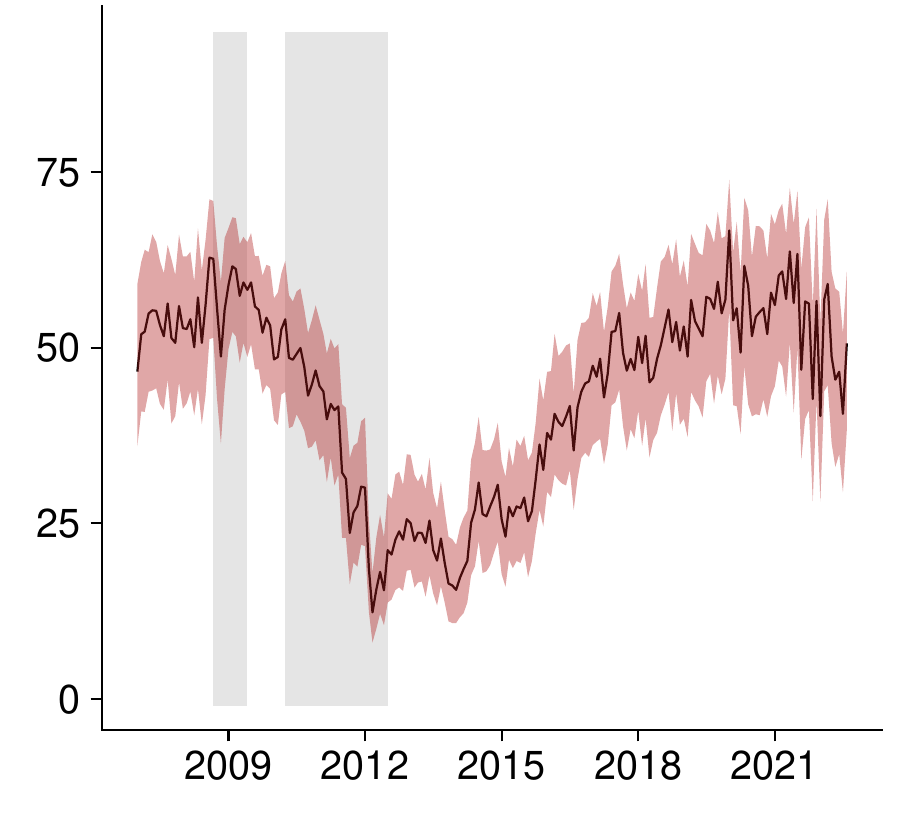}
\end{minipage}

\begin{minipage}{\linewidth}
\vspace{2pt}
\scriptsize \emph{Note:} This index indicates the share of spillovers for each country according to \cite{diebold2009measuring} and is estimated based on an expanding window. The solid line is the posterior median alongside the $68\%$ posterior credible set. The grey shaded area depicts the period of the ESDC.
\end{minipage}
\end{figure}
\end{landscape}

\clearpage

\begin{landscape}
\begin{figure}[htbp!]
\caption{Spillover index for each country and the 12-months forecast horizon. \label{fig:DYindex_detail_12_2}}

\begin{minipage}{\textwidth}
\centering
\small \textit{\textbf{Borrow short}}
\end{minipage}

\begin{minipage}{0.21\textwidth}
\centering
\scriptsize \textit{AT}
\end{minipage}
\begin{minipage}{0.21\textwidth}
\centering
\scriptsize \textit{BE}
\end{minipage}
\begin{minipage}{0.21\textwidth}
\centering
\scriptsize \textit{DE}
\end{minipage}
\begin{minipage}{0.21\textwidth}
\centering
\scriptsize \textit{FI}
\end{minipage}
\begin{minipage}{0.21\textwidth}
\centering
\scriptsize \textit{FR}
\end{minipage}
\begin{minipage}{0.21\textwidth}
\centering
\scriptsize \textit{LU}
\end{minipage}
\begin{minipage}{0.21\textwidth}
\centering
\scriptsize \textit{NL}
\end{minipage}

\begin{minipage}{0.21\textwidth}
\centering
\includegraphics[scale=.24]{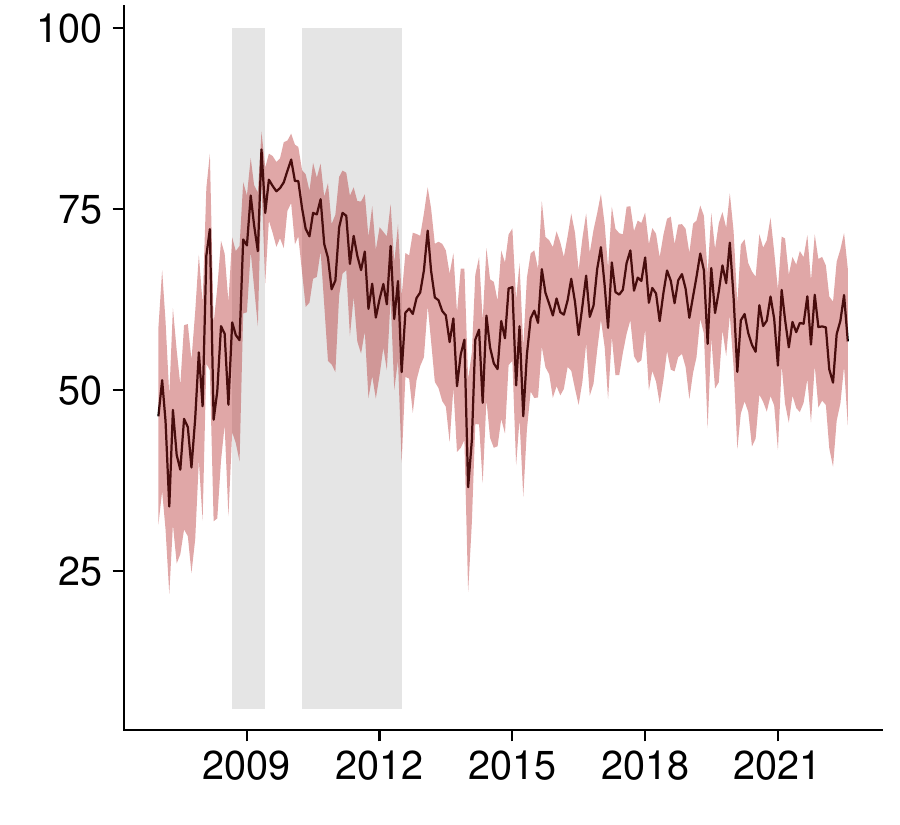}
\end{minipage}
\begin{minipage}{0.21\textwidth}
\centering
\includegraphics[scale=.24]{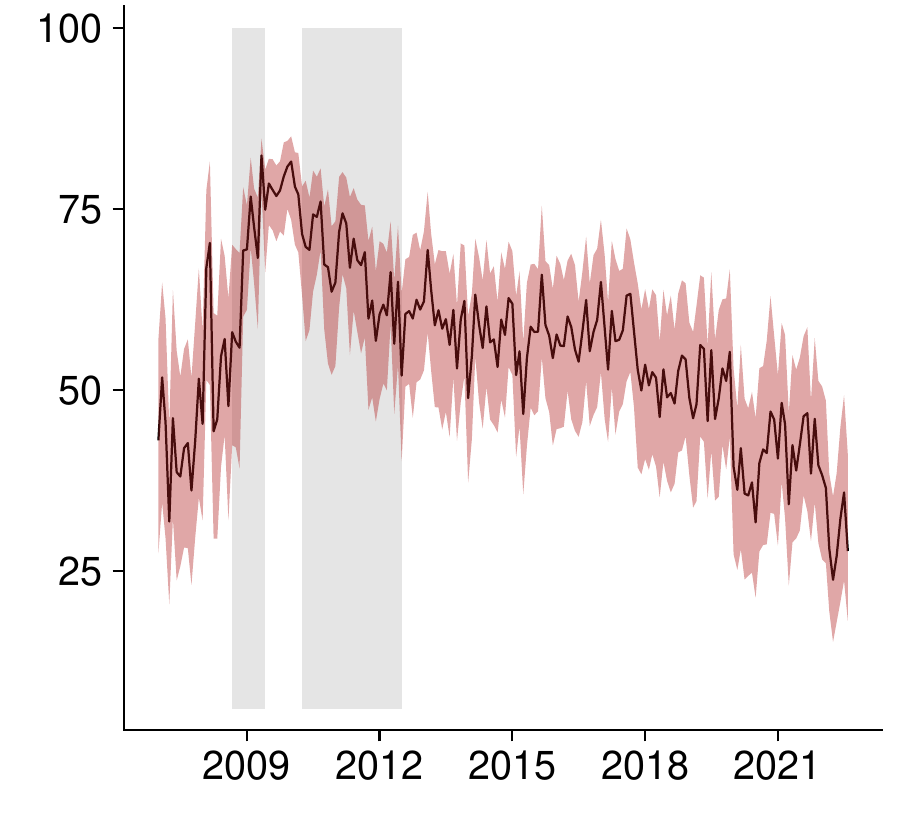}
\end{minipage}
\begin{minipage}{0.21\textwidth}
\centering
\includegraphics[scale=.24]{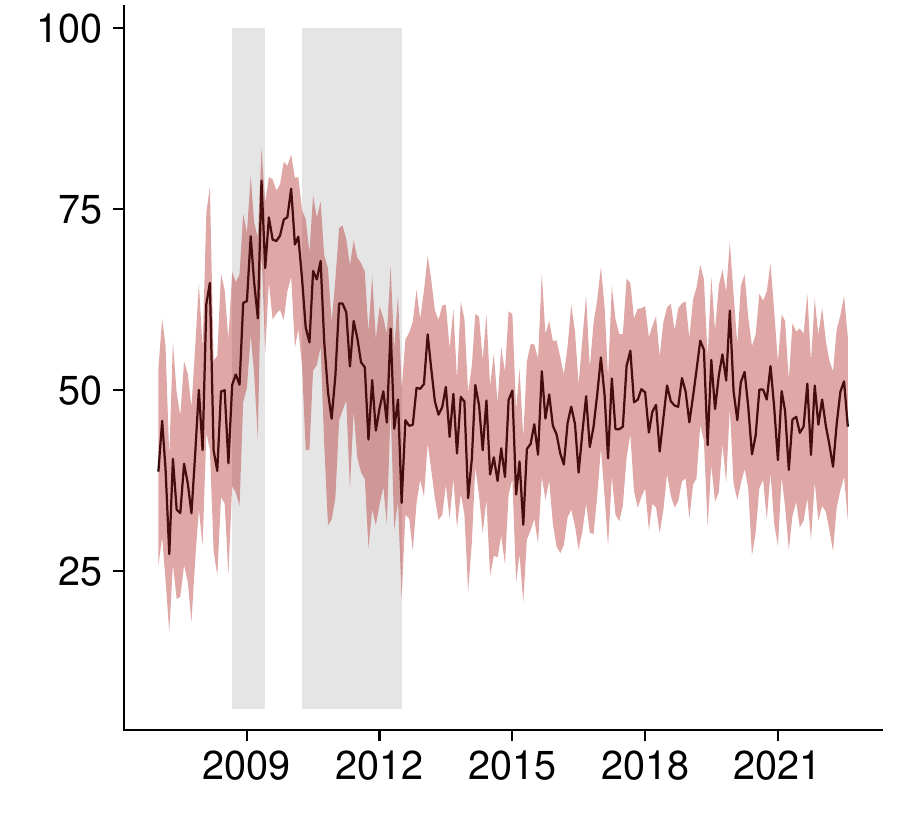}
\end{minipage}
\begin{minipage}{0.21\textwidth}
\centering
\includegraphics[scale=.24]{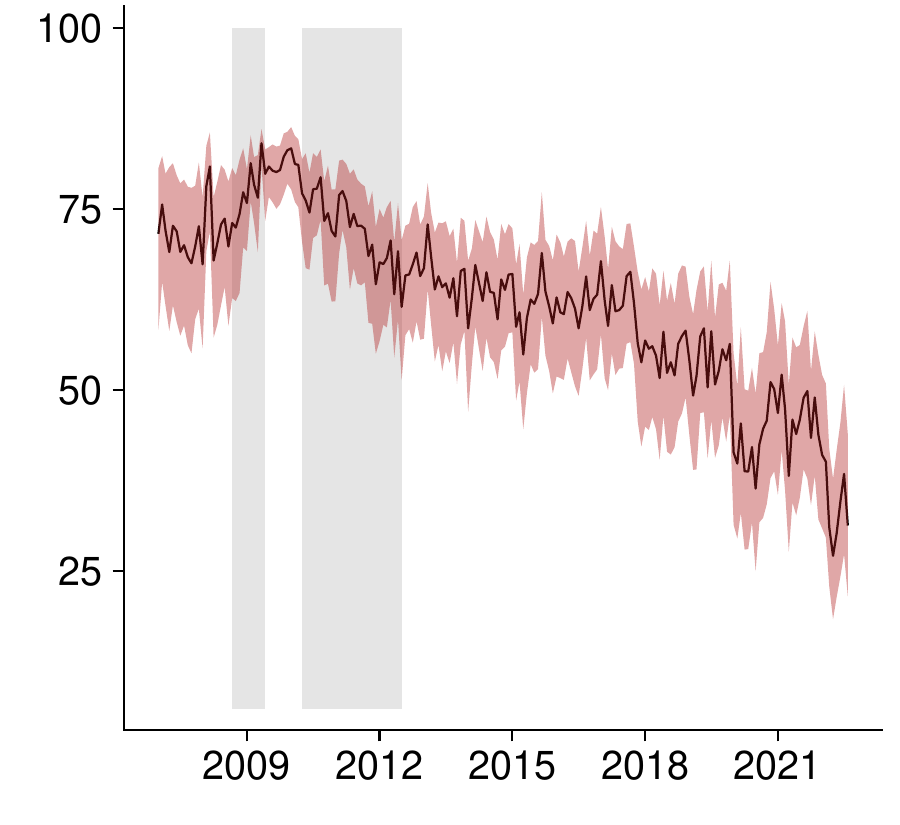}
\end{minipage}
\begin{minipage}{0.21\textwidth}
\centering
\includegraphics[scale=.24]{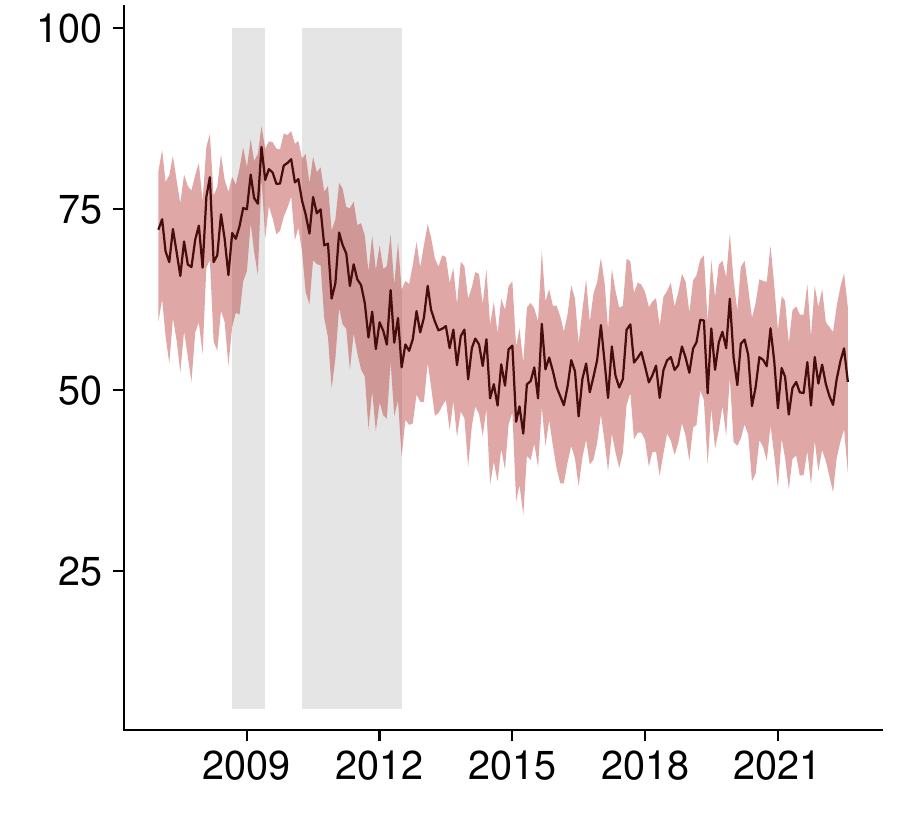}
\end{minipage}
\begin{minipage}{0.21\textwidth}
\centering
\includegraphics[scale=.24]{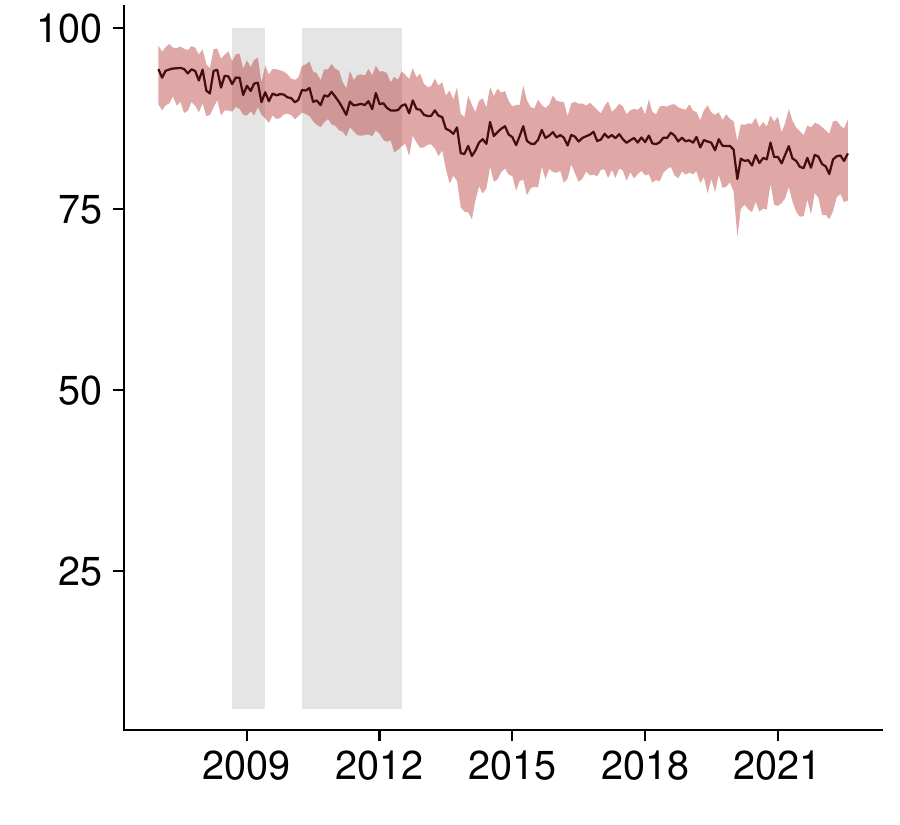}
\end{minipage}
\begin{minipage}{0.21\textwidth}
\centering
\includegraphics[scale=.24]{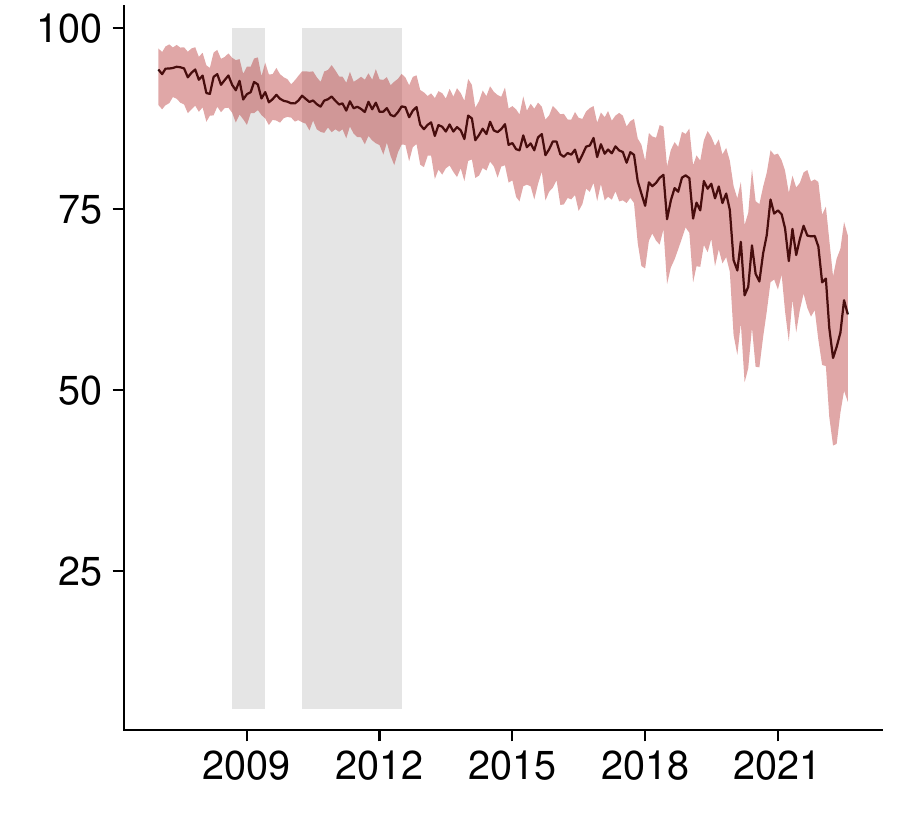}
\end{minipage}

\begin{minipage}{0.24\textwidth}
\centering
\scriptsize \textit{GR}
\end{minipage}
\begin{minipage}{0.24\textwidth}
\centering
\scriptsize \textit{IE}
\end{minipage}
\begin{minipage}{0.24\textwidth}
\centering
\scriptsize \textit{IT}
\end{minipage}
\begin{minipage}{0.24\textwidth}
\centering
\scriptsize \textit{PT}
\end{minipage}
\begin{minipage}{0.24\textwidth}
\centering
\scriptsize \textit{ES}
\end{minipage}

\begin{minipage}{0.24\textwidth}
\centering
\includegraphics[scale=.24]{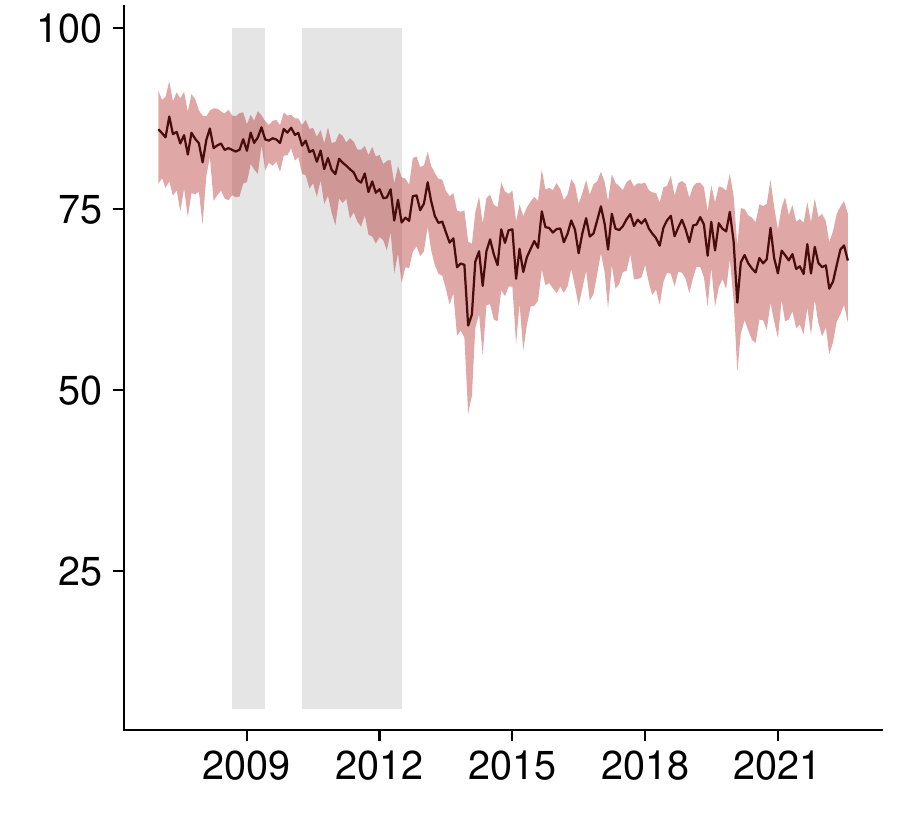}
\end{minipage}
\begin{minipage}{0.24\textwidth}
\centering
\includegraphics[scale=.24]{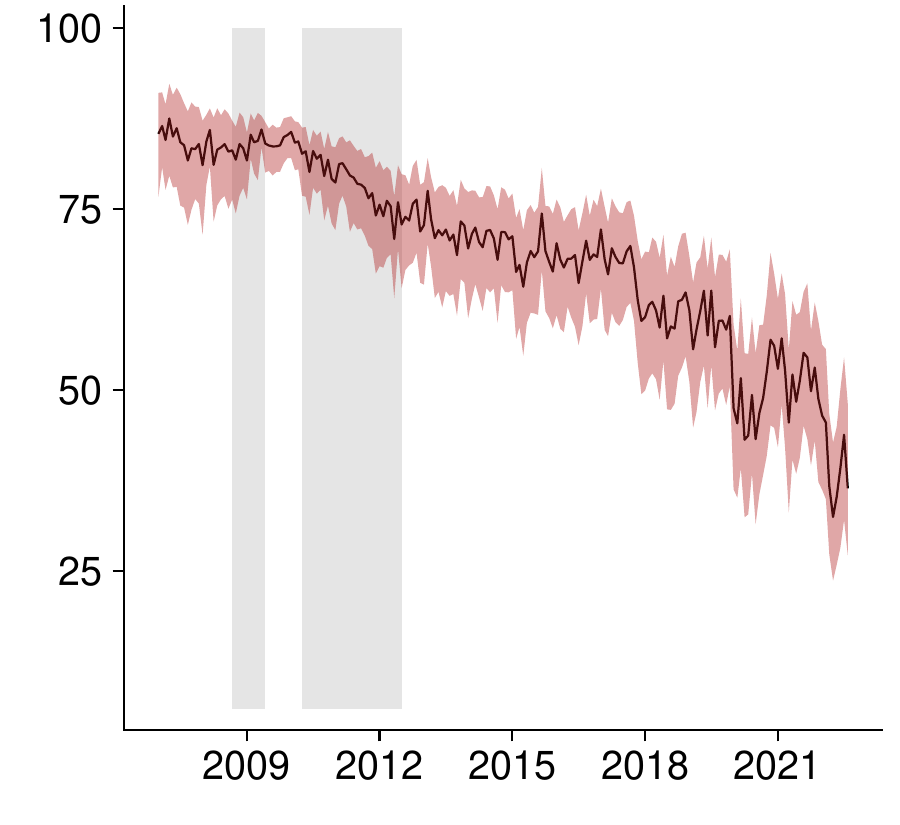}
\end{minipage}
\begin{minipage}{0.24\textwidth}
\centering
\includegraphics[scale=.24]{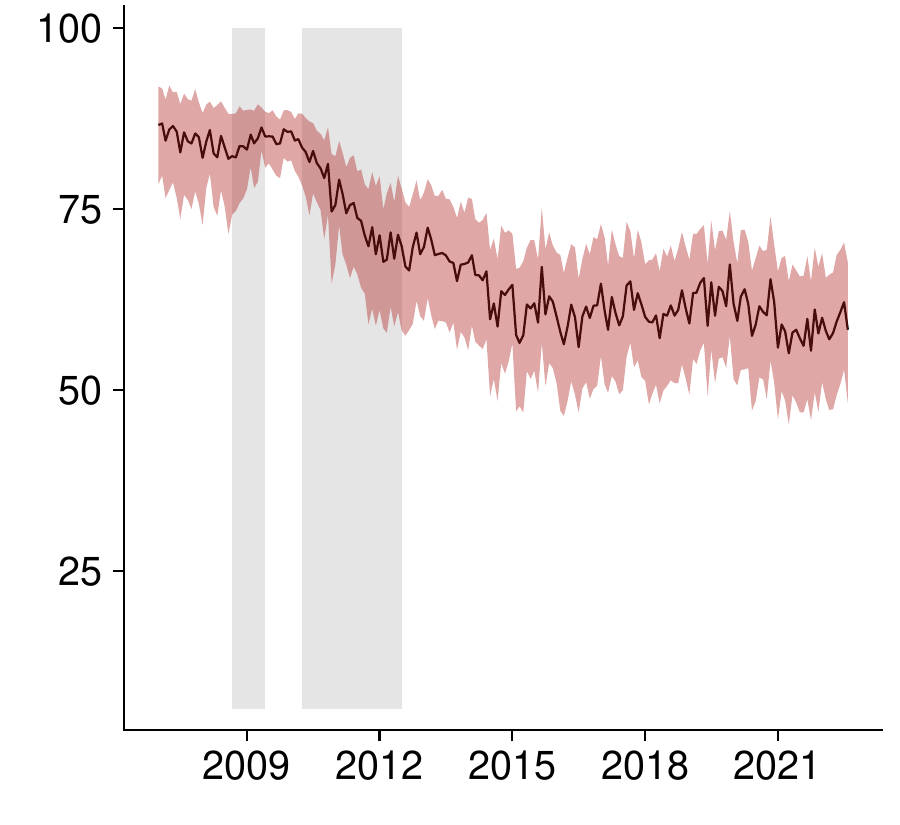}
\end{minipage}
\begin{minipage}{0.24\textwidth}
\centering
\includegraphics[scale=.24]{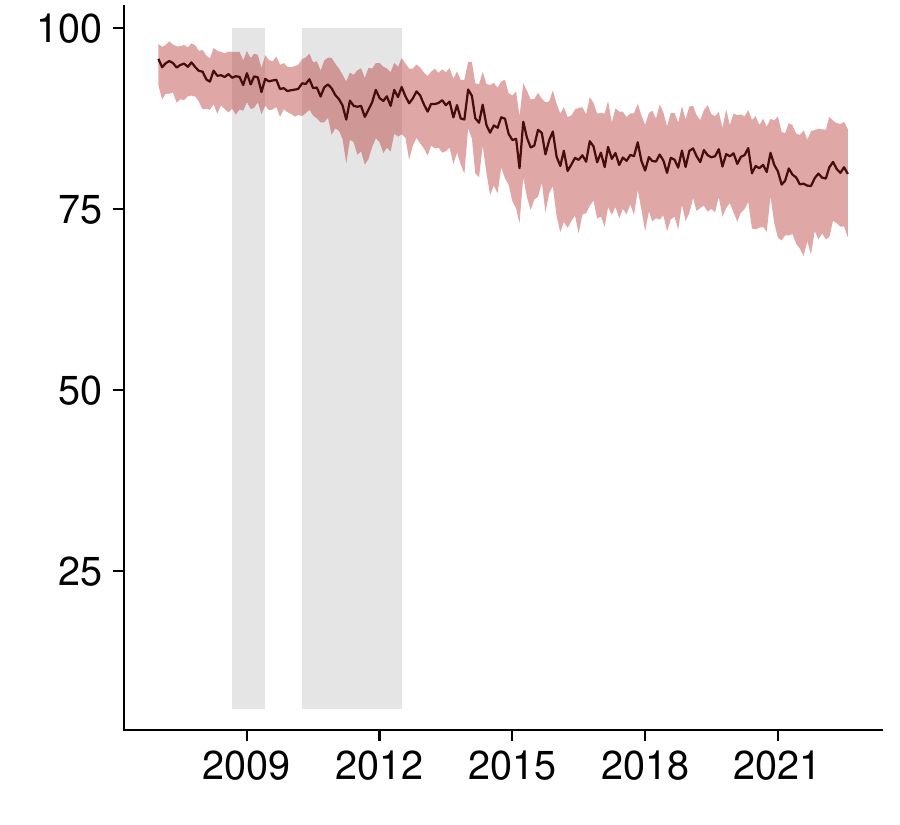}
\end{minipage}
\begin{minipage}{0.24\textwidth}
\centering
\includegraphics[scale=.24]{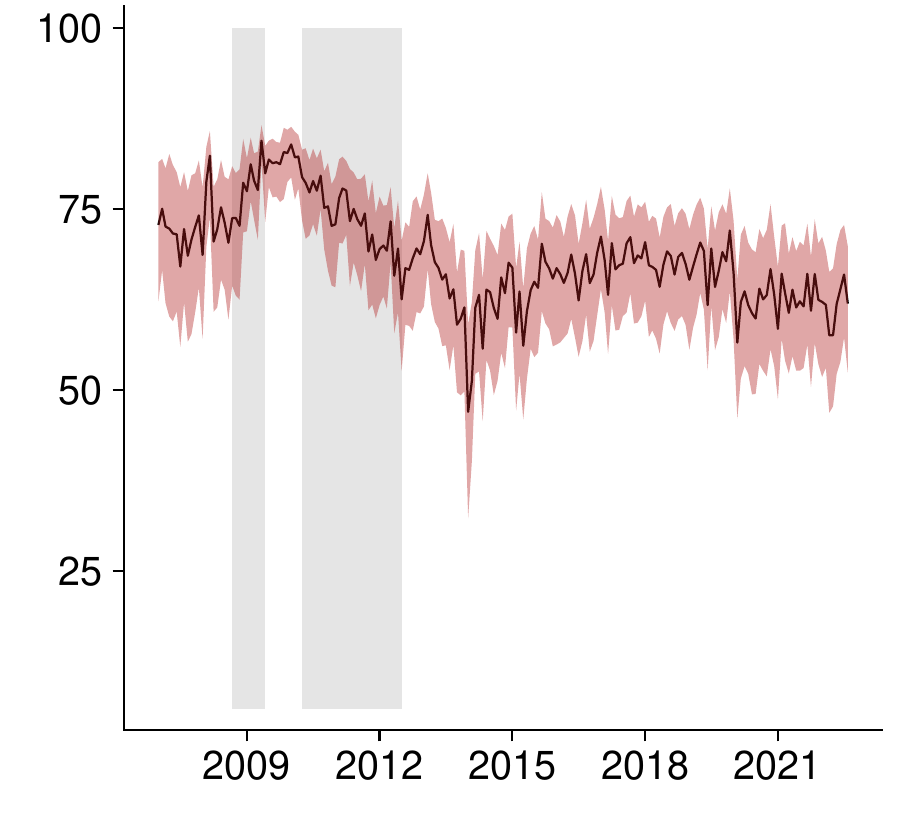}
\end{minipage}

\begin{minipage}{\textwidth}
\centering
\small \textit{\textbf{TARGET2}}
\end{minipage}

\begin{minipage}{0.21\textwidth}
\centering
\scriptsize \textit{AT}
\end{minipage}
\begin{minipage}{0.21\textwidth}
\centering
\scriptsize \textit{BE}
\end{minipage}
\begin{minipage}{0.21\textwidth}
\centering
\scriptsize \textit{DE}
\end{minipage}
\begin{minipage}{0.21\textwidth}
\centering
\scriptsize \textit{FI}
\end{minipage}
\begin{minipage}{0.21\textwidth}
\centering
\scriptsize \textit{FR}
\end{minipage}
\begin{minipage}{0.21\textwidth}
\centering
\scriptsize \textit{LU}
\end{minipage}
\begin{minipage}{0.21\textwidth}
\centering
\scriptsize \textit{NL}
\end{minipage}

\begin{minipage}{0.21\textwidth}
\centering
\includegraphics[scale=.24]{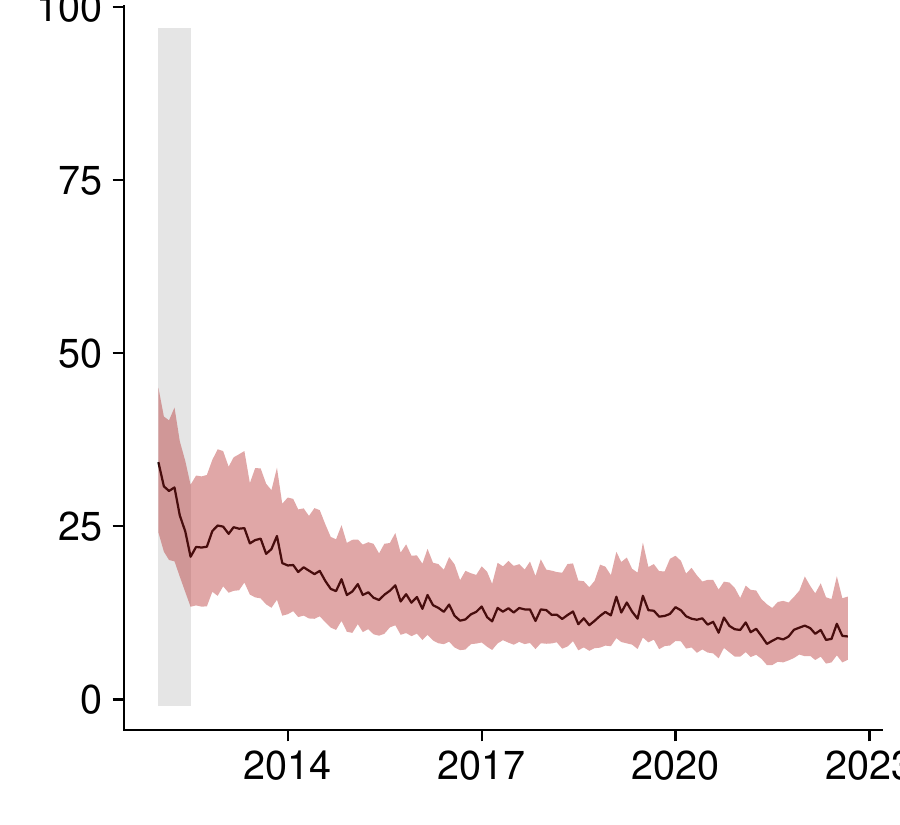}
\end{minipage}
\begin{minipage}{0.21\textwidth}
\centering
\includegraphics[scale=.24]{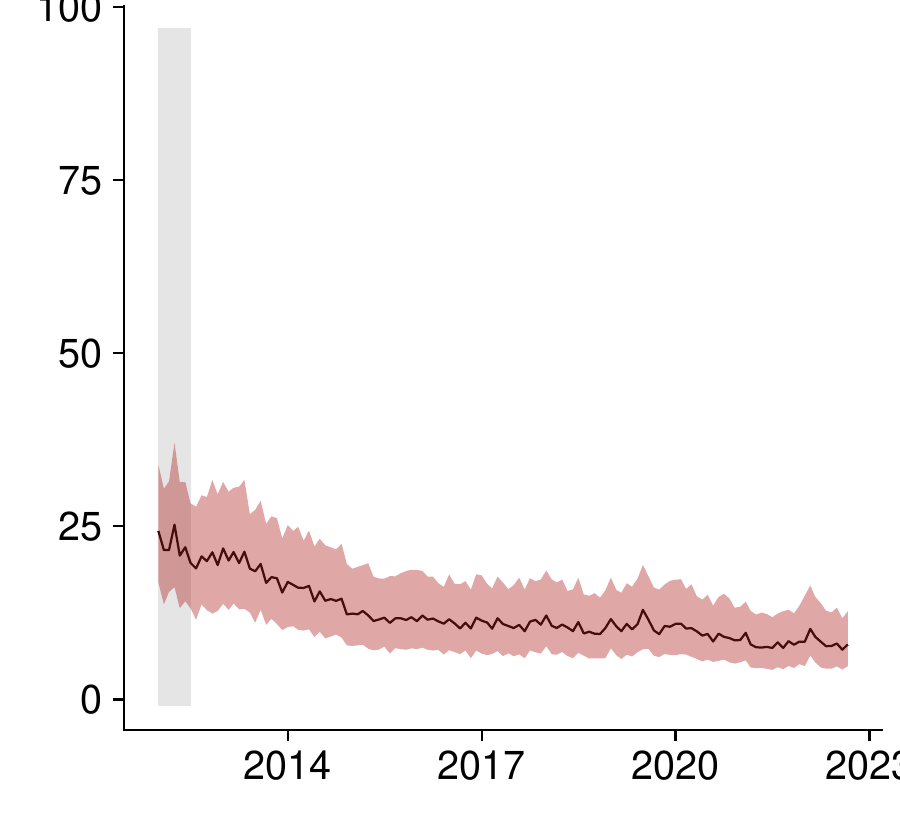}
\end{minipage}
\begin{minipage}{0.21\textwidth}
\centering
\includegraphics[scale=.24]{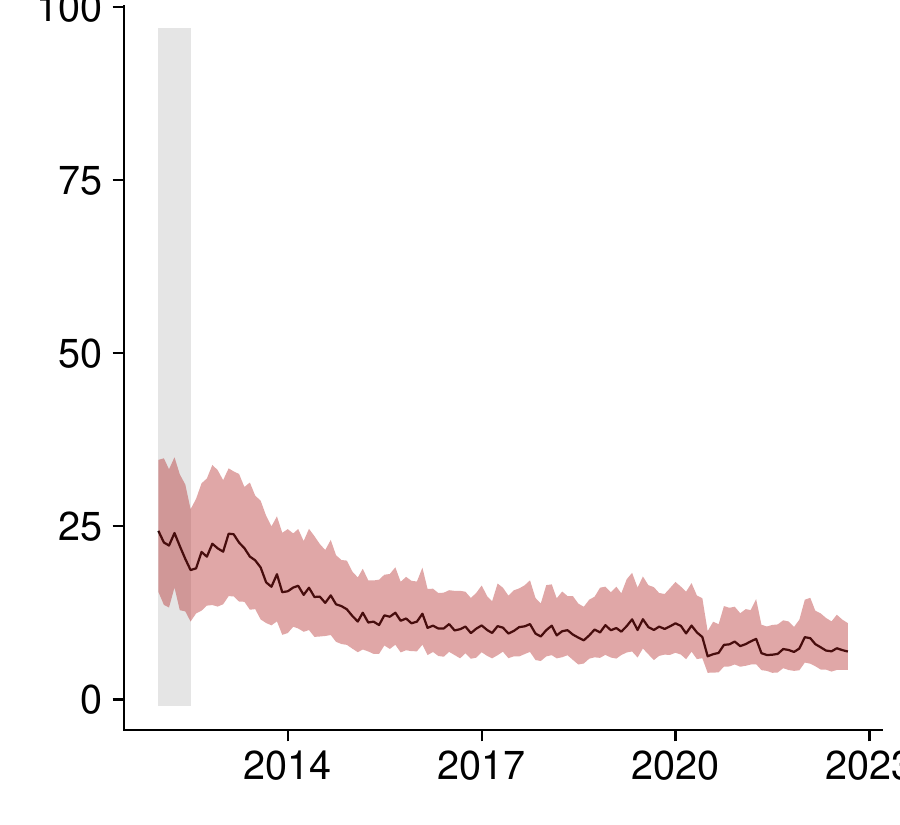}
\end{minipage}
\begin{minipage}{0.21\textwidth}
\centering
\includegraphics[scale=.24]{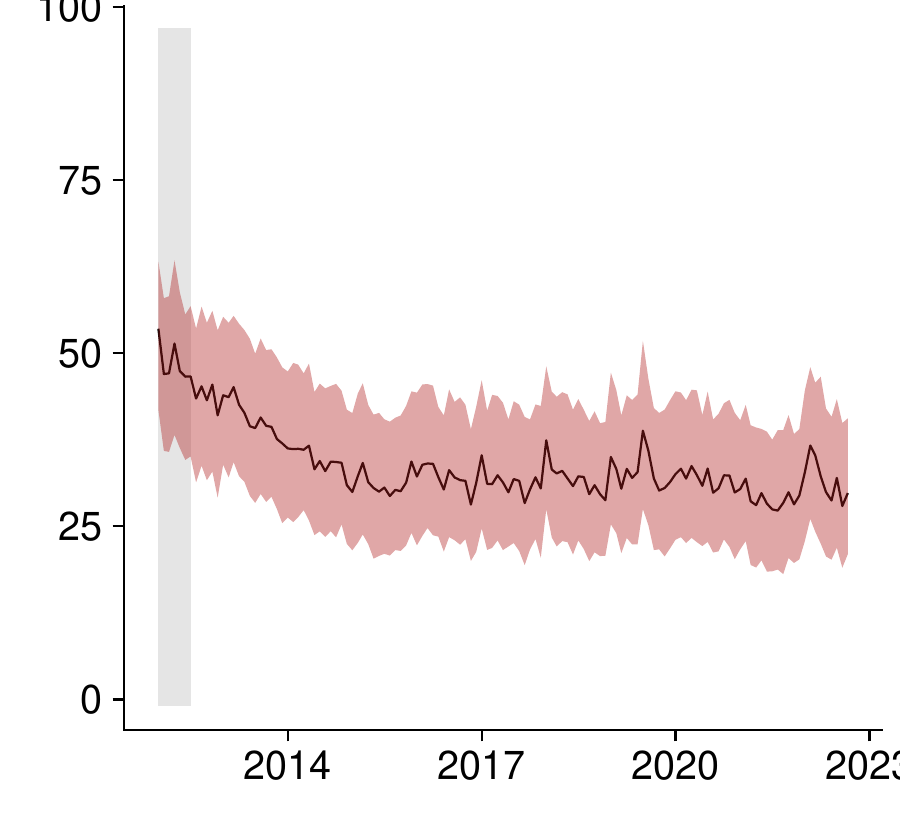}
\end{minipage}
\begin{minipage}{0.21\textwidth}
\centering
\includegraphics[scale=.24]{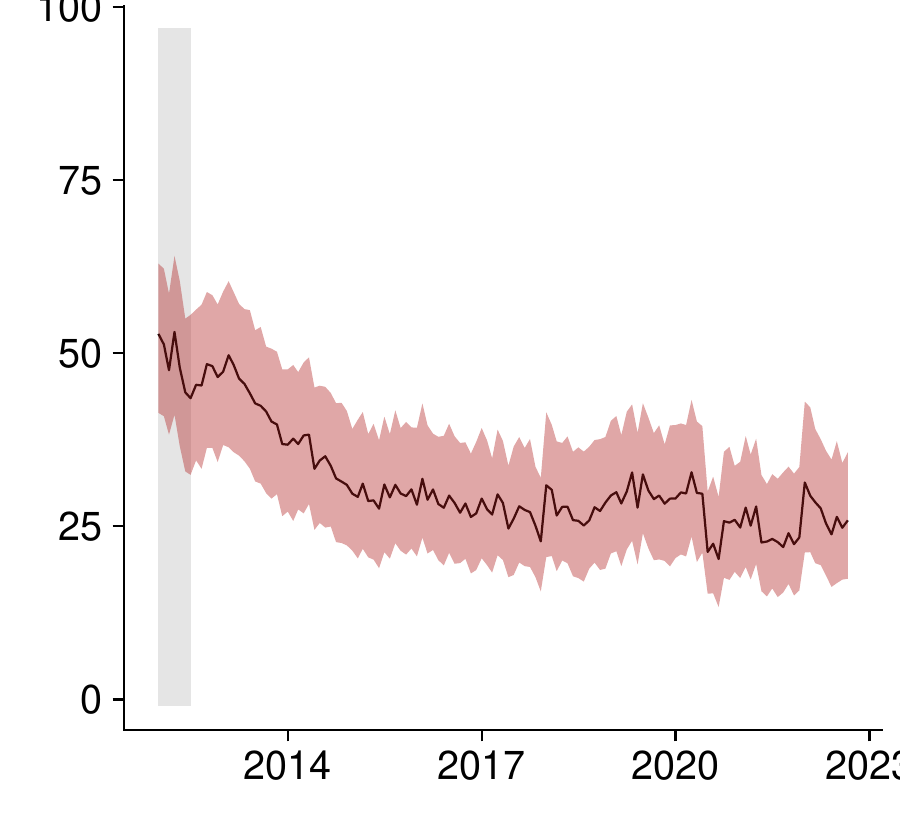}
\end{minipage}
\begin{minipage}{0.21\textwidth}
\centering
\includegraphics[scale=.24]{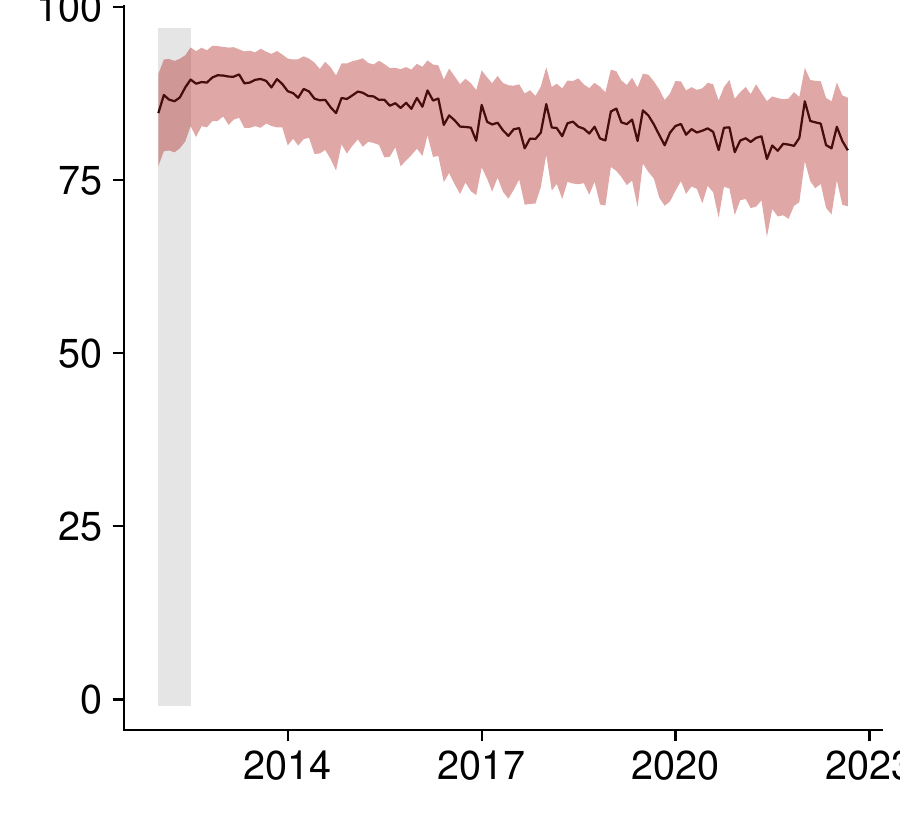}
\end{minipage}
\begin{minipage}{0.21\textwidth}
\centering
\includegraphics[scale=.24]{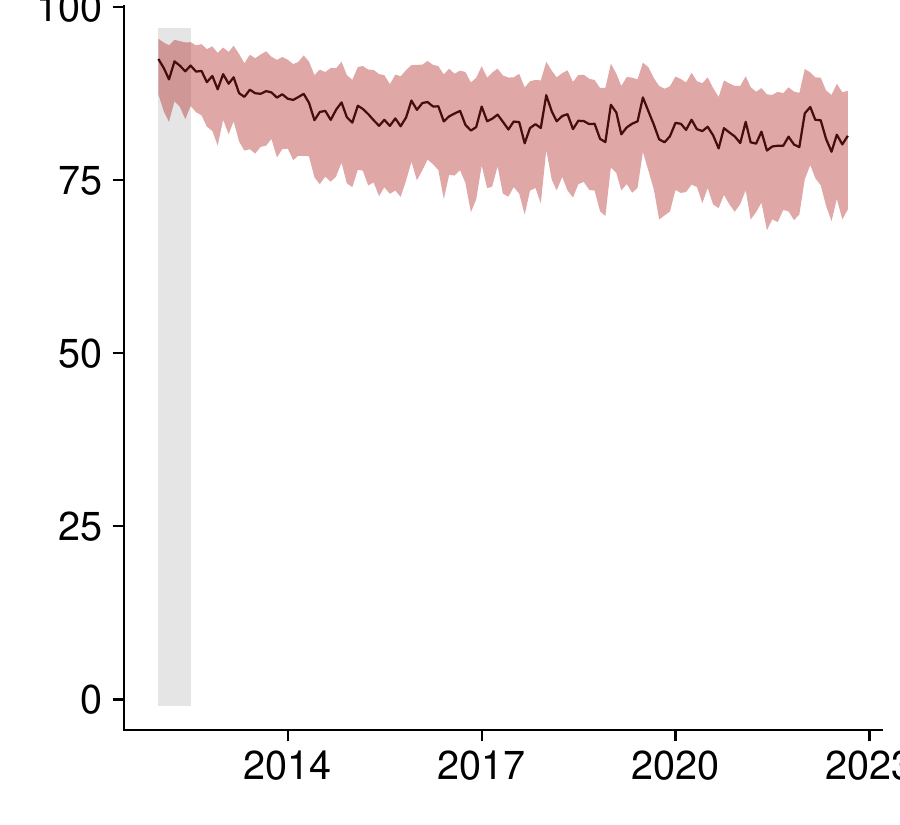}
\end{minipage}

\begin{minipage}{0.24\textwidth}
\centering
\scriptsize \textit{GR}
\end{minipage}
\begin{minipage}{0.24\textwidth}
\centering
\scriptsize \textit{IE}
\end{minipage}
\begin{minipage}{0.24\textwidth}
\centering
\scriptsize \textit{IT}
\end{minipage}
\begin{minipage}{0.24\textwidth}
\centering
\scriptsize \textit{PT}
\end{minipage}
\begin{minipage}{0.24\textwidth}
\centering
\scriptsize \textit{ES}
\end{minipage}

\begin{minipage}{0.24\textwidth}
\centering
\includegraphics[scale=.24]{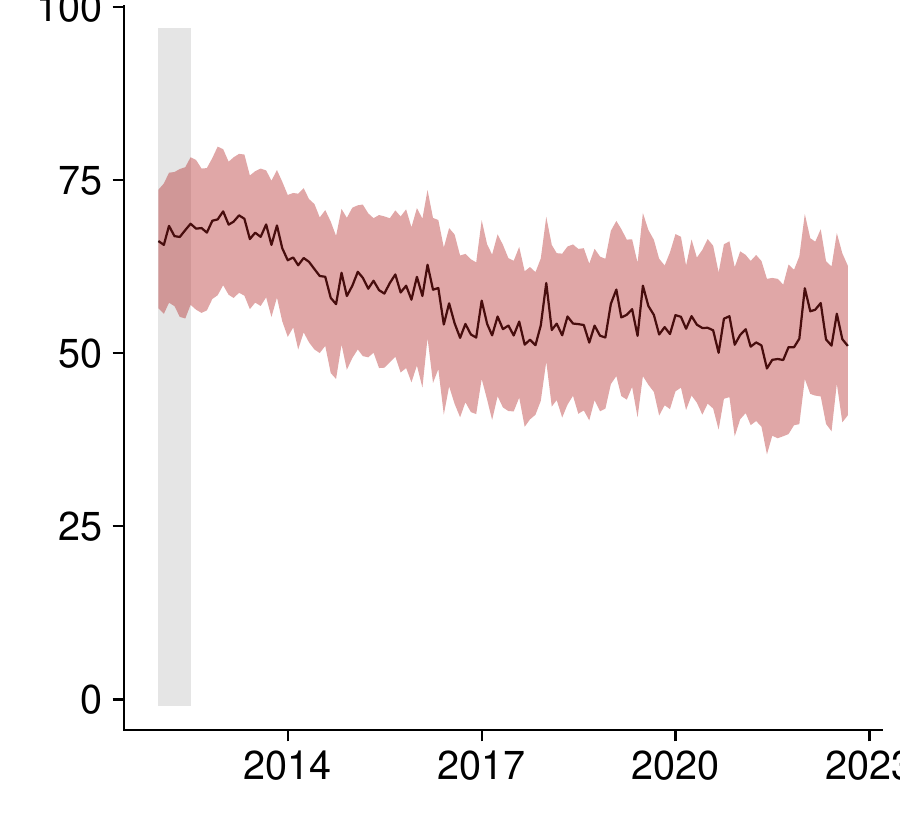}
\end{minipage}
\begin{minipage}{0.24\textwidth}
\centering
\includegraphics[scale=.24]{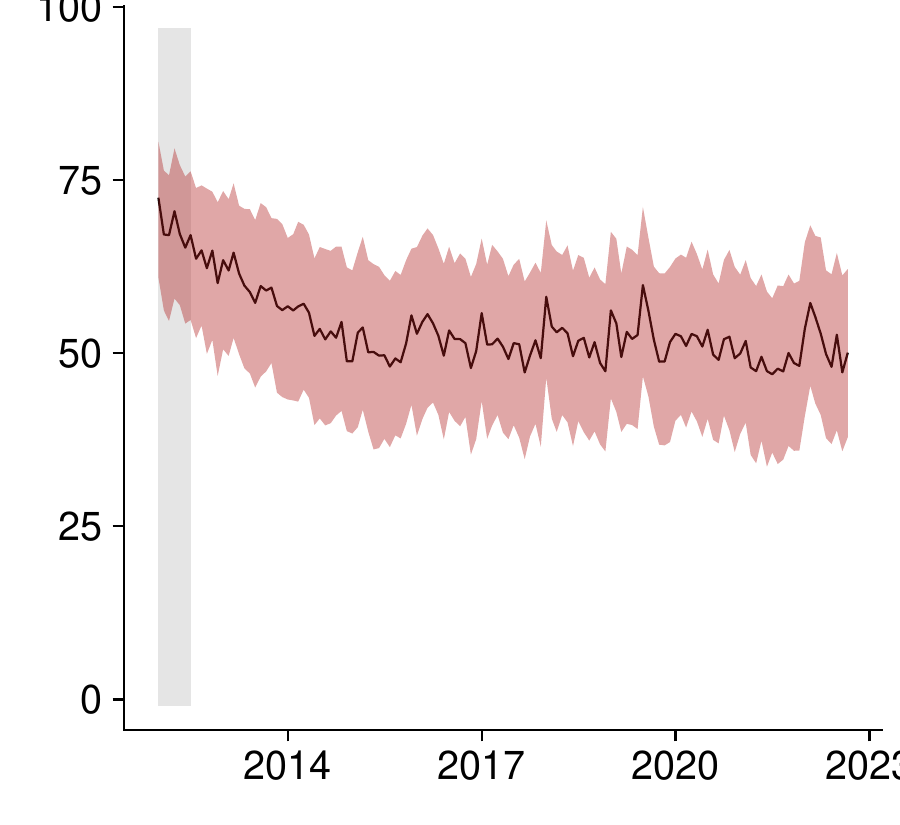}
\end{minipage}
\begin{minipage}{0.24\textwidth}
\centering
\includegraphics[scale=.24]{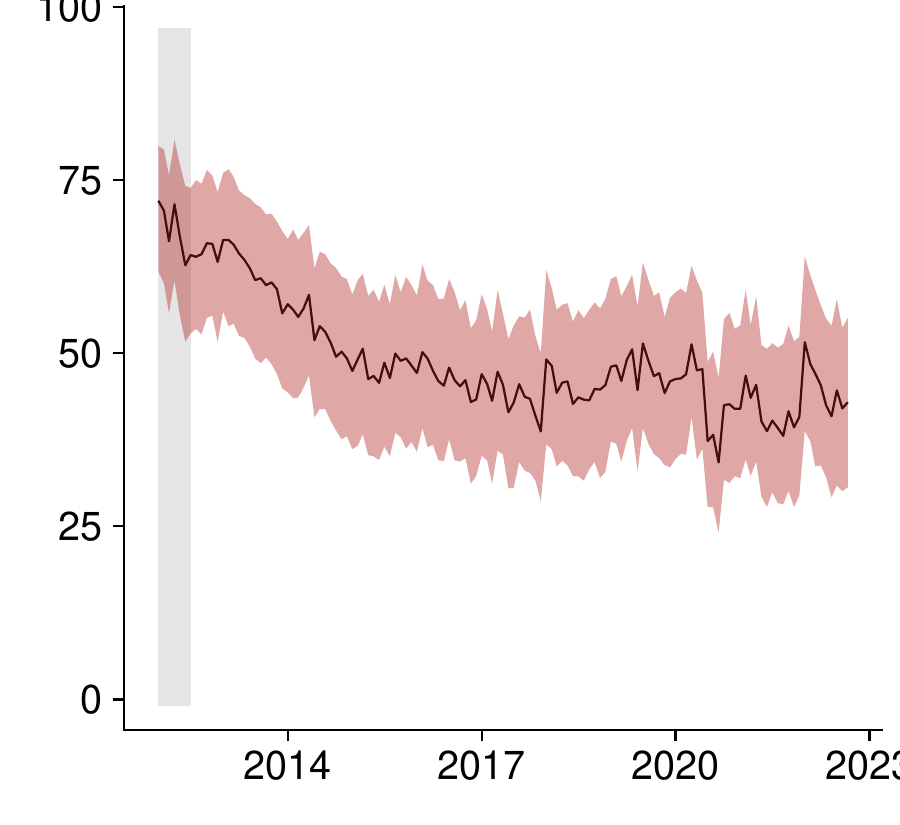}
\end{minipage}
\begin{minipage}{0.24\textwidth}
\centering
\includegraphics[scale=.24]{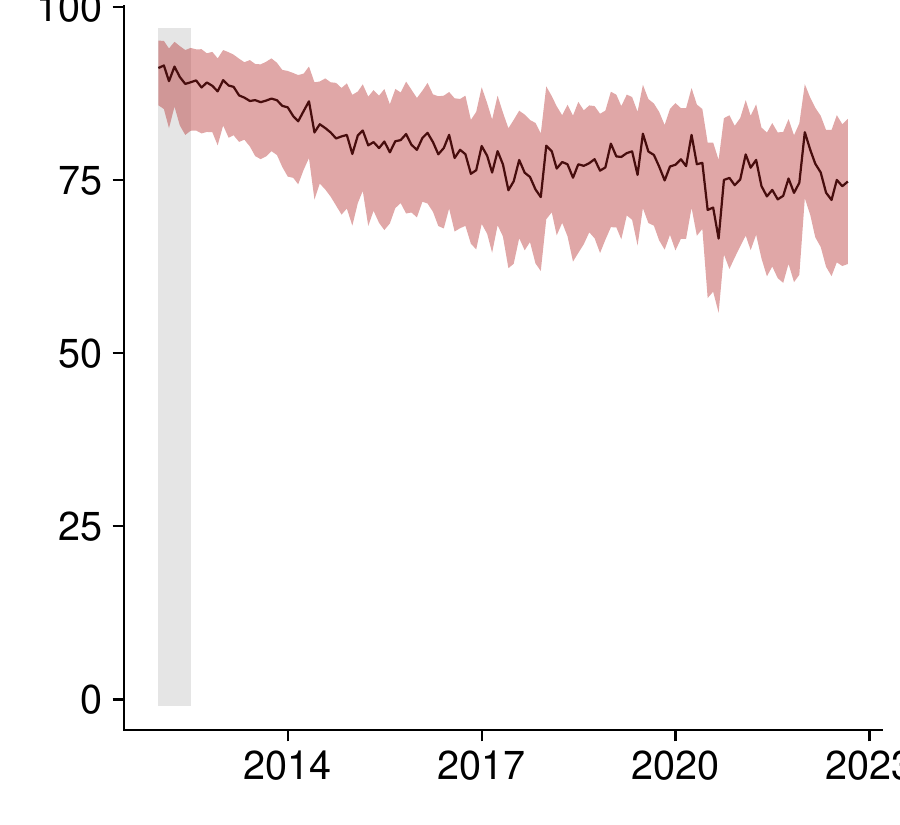}
\end{minipage}
\begin{minipage}{0.24\textwidth}
\centering
\includegraphics[scale=.24]{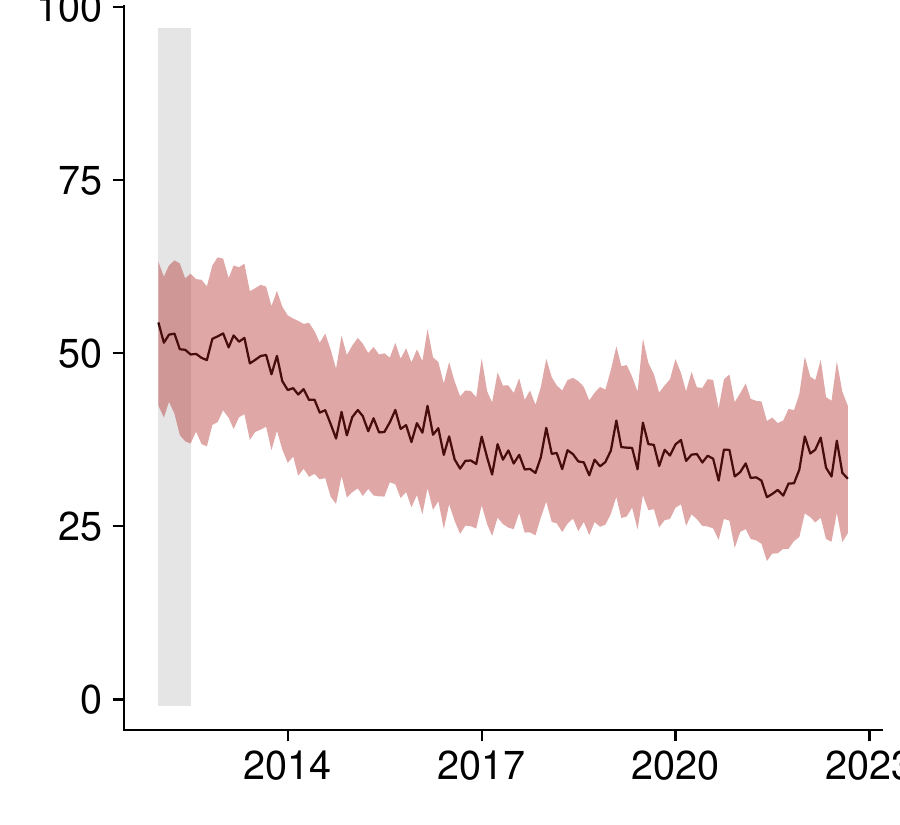}
\end{minipage}

\begin{minipage}{\linewidth}
\vspace{2pt}
\scriptsize \emph{Note:} This index indicates the share of spillovers for each country according to \cite{diebold2009measuring} and is estimated based on an expanding window. The solid line is the posterior median alongside the $68\%$ posterior credible set. The grey shaded area depicts the period of the ESDC.
\end{minipage}
\end{figure}
\end{landscape}

\clearpage

\begin{figure}[htbp!]
\caption{Clustering of GFEVD using a generalized block model for 12-months forecast horizon.}\label{fig:blockmodel12}
\begin{minipage}{1\linewidth}~\\
\centering \textbf{Government bond yields}
\end{minipage}\\
\begin{minipage}[b]{.24\linewidth}
Pre-GFC
\centering \includegraphics[scale=0.22]{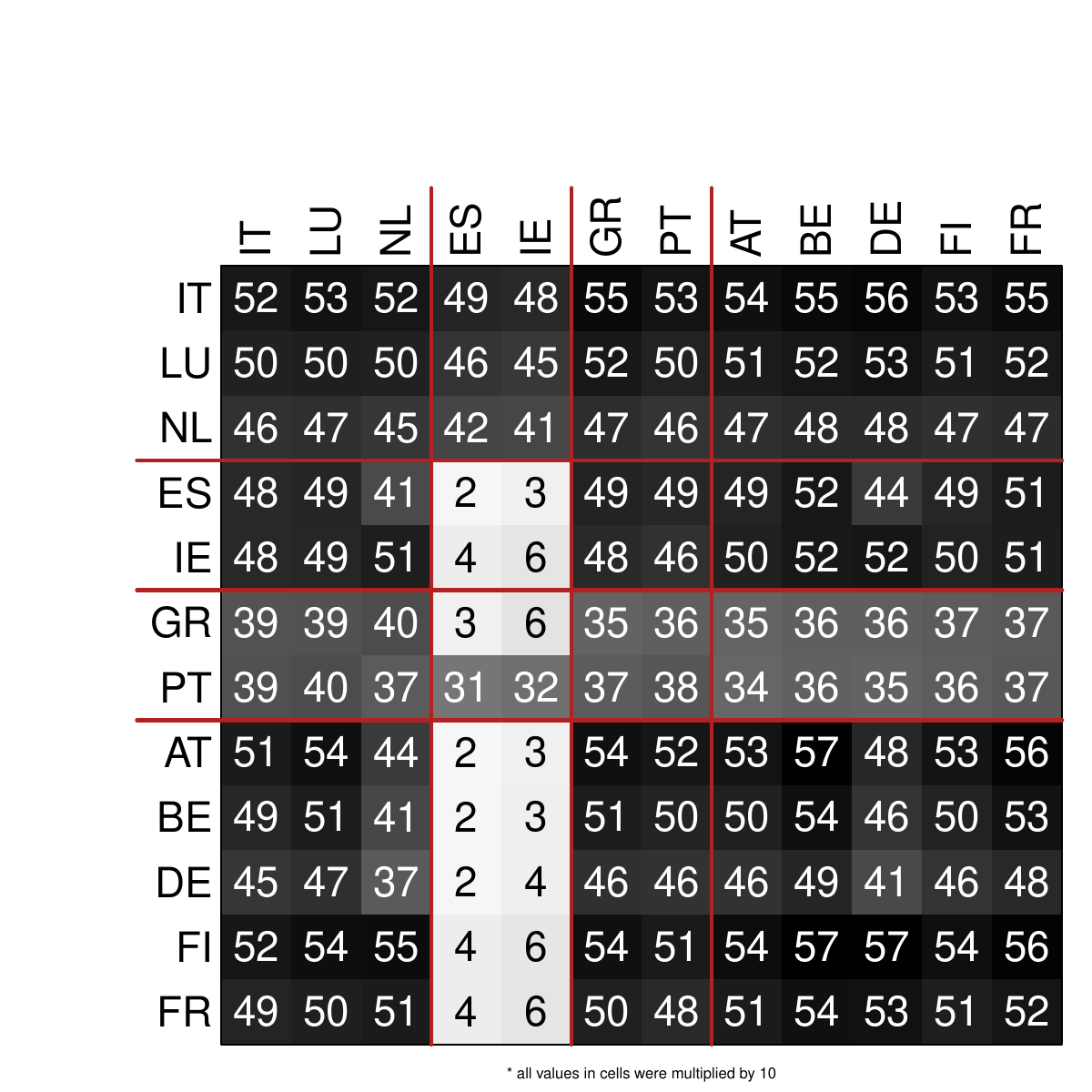}
\end{minipage}%
\begin{minipage}[b]{.24\linewidth}
GFC
\centering \includegraphics[scale=0.22]{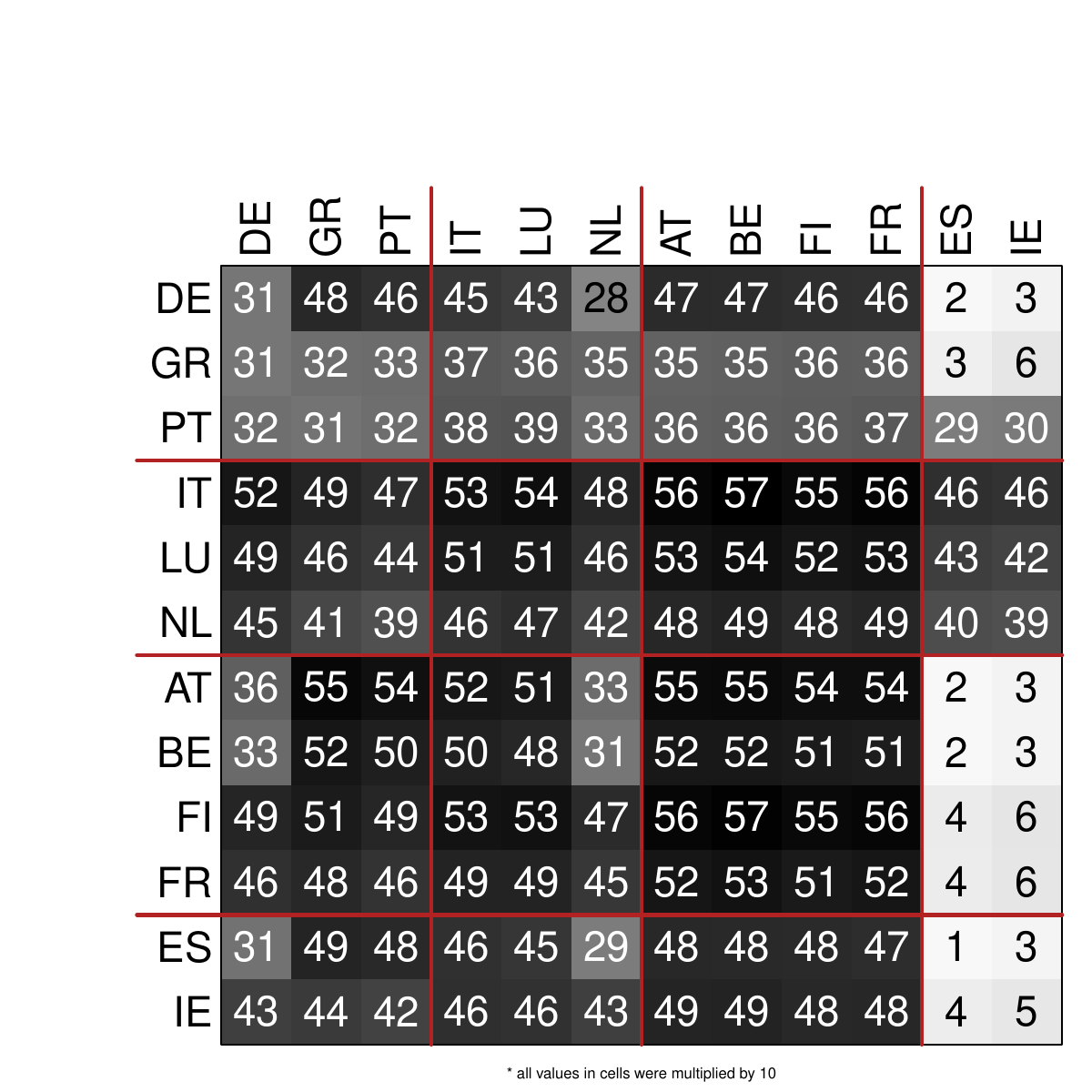}
\end{minipage}
\begin{minipage}[b]{.24\linewidth}
ESDC
\centering \includegraphics[scale=0.22]{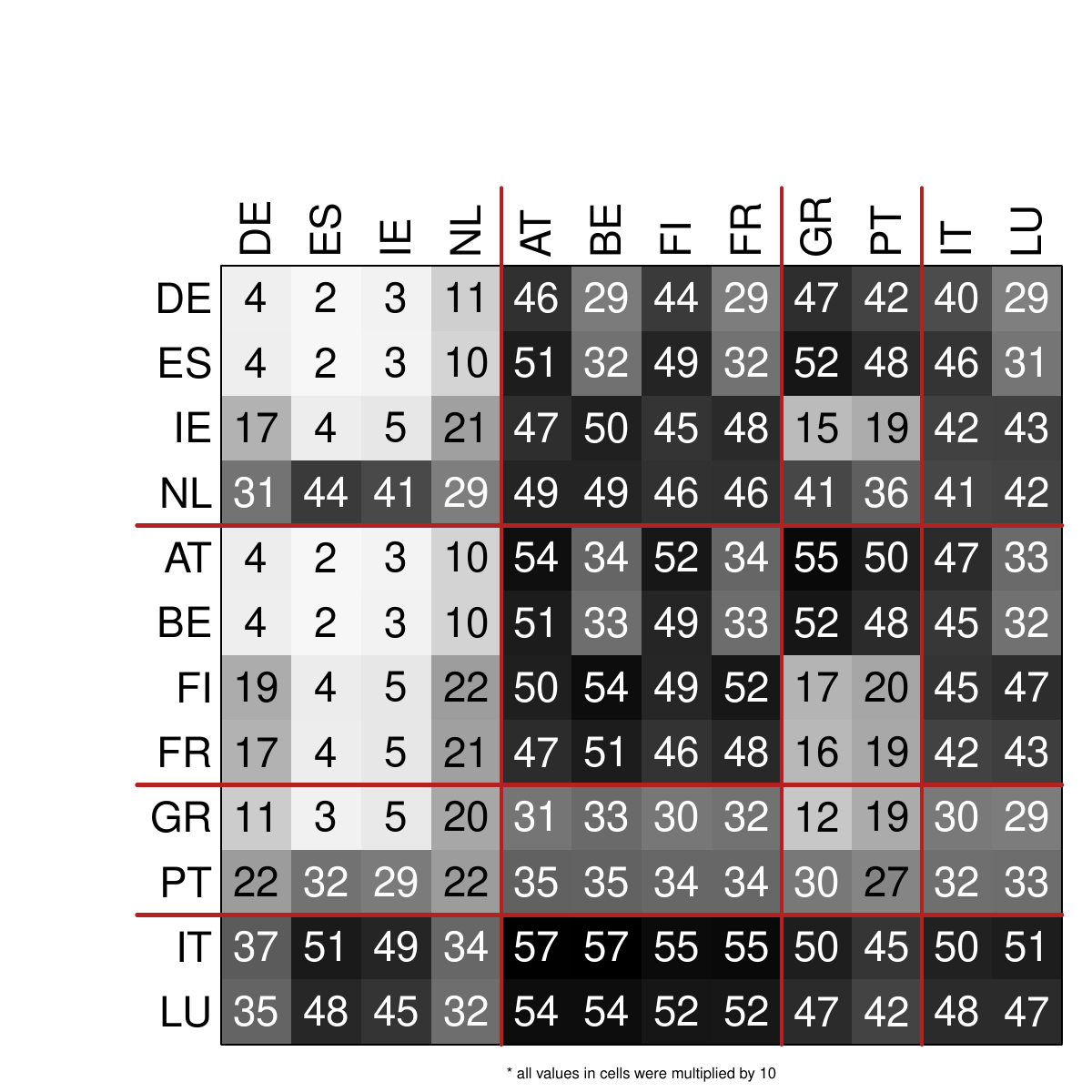}
\end{minipage}
\begin{minipage}[b]{.24\linewidth}
Post-ESDC
\centering \includegraphics[scale=0.22]{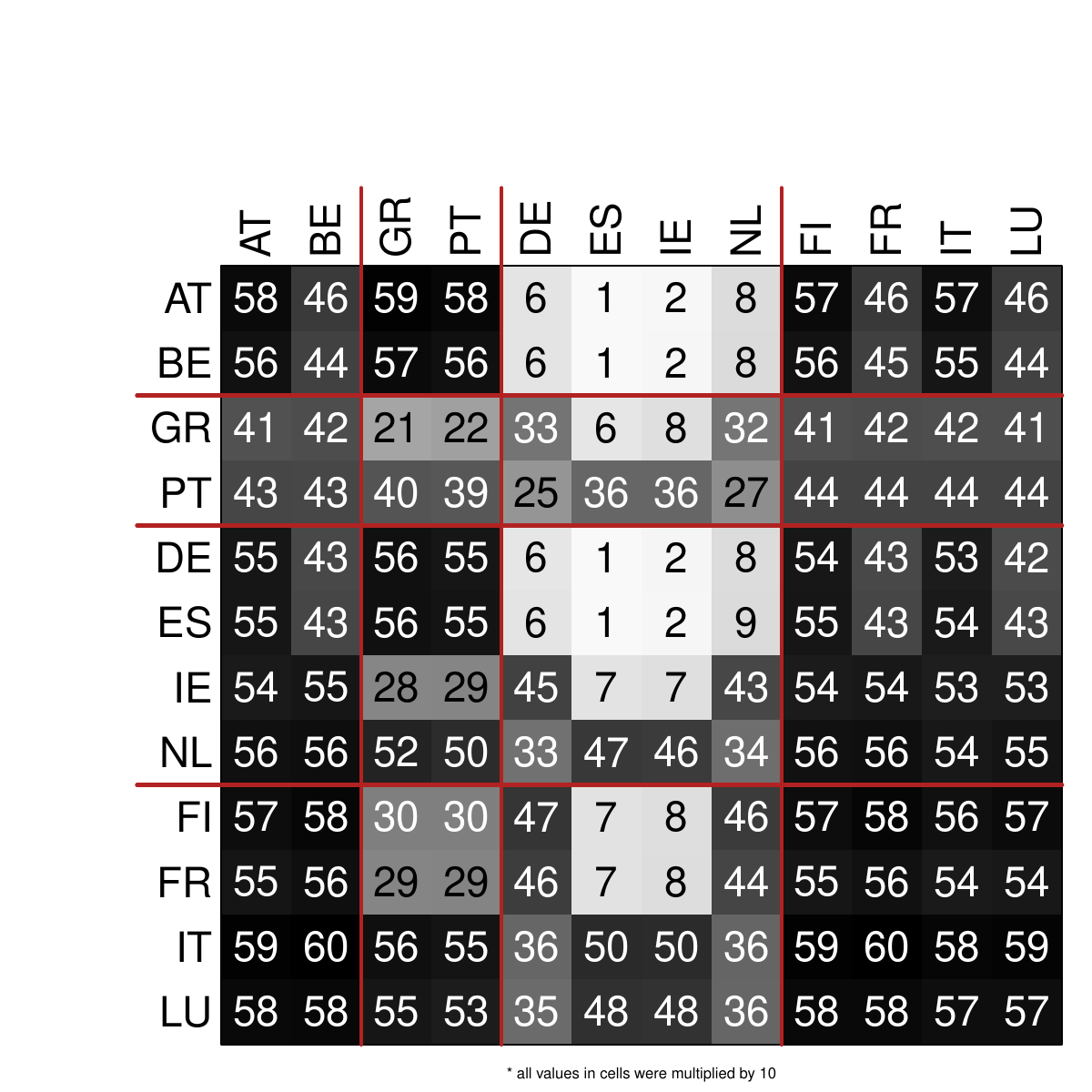}
\end{minipage}\\
\begin{minipage}{1\linewidth}~\\
\centering \textbf{Borrow long}
\end{minipage}\\
\begin{minipage}[b]{.24\linewidth}
Pre-GFC
\centering \includegraphics[scale=0.22]{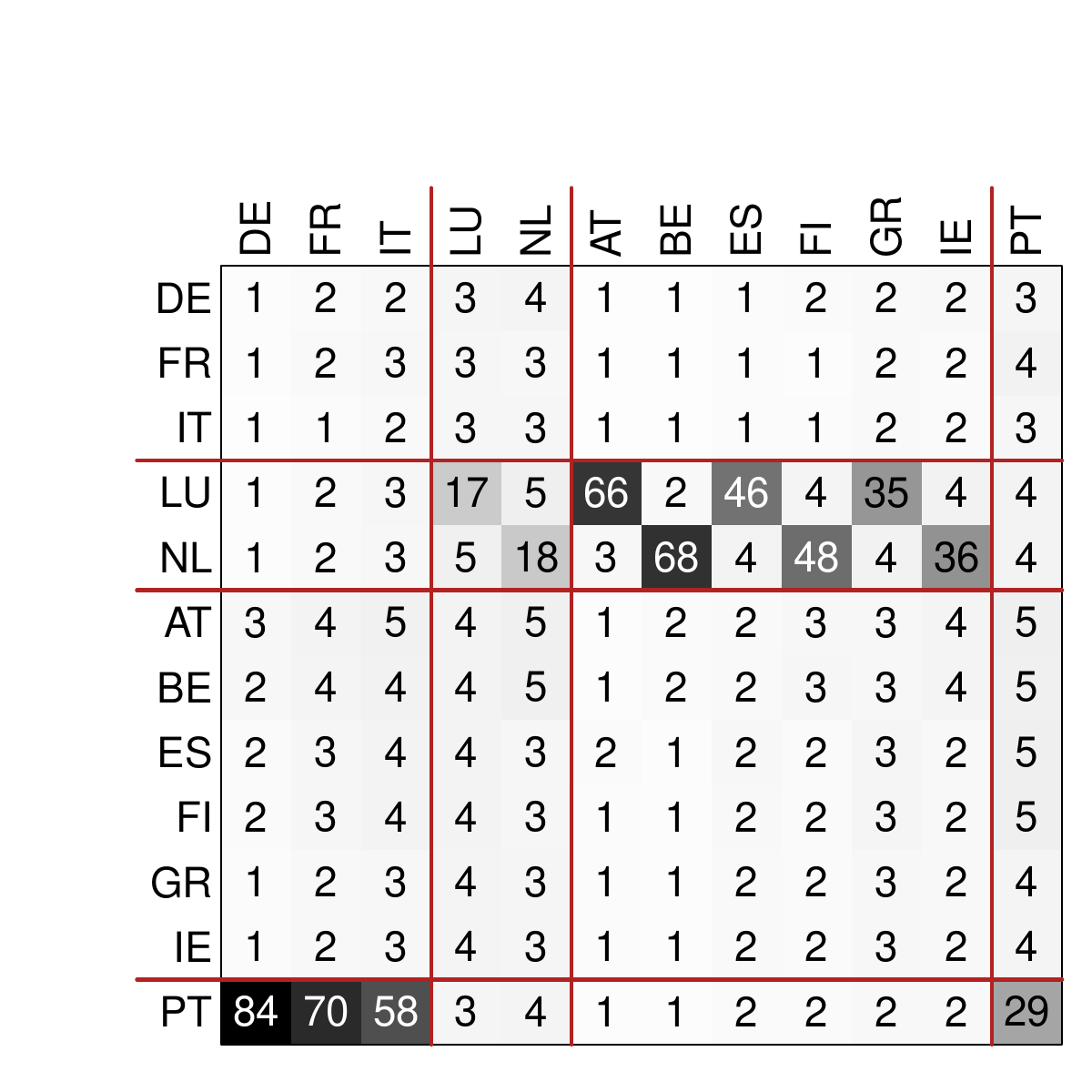}
\end{minipage}%
\begin{minipage}[b]{.24\linewidth}
GFC
\centering \includegraphics[scale=0.22]{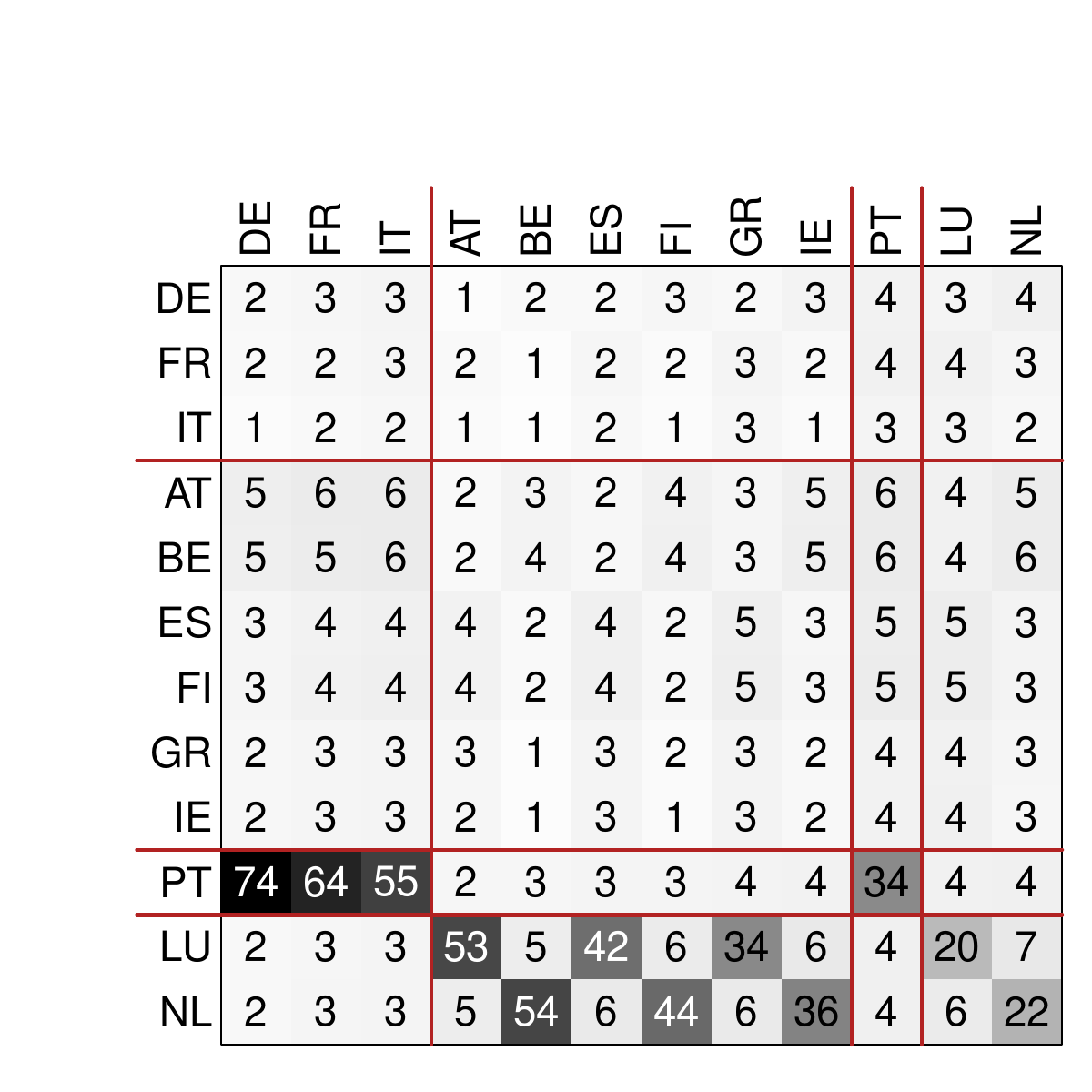}
\end{minipage}
\begin{minipage}[b]{.24\linewidth}
ESDC
\centering \includegraphics[scale=0.22]{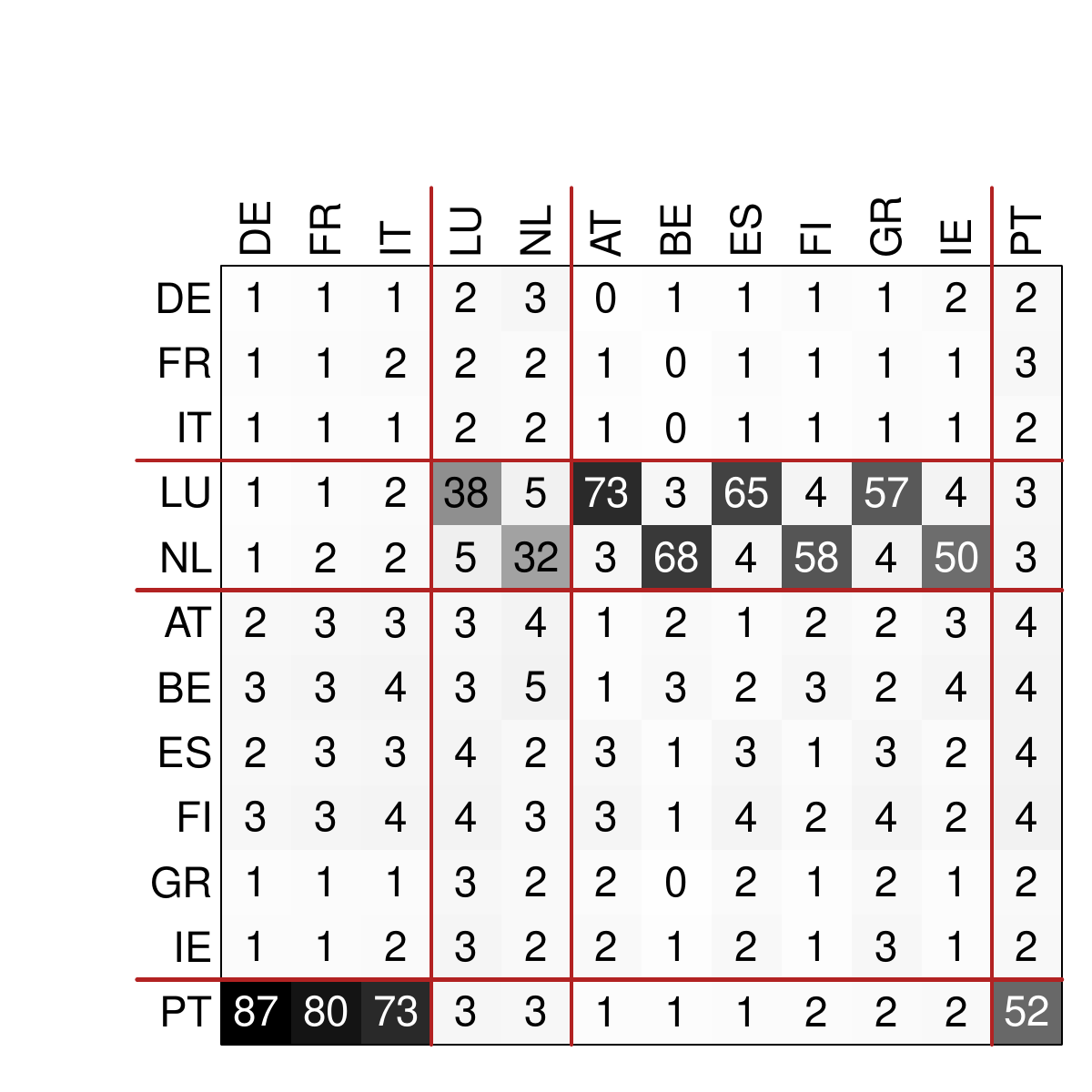}
\end{minipage}
\begin{minipage}[b]{.24\linewidth}
Post-ESDC
\centering \includegraphics[scale=0.22]{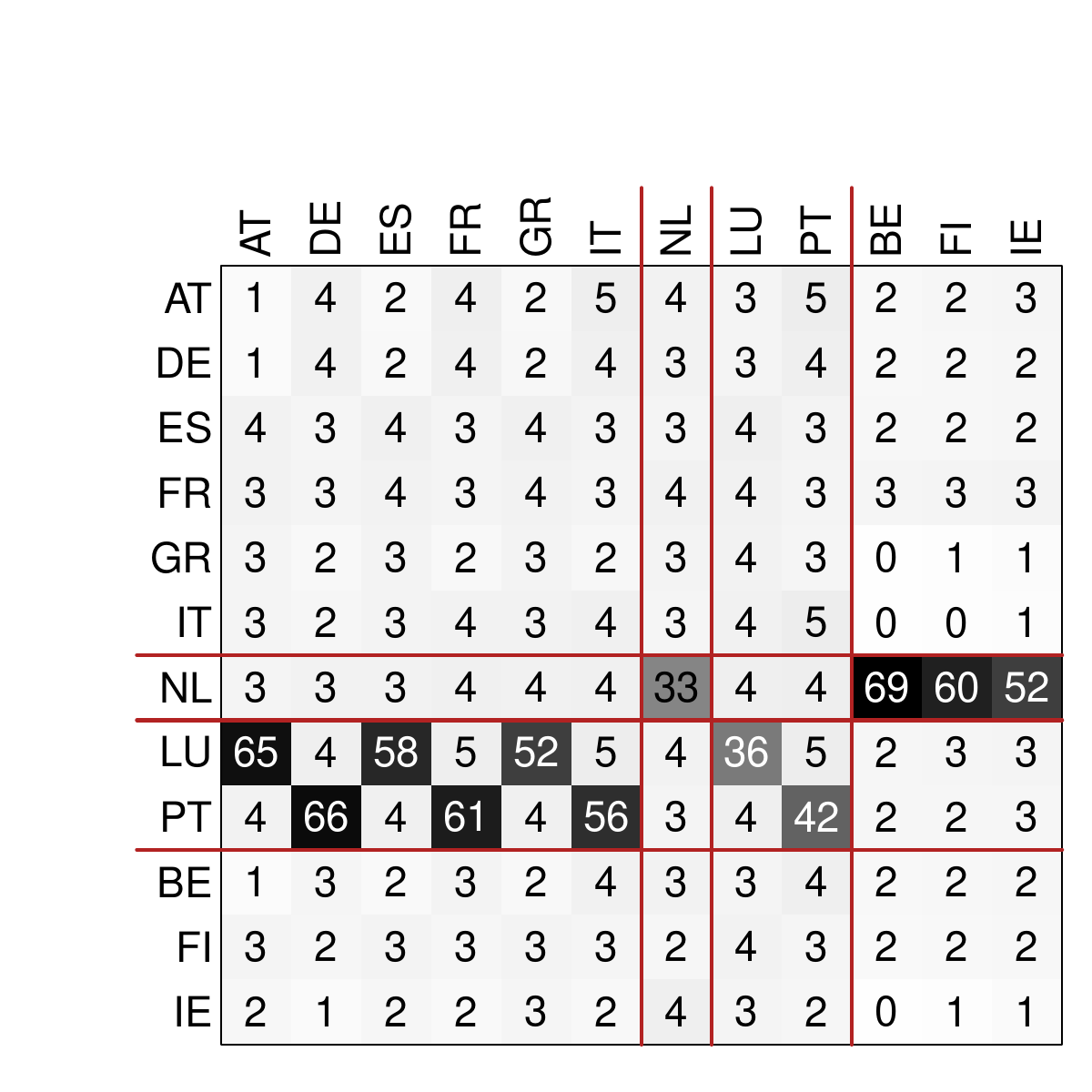}
\end{minipage}\\
\begin{minipage}{1\linewidth}~\\
\centering \textbf{Borrow short}
\end{minipage}\\
\begin{minipage}[b]{.24\linewidth}
Pre-GFC
\centering \includegraphics[scale=0.22]{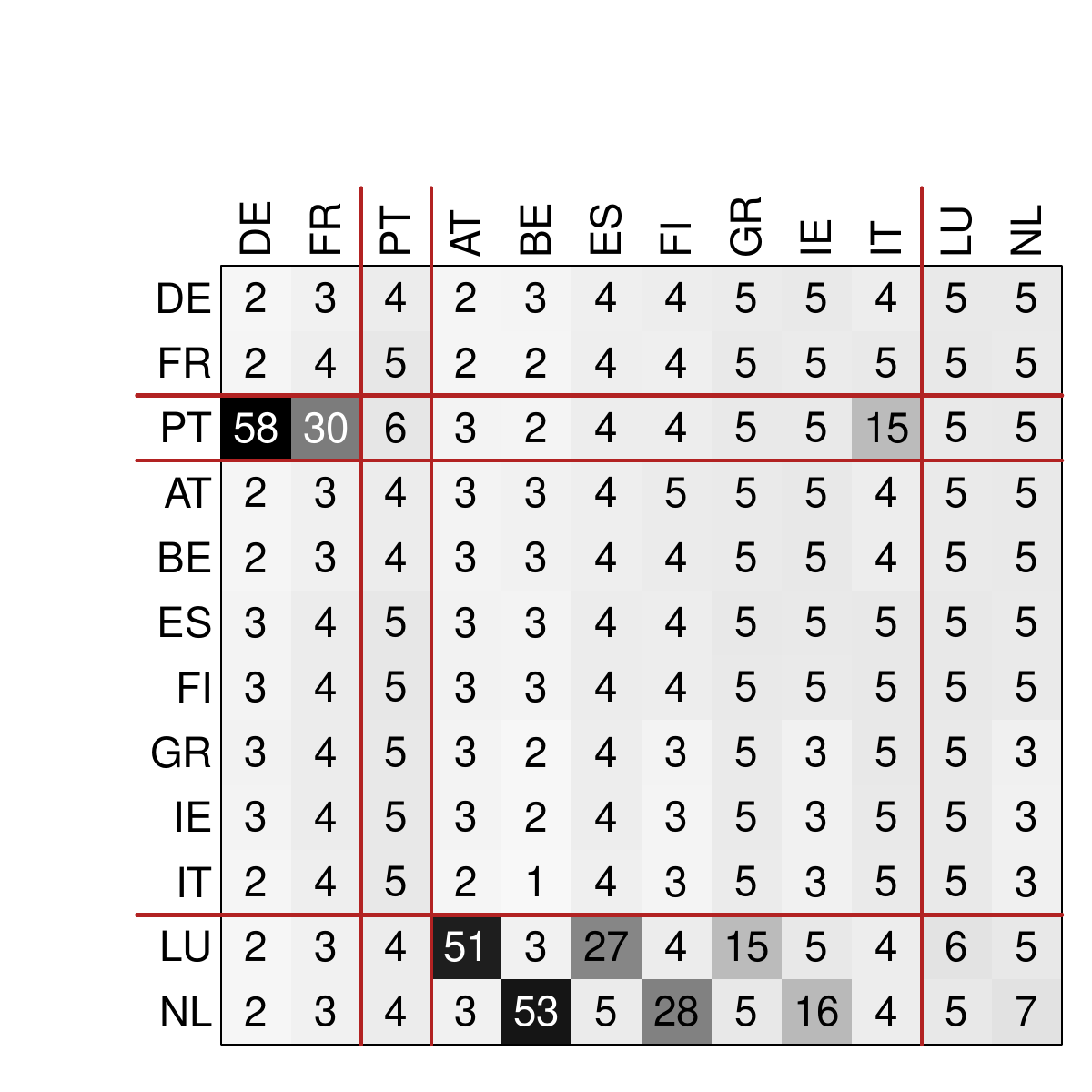}
\end{minipage}%
\begin{minipage}[b]{.24\linewidth}
GFC
\centering \includegraphics[scale=0.22]{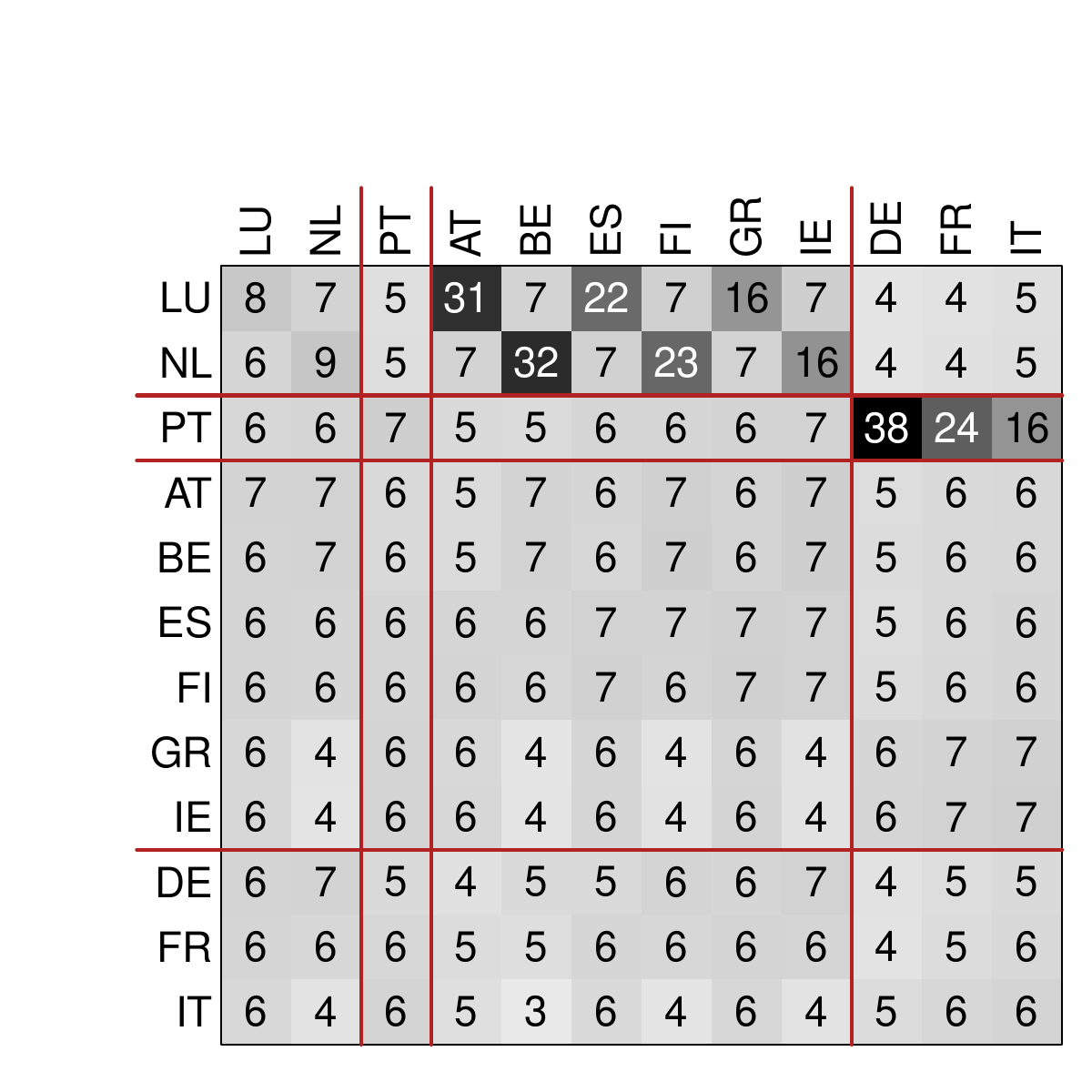}
\end{minipage}
\begin{minipage}[b]{.24\linewidth}
ESDC
\centering \includegraphics[scale=0.22]{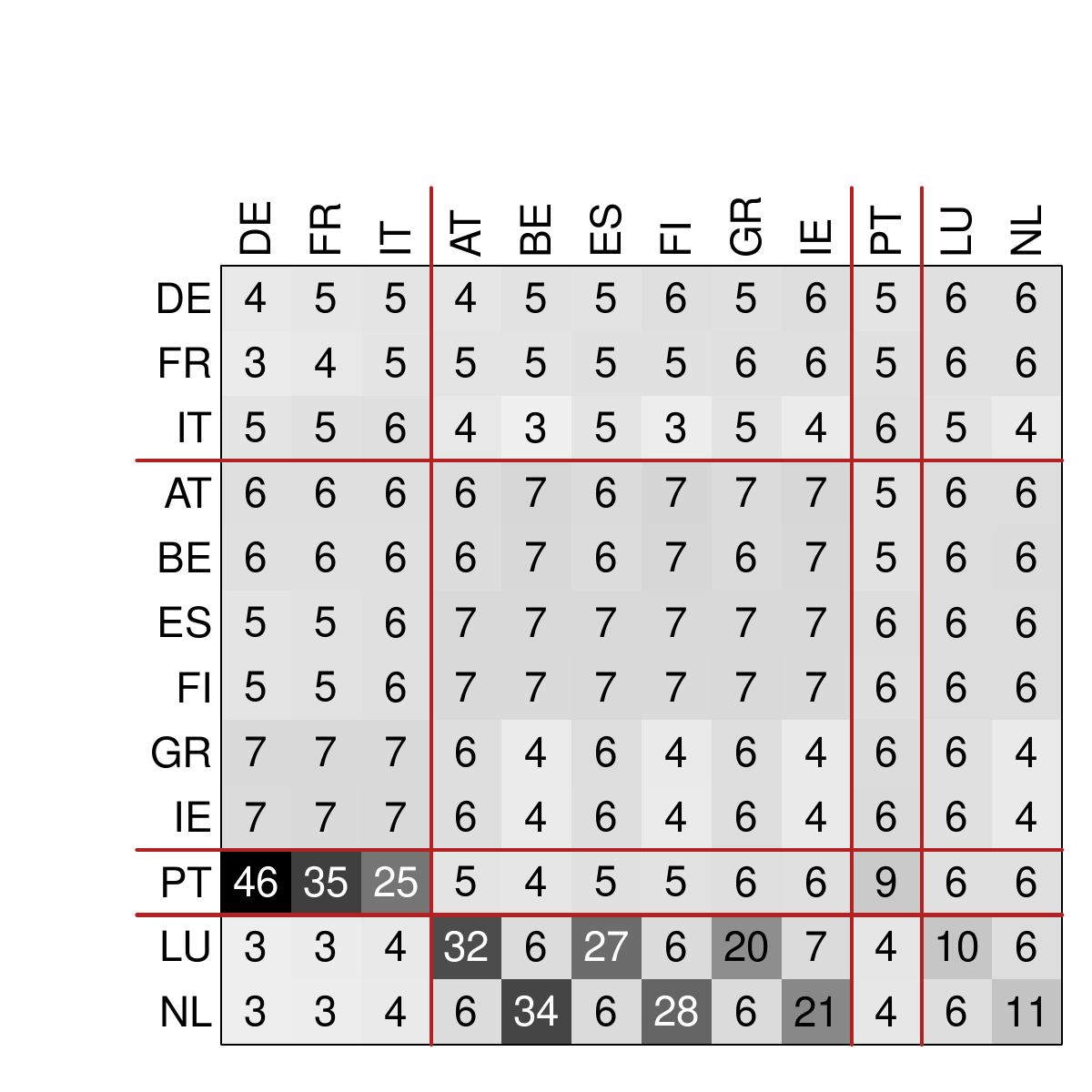}
\end{minipage}
\begin{minipage}[b]{.24\linewidth}
Post-ESDC
\centering \includegraphics[scale=0.22]{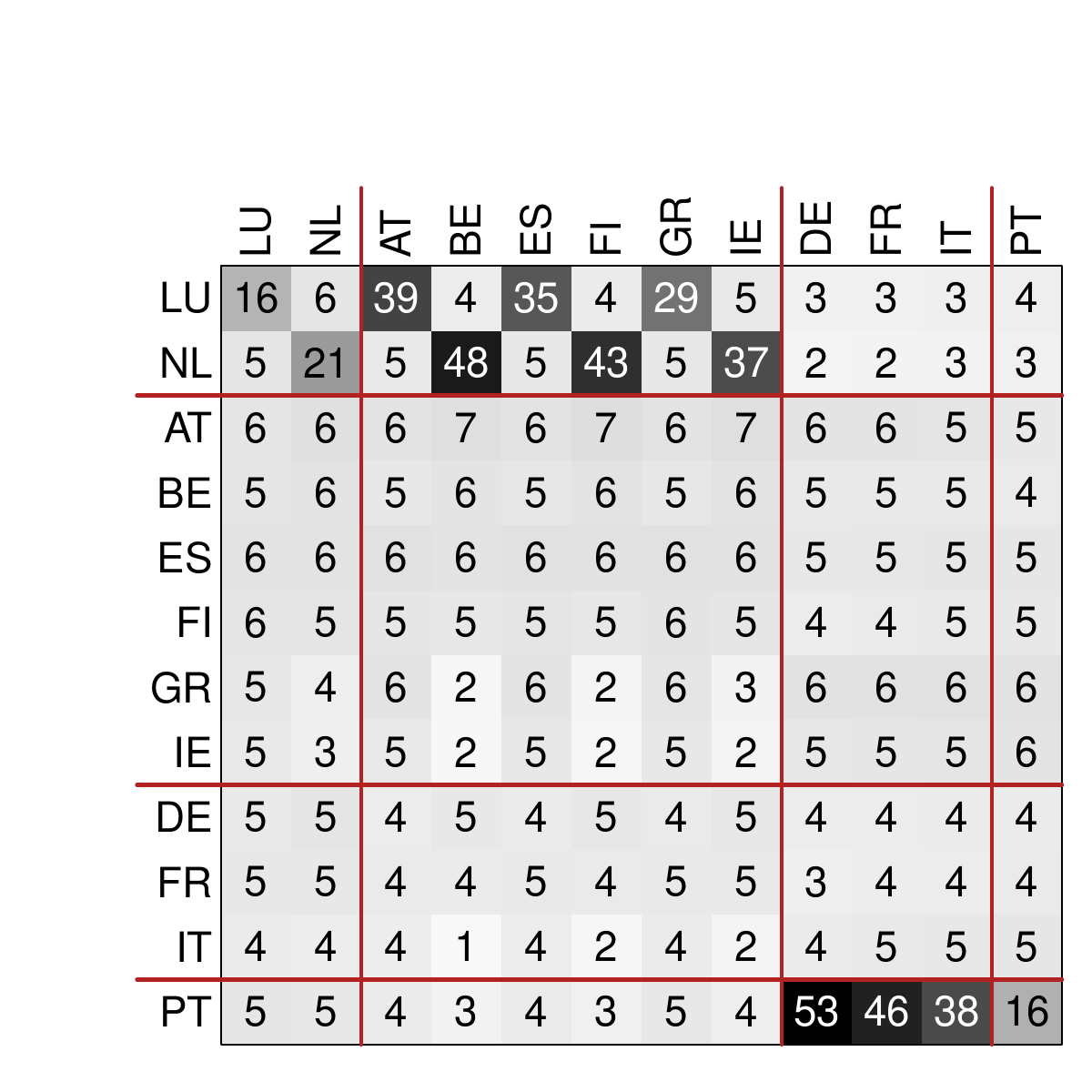}
\end{minipage}\\
\begin{minipage}{1\linewidth}~\\
\centering \textbf{TARGET2}
\end{minipage}\\
\begin{minipage}[b]{.24\linewidth}
\centering 
\end{minipage}%
\begin{minipage}[b]{.24\linewidth}
\centering 
\end{minipage}
\begin{minipage}[b]{.24\linewidth}
ESDC
\centering \includegraphics[scale=0.22]{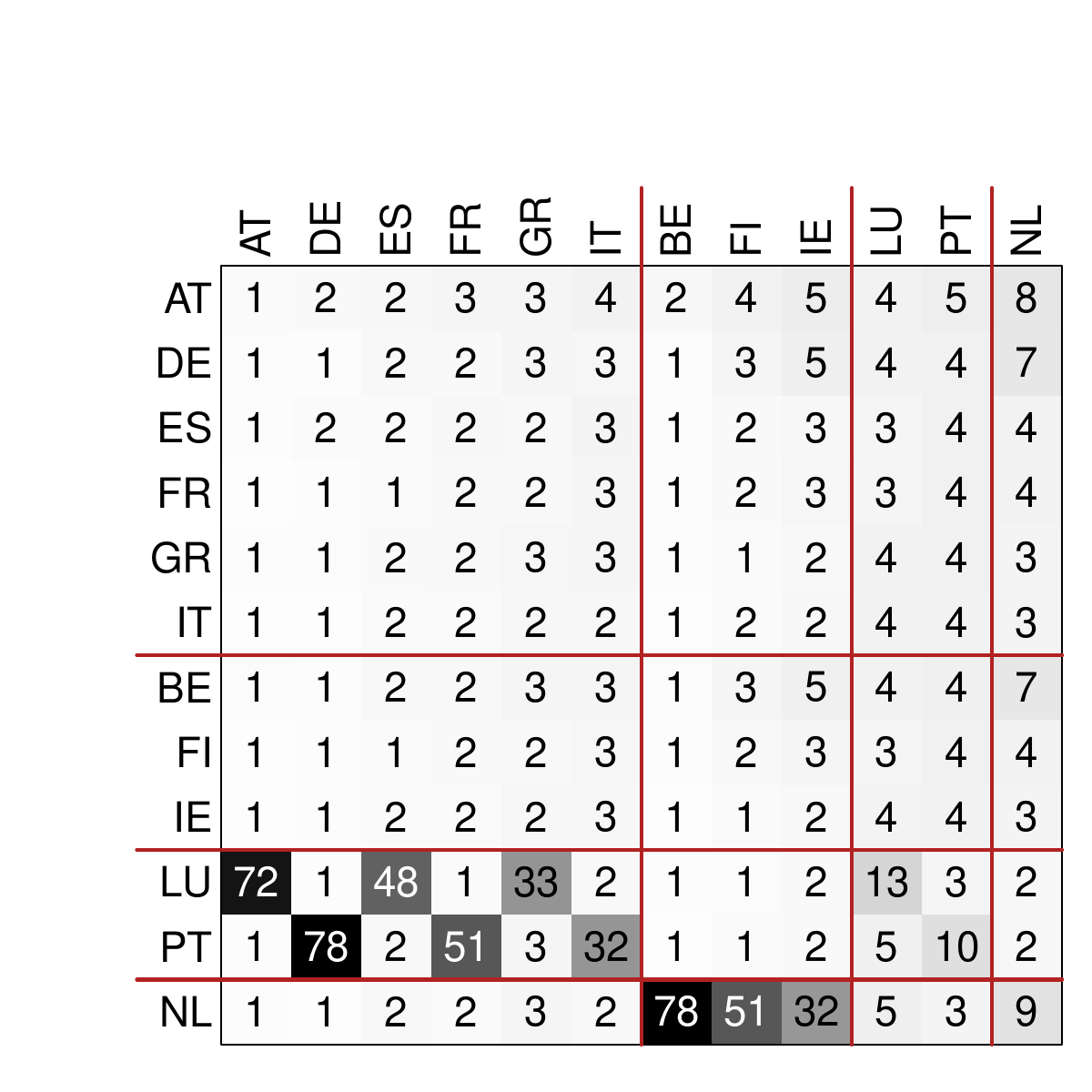}
\end{minipage}
\begin{minipage}[b]{.24\linewidth}
Post-ESDC
\centering \includegraphics[scale=0.22]{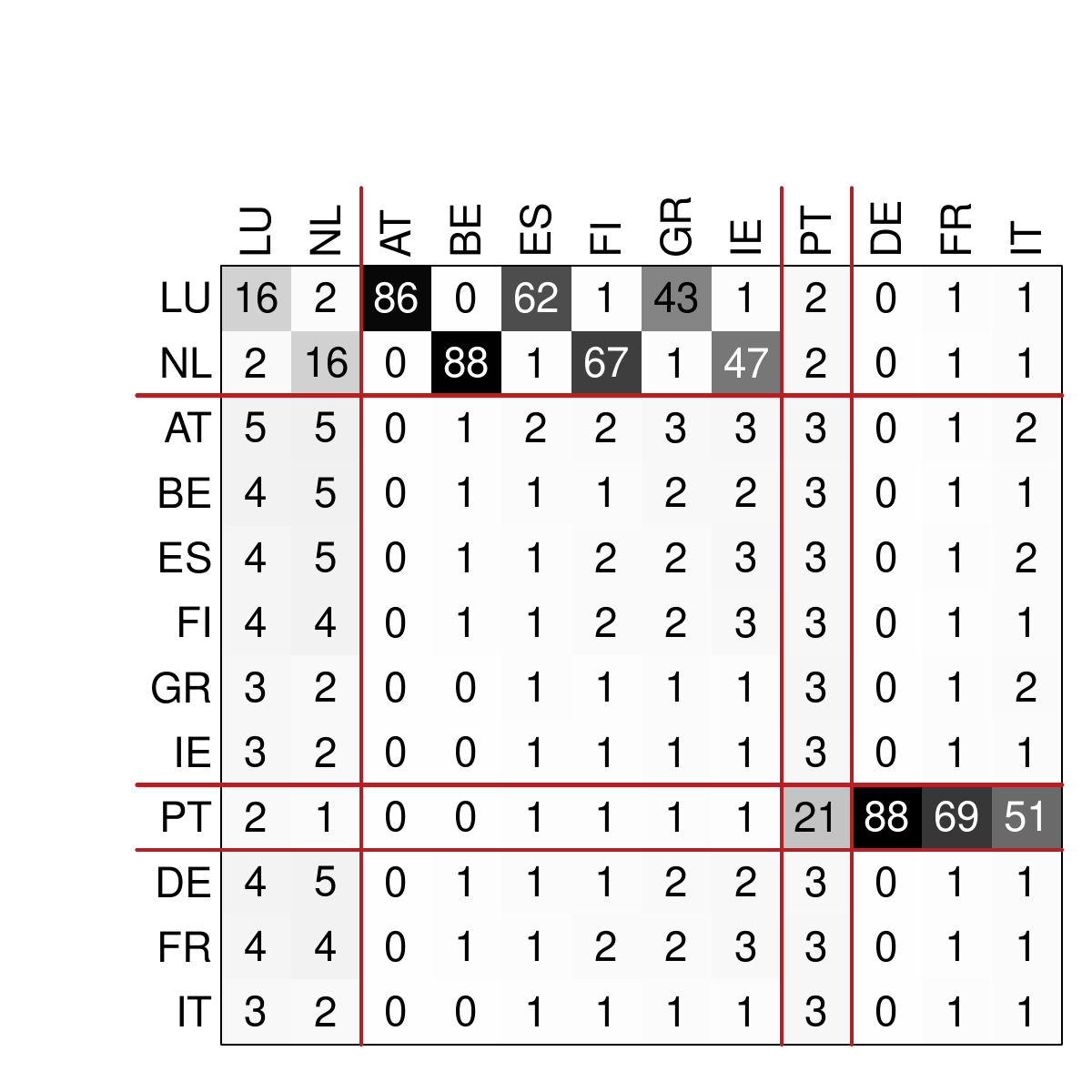}
\end{minipage}\\

\begin{minipage}{\linewidth}
\footnotesize \textit{Note:}  The figure shows results of a generalized blockmodel that clusters the GFEVD at the 12-months horizon. The pre-GFC period is defined as running from 2007m01 to 2008-08, the GFC runs from 2008m09 to 2009m06, ESDC from 2010m04 to 2012m07 and the post-ESDC from 2012m08 to 2022m05.
  \end{minipage}%
\end{figure}

\clearpage
\begin{table}[!htbp] \centering 
  \caption{Taylor rule specifications with interest rate smoothing: 2007m1 to 2022m5}\label{tbl:TRint}
  \footnotesize
\begin{tabular}{@{\extracolsep{5pt}}lcccccc} 
\\[-1.8ex]\hline 
\hline \\[-1.8ex] 
 & \multicolumn{6}{c}{\textit{Dependent variable:}} \\ 
\cline{2-7} 
\\[-1.8ex] & \multicolumn{5}{c}{$i_s$: euribor3m} & $i_s$: SSR \\ 
\\[-1.8ex] & (1) & (2) & (3) & (4) & (5) & (6)\\ 
\hline \\[-1.8ex] 
 $i_{s, t-1}$ & 0.959$^{***}$ & 0.951$^{***}$ &  0.961$^{***}$ & 0.948$^{***}$ & 0.942$^{***}$ & 0.928$^{***}$ \\ 
  & (0.014) & (0.015) & (0.014) & (0.016) & (0.017) & (0.035)  \\ 
    & & & & & & \\ 
 $y_{gap}$ & 0.040$^{***}$ & 0.037$^{***}$ & 0.039$^{***}$ & 0.038$^{***}$ & 0.035$^{***}$ & 0.010 \\ 
  & (0.007) & (0.007) & (0.007) & (0.007) & (0.007) & (0.016) \\ 
  & & & & & & \\ 
 $Dp_{yoy}$ & $-$0.068$^{***}$ & $-$0.069$^{***}$ & $-$0.065$^{***}$ & $-$0.064$^{***}$ & $-$0.064$^{***}$ & $-$0.031 \\ 
  & (0.012) & (0.012) & (0.013) & (0.012) & (0.012) & (0.027) \\ 
  & & & & & & \\ 
 $m2_{mom}$ & 0.068$^{**}$ & 0.066$^{**}$ & 0.069$^{**}$ & 0.059$^{*}$ & 0.061$^{*}$ & $-$0.140$^{**}$ \\ 
  & (0.031) & (0.031) & (0.031) & (0.031) & (0.032) & (0.067) \\ 
  & & & & & & \\ 
 $pcom$ & 0.008$^{***}$ & 0.008$^{***}$ & 0.008$^{***}$ & 0.007$^{***}$ & 0.007$^{***}$ & 0.008$^{***}$ \\ 
  & (0.001) & (0.001) & (0.001) & (0.001) & (0.001) & (0.002) \\ 
  & & & & & & \\ 
 $gb_{med}$ & $-$0.010 & $-$0.005 & $-$0.011$^{*}$ & $-$0.011$^{*}$ & $-$0.004 & $-$0.005 \\ 
  & (0.006) & (0.007) & (0.006) & (0.006) & (0.007) & (0.015) \\ 
  & & & & & & \\ 
 $gb_{long}$ & 0.063$^{***}$ & 0.071$^{***}$ & 0.060$^{***}$ & 0.071$^{***}$ & 0.074$^{***}$ & 0.113$^{**}$ \\ 
  & (0.013) & (0.014) & (0.013) & (0.013) & (0.015) & (0.046) \\ 
  & & & & & & \\ 
 $S_{gb}(1)$ &  & 0.002 &  &  & 0.002$^{*}$ & 0.007$^{**}$ \\ 
  &  & (0.001) &  &  & (0.001) & (0.003) \\ 
  & & & & & & \\ 
 $S_{bl}(1)$ &  &  & $-$0.0003 &  & $-$0.0004 & 0.001 \\ 
  &  &  & (0.0003) &  & (0.0004) & (0.001) \\ 
  & & & & & & \\ 
 $S_{bs}(1)$ &  &  &  & $-$0.001$^{*}$ & $-$0.001 & $-$0.003 \\ 
  &  &  &  & (0.001) & (0.001) & (0.002) \\ 
  & & & & & & \\ 
 Constant & $-$0.035$^{***}$ & $-$0.044$^{***}$ & $-$0.034$^{***}$ & $-$0.027$^{***}$ & $-$0.039$^{***}$ & $-$0.061$^{***}$ \\ 
  & (0.005) & (0.008) & (0.005) & (0.007) & (0.010) & (0.021) \\ 
  & & & & & & \\ 
\hline \\[-1.8ex] 
Observations & 185 & 185 & 185 & 185 & 185 & 185 \\ 
R$^{2}$ & 0.995 & 0.995 & 0.995 & 0.995 & 0.995 & 0.986 \\ 
Adjusted R$^{2}$ & 0.995 & 0.995 & 0.995 & 0.995 & 0.995 & 0.986 \\ 
Residual Std. Error & 0.001 & 0.001  & 0.001  & 0.001 & 0.001  & 0.002 \\ 
F Statistic & 4,887$^{***}$  & 4,306$^{***}$  & 4,265$^{***}$& 4,325$^{***}$  & 3,478$^{***}$  & 1,261$^{***}$ \\ 
\hline 
\hline \\[-1.8ex] 
\textit{Note:}  & \multicolumn{6}{r}{$^{*}$p$<$0.1; $^{**}$p$<$0.05; $^{***}$p$<$0.01} \\ 
\end{tabular} 
\end{table} 
\clearpage

\setcounter{equation}{0}
\setcounter{table}{0}
\setcounter{figure}{0}
\renewcommand\theequation{B.\arabic{equation}}
\renewcommand\thetable{B.\arabic{table}}
\renewcommand\thefigure{B.\arabic{figure}}

\section{Robustness}\label{sec:robust}
In addition to the PVAR model estimated in \autoref{sec:pvar},  we estimated a Bayesian vector autoregression (BVAR) with horseshoe shrinkage prior and stochastic volatility as well as a global vector autoregression (GVAR).

In contrast to the PVAR approach, the GVAR takes into account the dynamic interdependencies between economies by introducing pre-specified weights that measure the strength of the cross-country links \citep{pesaran2004gvar}. In this case, we define the $M$-dimensional vector $\bm y^{\ast}_{it} = \sum_{j=1}^N w_{ij} \bm y_{jt}$ with $w_{ij}$ denoting the pre-specified weights for which we assume that $w_{ii} = 0$, $w_{ij} \geq 0$ and $\sum_{j=1}^N w_{ij} = 1$ (for $i,j = 1,\dots,N$). To get a sparse representation of our system we introduce a Normal-Gamma shrinkage prior in the spirit of \cite{brown2010inference}. Similar to the PVAR, the GVAR allows us to jointly model and estimate the bond yield dynamics for a large cross-section of economies. This results in superior predictive performance due to the large underlying information set and the consideration of dynamic interdependencies. In the empirical application, we follow the bulk of the literature \citep{pesaran2004gvar,cuaresma2016forecasting,feldkircher2016gvar,pfarrhofer2022measuring} and use trade weights to measure to strength of economic activity between countries. More specifically, we use  annual trade flows from the Direction of Trade Statistics (DOTS) of the IMF, averaged over the sample period.

\begin{figure}[!htbp]
\caption{Overall spillover index of bond yields for the 1-month forecast horizon estimated with different modeling approaches. \label{fig:DYindex_ltir_models}}

\begin{minipage}{\textwidth}
\centering
\vspace{5pt}
\small (a) \textit{Constant parameter specification}
\vspace{2pt}
\end{minipage}

\begin{minipage}{0.49\textwidth}
\centering
\small \textit{BVAR}
\end{minipage}
\begin{minipage}{0.49\textwidth}
\centering
\small \textit{GVAR}
\end{minipage}

\begin{minipage}{0.49\textwidth}
\centering
\includegraphics[scale=.3]{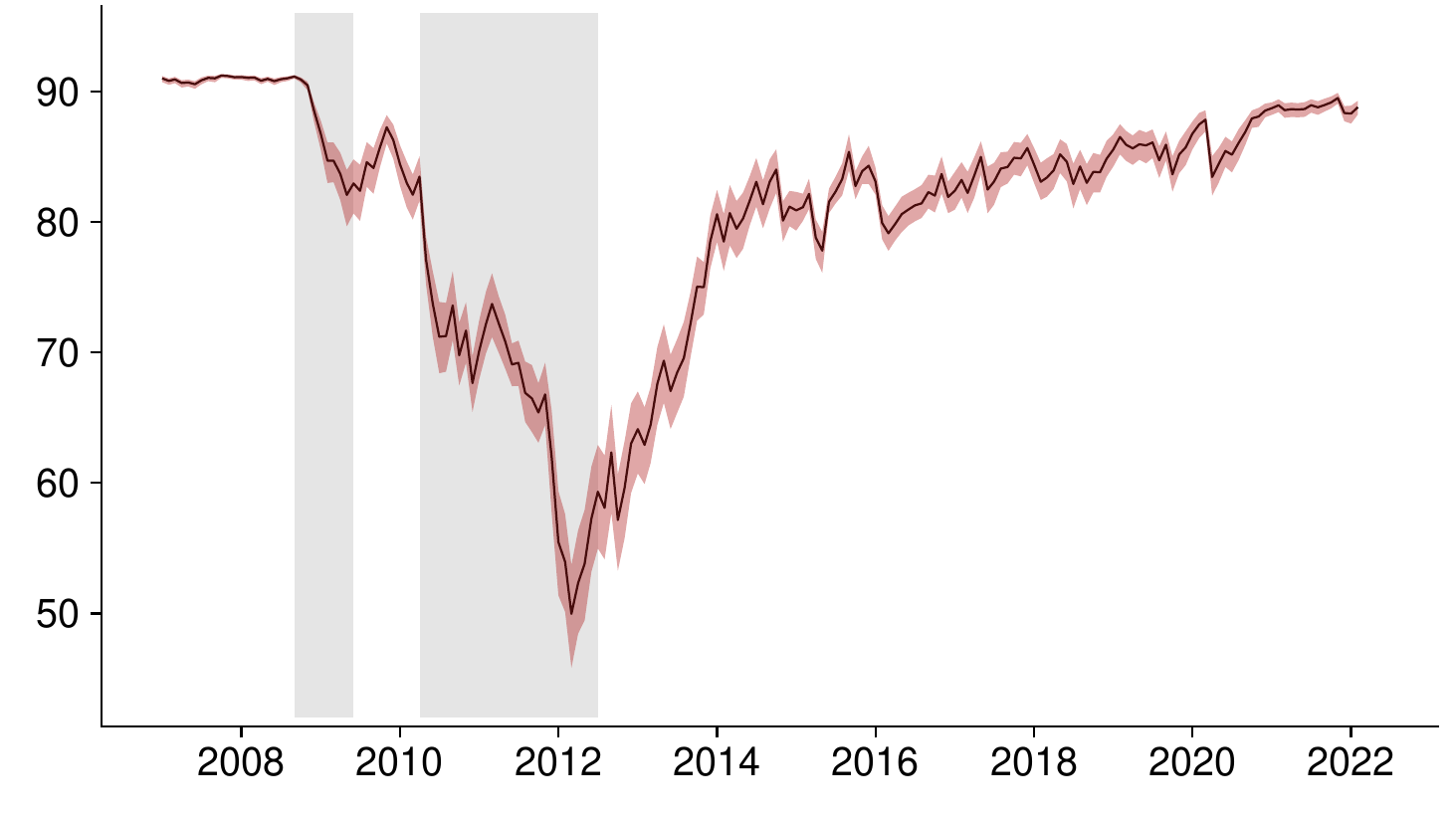}
\end{minipage}
\begin{minipage}{0.49\textwidth}
\centering
\includegraphics[scale=.3]{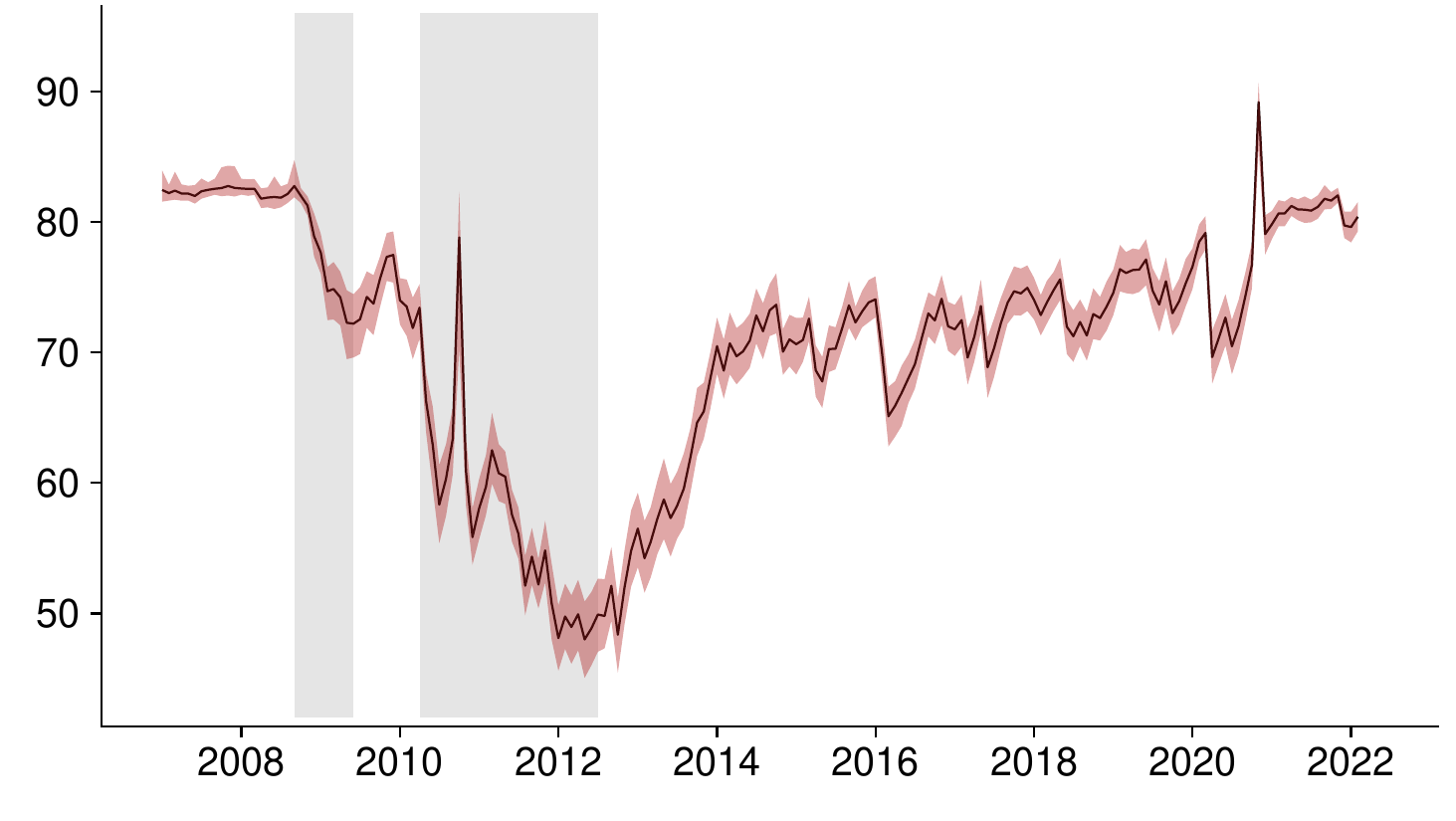}
\end{minipage}

\begin{minipage}{\textwidth}
\centering
\vspace{5pt}
\small (b) \textit{Time-varying parameter specification}
\vspace{2pt}
\end{minipage}

\begin{minipage}{0.49\textwidth}
\centering
\small \textit{BVAR}
\end{minipage}
\begin{minipage}{0.49\textwidth}
\centering
\small \textit{GVAR}
\end{minipage}

\begin{minipage}{0.49\textwidth}
\centering
\includegraphics[scale=.3]{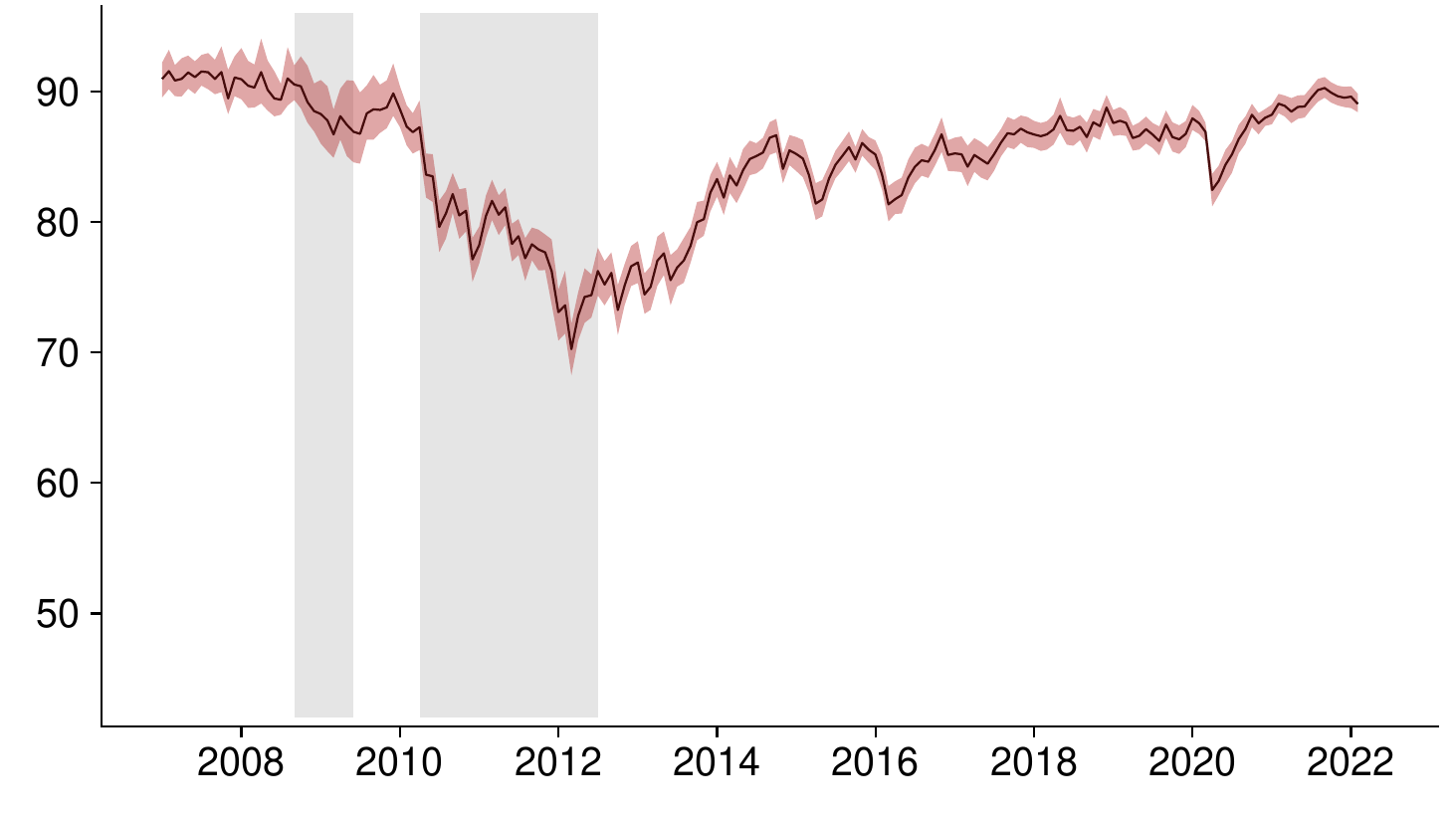}
\end{minipage}
\begin{minipage}{0.49\textwidth}
\centering
\includegraphics[scale=.3]{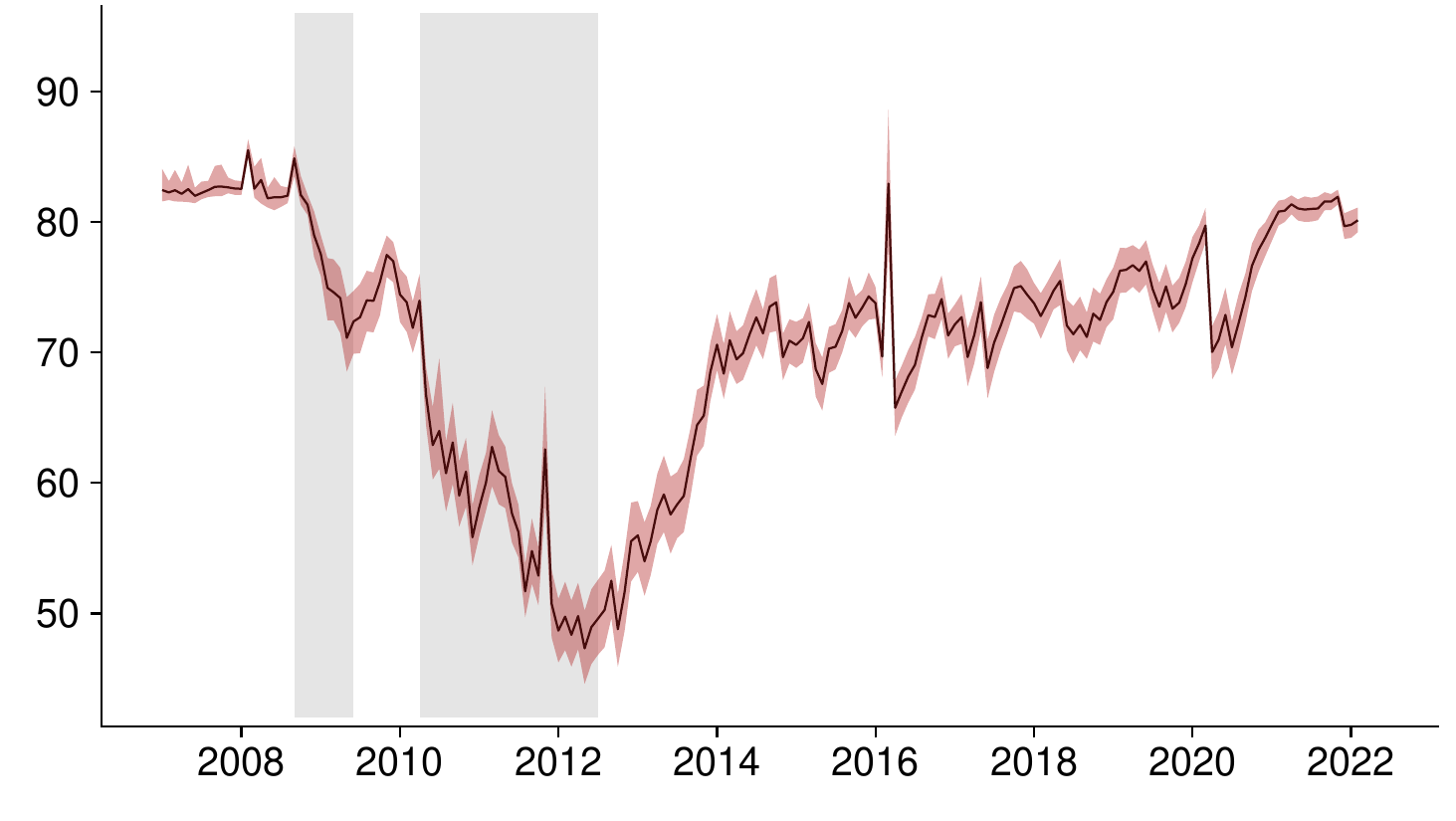}
\end{minipage}

\begin{minipage}{\textwidth}
\vspace{2pt}
\scriptsize \emph{Note:} This index indicates the share of spillovers across countries (averaged over all countries) according to \cite{diebold2009measuring} and is estimated based on an expanding window. The solid line is the posterior median alongside the $68\%$ posterior credible set. The grey shaded area depicts the periods of the GFC and ESDC.
\end{minipage}
\end{figure}

\autoref{fig:DYindex_ltir_models} shows the overall DY spillover index estimated for the 1-month forecast horizon with different modeling approaches. All models yield similar results, i.e., a significant decline in spillovers until the end of the ESDC and a reversing upward trend in the consecutive periods. \autoref{tab:eval_h1} shows the point forecasts performance of the different models for the 1-month ahead forecast horizon. We evaluate the performance of the different modeling approaches based on the root mean squared error relative to the BVAR with constant parameters. The analysis reveals the PVAR as a competitive forecasting tool for most economies and the chosen forecast horizon.  

\begin{table}[!htbp]
{\tiny
\caption{Point forecast performance based on relative RMSEs (1-month ahead). \label{tab:eval_h1}} 
\begin{center}
\begin{tabular*}{\textwidth}{l @{\extracolsep{\fill}} cccccc}
\toprule
\multicolumn{1}{c}{\bfseries Country}&\multicolumn{1}{c}{\bfseries }&\multicolumn{5}{c}{\bfseries Model}\tabularnewline
\cline{1-1} \cline{3-7}
\multicolumn{1}{c}{}&\multicolumn{1}{c}{}&\multicolumn{2}{c}{BVAR}&\multicolumn{2}{c}{GVAR}&\multicolumn{1}{c}{PVAR}\tabularnewline
\multicolumn{1}{c}{}&\multicolumn{1}{c}{}&\multicolumn{1}{c}{constant}&\multicolumn{1}{c}{time-varying}&\multicolumn{1}{c}{constant}&\multicolumn{1}{c}{time-varying}&\multicolumn{1}{c}{constant}\tabularnewline
\midrule
   AT&   &  \shadecell 0.922&   0.992&   1.020&   1.004&   \textbf{0.975}\tabularnewline
   BE&   &  \shadecell 0.963&   1.008&   1.002&   1.009&   \textbf{0.996}\tabularnewline
   DE&   &   0.887&   0.995&   1.005&   1.010&   \textbf{0.959}\tabularnewline
   ES&   & \shadecell  0.933&   1.019&   \textbf{1.002}&   1.004&   1.069\tabularnewline
   FI&   & \shadecell  0.671&   1.013&   1.002&   1.005&   \textbf{0.992}\tabularnewline
   FR&   & \shadecell  0.904&   0.999&   0.999&   0.999&   \textbf{0.960}\tabularnewline
   GR&   & \shadecell  1.918&   1.024&   0.986&   \textbf{0.967}&   1.013\tabularnewline
   IE&   &  \shadecell 1.448&   \textbf{0.997}&   1.005&   0.997&   1.047\tabularnewline
   IT&   &  \shadecell 0.935&   1.014&   \textbf{1.000}&   1.005&   1.082\tabularnewline
   LU&   & \shadecell  0.828&   1.017&   1.003&   1.009&   \textbf{0.965}\tabularnewline
   NL&   &  \shadecell 0.871&   1.004&   1.009&   1.012&   \textbf{0.961}\tabularnewline
   PT&   & \shadecell  1.371&   1.007&   \textbf{0.999}&   1.005&   1.028\tabularnewline
   \midrule
   Core &   & \shadecell 0.756&  0.879&  0.880&  0.881&  \textbf{0.851}\tabularnewline
   Periphery &   & \shadecell 1.321&  1.012&  0.998&  \textbf{0.996}&  1.048\tabularnewline
   Total &   & \shadecell 1.054&  1.007&  1.003&   \textbf{1.002}&  1.004\tabularnewline
\bottomrule
\end{tabular*}\end{center}}
\begin{minipage}{\textwidth}
\vspace{-12pt}
\tiny \emph{Note:} The table shows root mean squared errors (RMSEs) relative to the BVAR with constant parameters. In bold we mark the best performing model for each country. The grey shaded area gives the actual RMSE scores of our benchmark (BVAR with constant parameters). Results are averaged across the hold-out.
\end{minipage}
\end{table}

\end{appendices}
\end{document}